\documentclass[%
 pre,%
 amsmath,amssymb,
 reprint,singlecolumn,%
]{revtex4-2}

\usepackage{epsdice}
\usepackage{epigraph}
\usepackage{skak}
\usepackage[T1]{fontenc}
\usepackage[utf8]{inputenc}
\usepackage{amsmath}
\usepackage{amssymb}
\usepackage{graphicx}
\usepackage{yfonts}
\usepackage{mathtools}

\usepackage[
  citecolor=blue,
  colorlinks,
  linkcolor=blue,
  urlcolor=blue,
]{hyperref}
\usepackage[dvipsnames]{xcolor}
\usepackage{dcolumn}
\usepackage{bm}
\newcommand{\avg}[1]{\left\langle #1 \right\rangle}

\usepackage{graphicx}
\usepackage{capt-of}
\usepackage{tikz}
\usetikzlibrary{arrows.meta,decorations.markings,patterns}
\usetikzlibrary{decorations.pathreplacing}\newcommand{\ring}[4]{%
  \begin{tikzpicture}[baseline=(c.center),x=1cm,y=1cm,scale=0.63]
    \node (c) at (0,0) {};
    \draw[thin] (0,0) circle (1);
    \foreach \a/\o in {90/#1,0/#2,-90/#3,180/#4}{%
      \begin{scope}[shift={(\a:1)}]
        \draw[fill=white] (0,0) circle (0.3);
        \ifnum\o=1
          \draw (-0.212,-0.212) -- (0.212,0.212);
          \draw (-0.212,0.212) -- (0.212,-0.212);
        \fi
      \end{scope}}
  \end{tikzpicture}}
\newcommand{\bs}[1]{\boldsymbol{#1}}
\newcommand{\mc}[1]{\mathcal{#1}}

\newcommand{\kB}{k_{\mathrm{B}}}
\newcommand{\scutint}{\mathrel{\rlap{\lower-.00ex\hbox{\hskip.25em$\scriptstyle {{S}}$}}{\int}}}

\usepackage{tikz}
\usetikzlibrary{arrows.meta,positioning,calc,fit,backgrounds}
\usetikzlibrary{arrows.meta}

\newcommand{\spin}[3]{%
  \ifnum#3>0
    \draw[-{Stealth[length=4pt]},thick] (#1,#2-0.25) -- (#1,#2+0.25);
  \else
    \draw[-{Stealth[length=4pt]},thick] (#1,#2+0.25) -- (#1,#2-0.25);
  \fi}
\newcommand{\gchain}[5]{%
  \spin{#1}{#2}{#3}
  \spin{#1+0.55}{#2}{#4}
  \spin{#1+1.1}{#2}{#5}
  \draw[dashed,gray] (#1+0.3,#2-0.38) rectangle (#1+0.8,#2+0.38);}

\newcommand{\kchain}[6]{%
  \spin{#1}{#2}{#3}
  \spin{#1+0.55}{#2}{#4}
  \spin{#1+1.1}{#2}{#5}
  \spin{#1+1.65}{#2}{#6}
  \draw[dashed,gray] (#1+0.3,#2-0.38) rectangle (#1+1.35,#2+0.38);}
\providecommand{\kB}{k_B}
\usepackage{color}
\newcommand{\fivechain}[5]{\tikz[baseline=-0.6ex,scale=0.8]{%
  \spin{0}{0}{#1}\spin{0.55}{0}{#2}\spin{1.1}{0}{#3}%
  \spin{1.65}{0}{#4}\spin{2.2}{0}{#5}%
  \draw[dashed,gray] (0.85,-0.38) rectangle (1.35,0.38);}}
  
\makeatletter
\renewcommand{\bibnumfmt}[1]{[\ifnum\c@NAT@ctr<10 0\fi#1]}
\renewcommand{\citenumfont}[1]{\two@digits{#1}}
\makeatother

\begin{document}
\definecolor{deepsea}{rgb}{0,0.25,0}

\preprint{AIP/123-QED}

\title{\textit{{\color{deepsea}Nonequilibrium Statistical Mechanics:  Physical Concepts \& Mathematical Tools}}}
\author{Sagar Chakraborty}
\email{sagarc@iitk.ac.in}
\affiliation{
  Department of Physics,
  Indian Institute of Technology Kanpur,
  Uttar Pradesh 208016, India
}
\date{\today}
  %
%
\begin{abstract}
As anything not in equilibrium is, by definition, nonequilibrium, there is no universally agreed-upon set of topics that can constitute an exhaustive presentation of nonequilibrium statistical mechanics. At the same time, nonequilibrium statistical mechanics has transcended traditional physics to become a powerful language for complex systems. Prompted by an enthusiastic curiosity about complex systems, these cursory lecture notes attempt to elucidate some introductory concepts and techniques used in classical nonequilibrium statistical mechanics, which find successful applications not only in foundational physics, but also across other fields such as biology, ecology, economics, and sociology. Except for the specific synthesis and presentation of the material, which assumes basic familiarity with equilibrium statistical mechanics, the author claims no originality.
\begin{center}
	$\symking\symqueen\symrook\symknight\symbishop\sympawn\epsdice{1}\,\epsdice{2}\,\epsdice{3}\,\epsdice{4}\,\epsdice{5}\,\epsdice{6}\,\spadesuit\heartsuit\clubsuit\diamondsuit$
\end{center}
\end{abstract}

%
\maketitle
\tableofcontents
\begin{center}
	$\symking\symqueen\symrook\symknight\symbishop\sympawn\epsdice{1}\,\epsdice{2}\,\epsdice{3}\,\epsdice{4}\,\epsdice{5}\,\epsdice{6}\,\spadesuit\heartsuit\clubsuit\diamondsuit$
\end{center}
\newpage
\newpage
   \clearpage
   \onecolumngrid
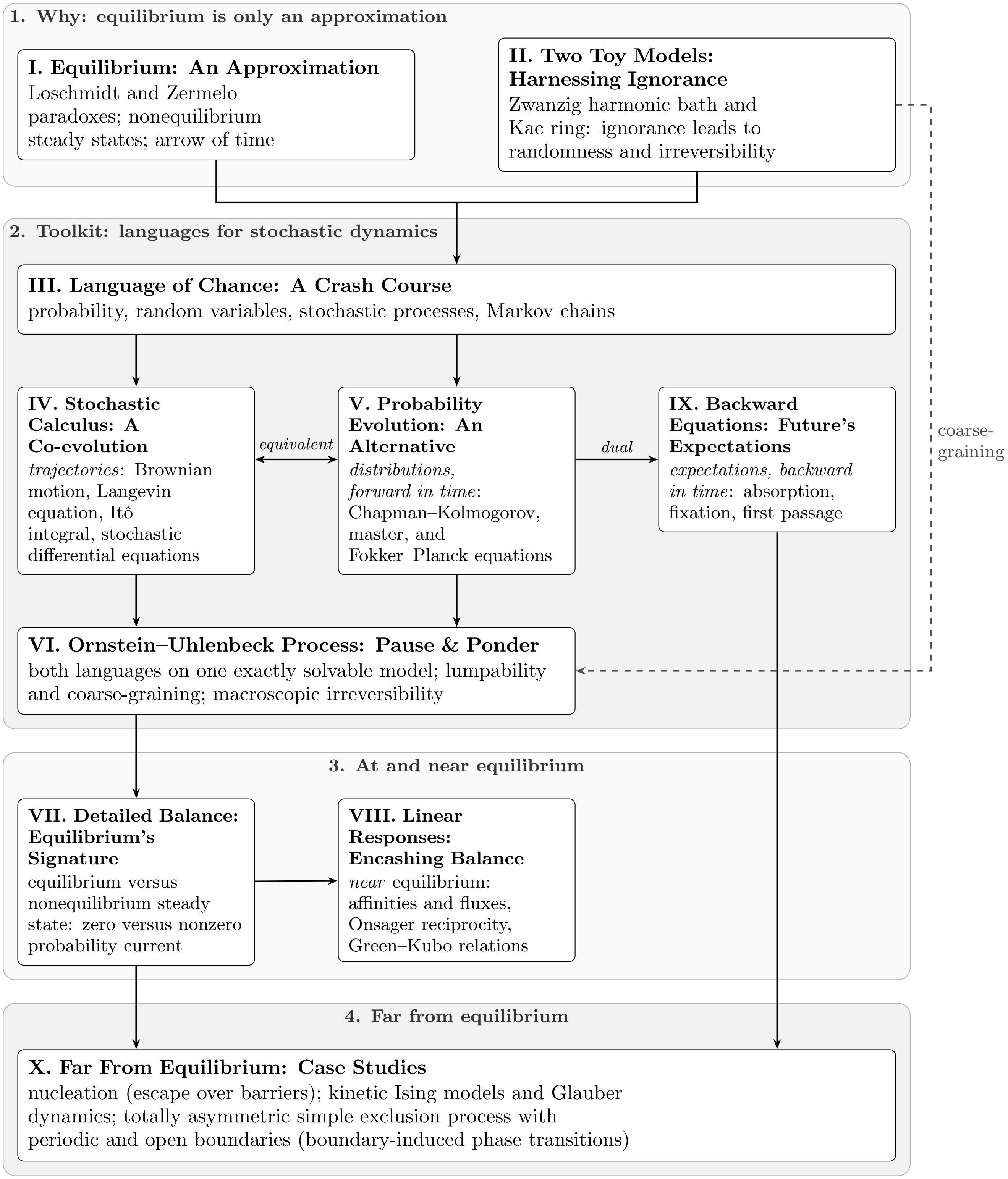
\begin{figure*}[p]
\centering
\begin{tikzpicture}[
  font=\normalsize,
  >={Stealth[length=5pt]},
  box/.style={draw, rounded corners=3pt, fill=white, align=left,
              execute at begin node={\hyphenpenalty=10000\exhyphenpenalty=10000},
              text width=#1, inner sep=5pt, minimum height=1.4cm},
  one/.style={box=3.75cm, font=\small},    
  two/.style={box=9.3cm},                 
  half/.style={box=6.525cm},                 
  full/.style={box=14.85cm, minimum height=1.2cm},
  tier/.style={draw=gray!60, rounded corners=6pt, fill=gray!#1, inner sep=7pt},
  tierlabel/.style={font=\bfseries\small, anchor=north west, text=black!80},
  req/.style={->, thick},
  link/.style={->, dashed, thick, gray!60!black},
]
\coordinate (A)  at (0,0);
\coordinate (B)  at (5.55,0);
\coordinate (C)  at (11.1,0);
\coordinate (AB) at (2.775,0);
\coordinate (M)  at (5.55,0);
\coordinate (H1) at (1.3875,0);
\coordinate (H2) at (9.7125,0);

\node[half] (I) at (H1)
  {\textbf{I. Equilibrium: An Approximation}\\[1pt]
   Loschmidt and Zermelo paradoxes; nonequilibrium steady states; arrow of time};
\node[half] (II) at (H2)
  {\textbf{II. Two Toy Models: Harnessing Ignorance}\\[1pt]
   Zwanzig harmonic bath and Kac ring: ignorance leads to randomness and irreversibility};

\node[full, below=1.6cm of II.south -| M] (III)
  {\textbf{III. Language of Chance: A Crash Course}\\[1pt]
   probability, random variables, stochastic processes, Markov chains};
\node[one, below=0.9cm of III.south -| A] (IV)
  {\textbf{IV. Stochastic Calculus: A Co-evolution}\\[1pt]
   \emph{trajectories}: Brownian motion, Langevin equation, It\^o integral, stochastic differential equations};
\node[one, below=0.9cm of III.south -| B] (V)
  {\textbf{V. Probability Evolution: An Alternative}\\[1pt]
   \emph{distributions, forward in time}: Chapman--Kolmogorov, master, and Fokker--Planck equations};
\node[one, below=0.9cm of III.south -| C] (IX)
  {\textbf{IX. Backward Equations: Future's Expectations}\\[1pt]
   \emph{expectations, backward in time}: absorption, fixation, first passage};
\node[two, below=0.9cm of IV.south -| AB] (VI)
  {\textbf{VI. Ornstein--Uhlenbeck Process: Pause \& Ponder}\\[1pt]
   both languages on one exactly solvable model; lumpability and coarse-graining; macroscopic irreversibility};

\node[one, below=1.45cm of VI.south -| A] (VII)
  {\textbf{VII. Detailed Balance: Equilibrium's Signature}\\[1pt]
   equilibrium versus nonequilibrium steady state: zero versus nonzero probability current};
\node[one, below=1.45cm of VI.south -| B] (VIII)
  {\textbf{VIII. Linear Responses: Encashing Balance}\\[1pt]
   \emph{near} equilibrium: affinities and fluxes, Onsager reciprocity, Green--Kubo relations};

\node[full, below=1.45cm of VII.south -| M, minimum height=1.4cm] (X)
  {\textbf{X. Far From Equilibrium: Case Studies}\\[1pt]
   nucleation (escape over barriers); kinetic Ising models and Glauber dynamics;
   totally asymmetric simple exclusion process with periodic and open boundaries
   (boundary-induced phase transitions)};

\draw[req] (III.south -| IV) -- (IV);
\draw[req] (III.south -| V) -- (V);
\coordinate (row2) at ($(IX.north)!0.5!(IX.south)$);
\draw[<->, thick] (IV.east |- row2) -- node[above, font=\footnotesize]{\emph{equivalent}} (V.west |- row2);
\draw[req] (V.east |- row2) -- node[above, font=\footnotesize]{\emph{dual}} (IX.west |- row2);
\draw[req] (IV.south) -- (IV.south |- VI.north);
\draw[req] (V.south) -- (V.south |- VI.north);
\draw[req] (VI.south -| A) -- (VII.north);
\draw[req] (VII) -- (VIII);
\draw[req] (VII.south) -- (VII.south |- X.north);
\draw[req] (IX.south) -- (IX.south |- X.north);

\coordinate (t1) at ($(I.north)+(0,0.55)$);
\coordinate (t2) at ($(III.north)+(0,0.55)$);
\coordinate (t3) at ($(VII.north)+(0,0.55)$);
\coordinate (t4) at ($(X.north)+(0,0.55)$);
\coordinate (r3) at (X.east |- VII.south);
\begin{scope}[on background layer]
  \node[tier=4,  fit=(I)(II)(t1)] (T1) {};
  \node[tier=10, fit=(III)(IV)(V)(IX)(VI)(t2)] (T2) {};
  \node[tier=4,  fit=(VII)(VIII)(t3)(r3)] (T3) {};
  \node[tier=10, fit=(X)(t4)] (T4) {};
\end{scope}
\node[tierlabel] at (T1.north west) {1. Why: equilibrium is only an approximation};
\node[tierlabel] at (T2.north west) {2. Toolkit: languages for stochastic dynamics};
\node[tierlabel, anchor=north] at (T3.north -| B) {3. At and near equilibrium};
\node[tierlabel, anchor=north] at (T4.north -| B) {4. Far from equilibrium};

\coordinate (m1) at ($(T1.south)!0.5!(T2.north)$);
\draw[thick] (I.south) -- (I.south |- m1) -- (II.south |- m1) -- (II.south);
\draw[req] (m1 -| III.north) -- (III.north);

\coordinate (gx) at ($(T1.east)+(0.35,0)$);
\draw[link] (II.east) -- (II.east -| gx) |- node[pos=0.3, right, font=\small, align=left]{coarse-\\graining} (VI.east);

\end{tikzpicture}
\caption{{\bf Roadmap of the lecture notes}: the conceptual flow from why equilibrium is only an approximation (tier 1), through the mathematical toolkit with its three complementary languages---viz., trajectories, probabilities evolving forward in time, and expectations evolving backward in time (tier 2)---and the behaviour at and near equilibrium (tier 3), to case studies far from equilibrium (tier 4). Roman numerals refer to sections; the boxes follow the logical rather than the numerical order. Solid arrows mark prerequisites, the double-headed arrow connects equivalent descriptions, and the dashed arrow marks an idea that returns later in a new guise.}
\label{fig:roadmap}
\end{figure*}
\newpage
\newpage
\clearpage
\onecolumngrid
\section{Equilibrium: An Approximation}
\epigraph{\textsf{There is nothing permanent except change.}}{Heraclitus}
\twocolumngrid

\indent {\color{deepsea}{\LARGE \textgoth{E}}}quilibrium in thermodynamic systems is a statistical reality but not, strictly speaking, 
everlasting. There is a mathematically unique way in which such an equilibrium is defined: 
as the macrostate that maximizes the accessible phase-space volume under the constraints 
of the system's conserved invariants, thereby maximizing the Boltzmann entropy.  Macroscopic thermodynamic equilibrium is thus conceptualized not as a frozen, static 
configuration in microscopic phase space, but as an ensemble-averaged manifestation 
characterized by the overwhelming geometric dominance of the maximum-entropy macrostate. 
Because phase-space hyper-volume scales exponentially with the system's massive degrees of 
freedom, the equilibrium macrostate physically monopolizes virtually the entirety of the 
accessible manifold. The apparent permanence of equilibrium is a direct consequence of 
this geometric asymmetry; a chaotic, time-reversible trajectory wanders dynamically from 
one microstate to another, yet it remains statistically trapped within the 
macroscopically identical configurations of the dominant volume over all physical 
observation timescales.

In fact, through the discussion of two fundamental paradoxes---Loschmidt's paradox 
and Zermelo's paradox---in the foundations of statistical mechanics, we shall be able to 
convince ourselves of the mathematical truth that a macroscopic thermodynamic system 
consisting of particles obeying deterministic, reversible Newtonian equations of motion 
can never remain strictly in equilibrium for all time. This follows from the dual 
realities that there always exists a non-zero probability that entropy may spontaneously 
decrease along a given phase-space trajectory, and almost every phase trajectory, 
over an infinite temporal horizon, returns arbitrarily close to its initial configuration.

Although equilibrium is mathematically always an approximation, it practically serves as 
a highly accurate description in many macroscopic situations, as evidenced by the 
astonishing success of equilibrium statistical mechanics over more than a century of 
applications. However, when a student of the subject ponders beyond this paradigm 
toward nonequilibrium processes, a unique, operational definition evades them; 
the statement that ``anything not in equilibrium is nonequilibrium'' remains a 
mere tautology.

In any case, two distinct nonequilibrium scenarios come to mind readily: a system (near equilibrium)
spontaneously relaxing toward equilibrium, and a system (far from equilibrium) continuously driven externally 
to prevent it from reaching equilibrium. A straightforward example of the latter is 
water being steadily pumped through a pipe. At any fixed spatial coordinate within the 
pipe, the local density and velocity profile remain statistically constant over time, 
yet the system cannot be classified as being in thermodynamic equilibrium; rather, it 
exists in a \emph{nonequilibrium steady state} (NESS) maintained by a continuous 
external pressure gradient. Such nonequilibrium steady states can give rise to intriguing macroscopic pattern 
formation, as exemplified by Rayleigh--B\'enard convection: When two horizontal plates 
are maintained at different temperatures, with the lower plate being hotter, a confined 
fluid between them can overcome internal viscosity and gravity to initiate macroscopic 
convection by virtue of the thermal gradient. This unstable buoyancy mechanism breaks 
the homogeneous spatial symmetry, resulting in spectacular stationary and oscillatory 
patterns of fluid flow.

Just to bring the contrast between thermodynamic equilibrium and a NESS to the fore, 
we highlight that a NESS is fundamentally characterized by the presence of a persistent, 
net macroscopic flux; in contrast, in true equilibrium, all net macroscopic fluxes 
are strictly zero. Consider, for example, a chemical reaction in a beaker where a reactant $R$ transforms 
into a product $P$ via first-order kinetics. In a state of true thermodynamic equilibrium, 
the ratio of concentrations remains constant ($[R]/[P] = \text{constant}$) because the 
system satisfies the principle of detailed balance: the forward reaction rate exactly 
balances the reverse rate ($k_f[R] = k_b[P]$), leaving the net chemical flux completely 
extinguished.  Interestingly, we can drive this exact same system into a nonequilibrium steady state by 
continuously feeding $R$ into the beaker and selectively removing $P$ via an external 
mass transport mechanism. Once a steady state is reached, the concentration ratio 
$[R]/[P]$ will once again appear constant over time, mimicking equilibrium. However, 
the microscopic rates no longer balance ($k_f[R] \neq k_b[P]$), and the constancy of the 
concentrations is instead maintained by a continuous, global mass flux traversing the 
boundaries of the system.

While nonequilibrium statistical mechanics originated as a core framework within 
theoretical physics, its underlying conceptual architecture extends far beyond traditional 
physical systems. When a system is liberated from the constraints of thermodynamic 
equilibrium, its mathematical machinery becomes universally applicable to any complex, 
many-body ensemble characterized by broken time-reversal symmetry and continuous external 
or internal driving. 

In biophysics, this framework underpins the study of {active matter}, where individual 
constituents---such as schooling fish, swarming bacteria, or intracellular molecular 
motors---convert ambient energy into autonomous mechanical work, manifesting collective 
nonequilibrium phase transitions. Similarly, in quantitative finance, economic markets 
are modelled not as static allocations, but as dynamic driven systems where the continuous 
global flux of capital and information establishes nonequilibrium steady states. 

Remarkably, this paradigm has also revolutionized computer science; modern generative 
artificial intelligence relies on algorithmic {diffusion models} that parameterize 
the time-reversed thermodynamic trajectories of stochastic noise to engineer highly structured, 
low-entropy data states. It is by no means an exaggeration to state that nonequilibrium statistical 
mechanics---along with nonlinear dynamics, information theory, network theory, and game theory---constitutes 
the definitive language of \emph{complex systems}. Ultimately, the transition from equilibrium to 
nonequilibrium statistical mechanics transforms the discipline from a localized study of matter 
and heat into a universal calculus of emergence, self-organization, and information processing 
across biological, financial, and technological scales.

Thus, nonequilibrium statistical mechanics beautifully illustrates how practical science actually operates: even without exact knowledge of the explicit mathematical laws governing the invisible microscopic world, the collective interactions of a system's constituent particles give rise to visible macroscopic, emergent phenomena. These phenomena can be adequately understood using models tailored precisely to their specific length and time scales. Crucially, this framework exposes the futility of radical reductionism---whether, e.g., in physics or biology---in an optimistic way. It opens new doors to comprehending real-world complexity through pragmatic, scale-specific models that satiate our endless curiosity for both `how' and `why', while spotlighting the universality that spans transdisciplinary phenomena.

Since the subject lacks a canonical syllabus, a word on the organization of these notes is in order; the conceptual flow is summarized in Fig.~\ref{fig:roadmap}. We first argue why equilibrium is only an approximation and why a stochastic description is natural (Secs.~I and II). We then develop the mathematical toolkit, built on the basic language of chance (Sec.~III), in three complementary languages: trajectories, governed by stochastic differential equations (Sec.~IV); probabilities evolving forward in time, governed by master and Fokker--Planck equations (Sec.~V); and expectations evolving backward in time, governed by the dual backward equations, which answer questions about the future such as absorption and first passage (Sec.~IX). The first two are brought together on the exactly solvable Ornstein--Uhlenbeck process (Sec.~VI). Equipped with these tools, we examine equilibrium and its neighbourhood: detailed balance as the signature of equilibrium (Sec.~VII), and linear response to small departures from it (Sec.~VIII). Finally, building on detailed balance and backward equations alike, we turn to case studies far from equilibrium (Sec.~X). Throughout, three questions are worth asking of every model: Is the dynamics Markov? Is it best described by its trajectories, its probabilities, or its expectations? And does it obey detailed balance, i.e., is its steady state an equilibrium or a NESS?

One final remark is in order before we turn to the two paradoxes mentioned earlier:
Through the study of nonequilibrium statistical mechanics, we are essentially
confronting one of the three fundamental sources of uncertainty in nature---the
triad being deterministic (chaotic), stochastic, and quantum---though, more often
than not, an intricate interplay among all three is present in varying degrees.
Nonequilibrium systems are conventionally characterized by an extraordinarily
large number of degrees of freedom, rendering our detailed microscopic ignorance
an inescapable practical constraint. This fundamental ignorance compels us to
interpret the system's dynamics as inherently unpredictable and random---or,
in a single word, \emph{stochastic}.

\subsection{Loschmidt's Paradox}
Within the paradigm of classical statistical mechanics, we may consider the particle dynamics to be given by some Newtonian dynamics where the interaction forces are derivable from some time-independent potential and so the particle dynamics is governed by a Hamiltonian $H(\bs{q},\bs{p})$; the exact form of the potential, however, is not precisely known. Since there can be many {microstates} corresponding to a given {macrostate}, a set of many phase points---constrained by the values of the thermodynamic variables at a given point of time---would correspond to a single macrostate (defined by the very values of the thermodynamics variables) at that time instant; this set is called an \emph{ensemble} of microstates, each of which is a copy of the macrostate.  One can define the phase space density $\rho(\bs{q},\bs{p},t)$ of the ensemble such that $\rho(\bs{q},\bs{p},t)\prod_{i=1}^Nd\bs{q}_id\bs{p}_i$ is the fraction of all possible microstates present in the elementary phase space volume $\prod_{i=1}^Nd\bs{q}_id\bs{p}_i$. As per the \emph{Liouville theorem} for the Hamiltonian systems, the time evolution in the phase space $\Omega$ is such that $d\rho/dt=0$. Thus, in the terminology of dynamical systems, Hamiltonian flow is a volume-preserving.

The microscopic dynamics, being Newtonian, is \emph{reversible}. 
In other words, since the invariance of the equation of motion under 
time reversal---i.e., $(\bs{q},\bs{p},t)\rightarrow(\bs{q},-\bs{p},-t)$---is 
manifest, the equilibrium phase space density satisfies 
$\rho(\bs{q},\bs{p}) = \rho(\bs{q},-\bs{p})$. In passing, 
we remark that the reversibility of Newton's laws generally depends on the 
nature of the force. While velocity-dependent forces that are non-dissipative---such 
as the magnetic Lorentz force---preserve reversibility under the simultaneous 
transformation of external magnetic fields $\bs{B} \rightarrow -\bs{B}$, 
a macroscopically dissipative force will render the corresponding Newton's 
equation explicitly irreversible. Similarly, Newtonian dynamics becomes 
irreversible if the force is explicitly time-dependent, unless the 
forcing protocol happens to be an even function of time.

Many paradoxes related to the microscopic statistical justification of the Second Law of Thermodynamics---entropy of
an isolated thermodynamic system can not decrease in
any spontaneous process---have been put forward over the years and they have actually helped in developing deeper insights into the subject matter. They mainly hover around the topic of the intriguing emergence of macroscopic or thermodynamic irreversibility, sometimes fancily known as the concept of thermodynamic \emph{arrow of time}. One of the earliest ones is \emph{Loschmidt's paradox} (due to time-reversibility of the Hamiltonian dynamics, increase in entropy with time cannot occur for all the initial phase points). However, this is not a true contradiction: the  Second Law of Thermodynamics does not claim to be applicable either for all phase points---the law is fundamentally statistical in origin.

Thus, the Loschmidt paradox is resolved by recognizing that the transition from microscopic reversibility to macroscopic irreversibility is a consequence of {probability and statistics}, rather than a flaw in Newtonian mechanics. While the fundamental laws of motion permit a system to spontaneously reorder itself and decrease its entropy, the number of disordered configurations vastly outnumbers the ordered ones. This statistical asymmetry renders entropy-decreasing trajectories astronomically improbable.

To understand this statistical behaviour we present analogy of a coin-toss experiment. Consider a macrostate defined by flipping $N = 100$ coins:
\begin{itemize}
    \item \emph{Low-Entropy State:} A highly ordered state where all 100 coins land on {Heads}. There is only $1$ unique configuration (microstate) that achieves this result.
    \item \emph{High-Entropy State:} A disordered state with an even split of {50 Heads and 50 Tails}. There are roughly ${}^{100}C_{50}\approx10^{29}$ different configurations that yield this same macrostate.
\end{itemize}
If we shake a box containing 100 ``Heads'' coins, it will naturally transition into a messy 
50/50 mix because the system randomly wanders into the overwhelmingly larger pool of disordered 
states. Mathematically, it is entirely possible for continued shaking to reverse this path and 
return all coins to ``Heads.'' However, the probability of hitting that single, perfectly 
ordered microstate by random chance is $1$ in $2^{100}$ (approximately $1$ in $10^{30}$).

When scaling this principle up to a macroscopic system containing Avogadro's number of 
particles ($N_A \sim 10^{23}$), statistical deviations from the average behaviour completely 
vanish. The microstate volume ratio between a macrostate where a gas is compressed into 
half a container versus one where it fills the entire volume scales as:
\begin{equation}
    \frac{\Omega_{\text{compressed}}}{\Omega_{\text{expanded}}} \sim \left(\frac{1}{2}\right)^{10^{23}}.
\end{equation}
This fraction is so infinitesimally close to zero that the compressed macrostate represents 
a zone of extreme scarcity in phase space. Consequently, a spontaneous reduction 
in entropy can only be triggered by selecting from an exceptionally rare subset of 
specially engineered initial phase points whose molecular momenta are inverted with 
absolute, conspiratorial precision.

Ultimately, the arrow of time is dictated not by the time-symmetric Newtonian equations, 
but by the universe's boundary conditions. Modern physics relies on the \emph{Past Hypothesis}: 
the assertion that the universe was born in a macrostate of exceptionally low entropy at 
the Big Bang. Because our system is historically bottlenecked by this initial scarcity, 
any forward-evolving microscopic trajectory has nowhere to go but to expand out of this 
unlikely state and into the vastly more abundant, high-entropy regions of phase space. 
Macroscopic irreversibility is therefore an inescapable certainty driven by an asymmetric 
cosmological initial condition.

\subsection{Zermelo's Paradox}
For the thermodynamic system under consideration, another mathematical feature of the 
phase space dynamics is \emph{Poincar\'e recurrence}: after a finite (but typically 
astronomical) duration, almost every microstate on the constant-energy manifold returns arbitrarily close to its initial point in phase space. Thus, we 
face another challenge to understanding omnipresent macroscopic irreversibility 
through deterministic Hamiltonian dynamics. This is known as \emph{Zermelo's paradox}: 
due to the inevitability of Poincar\'e recurrences, the entropy of a closed system 
cannot monotonically increase for all time. However, this also is not a true 
contradiction; after all, the Second Law of Thermodynamics is a statistical principle valid over 
physical observation timescales, rather than a rigid theorem intended to apply across 
infinite temporal horizons.

Let us now understand the Poincar\'e recurrence. We note that the Hamiltonian dynamics on the phase space can be mathematically viewed as a triplet $(\mathcal{M},\mu, f^t)$ where $f^t$ is a one-parameter (parameter being $t$) measure-preserving flow on space (manifold) $\mathcal{M}$ with $\mu$ as its measure.  The notion of \emph{measure} generalizes geometric notions---such as length, area, and volume---alongside physical properties like mass and event probabilities, assigning a non-negative ``size'' to subsets of a space. While we present the formal technical definition below, for the sake of our phase-space discussion, a measure can conveniently be viewed as a tool to determine the volume of any phase-space subset (i.e., a collection of microstates).

\emph{Definition of measure}: For a space $\mathcal{M}$, a function $\mu$ defined on a ($\sigma$-algebra of) subsets of $M$ is a \emph{measure} if it assigns a non-negative real number to subsets $A, B \subseteq \mathcal{M}$ according to following three conditions:
\begin{enumerate}
    \item \emph{Non-negativity:} The size of any subset $A \subseteq \mathcal{M}$ can never be negative: $\mu(A) \ge 0$.
    \item \emph{Empty set is zero:} The empty set $\emptyset \subseteq \mathcal{M}$ has measure zero: $\mu(\emptyset) = 0$.
    \item \emph{Countable Additivity ($\sigma$-additivity):} For any two disjoint (non-overlapping) subsets $A, B \subseteq \mathcal{M}$ (where $A \cap B = \emptyset$), the measure of their union is the sum of their individual measures:  $\mu(A \cup B) = \mu(A) + \mu(B)$.
    More generally, for any countable sequence of pairwise disjoint subsets $\{A_i\}_{i=1}^{\infty} \subseteq \mathcal{M}$:
    \begin{equation*}
        \mu\left( \bigcup_{i=1}^{\infty} A_i \right) = \sum_{i=1}^{\infty} \mu(A_i).
    \end{equation*}
\end{enumerate}
A normalized measure with $\mu(\mathcal{M})=1$ is known as \emph{probability measure} in which context $\mathcal{M}$ is known as \emph{sample space} and its subsets ($A_i$s) are called \emph{events}.

Going back to the triplet, the measure-preservation essentially implies: $\mu(f^{-t} (A))=\mu(A)\,\forall A\subseteq \mathcal{M}$. Recall for the phase space dynamics, Liouville theorem guarantees the measure-preservation.

Now we are ready to state the \emph{Poincar\'e recurrence theorem} where we assume, for the sake of convenience (with no loss of generality), $t$ to be discrete parameter and hence denoted by $n$:
\begin{enumerate}
\item[]
\emph{Theorem:} Let $(\mathcal{M}, \mu, f)$ be a measure-preserving dynamical system with $\mu(\mathcal{M}) < \infty$. For any measurable subset $A \subseteq \mathcal{M}$ with $\mu(A) > 0$, there exists a point $a \in A$ and a positive integer $n\ge 1$ such that
\begin{equation}
    f^n(a) \in A.
\end{equation}
Furthermore, this recurrence holds for almost all points in $A$---i.e., for every point in $A$ except possibly a subset of measure zero:
\begin{equation}
    \mu\Big(\{ a \in A \mid f^n(a) \notin A \; \forall n \ge 1 \}\Big) = 0.
\end{equation}
\end{enumerate}

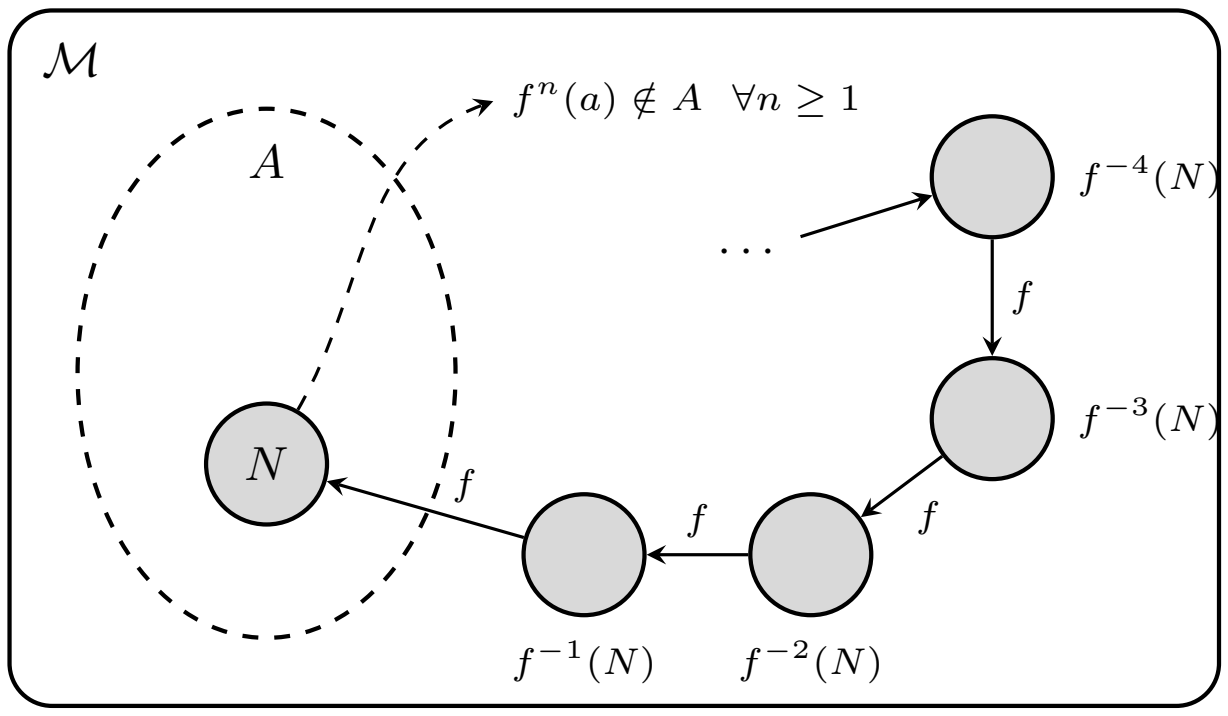
\begin{figure}[htbp]
  \small
  \centering
  \begin{tikzpicture}[
    blob/.style={circle, draw, thick, fill=gray!30, minimum size=8mm, inner sep=0pt},
    edge/.style={->, >=stealth, semithick},
    lab/.style={font=\scriptsize}
  ]
    \draw[thick, rounded corners=8pt] (0,0) rectangle (8,4.6);
    \node[anchor=north west] at (0.1,4.5) {$\mathcal{M}$};
    \draw[thick, dashed] (1.7,2.2) ellipse (1.25cm and 1.75cm);
    \node at (1.7,3.6) {$A$};

    \node[blob] (N)  at (1.7,1.6) {$N$};
    \node[blob] (N1) at (3.8,1.0) {};
    \node[blob] (N2) at (5.3,1.0) {};
    \node[blob] (N3) at (6.5,1.9) {};
    \node[blob] (N4) at (6.5,3.5) {};
    \node (D) at (4.9,3.0) {$\cdots$};

    \node[lab, below=1pt] at (N1.south) {$f^{-1}(N)$};
    \node[lab, below=1pt] at (N2.south) {$f^{-2}(N)$};
    \node[lab, right=1pt] at (N3.east)  {$f^{-3}(N)$};
    \node[lab, right=1pt] at (N4.east)  {$f^{-4}(N)$};

    \draw[edge] (D) -- (N4);
    \draw[edge] (N4) -- node[lab, right] {$f$} (N3);
    \draw[edge] (N3) -- node[lab, below right=-1pt] {$f$} (N2);
    \draw[edge] (N2) -- node[lab, above] {$f$} (N1);
    \draw[edge] (N1) -- node[lab, above, pos=0.3] {$f$} (N);

    \draw[edge, dashed] (N) to[out=60, in=200] (3.2,4.0)
      node[lab, right] {$f^n(a)\notin A\ \ \forall n\ge1$};
  \end{tikzpicture}
  \caption{\small Idea behind the proof: The set $N\subseteq A$ (shaded) contains the
  points of $A$ that never return to $A$. Its preimages $f^{-k}(N)$, $k\ge0$, are
  pairwise disjoint, and since $f$ is measure preserving they all have the same
  measure $\mu(N)$ (drawn as equal-size disks). A space of finite measure cannot
  contain infinitely many disjoint sets of equal positive measure, so $\mu(N)=0$.}
  \label{fig:poincare}
\end{figure}

\emph{Proof:} Define $N \subseteq A$ to be the set of points in $A$ that never return to $A$ under positive iterates of $f$:
\begin{equation}
    N = \{ a \in A \mid f^n(a) \notin A \; \forall n \ge 1 \}.
\end{equation}

Consider the sequence of preimages $\{ f^{-k}(N) \}_{k=0}^{\infty} = \{ N, f^{-1}(N), f^{-2}(N), \dots \}$. We assert that these sets are pairwise disjoint (see Fig.~\ref{fig:poincare}).

Suppose for \emph{contradiction} that $f^{-i}(N) \cap f^{-j}(N) \neq \emptyset$ for integers $0 \le i < j$. Then there exists a point $p \in \mathcal{M}$ such that $f^i(p) \in N$ and $f^j(p) \in N$. Setting $a = f^i(p) \in N$, we observe that:
\begin{equation}
    f^{j-i}(a) = f^{j-i}(f^i(p)) = f^j(p) \in N \subseteq A.
\end{equation}
Since $j - i \ge 1$, this implies that $a \in N$ returns to $A$ after $j - i$ steps, directly contradicting the definition of $N$. Thus, the preimages must be pairwise disjoint.

Because $f$ is measure-preserving, we have $\mu(f^{-k}(N)) = \mu(N)$ for all $k \ge 0$. Summing over the disjoint union yields:
\begin{equation}
    \mu(\mathcal{M}) \ge \mu\left( \bigcup_{k=0}^{\infty} f^{-k}(N) \right) = \sum_{k=0}^{\infty} \mu(f^{-k}(N)) = \sum_{k=0}^{\infty} \mu(N).
\end{equation}
Since $\mu(\mathcal{M}) < \infty$, the infinite series $\sum_{k=0}^{\infty} \mu(N)$ can only converge if $\mu(N) = 0$. Hence, almost every point $a \in A$ returns to $A$. 

In fact, almost every point returns infinitely often. For any $k\ge1$, the map $f^k$ is also measure preserving, so the same argument applied to $f^k$ shows that
\begin{equation}
    N_k = \{ a \in A \mid f^{mk}(a) \notin A \; \forall m \ge 1 \}
\end{equation}
has $\mu(N_k)=0$. Every point of $A$ that does not return to $A$ at any time $n\ge k$ belongs to $N_k$. A point that returns to $A$ only finitely many times lies in $N_k$ for some $k$, and hence in $\bigcup_{k\ge1}N_k$, which is a countable union of sets of measure zero and therefore has measure zero.

While we did not initially require $\mu$ to be a probability measure, in what follows 
we shall fix it to be so. Now, the resolution of Zermelo's paradox dawns when the 
solution to the question---``What is the expected number of steps $n$ for a trajectory 
originating in $A$ to return inside $A$?''---is searched for. There is a precise 
mathematical answer to this in the form of \emph{Kac's Lemma}, provided the system is 
additionally ergodic. A system is \emph{ergodic} if, for any measurable set 
$A \subseteq \mathcal{M}$, the condition $T^{-1}(A) = A$ implies $\mu(A) \in \{0,1\}$ 
[in other words, if a region is completely trapped or unchanged by the flow of time, 
that region must either be nothing ($\mu=0$) or everything ($\mu=1$)]. Under this 
condition, Kac's lemma dictates that, on average, the recurrence time to a region 
is inversely proportional to its measure. 

Naturally, if one has one mole of compressed air in a bottle inside a closed room, 
then after opening the bottle, one will need to wait for an astronomically long time 
to get almost all the air magically back into the bottle. This is because (ignoring 
velocity distribution differences between the bottle and the room), the measure of the 
phase-space subset corresponding to the bottle state is the ratio of the volume of the 
bottle to that of the room, raised to the power of Avogadro's number of particles!
\begin{center}
	$\symking\symqueen\symrook\symknight\symbishop\sympawn\epsdice{1}\,\epsdice{2}\,\epsdice{3}\,\epsdice{4}\,\epsdice{5}\,\epsdice{6}\,\spadesuit\heartsuit\clubsuit\diamondsuit$
\end{center}
\newpage
\newpage
\clearpage
\onecolumngrid
\section{Two Toy Models: Harnessing Ignorance}
\epigraph{\textsf{All models are wrong, but some are useful.}}{George E. P. Box}
\twocolumngrid

\indent {\color{deepsea}{\LARGE \textgoth{A}}}s discussed earlier, the philosophical framework we are dealing with is emergentism (specifically, epistemic and ontological emergence), paired with scientific pragmatism (or instrumentalism).

This philosophical framework finds its ultimate, rigorous validation in two seminal toy models 
of nonequilibrium statistical mechanics: the Zwanzig harmonic bath model and the Kac ring model. 
Together, these models serve as a literal mathematical execution of the principle that 
macroscopic irreversibility, dissipation, and stochasticity do not require fundamentally 
random or irreversible microscopic laws. Instead, they demonstrate that the thermodynamic arrow 
of time emerges naturally from entirely deterministic, time-reversible Hamiltonian dynamics 
whenever a focal system is projected onto a coarse-grained space. Crucially, both toy models formalize 
the precise mechanism by which human ignorance---far from destroying predictive power---acts as the 
essential mathematical catalyst that transforms microscopic determinism into macroscopic probability. Ultimately, these two models present a unified front against radical reductionism. 

The Zwanzig harmonic bath model (introduced by Robert Zwanzig in 1973) achieves this by explicitly 
coupling a mesoscopic system particle to an environment consisting of an infinite collection of 
deterministic harmonic oscillators. In this framework, `noise' and `randomness' are stripped 
of any intrinsic status in the universe; they are revealed to be the direct mathematical consequence 
of our microscopic ignorance regarding the bath's exact initial conditions. By integrating out the 
unobservable bath degrees of freedom and replacing our lack of exact knowledge with a statistical 
ensemble average, a smooth friction term and a fluctuating stochastic force emerge from purely 
Newtonian trajectories. The deep philosophical insight of the Zwanzig model is that subjective 
ignorance yields objectively correct macroscopic dynamics, providing a first-principles derivation 
of the \emph{fluctuation--dissipation theorem/relation}.

In a structurally distinct yet philosophically complementary manner, the Kac ring model 
(introduced by Mark Kac in 1956) explores the manifestation of ignorance through spatial rather 
than environmental complexity. While Zwanzig relies on an infinite environmental bath to derive 
dissipation, the Kac model employs a discrete, time-reversible toy universe to demonstrate how an 
irreversible approach to equilibrium arises from our systematic ignorance of microscopic 
correlations. By invoking a statistical assumption akin to Boltzmann's \emph{Stosszahlansatz} 
(the hypothesis of molecular chaos), the model deliberately discards the trackless, exact locations 
of scattering markers in favour of a singular, macroscopic density parameter. Averaging over this 
structural ignorance effectively erases the system's exact microscopic memory, forcing a 
spontaneously emergent arrow of time to materialize.

\subsection{Zwanzig Harmonic Bath Model}
Consider a focal particle (system) described by coordinate $x$ and conjugate momentum $p$ in a potential $U(x)$. The heat bath consists of independent harmonic oscillators (with mass $m_j = 1$, $j$ denoting $j$th oscillator) defined by coordinates $q_j$ and conjugate momenta $p_j$. The harmonic bath is explicitly assumed to be at a fixed, constant temperature, $T$.  The system and the bath oscillators are coupled via a linear interaction, where the coupling force depends on their relative coordinates. The motion of the focal particle is generally an out-of-equilibrium (nonequilibrium) process, though it can asymptotically relax into an equilibrium state over time as we shall see. In passing, we note that the central particle is described by a single coordinate and a single momentum; the interaction with the bath oscillators is also treated along this single axis. This one-dimensional setup is adopted primarily to simplify the algebra.

The total Hamiltonian, thus, is given by:
\begin{equation}
H = \frac{p^2}{2m} + U(x) + \sum_{j} \left[ \frac{p_j^2}{2} + \frac{\omega_j^2}{2} \left( q_j - \frac{\gamma_j}{\omega_j^2} x \right)^2\right],\label{eq:Hz}
\end{equation}
where $\omega_j$ is the frequency of the $j^{\text{th}}$ oscillator and $\gamma_j$ measures the coupling strength between the focal particle and the $j^{\text{th}}$ oscillator. Using Hamilton's equations ($\dot{q} = \frac{\partial H}{\partial p}$ and $\dot{p} = -\frac{\partial H}{\partial q}$), the equation of motion for the focal particle is found to be
\begin{subequations}
\begin{align}
\dot{x} &= \frac{p}{m}, \\
\dot{p}&= -\frac{\partial H}{\partial x} = -\frac{dU}{dx} + \sum_{j} \gamma_j \left( q_j - \frac{\gamma_j}{\omega_j^2} x \right).\label{eq:fpeq2}
\end{align}
\label{eq:fpeq}
\end{subequations}
Clearly, the system and the bath are made to interact via a linear coupling of the form $-\gamma_jxq_j$ in the Hamiltonian [just expand the square in Eq.~(\ref{eq:Hz})]; the coupling appears as a superposition of linear forcing terms---$j$th oscillator contributes $\gamma_j q_j$ force---in momentum equation~(\ref{eq:fpeq2}).

Similarly, for bath oscillators, we compactly arrive at
\begin{align}
 \ddot{q}_j + \omega_j^2 q_j = \gamma_j x(t).\label{eq:bath}
\end{align}
We may solve it using the Green's function method. Consider
\begin{equation}
\left( \frac{d^2}{d\tau^2} + \omega_j^2 \right) G(\tau) = \delta(\tau).
\end{equation}
Using causality---$G(\tau) = 0$ for $\tau < 0$---and matching boundary conditions at $\tau = 0$, we get:
\begin{equation}
G(t - t') = \frac{1}{\omega_j} \sin[\omega_j (t - t')].
\end{equation}
Therefore, the convolution of the Green's function with the source term ($\gamma_j x(t)$) yields the general solution for $q_j(t)$ as
\begin{eqnarray}
q_j(t) =& q_j(0) \cos(\omega_j t) + \frac{p_j(0)}{\omega_j} \sin(\omega_j t) \nonumber\\
 &+\int_{0}^{t} \frac{\sin[\omega_j(t - t')]}{\omega_j} \gamma_j x(t') \, dt'.
\end{eqnarray}

Integrating the last term by parts and substituting $\frac{dx}{dt'} = \frac{p(t')}{m}$, we obtain:
\begin{eqnarray}
q_j(t) &-& \frac{\gamma_j}{\omega_j^2}x(t) = \left( q_j(0) - \frac{\gamma_j}{\omega_j^2}x(0) \right) \cos(\omega_j t) \nonumber\\
&+& \frac{p_j(0)}{\omega_j} \sin(\omega_j t) - \int_{0}^{t} \frac{\cos[\omega_j(t - t')]}{\omega_j^2} \gamma_j \frac{p(t')}{m} \, dt'.\qquad
\end{eqnarray}

Substituting $q_j(t) - \frac{\gamma_j}{\omega_j^2}x(t)$ back into the focal particle's momentum equation [Eq.~(\ref{eq:fpeq2})], we get \emph{generalized Langevin equation}:
\begin{equation}
\dot{p} = -\frac{dU}{dx} - \int_{0}^{t} K(t - t') \frac{p(t')}{m} \, dt' + f(t),\label{eq:fpdot}
\end{equation}
where memory function
\begin{equation}
K(t - t') = \sum_{j} \frac{\gamma_j^2}{\omega_j^2} \cos[\omega_j(t - t')],
    \end{equation}
and noise term    
\begin{eqnarray}
    f(t) = &&\sum_{j} \gamma_j \left( q_j(0) - \frac{\gamma_j}{\omega_j^2}x(0) \right) \cos(\omega_j t)\nonumber\\
     &&+ \sum_{j} \gamma_j \frac{p_j(0)}{\omega_j} \sin(\omega_j t).
    \end{eqnarray}

Although the equation of motion for $p(t)$ remains formally deterministic, the force $f(t)$ is governed by the initial conditions of a vast number of bath variables---typically on the order of Avogadro's number. Since precisely specifying these initial values is practically impossible, a statistical framework is required. In this approach, $f(t)$ is treated as a stochastic force defined by its moments, which are derived from the initial distribution of the bath's coordinates and momenta. Assuming the heat bath is in thermal equilibrium at temperature $T$, these variables follow a Maxwell--Boltzmann (Gaussian) distribution, which implies 
\begin{equation}
P_{\text{eq}}(\{q_i\},\{p_i\}\mid x) \propto \prod_{j} \exp \left( -\frac{\frac{p_j^2}{2} + \frac{\omega_j^2}{2}\left(q_j - \frac{\gamma_j}{\omega_j^2}x\right)^2}{k_B T} \right).
\end{equation}
This leads to the following thermal averages:
\begin{subequations}
  \begin{eqnarray*}
\langle p_i(0) \rangle &=& 0, \\
\left\langle \left( q_i(0) - \frac{\gamma_i}{\omega_i^2} x(0) \right)\right\rangle &=&0,\\
\left\langle p_i(0) \left( q_j(0) - \frac{\gamma_j}{\omega_j^2} x(0) \right)\right\rangle &=&0,\\
\langle f(t) \rangle &=& 0, \\
\langle p_i(0) p_j(0) \rangle &=& k_B T \delta_{ij}, \\
\left\langle \left( q_i(0) - \frac{\gamma_i}{\omega_i^2} x(0) \right) \left( q_j(0) - \frac{\gamma_j}{\omega_j^2} x(0) \right) \right\rangle &=& \frac{k_B T}{\omega_i^2} \delta_{ij}.
    \end{eqnarray*}
\end{subequations}

Most interestingly, we get a fluctuation--dissipation relation/theorem on evaluating $\langle f(t) f(t') \rangle$ using $\cos(\omega_i t)\cos(\omega_i t') + \sin(\omega_i t)\sin(\omega_i t') = \cos[\omega_i(t - t')]$:
\begin{equation}
\langle f(t) f(t') \rangle = k_B T K(t - t').
\end{equation}
This is an interesting relation which says that  [see Eq.~(\ref{eq:fpdot})] in the dynamics of the focal particle is charaterized by a balance between the noisy kicks (via $f(t)$) and the damping (via memory kernel). This becomes clearer and simpler, when we take a limiting case where the autocorrelation is delta-correlated. To this end, we assume a continuous spectrum of frequencies $P(\omega) = \frac{\Gamma \omega^2}{\pi \gamma^2}$ as $\omega_{\text{max}} \to \infty$, so that
\begin{equation}
K(t - t') = \lim_{\omega_{\text{max}} \to \infty} \frac{\Gamma}{\pi} \int_{0}^{\omega_{\text{max}}} \cos[\omega(t - t')] \, d\omega = \Gamma \delta(t - t').
\end{equation}
Substituting this Dirac delta function back into the integral memory term of Eq.~(\ref{eq:fpdot}) requires careful evaluation at the integration boundary $t' = t$:
\begin{equation}
\int_0^t \Gamma \delta(t - t') \frac{p(t')}{m} \, dt' = \frac{\Gamma}{2m} p(t),
\end{equation}
where the factor of $1/2$ arises because the Dirac delta singularity lands directly on the upper limit of integration ($t'=t$). Next, 
setting $-\frac{dU}{dx} = 0$ for a free particle (or isolating momentum dynamics), the equation of motion reduces to:
\begin{equation}
\dot{p}(t) = -\frac{\Gamma}{2m} p(t) + f(t),\label{eq:le}
\end{equation}
with the noise correlation simplifying to that of the \emph{white noise}:
\begin{equation}
\langle f(t) f(t') \rangle = \Gamma k_B T \delta(t - t').\label{eq:mfdr}
\end{equation}
Eq.~(\ref{eq:le}) is the standard \emph{Langevin equation} and Eq.~(\ref{eq:mfdr}) is a fluctuation--dissipation relation.

The fluctuation--dissipation relation establishes that random noise ($f$) and frictional drag ($\Gamma$) are inseparable consequences of the same physical coupling between the particle and the bath oscillators. Mathematically, this link is expressed by the fact that the autocorrelation of the stochastic force is directly proportional to the friction (memory) kernel (here reduced to $\delta$-function), scaled by the thermal energy $k_B T$. Because both quantities share the same underlying structure and time-scales, this relation ensures a perfect energy balance: the agitation provided by thermal fluctuations is exactly countered by the energy lost to dissipation. Ultimately, this fundamental symmetry guarantees that the system maintains thermal equilibrium at the designated temperature and satisfies the \emph{equipartition theorem} (i.e., in thermal equilibrium, each degree of freedom holds $k_BT/2$ amount of energy on average).

To check the equipartition, we solve the Langevin equation (a \emph{stochastic differential equation}) using the integrating factor method:
\begin{equation}
    p(t_1) = e^{-\frac{\Gamma}{2m} t_1} p(0) + e^{-\frac{\Gamma}{2m} t_1} \int_0^{t_1} e^{\frac{\Gamma}{2m} s_1} f(s_1) \, ds_1.\label{eq:pt1}
\end{equation}
To find the correlation $\langle p(t_1) p(t_2) \rangle$, we multiply two such expressions and take the ensemble average. Since all cross terms between the initial momentum and the noise average to zero, we get:
\begin{eqnarray}
    \langle &&p(t_1) p(t_2) \rangle = e^{-\frac{\Gamma}{2m}(t_1+t_2)} \langle p(0)^2 \rangle\nonumber\\
     &&+ e^{-\frac{\Gamma}{2m}(t_1+t_2)} \int_0^{t_2} \int_0^{t_1} e^{\frac{\Gamma}{2m}(s_1+s_2)} \langle f(s_1) f(s_2) \rangle \, ds_1 \, ds_2.\qquad\label{eq:p1p2}
\end{eqnarray}

We evaluate the double integral term explicitly. Using the fluctuation--dissipation relation, $\langle f(s_1) f(s_2) \rangle = \Gamma k_B T \delta(s_1 - s_2)$. Assuming $t_1 < t_2$ (without any loss of generality):
\begin{align}
    I &= \int_0^{t_2} \int_0^{t_1} e^{\frac{\Gamma}{2m}(s_1+s_2)} \Gamma k_B T \delta(s_1 - s_2) \, ds_1 \, ds_2.
\end{align}
Evaluating the inner integral over $s_1$, the delta function $\delta(s_1 - s_2)$ constrains $s_1 = s_2$. This is non-zero only if $s_2$ is within the range $[0, t_1]$:
\begin{align}
    I &= \Gamma k_B T \int_0^{t_1} e^{\frac{\Gamma}{m}s_2} \, ds_2  = m k_B T \left( e^{\frac{\Gamma}{m}t_1} - 1 \right).
\end{align}
Substituting this back into the expression~(\ref{eq:p1p2}), we arrive at
\begin{align}
    \langle p(t_1) p(t_2) \rangle &=& \left( \langle p(0)^2 \rangle - m k_B T \right) e^{-\frac{\Gamma}{2m}(t_1+t_2)} \nonumber\\
    &&+ m k_B T e^{-\frac{\Gamma}{2m}(t_2-t_1)}.
\end{align}
We observe that in the limit $t_1=t_2=t\rightarrow\infty$: 
\begin{align}
    \left\langle \frac{p(t)^2}{2m} \right\rangle \rightarrow  \frac{k_B T}{2},
\end{align}
in line with the equipartition theorem.

From Eq.~(\ref{eq:pt1}), we also observe that in long time limit,  $\left\langle {p(t)} \right\rangle \rightarrow  0$. We, thus, make an educated guess that the probability distribution, $ P(p)$, of momentum of the focal particle after equilibration is Gaussian with mean $0$ and variance $mk_BT$, i.e.,
\begin{equation}
    P(p) \sim e^{-p^2/(2mk_BT)}.\label{eq:exp}
\end{equation}
This is actually true because one can show that every \emph{cumulant} of order 3 and higher, calculate using  Eq.~(\ref{eq:pt1}), are exactly zero.


We digress a bit to remark that while the Langevin equation describes the time evolution of the {random variable} $p(t)$, the \emph{Fokker--Planck equation} describes the time evolution of the {probability density distribution} $P(p, t)$ in momentum space. For a general stochastic differential equation of the form:
\begin{equation}
\dot{p}(t) = A(p, t) + B(p, t) f(t)
\end{equation}
with noise correlation $\langle f(t) f(s) \rangle = 2 C \delta(t - s)$, a corresponding Fokker--Planck equation exists. It is
\begin{equation}
\frac{\partial P(p, t)}{\partial t} = -\frac{\partial}{\partial p} \left[ \left( A + C B \frac{\partial B}{\partial p} \right) P \right] + C \frac{\partial^2 (B^2 P)}{\partial p^2}.
\end{equation}
The `drift correction term'---$C B \frac{\partial B}{\partial p}$---vanishes in our case because $B(p, t) = 1$.

Thus, corresponding to Eq.~(\ref{eq:le}), we have
\begin{equation}
\frac{\partial P}{\partial t} = -\frac{\partial}{\partial p} \left( -\frac{\Gamma}{2m} p P \right) + \frac{\Gamma k_B T}{2} \frac{\partial^2 P}{\partial p^2}.
\end{equation}
This Fokker--Planck equation can be rewritten as a continuity equation in momentum space:
\begin{equation}
\frac{\partial P}{\partial t} = -\frac{\partial J}{\partial p},
\end{equation}
where the probability current $J(p, t)$ is defined as:
\begin{equation}
J = -\frac{\Gamma}{2m} p P - \frac{\Gamma k_B T}{2} \frac{\partial P}{\partial p}.
\end{equation}

In equilibrium, the probability distribution must be stationary ($\frac{\partial P}{\partial t} = 0$), and the net global probability current vanishes ($J = 0$):
\begin{equation}
-\frac{\Gamma}{2m} p P - \frac{\Gamma k_B T}{2} \frac{\partial P}{\partial p} = 0 \implies P(p) \sim e^{-p^2/(2mk_BT)}.
\end{equation}
This recovers the expected Maxwell--Boltzmann momentum distribution at thermal equilibrium (recall Eq.~\ref{eq:exp}).

\subsection{Kac Ring Model}
Unlike the Zwanzig model, which uses an infinite continuum of oscillators, the Kac ring model uses a discrete, finite, and deterministic system to illustrate the transition from non-equilibrium to a pseudo-equilibrium--- a state where a system appears completely stable and unchanging on a macroscopic level, but does not satisfy the strict fundamental laws of true thermodynamic equilibrium.

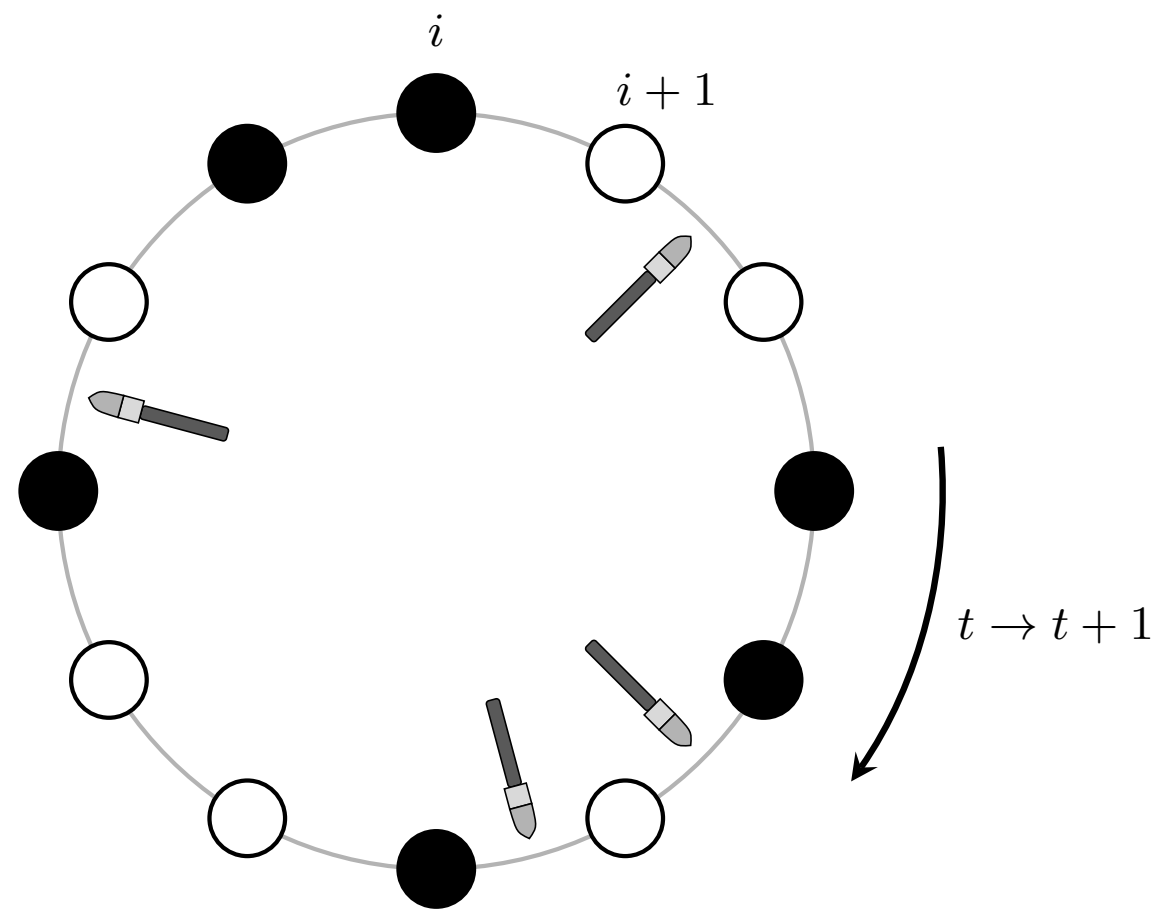
\begin{figure}[h]
\centering
\begin{tikzpicture}[
    ball/.style = {circle, draw=black, thick, minimum size=5mm, inner sep=0pt},
    >=stealth,
    brush/.pic = {
      \filldraw[draw=black, line width=0.3pt, fill=gray!70!black, rounded corners=0.6pt]
            (-0.95,-0.045) rectangle (-0.36,0.045);                 
      \filldraw[draw=black, line width=0.3pt, fill=gray!30]
            (-0.36,-0.075) rectangle (-0.22,0.075);                 
      \filldraw[draw=black, line width=0.3pt, fill=gray!60]
            (-0.22,-0.075) -- (-0.22,0.075)
            .. controls (-0.06,0.07) .. (0,0)
            .. controls (-0.06,-0.07) .. (-0.22,-0.075) -- cycle;   
    }]
  \path (-4.0,0) -- (6.4,0);
  \draw[gray!60, thick] (0,0) circle (2.5);
  \foreach \k in {1,4,5,9}{
    \pic[rotate={90-30*\k-15}] at ({90-30*\k-15}:2.38) {brush};
  }
  \foreach \k/\c in {0/black,1/white,2/white,3/black,4/black,5/white,
                     6/black,7/white,8/white,9/black,10/white,11/black}{
    \node[ball, fill=\c] at ({90-30*\k}:2.5) {};
  }
  \node at (90:3.05) {$i$};
  \node at (60:3.05) {$i+1$};
  \draw[->, very thick] (5:3.35) arc[start angle=5, end angle=-35, radius=3.35];
  \node[right] at (-15:3.45) {$t \to t+1$};
\end{tikzpicture}
\caption{\small The Kac ring with $N=12$ sites and $M=4$ markers (paint brushes)
placed on edges between neighbouring sites. At each time step every ball moves
one site clockwise; a ball crossing a marked edge is repainted
(black $\leftrightarrow$ white).}
\label{fig:kac_ring}
\end{figure}

In the Kac Ring model (see Fig.~\ref{fig:kac_ring}), we have $N$ sites arranged in a circle. Each site $i$ contains a ball that is either black or white. There are $M<N$ fixed markers uniformly, randomly placed on the edges between sites---each edge has at most one marker. As time progresses ($t \to t+1$), every ball moves one step clockwise. If a ball passes a marker, it changes colour (black $\leftrightarrow$ white); otherwise, it retains its colour.

Let $B(t)$ and $W(t)$, respectively, be the total number of black and the total number of white balls at time $t$. We define the macrostate as being characterized by macroscopic variable
\begin{equation}
    \Delta(t) \equiv B(t) - W(t).
\end{equation}
Since the total number of balls is constant, we have the constraint 
\begin{equation}
B(t) + W(t) = N.\label{eq:kN}
\end{equation}

To track the evolution of the macrostate, let $b(t)$ and $w(t)$, respectively, be the number of black balls and the number of white balls in front of a marker at time $t$. Of course, following constraint holds:
\begin{equation}
b(t) + w(t) = M.\label{eq:kM}
\end{equation}
At the next time step:
\begin{itemize}
    \item The number of black balls becomes: $B(t+1) = B(t) - b(t) + w(t)$. This relation essentially tracks the loss of black balls that hit a marker, and gain from white balls hitting a marker.
    \item Similarly, the number of white balls becomes: $W(t+1) = W(t) - w(t) + b(t)$.
\end{itemize}
Consequently,
\begin{align}
    \Delta(t+1) &= B(t+1) - W(t+1)= \Delta(t) - 2[b(t) - w(t)].\label{eq:maev}
\end{align}
Because we are ignorant about the exact marker locations, just by using the information in Eq.~(\ref{eq:kN}) and Eq.~(\ref{eq:kM}), one cannot predict $\Delta(t)$ [equivalently, $W(t)$ or $B(t)$] precisely at all times---it is the standard closure problem.

To do something about it, we apply an analogous version of \emph{Stosszahlansatz}, assuming the markers are uniformly distributed with density $\mu = M/N$ and are uncorrelated with the ball colours. Effectively, our ansatz is that
\begin{equation}
    (b(t),w(t)) = (\mu B(t),  \mu W(t)).
\end{equation}
Substituting these into Eq.~ (\ref{eq:maev}), we get
\begin{align}
    \Delta(t+1) = (1 - 2\mu)\Delta(t).\label{eq:kme}
\end{align}
Iterating this gives the exponential decay 
\begin{align}
\Delta(t) = (1 - 2\mu)^t \Delta(0) \quad\forall t\label{eq:mchaos}
\end{align}
as the solution.

As a side note: Eq.~(\ref{eq:kme}) is like a \emph{(Markovian) Master Equation} because it treats the colour changes as random transitions based on a density $\mu$ to describe the average behaviour of the system; it also has the defining property of being memoryless (Markovian) as it assumes that the probability of the next transition depends only on the current state, not on the history of the system---at every step, the probability of a ball hitting a marker is a constant $\mu$, regardless of whether that ball hit a marker in the previous step.

Fortunately, the Kac model can be solved exactly using microscopic variables. By verifying the predictions made at the macroscopic level against this exact solution, we can clearly understand where our initial `feigning of ignorance' regarding microscopic correlations has led us.
To treat the system exactly, we introduce:
\begin{itemize}
    \item $\eta_i(t) = +1$ (black ball at site $i$ at time $t$) or $-1$ (white ball at site $i$ at time $t$),
    \item $\epsilon_i = -1$ (marker exists between site $i$ and $i+1$) or $+1$ (no marker between site $i$ and $i+1$).
\end{itemize}
In these notations, we may express the macroscopic variable under scrutiny as:
\begin{equation}
\Delta(t)  = \sum_i \eta_i(t),\label{eq:kdel}
\end{equation}
summing over all sites $i$. A complete microstate of the Kac ring system at time $t$ is specified by $\{\eta_1(t), \eta_2(t), \ldots, \eta_N(t)\}$. The total number of possible microstates is $2^N$. It is important to note that the marker configuration $\{\epsilon_1, \epsilon_2, \ldots, \epsilon_N\}$ is not part of a microstate. Instead, the markers constitute the \emph{quenched disorder} (fixed external structure) of the system.

The transformation for a single ball moving from $i$ to $i+1$ is \begin{equation}\eta_{i+1}(t+1) = \epsilon_i \eta_i(t).\label{eq:kacre}\end{equation} 
We can trace the state of site $i$ at time $t$ back to the initial state at $t=0$ through the following intermediate iterations:
\begin{itemize}
    \item For $t=1$: $\eta_i(1) = \epsilon_{i-1} \eta_{i-1}(0)$.
    \item For $t=2$: $\eta_i(2) = \epsilon_{i-1} \eta_{i-1}(1) = \epsilon_{i-1} \epsilon_{i-2} \eta_{i-2}(0)$.
    \item By mathematical induction: \begin{equation}\eta_i(t) = \left( \prod_{k=1}^{t} \epsilon_{i-k} \right) \eta_{i-t}(0).\end{equation}
\end{itemize}
Note we only know that $M$ markers are uniformly distributed, we do not know their exact locations---many marker configurations are possible. Therefore, taking the average in Eq.~(\ref{eq:kdel}) over such different configurations and microstates, while assuming that the initial configuration of balls $\{\eta(0)\}$ is independent of the distribution of markers $\{\epsilon\}$, yields
\begin{equation}
    \langle \Delta(t) \rangle = \left\langle \sum_i \left( \prod_{k=1}^{t} \epsilon_{i-k} \right) \eta_{i-t}(0) \right\rangle = \langle \epsilon_1 \epsilon_2 \dots \epsilon_t \rangle    \langle \Delta(0) \rangle. \label{eq:savdel}
\end{equation}
The last equality is result of the fact that the markers being placed uniformly, no single site is special: the probability of having a marker at site $i-t$ is the same as the probability at site $t$.The term $\langle \epsilon_1 \dots \epsilon_t \rangle$ represents the average value of the product of $t$ marker variables.

Before proceeding further, let us clarify the averaging process above. The notation $\langle \cdots \rangle$ in Eq.~(\ref{eq:kdel}) actually conceals two fundamentally different averaging procedures---(i) standard microcanonical ensemble average over ball configurations (at fixed markers) and (ii) disorder average over marker configurations (at fixed initial balls). The former means that we divide the sum of all $\Delta$s by total number of possible microstates ($2^N$) and the latter averaging process is denoted by ${\mathbb {E}}_{\{\epsilon\}}[\cdots]$. Therefore,
\begin{equation}
    \langle \Delta(t) \rangle \equiv \mathbb{E}_{\{\epsilon\}} \left[ \frac{1}{2^N} \sum_{\{\eta(0)\}} \Delta(t;\{\eta(0)\}, \{\epsilon\}) \right].
\end{equation}
This is justified by the aforementioned {independence} assumption: the initial ball configuration $\{\eta(0)\}$ is generated independently of the marker placement $\{\epsilon\}$.

Now, first we consider the case where $t<N$. The probability of encountering exactly $j$ markers in $t$ steps is given by the binomial distribution ${}^t C_j \mu^j (1-\mu)^{t-j}$. Thus, we have
\begin{equation}
    \langle \epsilon_1 \dots \epsilon_t \rangle = \sum_{j=0}^{t} {}^t C_j (-1)^j \mu^j (1 - \mu)^{t-j}.
\end{equation}
Note that the term \((-1)^j\) represents the net sign change or the total mathematical value of the product \(\epsilon_1 \dots \epsilon_t\) when a particle encounters exactly \(j\) markers over \(t\) steps. (Because the particles in the Kac ring model do not interact with each other, we can solve the entire system by looking at just one representative particle at a time.) Using the Binomial Theorem $(a + b)^t = \sum {}^t C_j a^j b^{t-j}$ where $a = -\mu$ and $b = (1-\mu)$:
\begin{equation} 
    \langle \epsilon_1 \dots \epsilon_t \rangle = [-\mu + (1 - \mu)]^t = (1 - 2\mu)^t,\label{eq:mutn}
\end{equation}
which implies for Eq.~(\ref{eq:savdel}) that
\begin{equation}
    \langle \Delta(t) \rangle = (1 - 2\mu)^t   \langle \Delta(0) \rangle \quad  (0< t<N). \label{eq:savdel}
\end{equation}
This confirms that,  for $t < N$, the statistical average matches solution of the earlier `Master Equation' derivation [see Eq.~(\ref{eq:kme})].

Next, for the time range $N < t < 2N$, we have to be careful as repetitions of the microstate comes into play. Letting $ t = N + s$ ($0 < s < N$), the average over the product of $\epsilon$s in Eq.~(\ref{eq:savdel}) can be recast as:
\begin{equation}
\langle \mathcal{\epsilon}_1, \dots, \mathcal{\epsilon}_{N+s} \rangle = \left\langle \frac{[\mathcal{\epsilon}_1\dots\mathcal{\epsilon}_{N+s}]^2 [\mathcal{\epsilon}_{N+s+1} \dots\mathcal{\epsilon}_{2N}]}{[\mathcal{\epsilon}_1\dots\mathcal{\epsilon}_{N+s}] [\mathcal{\epsilon}_{N+s+1} \dots\mathcal{\epsilon}_{2N}]} \right\rangle.
\end{equation}
The square term in the numerator must be identically 1 and so is the denominator because every marker-variable appears twice. Therefore, we obtain
\begin{eqnarray}
\langle \mathcal{\epsilon}_1\dots \mathcal{\epsilon}_{N+s} \rangle &=& \langle \mathcal{\epsilon}_{N+s+1} \dots \mathcal{\epsilon}_{2N} \rangle \nonumber\\
&=&\langle \mathcal{\epsilon}_{N+s+1-t} \dots \mathcal{\epsilon}_{2N-t} \rangle\rangle \nonumber\\
&=& \langle \mathcal{\epsilon}_1 \dots \mathcal{\epsilon}_{2N-t} \rangle.
\end{eqnarray}
In the last step, we have used the equality: $N-s=2N-t$. $\langle \mathcal{\epsilon}_1 \dots \mathcal{\epsilon}_{2N-t} \rangle$ can now be evaluated exactly as in Eq.~(\ref{eq:mutn}) since $2N-t<N$. Consequently, we have
\begin{equation}
 \langle \Delta(t) \rangle = (1 - 2\mu)^{2N-t} \langle \Delta(0) \rangle \quad (N<t<2N).\label{eq:N2N}
\end{equation}

Finally, when $t = kN$ where $k$ is a positive integer, each ball completes exactly $k$ full circuits around the ring. We need to consider Eq.~(\ref{eq:savdel}) in the following light. Since the ring has period $N$,    $\langle \epsilon_1 \cdots \epsilon_{kN} \rangle = \langle (\epsilon_1 \cdots \epsilon_N)^k \rangle = \langle [(-1)^M]^k \rangle = (-1)^{kM}$ implying
\begin{equation}
    \langle \Delta(t) \rangle = (-1)^{kM} \langle \Delta(0) \rangle\quad (t=kN).\label{eq:0kN}
\end{equation}
Therefore, depending on whether $M$ is even or odd, the recurrence period ($\tau_{\rm rec}$) is $N$ or $2N$.

Comparing the coarse-grained solution via Stosszahlansatz in Eq.~(\ref{eq:mchaos}) with the exact solutions in Eq.~(\ref{eq:savdel}), Eq.~(\ref{eq:N2N}), and Eq.~(\ref{eq:0kN}), we arrive at the following key conclusions:

\begin{enumerate}
\item The macroscopic variable $\Delta$ in Eq.~(\ref{eq:mchaos}) should be recognized as an ensemble averaged quantity.
\item For times $t \ll N$, the \textit{Stosszahlansatz} prediction agrees perfectly with the exact solution. This validates the coarse-graining approach under two primary conditions: first, when balls sample only a small fraction of the ring (ensuring the system is far from Poincar\'e recurrence); and second, when correlations between ball colours and marker positions remain negligible (maintaining statistical independence). Under these conditions, the system exhibits an apparent macroscopic relaxation toward a statistical `equilibrium'---a configuration where the number of black and white balls is roughly equal---behaving as if it were irreversible on short time scales. The relaxation timescale (the time at which the system decays to $1/e$ of its initial value) is given by $\tau_{\text{rel}} \sim 1/(2\mu) = N/(2M)$. For $\mu \ll 1$ (representing sparse markers), this creates a clear temporal separation between the regime of apparent equilibration and the eventual recurrence of the initial state.
\item At integral multiples of $N$, the Stosszahlansatz completely fails. It predicts $\langle \Delta(0) \rangle \to 0$ as $N \to \infty$ which is in contrast with the exact result's prediction---$\langle \Delta(0) \rangle = \mathcal{O}(\langle \Delta(0) \rangle)$. The discrepancy arises because balls traverse the {exact same} marker configuration $k$ times, creating perfect correlations that violate the Stosszahlansatz assumption.

\item For $N < t < 2N$, the exact solution exhibits \emph{anti-relaxation}:
\begin{equation}
    \langle \Delta(t) \rangle \to \langle \Delta(0) \rangle \quad \text{as } t \to 2N^-.
\end{equation}
The system spontaneously `un-relaxes', demonstrating Poincar\'e recurrence. This violates the apparent irreversibility suggested by the master equation, revealing the fundamentally reversible microscopic dynamics.

\item The Kac ring illustrates a crucial distinction between pseudo-equilibrium vs. true thermodynamic equilibrium:
\begin{itemize}
    \item {Pseudo-equilibrium:} A macroscopically stationary state that appears stable for $\tau_{\text{rel}} \ll t \ll \tau_{\text{rec}}$ but eventually exhibits recurrence.
    \item {True thermodynamic equilibrium:} Requires the thermodynamic limit ($N \to \infty$) where $t_{\text{rec}} \to \infty$, making recurrence physically irrelevant.
\end{itemize}
For finite $N$, the Kac model never reaches true equilibrium, only a pseudo-equilibrium that persists for exponentially long (but finite) times.

\item The model is fundamentally {non-ergodic}: Each trajectory visits at most $\tau_{\rm rec}$ (either $N$ or $2N$) distinct microstates even though the total number of accessible microstates is $2^N$ (each ball can be black or white). Therefore, the phase space exploration is incomplete: no trajectory samples all possible microstates. Naturally, time averages and ensemble averages are {not} equivalent.
This demonstrates that ergodicity is {not necessary} for statistical mechanics to work---typicality and large $N$ are sufficient.
\item Despite non-ergodicity, the model exhibits typical behaviour through concentration of measure. One can compute the variance:
\begin{equation}
    \langle \Delta(t)^2 \rangle - \langle \Delta(t) \rangle^2 \sim N.
\end{equation}
This implies $\sigma \sim \sqrt{N}$, so the relative fluctuations are:
\begin{equation}
    \frac{\sigma}{\Delta_{\max}} \sim \frac{\sqrt{N}}{N} = \frac{1}{\sqrt{N}} \to 0 \quad \text{as } N \to \infty.
\end{equation}
The distribution becomes sharply peaked around the mean, making $\langle \Delta(t) \rangle$ a \emph{typical} value for almost all microscopic realizations. This is typicality in action: most microstates yield similar macroscopic behaviour.
\item The microscopic dynamics are completely reversible:  The transformation $\eta_{i+1}(t+1) = \epsilon_i \eta_i(t)$ is deterministic and invertible with perfect recurrence at $t = t_{\text{rec}}$, demonstrating time-reversibility. Yet the {macroscopic} description appears irreversible for $t \ll t_{\text{rec}}$. This irreversibility is not fundamental but emerges from coarse-graining (tracking only $\Delta(t)$, not individual ball positions) and ignoring microscopic correlations. This is the resolution of Zermelo's paradox in this model.
\item The Kac ring model elegantly resolves Loschmidt's paradox by demonstrating how macroscopic irreversibility emerges from perfectly time-reversible microscopic dynamics. The reversibility can be explicitly seen by multiplying Eq.~\eqref{eq:kacre} with $\epsilon_i$ to get   backward-in-time evolution equation: $ \eta_i(t) = \epsilon_i \eta_{i+1}(t+1)$, since $\epsilon_i^2=1$. Thus, knowing the complete microstate at time $t+1$ and the quenched marker configuration $\{\epsilon\}$ allows one to compute the exact microstate at time $t$ without any loss of information: Microscopically, stepping backward in time obeys the exact same dynamical rules as stepping forward. The apparent irreversible decay toward equilibrium is an artefact of coarse-graining and the Stosszahlansatz, which assume statistical independence and ignore the continuous buildup of hidden microscopic correlations between the balls and the markers. If the dynamics were reversed, the system would only ``un-relax'' to its initial state if these exact, delicate correlations were perfectly preserved; failing to track them---as macroscopic observation inevitably does---breaks the time-symmetry and produces a statistical arrow of time.
\end{enumerate}

\begin{center}
	$\symking\symqueen\symrook\symknight\symbishop\sympawn\epsdice{1}\,\epsdice{2}\,\epsdice{3}\,\epsdice{4}\,\epsdice{5}\,\epsdice{6}\,\spadesuit\heartsuit\clubsuit\diamondsuit$
\end{center}
\newpage
\newpage
\clearpage
\onecolumngrid
\section{Language of Chance: A Crash Course}
\epigraph{\textsf{To us, probability is the very guide of life.}}{Joseph Butler}
\twocolumngrid

\indent {\color{deepsea}{\LARGE \textgoth{T}}}he history of statistical mechanics is defined by the struggle to reconcile the time-reversal symmetry of microscopic laws with the stubborn irreversibility of the macroscopic world. Loschmidt's reversibility objection and Zermelo's recurrence paradox highlight the logical gaps in a purely deterministic framework, yet it is through the Kac ring model that we see how \emph{statistical} trends toward equilibrium can emerge from discrete, reversible steps. Further refinement through the Zwanzig harmonic bath model demonstrates that when a system is coupled to a vast environment, the resulting dissipation and loss of memory necessitate a transition from classical trajectories to a \emph{probabilistic} description. These models collectively prove that the bridge between the individual and the ensemble is built upon the mathematics of \emph{chance}. We must have realized by now that survival in this nonequilibrium complex world is an adventure born of our own \emph{ignorance}, where every realistic system is shrouded in uncertainty. To gain any true knowledge in such a landscape, we must move beyond the illusion of certainty and learn to characterize, quantify, and interpret these uncertainties through the rigorous `language of chance' comprising the tools of probability, random variables, and stochastic processes.

\subsection{Probability}
A probabilistic system is mathematically described by a triple $(\mathcal{S},\mathcal{E},P)$: $\mathcal{S}$ is the \emph{sample space}---set of all possible outcomes, $\mathcal{E}$ is the \emph{event space}---set of all possible (measurable) sets of outcomes, and $P:\mathcal{E}\rightarrow[0,1]$ is the \emph{probability measure or probability function} that obeys \emph{Kolmogorov's axioms}:
\begin{enumerate}
\item \emph{Normalization:} $P(\mathcal{S})=1$.
\item \emph{Non-negativity:} $P(E)\ge0$ $\forall E\in\mathcal{E}$.
\item \emph{Finite additivity:} $P(\cup_{i=1}^nE_i)=\sum_{i=1}^nP(E_i)$ for any finite sequence ($\{E_i\}_{i=1}^n$) of mutually disjoint events.
\end{enumerate}
For an infinite sample space, a fourth axiom---\emph{Continuity at zero}---needs to be added: If $E_1\supseteq E_2\supseteq E_3\supseteq\cdots$ is a decreasing sequence of events that tends to empty set, then $\lim_{i\rightarrow\infty}P(E_i)=0$. Equivalently, rather than adding the fourth axiom, the third axiom may be modified for an infinite sample space; it becomes: (\emph{Countable additivity}) $P(\cup_{i=1}^\infty E_i)=\sum_{i=1}^\infty P(E_i)$ for any infinite sequence, $\{E_i\}_{i=1}^\infty$, of mutually disjoint events. By taking almost all events as empty set, the countable additivity can be seen to imply finite additivity; and the `continuity at zero' axiom can be obtained from the countable additivity by choosing events, $E_i=F_{i-1}\backslash F_i$ where $F_i\equiv \cup_{k=1}^{\infty}E_k\backslash \cup_{k=1}^{i}E_k$ ($\{F_i\}$ is a decreasing sequence of events approaching empty set). 

Observe that if $\mathcal{S}$ is countably infinite (or finite) and $\mathcal{E}$ is taken to be the power set $2^\mathcal{S}$, then $\mathcal{E}$ satisfies the following properties:
\begin{enumerate}
\item  $\mathcal{S}\in\mathcal{E}$.
\item  If $E\in\mathcal{E}$, then $E^c\in\mathcal{E}$.
\item If $E_i\in\mathcal{E}\,\, \forall i\in\{1,2,\cdots,\infty\}$, then $\cup_{i=1}^\infty E_i\in\mathcal{E}$ (or, for finite $\mathcal{S}$: if $E_i\in\mathcal{E}\,\, \forall i\in\{1,2,\cdots,n\}$, then $\cup_{i=1}^nE_i\in\mathcal{E}$). 
\end{enumerate}
Note that it follows that the impossible event $\emptyset\in\mathcal{E}$ (for which $P(\emptyset)=0$) and $\cap_{i=1}^\infty E_i\in\mathcal{E}$ (or, for finite $\mathcal{S}$, $\cap_{i=1}^nE_i\in\mathcal{E}$). In technical jargon, $2^\mathcal{S}$ is a \emph{sigma algebra} (or \emph{algebra} if $\mathcal{S}$ is finite; all sigma algebras are algebras). In simpler words, a sigma algebra is just a set of sets that is closed under complementation (denoted by superscript `$c$'), countable union, and countable intersection.

Interestingly, there may exist smaller sigma algebra $\mathcal{E}'\subset 2^\mathcal{S}$ for countable sample space: e.g., for the sample space $\mathcal{S}=\{1,2,3,4\}$, $\mathcal{E}'=\{\emptyset,\{1,3\},\{2,4\},\mathcal{S}\}$ is a sigma algebra. However, it doesn't make sense to work with such smaller non-power sets in the probability theory as they do not exhaust all possible events.

Problem arises when one deals with uncountably infinite sample space: There are events with which one cannot associate probability measure. A classic example is that of the Vitali set. How should probability space be defined then? In this case, smaller non-power sets come to rescue. If the event space $\mathcal{E}$ has to be taken as the set of measurable sets of outcomes, we should take the one that is a sigma algebra even though it may not be the power set. For example, Borel sigma algebra on the sample space $[0,1]\subset\mathbb{R}$ is a set of all the open intervals $(a,b)$ with $0\le a\le b\le 1$, all their complements, all their finite or countably infinite unions, and all complements of these unions.

The probabilistic system has to be further supplemented with the notion of \emph{conditional probability}---$P(E_i|E_j)$ for any two events in the event space---that describes the probability of event $E_i$ given $E_j$ has occurred; there need not be any causal connection between the two events. It can either be introduced as a definition in the restricted sample space:
\begin{equation}
P(E_i|E_j)=\frac{P(E_i\cap E_j)}{P(E_j)}
\end{equation}
(termed \emph{ratio formula}) with $P(E_j)\ne0$ or as an independent axiom for the conjunction:
\begin{equation}\label{eq:indt}
{P(E_i\cap E_j)}=P(E_i|E_j){P(E_j)},
\end{equation}
where $E_i$ and $E_j$ need not occur at the same time.

Naturally, $E_i$ and $E_j$ can be said to be \emph{independent} if 
\begin{equation}
P(E_i|E_j)={P(E_i)};
\end{equation}
i.e., in the light of Eq.~(\ref{eq:indt}), if
\begin{equation}
{P(E_i\cap E_j)}=P(E_i){P(E_j)}.
\end{equation}
A related concept is that of \emph{conditional independence}: Two events, say $E_i$ and $E_j$, are known as conditionally independent given an event $E_k$ if 
\begin{equation}
P(E_i\cap E_j|E_k)=P(E_i|E_k)P(E_j|E_k),
\end{equation}
or equivalently, if
\begin{equation}
P(E_i|E_j,E_k)={P(E_i|E_k)}.
\end{equation}
Here, $P(E_i|E_j,E_k)$ reads as probability of $E_i$ given both $E_j$ and $E_k$. It is neither necessary nor sufficient condition for independence between two events $E_i$ and $E_j$. 

It is easy to see that a relation---known as \emph{Bayes' theorem}---trivially follows from the definition of the conditional probability:
\begin{eqnarray}
&&P(E_i|E_j)=\frac{P(E_j|E_i){P(E_i)}}{P(E_j)}\nonumber\\
\implies&&P(E_i|E_j)=\frac{P(E_j|E_i){P(E_i)}}{P(E_j|E_i){P(E_i)}+P(E_j|E^c_i){P(E^c_i)}}.\qquad
\end{eqnarray}
Note that the expanded form of denominator as been in the last expression is an application of what is known as the \emph{law of total probability}: If $\{E_1,E_2,\cdots,E_n\}$ is a partition of a sample space, then any event $E$ satisfies the relation, $P(E)=\sum_{i=1}^nP(E\cap E_i)$. Here \emph{partition} means a set of non-overlapping, non-empty subsets of the sample space such that their union is the sample space itself.

Finally, we must be aware that what is meant by the word `probability' is not unanimously agreed upon. Broadly, there are three formulations of probability: \emph{formalist}, \emph{subjectivist}, and \emph{objectivist}. Although all three adopt Kolmogorov's three axioms and the ratio formula, probability carries distinctly different interpretations across these views.
\begin{enumerate}
\item From the formalist viewpoint, probability is nothing more than a notation used within an axiomatic structure. One can use consistent syntax to pose and prove various propositions unambiguously while remaining entirely within the paradigm of that structure. Probability exists independently of the physical world---living or non-living. Beyond the pure mathematical interest of generating consistent abstract structures, it serves as a conceptual framework that can be applied in practice to model real-world phenomena.
\item The subjectivist viewpoint is inherently personal. It does not require events to be repeatable; an individual can assign a probability to any single event and communicate it numerically. Naturally, for the exact same event, the assigned probability may differ between observers depending on their knowledge of the system's state. Crucially, this implies that randomness is not an intrinsic property of the event, but rather a reflection of the observer's state of knowledge. A subjectivist dealing with uncertainty seeks to make logically coherent inferences.
\item An objectivist view of probability holds that probabilities are real, objective facts about the world or logic, existing independently of human belief, bias, or personal opinion. Unlike subjectivism---which treats probability as an observer's personal degree of belief---objectivism asserts that for any well-defined event or condition, there is a single, uniquely correct probability value. Note, however, that in this formulation, probability itself is not directly observable.
\end{enumerate}
\emph{Frequentism}---the most common objective formulation in physics---is \emph{empirical}: probabilities are defined in terms of actual physical events by calculating an event's relative frequency of occurrence. While observed frequencies naturally fluctuate depending on the number of trials, a stronger variation---\emph{hypothetical frequentism}---defines probability as the limiting frequency. This approach faces severe theoretical hurdles: the domain over which the limit frequency is defined may fail to form an algebra of sets (a field), thereby violating countable additivity. Furthermore, the limiting relative frequency can depend on the ordering of trials and may not mathematically exist. Last but not least, frequentism suffers from the \emph{reference class problem}, yielding only conditional probabilities. That is, the same event can have multiple probabilities depending on the reference class to which it is assigned (e.g., asking "What is the probability that a cow will die at age 15?" depends entirely on the specific class of cows being considered).

\subsection{Random Variable}
A single-valued function $X(s)$ is called a \emph{random variable} (actually a `function' in traditional sense!) if it maps each outcome $s$ of sample space $\mathcal{S}$ to a real number; two distinct outcomes may be mapped to same real number but not vice versa. One can then use the random variable to define events. 

\emph{Cumulative distribution function} (cdf) $F_X(x)$ of $X$ is defined as
\begin{eqnarray}
F_X(x)=P(X\le x),\, x\in(-\infty,+\infty).
\end{eqnarray}
If a cdf changes values only in a countable number of jumps and is otherwise constant between two subsequent jumps, the corresponding $X$ is a \emph{discrete random variable}. Essentially, a discrete random variable is the one for which $\{X(s): s\in\mathcal{S}\}$ is finite or countably infinite. On the other hand, a random variable is \emph{continuous} if $\{X(s): s\in\mathcal{S}\}$ is either a single interval or a set of disjoint intervals of real line. Note that for a continuous random variable, $F_X(x)$ is continuous and its piecewise continuous derivative exists everywhere except maybe at a finite number of points. One can construct a \emph{mixed} random variable that has a cdf with properties of both the discrete and the continuous random variables. The notion of \emph{percentile} is defined via cdf: The $u\,(\in[0,1])$ percentile of $X$ is the smallest real number $x_u$ that satisfies $u=F_X(x_u)$.

For the discrete random variable, one can define \emph{probability mass function} (pmf):
\begin{eqnarray}
p_X(x)=P(X=x),
\end{eqnarray}
where
\begin{eqnarray}
P(X=x_i)=F_X(x_i)-F_X(x_{i-1}).
\end{eqnarray}
Here subscript $i$ denotes the points of jumps for the cdf.

Equivalently for the continuous random variable, one can define \emph{probability density function} (pdf):
\begin{eqnarray}
p_X(x)=\frac{d}{dx}F_X(x).
\end{eqnarray}
One can invert this to get,
\begin{eqnarray}
F_X(x)=P(X\le x)=\int_{-\infty}^{x}p_X(x)dx.
\end{eqnarray}
Although we are using the same symbol for pmf and pdf, it should not be a source of confusion as the context will make it clear.

It is straightforward to define \emph{conditional cdf} of an event given some event $E$:
\begin{eqnarray}
F(x|E)=P(X\le x|E)=\frac{P(\{s:X(s)\le x\}\cap E)}{P(E)}.\quad
\end{eqnarray}
Using this definition, the corresponding definitions for \emph{conditional pmf} or \emph{conditional pdf} follow in analogy with what was done before for the unconditional cases.

An all important concept in the theory of random variable is that of the functions of random variables. Consider a function $f(x)$. Notationally, 
\begin{equation}
Y=f(X)
\end{equation}
defines a new random variable such that for any subset $\mathcal{I}$ with $f(x)\le y$ and $\mathcal{I}\subseteq({\rm range~of~}X)\subseteq\mathbb{R}$ ,
\begin{equation}
\{s:Y=f(X(s))\le y\}=\{s:X(s)\in \mathcal{I}\}.
\end{equation}
Naturally, for continuous random variable,
\begin{eqnarray}
F_Y(y)=P(Y\le y)=P(X\in \mathcal{I})=\int_\mathcal{I}p_X(x)dx\,.\quad
\end{eqnarray}
The pdf of $Y$ can then be obtained through,
\begin{equation}\label{eq:conversionp}
p_Y(y)=\sum_i\frac{p_X(x_i)}{|f'(x_i)|}\,,
\end{equation}
where $x_i$'s are the real roots of $y=f(x)$.
In passing, we note that in case of the discrete random variable the pmf $p_X(x)$ under transformation $y=f(x)$ rather trivially implies
\begin{equation}
p_Y(y)=\sum_{x\in f^{-1}(y)}p_X(x).
\end{equation}

The \emph{expectation} of a function of a random variable is given by
\begin{equation}
\mathbb{E}[Y]=\mathbb{E}[f(X)]=\sum_if(x_i)p_X(x_i)\,{\rm and}\,\int_{-\infty}^{+\infty}f(x)p_X(x)dx,
\label{eq:expectation}
\end{equation}
for discrete and continuous cases, respectively. Expectations of the following three particular functions are useful:
\begin{enumerate}
\item \emph{Moment generating function}:
\begin{equation}
M_X(t)\equiv \mathbb{E}[e^{tX}],\,t\in\mathbb{R}.
\end{equation}
$M_X(t)$ need not exist as the integral or the sum may not converge for all $t$. The beauty of $M_X(t)$ is that $n$th \emph{moment} $m_n$ ($n\in\mathbb{N}$)---defined as $E(X^n)$---can be obtained from it through
\begin{equation}
m_n\equiv \mathbb{E}[X^n]=\left.\frac{d^n}{dt^n}M_X(t)\right|_{t=0}.
\end{equation}
An important result is that if two random variables have same moment generating functions, then they necessarily have same probability distribution. Sometimes it is convenient to work with \emph{central moments} $\mu_n$ ($n=2,3,\cdots$) defined as
\begin{equation}
\mu_n\equiv \mathbb{E}[(X-m_1)^n],
\end{equation}
that measures deviation of the random variables from the \emph{mean} $m_1$ (mostly commonly denoted by $\mu$ with no subscript; hence no confusion with the central moments). Mean measures the \emph{central tendency} of a probability distribution, i.e., it is a measure that locates where the data are concentrated. Two other widely used measures of the central tendency are: \emph{median} and \emph{mode}. Median is value of the random variable such that the probabilities of getting smaller and greater values are equal; and mode is the most probable value of the probability distribution. Note that median is $0.5$ percentile of $X$.
\item \emph{Characteristic function}:
\begin{equation}
\phi_X(t)\equiv \mathbb{E}[e^{\iota tX}],\,t\in\mathbb{R},\quad (\iota\equiv\sqrt{-1}).
\end{equation}
Unlike $M_X(t)$, $\phi_X(t)$ always exists. One can prove that if two random variables have same characteristic functions, then they necessarily have same probability distribution.
\item \emph{Cumulant generating function}: The cumulant generating function
\begin{equation}
K_X(t)\equiv\ln \mathbb{E}[e^{tX}]=\sum_{n=1}^\infty\kappa_n\frac{t^n}{n!},\,t\in\mathbb{R},
\end{equation}
yields \emph{cumulants} $\kappa_n$ through the following relation:
\begin{equation}
\kappa_n=\left.\frac{d^n}{dt^n}K_X(t)\right|_{t=0}.
\end{equation}
The cumulants are related to moments and central moments. In fact, the first cumulant $\kappa_1$ is the mean ($\mu$); and the next two, $\kappa_2$ and $\kappa_3$, are respectively equal to the central moments $\mu_2$ (usually denoted by $\sigma^2$, commonly known as the \emph{variance}) and $\mu_3$ (whose scaled form---$\mu_3/\sigma^3$---is called \emph{skewness}). Square root of the variance is called \emph{standard deviation} and is denoted as $\sigma$, or $\sigma_X$, if the corresponding random variable needs to be emphasized. Positive skewness means the probability distribution is skewed towards right and negative means the distribution is skewed towards left. The fourth cumulant $\kappa_4$ is not equal to any central moment but is related to $\mu_4$ as follows: $\kappa_4=\mu_4-3\sigma^4$. The scaled version of $\kappa_4$ is known as \emph{kurtosis} $\kappa\equiv\kappa_4/\sigma^4$. A positive kurtosis makes the distribution \emph{leptokurtic} meaning that the distribution's tail is heavier than in a \emph{Gaussian} (also called \emph{normal}) distribution. For Gaussian distribution, notationally,
\begin{equation}
\phantom{xxxx}X \sim \mathcal{N}(\mu, \sigma^2)\implies p_X(x) = \frac{1}{\sqrt{2\pi\sigma^2}}e^{-\frac{(x-\mu)^2}{2\sigma^2}}.\label{eq:gdnd}
\end{equation}
 Whereas a negative value of kurtosis indicates that distribution is \emph{platykurtic} meaning that the tail is lighter than in a Gaussian distribution. (Near zero value for kurtosis means that the distribution is \emph{mesokurtic}. Interestingly, the ``lightness'' or ``heaviness'' in these Greek names do not refer to the tails; rather they refer to the region around the peak: Because of the normalization of probability distribution, while a heavier tail implies a lighter weight in the peak region, a lighter tail implies more weight around the peak region.)
\end{enumerate}

Let us exemplify the above three functions using Gaussian distribution. First, consider $Y \sim \mathcal{N}(0,1)$---\emph{standard} Gaussian/normal distribution. For this
\begin{align}
\mathbb{E}[e^{\sigma t Y}] &= \int_{-\infty}^{\infty} e^{\sigma t y} \cdot \frac{1}{\sqrt{2\pi}} e^{-y^2/2} \, dy = e^{\sigma^2 t^2/2}.\label{eq:mgfg0}
\end{align}
Since $X = \mu + \sigma Y$,
\begin{align}
M_X(t) = \mathbb{E}[e^{t(\mu + \sigma Y)}]= e^{\mu t} \mathbb{E}[e^{\sigma t Y}]= \exp\left(\mu t + \frac{\sigma^2 t^2}{2}\right).\label{eq:mgfg}
\end{align}
Similarly, since for standard normal $Y$:
\begin{align}
\mathbb{E}[e^{\iota\sigma t Y}] = \int_{-\infty}^{\infty} e^{\iota\sigma t y} \cdot \frac{1}{\sqrt{2\pi}} e^{-y^2/2} \, dy= e^{-\sigma^2 t^2/2},
\end{align}
we have
\begin{equation}
\phi_X(t) = \exp\left(\iota\mu t - \frac{\sigma^2 t^2}{2}\right).
\end{equation}
Finally, from Eq.~(\ref{eq:mgfg}), we get
\begin{align}
K_X(t) = \ln M_X(t) &= \ln\left[\exp\left(\mu t + \frac{\sigma^2 t^2}{2}\right)\right]= \mu t + \frac{\sigma^2 t^2}{2}.
\end{align}
The cumulants are $\kappa_1 = \mu$, $\kappa_2 = \sigma^2$, and $\kappa_n = 0$ for $n \geq 3$, confirming that the Gaussian distribution $\mathcal{N}(\mu,\sigma^2)$ has only two non-zero cumulants.

Let us mention three extremely important practical results concerning sequences of \emph{i.i.d.} (independent, identically distributed) random variables. While the mathematically precise meaning of the term i.i.d. is defined later but the idea should be clear from the following intention that we have: Our intention is to say something concrete about a sample that contains a sequence ${X_1,X_2,\cdots,X_N}$ (called \emph{sample vector} or \emph{random sample};  to be denoted as $\{X_i\}_{i=1}^{i=N}$ or simply $\{X_i\}$) of $N$ observations for a particular experimental setup or phenomenon. Because we are interested in a particular fixed experimental setup, it is natural to assume that all the observations are modelled by identical random variables. Furthermore, assume that each observation is randomly generated independent of the other observations. One can define the \emph{sample mean}, $\bar{X}_N\equiv(\sum_i^NX_i)/N$, for the sample. If many such samples are collected, what is the statistical behaviour of the sample mean?
\begin{enumerate}
\item \emph{(Strong) Law of large numbers}: If each of the i.i.d. random variables have a finite mean $\mu$, then for any positive $\epsilon$,
\begin{equation}
P\left(\lim_{N\rightarrow\infty}|\bar{X}_N-\mu|>\epsilon\right)=0.
\end{equation}
It essentially says that the sequence of sample means converges to the mean of random variable $X_i$ with full certainty. The weak law---$\lim_{N\rightarrow\infty}P\left(|\bar{X}_N-\mu|>\epsilon\right)=0$---can be inferred from the strong law. If $\mathbb{E}[X_i]=\infty$, the strong law does not hold but using the idea of conditionally convergent series, sometimes a finite expectation value may be defined and the weak law may be seen to be obeyed with convergence towards the redefined expectation value.
\item \emph{Central limit theorem} (CLT): If each of the i.i.d. random variables have a finite mean $\mu$ and a finite variance $\sigma^2$, then 
\begin{equation}
\lim_{N\rightarrow\infty}\left(\frac{\bar{X}_N-\mu}{\sigma/\sqrt{N}}\right)\xrightarrow[]{\text{distribution}}\mathcal{N}(0,1),
\end{equation}
i.e.,  asymptotic distribution of the random variable, $\bar{X}_N$, is Gaussian with mean $\mu$ and variance $\sigma^2/N$.

The CLT provides more detailed description about the sample mean than the laws of large numbers. It allows to approximately find a probability of $P(\bar{X}_N>x)$ because we know that we just have to use $\mathcal{N}(0,1)$. However, such an approximation by the CLT will not be very accurate in the tails---where $x$ is quite distant from $\mu$---if $N$ is not large enough. Moreover, the CLT does not furnish any detail about the convergence of the tail probabilities with the sample size. This is where a theorem from the theory of large deviations comes to the rescue.
\item \emph{Cram\'er's theorem}: If each of the i.i.d. random variables have a finite mean $\mu$ and a cumulant generating function $K_{X_i}(t)=\ln \mathbb{E}[e^{tX_i}]$, then $\bar{X}_N$ satisfies the \emph{large deviation principle}, i.e.,
\begin{equation}
\lim_{N\rightarrow\infty}\left[-\frac{1}{N}\ln P(\bar{X}_N\ge x)\right]=I(x)\,\,\forall x>\mu,
\end{equation}
where $I(x)$---the Legendre--Fenchel transform of $K_X(t)$, i.e., $I(x)\equiv\sup_t(tx-K_{X_i}(t))$---is called \emph{rate function} or \emph{Cram\'er's function}. Sometimes, this is less rigorously presented as 
\begin{equation}
P(\bar{X}_N\ge x)\approx e^{-NI(x)},
\end{equation}
where `$\approx$' sign means that only the dominant exponential term has been kept and any sub-exponential terms have been ignored (hence the limit in the rigorous definition).
\end{enumerate}

Readers must have already realized a technical point: $\bar{X}_N$ is defined on combined sample space---$\mathcal{S}^N\equiv\mathcal{S}_1\times \mathcal{S}_2\times\cdots\times\mathcal{S}_N$---where $\mathcal{S}_i$ is the sample space corresponding to the random variable $X_i$. Thus, by independent random variable we mean that $X_i(s_1,s_2,\cdots,s_N)=X_i(s_i)$, where $s_i\in\mathcal{S}_i$. Moreover, if $\mathcal{S}_i=\mathcal{S}_j=\mathcal{S}$ $\forall i,j$ and $X_i(s_i)=X(s_i)$ $\forall i$ (where $X$ is some random variable defined on $\mathcal{S}$), then $X_i$'s can be said to be identically distributed. This way of looking at multiple random variables begs the question what if two or more random variables are not independent or not identical; more generally, how to generalize the concepts in single random variable theory to multiple random variables. 

\subsection{Multiple Random Variables}
While all the preceding ideas can be generalized to the case of multiple random variables, some newer concepts emerge. Let us list them. For convenience, we shall only work with the two random variables unless specified otherwise; further generalization is straightforward but notationally cumbersome.

First of all, we must realise that, for consistency of the concepts below, we should define a \emph{bivariate} or \emph{two-dimensional} random variable $(X,Y)$ on the same sample space. The \emph{joint cumulative distribution function} can be defined as
\begin{eqnarray}
F_{XY}(x,y)=P(X\le x,Y\le y);\,x,y\in(-\infty,+\infty).\qquad
\end{eqnarray}
The \emph{joint probability mass function} is
\begin{eqnarray}
p_{XY}(x,y)=P(X=x,Y=y),
\end{eqnarray}
and \emph{joint probability density function} is
\begin{eqnarray}
p_{XY}(x,y)=\frac{\partial^2}{\partial x\partial y}F_{XY}(x,y).
\end{eqnarray}
One defines \emph{marginal cumulative distribution function} as
\begin{eqnarray}
F_X(x)=\lim_{y\rightarrow\infty}F_{XY}(x,y)=P(X\le x,Y\le \infty).
\end{eqnarray}
Furthermore, \emph{conditional} pmf or pdf can now be defined as
\begin{equation}
p_{X|Y}(x|y)\equiv\frac{p_{XY}(x,y)}{p_Y(y)},\,p_Y(y)\ne0.
\end{equation}
Symmetry of $p_{XY}(x,y)$ in the arguments $x$ and $y$, allows to write \emph{Bayes' rule} for the random variables:
\begin{equation}
p_{X|Y}(x|y)\equiv\frac{p_{Y|X}(y|x)p_X(x)}{p_Y(y)},\,p_Y(y)\ne0.
\end{equation}

In the above notations, the random variables $X$ and $Y$ are defined to be independent if
\begin{eqnarray}
F_{XY}(x,y)=F_X(x)F_Y(y),
\end{eqnarray}
and then, for such variable, we find that 
\begin{eqnarray}
&&p_{XY}(x,y)=p_X(x)p_Y(y), \\
&&p_{X|Y}(x|y)=p_X(x)\,{\rm and}\,p_{Y|X}(y|x)=p_Y(y).
\end{eqnarray}
Here, $p_X(x)$ and $p_Y(y)$ are the marginal versions obtained after summing (in the case of pmf) or integrating (in the case of pdf) over all possible values of $y$ and $x$ respectively.

One can furthermore define $(i,j)$th moment as
\begin{equation}
m_{ij}\equiv \mathbb{E}[X^iY^j].
\end{equation}
Similarly, one can define bivariate central moments and cumulants; and also different generating functions. Two concepts are of particular interest: \emph{orthogonal} $X$ and $Y$, and \emph{uncorrelated} $X$ and $Y$. $X$ and $Y$ are said to be orthogonal if $m_{11}=0$. (This terminology will make intuitive sense when later we shall see an inner product structure in this context.)  The random variables are said to be uncorrelated (otherwise, they are correlated) if their \emph{covariance} ${\rm Cov}(X,Y)$ (or $\sigma_{XY}$), given by
\begin{equation}
\sigma_{XY}\equiv \mathbb{E}[(X-m_{10})(Y-m_{01})],
\end{equation}
is zero. Note that since
\begin{equation}
\sigma_{XY}= \mathbb{E}[XY]-\mathbb{E}[X]\mathbb{E}[Y],
\end{equation}
independent random variables are uncorrelated but the converse need not be true. It may be pointed out that (Pearson's) correlation coefficient,
\begin{equation}
 {\rm Corr}(X,Y)\equiv\frac{\sigma_{XY}}{\sigma_X\sigma_Y}\,,
\end{equation}
which ranges between $-1$ to $+1$, quantifies the degree to which the two random variables are linearly correlated:  ${\rm Corr}(X,Y)=1$ means completely linear increasing relationship, while ${\rm Corr}(X,Y)=-1$ means completely linear decreasing relationship.

It is useful to know a rather useful special multivariate random distribution: \emph{$n$-variate Gaussian (normal) distribution}. Denoting $\boldsymbol{X}$ as a column vector $[X_1\,X_2\,\cdots\,X_n]^T$ and its particular realization as $\boldsymbol{x}\equiv [x_1\,x_2\,\cdots\,x_n]^T$, we define $\{X_i\}_{i=1}^{i=n}$ an $n$-variate Gaussian (normal) random variable---represented as $\boldsymbol{X}\sim\mathcal{N}_n(\boldsymbol{\mu},{\sf C})$---if its joint pdf is
\begin{eqnarray}
p_{\boldsymbol{X}}(\boldsymbol{x})=\frac{1}{\sqrt{(2\pi)^n|\textsf{C}|}}e^{-\frac{1}{2}(\boldsymbol{x}-\boldsymbol{\mu})^T{\sf C}^{-1}(\boldsymbol{x}-\boldsymbol{\mu})},
\end{eqnarray}
where $\boldsymbol{\mu}$ is a column vector $[\mathbb{E}[X_1]\,\mathbb{E}[X_2]\,\cdots\,\mathbb{E}[X_n]]^T$ and ${\sf C}_{ij}\equiv\sigma_{X_iX_j}$ is the $(i,j)$-th element of the $n\times n$ matrix ${\sf C}$---the \emph{covariance matrix}---whose determinant has been denoted by $|{\sf C}|$ in the pdf. The set of such $n$ variables are equivalently termed \emph{jointly normal}, an apt term because the sum $\sum_{i=1}^{n}k_ix_i$ can be shown to be a normal random variable for any set of numbers, $k_i$'s. We remark that if additionally the variables are uncorrelated, then they are independent as well. Moreover, the central limit theorem extended to the case of multivariate random variables says: $\lim_{N\rightarrow\infty}\sqrt{N}\left({\bar{\boldsymbol{X}}_N-\boldsymbol{\mu}}\right)\xrightarrow[]{\text{distribution}}\mathcal{N}_n(\boldsymbol{0},{\sf C})$.

We end the discussion on multiple random variables by noticing a nice hidden geometry. Let $L^2$ be the space of square-integrable random variables defined as:
\[
L^2 = \left\{ X : S \to \mathbb{R} \ \middle\vert\ \ \mathbb{E}[X^2] < \infty \right\}
\]
where we consider two random variables to be equivalent if they are equal almost surely. For any two random variables $X$ and $Y$ belonging to the $L^2$ space, an \emph{inner product} can be defined:
\[
\langle X, Y \rangle = \mathbb{E}[XY].
\]
The angle brackets should not be confused with the averaging notation sometimes used in physics; the presence of a comma between the two arguments anyway prevents this confusion.

First, we need to ensure that the inner product maps to a finite real number ($|\langle X, Y \rangle| < \infty$). To this end, note that for any real constant $a \in \mathbb{R}$, $(X + aY)^2 \ge 0$. Expanding and applying the linearity of expectation yields:
\[
\mathbb{E}[(X + aY)^2] = \mathbb{E}[Y^2]a^2 + 2\mathbb{E}[XY]a + \mathbb{E}[X^2] \ge 0.
\]
This expression is a quadratic polynomial in $a$ of the form $Aa^2 + Ba + C \ge 0$. For this inequality to hold true for all $a$, the polynomial's discriminant ($B^2 - 4AC$) must be non-positive:
\[
(2\mathbb{E}[XY])^2 - 4\mathbb{E}[Y^2]\mathbb{E}[X^2] \le 0.
\]
Expanding the square, dividing the entire inequality by 4, and subsequently taking the square root of both sides delivers the famous \emph{Cauchy--Schwarz inequality}:
\[
\vert \mathbb{E}[XY] \vert \le \sqrt{\mathbb{E}[X^2]\mathbb{E}[Y^2]}.
\]
Now, since $X \in L^2$ and $Y \in L^2$, we have $\mathbb{E}[X^2] < \infty$ and $\mathbb{E}[Y^2] < \infty$. Their product is finite, proving via the Cauchy--Schwarz inequality that the inner product is \emph{well-defined} for all elements in the space.

Now, let us verify that this definition satisfies the core inner product axioms. Let $X, Y, Z \in L^2$ be random variables on the probability space $(\mathcal{S}, \mathcal{E}, P)$, and let $a, b \in \mathbb{R}$ be real constants. 
 \begin{enumerate}
 \item \emph{Symmetry (Commutativity):} Because the regular multiplication of real numbers is point-wise commutative ($X(s)Y(s) = Y(s)X(s)$ $\forall s \in \mathcal{S}$), their mathematical expectations must be identical:
\[
\langle X,Y \rangle = \mathbb{E}[XY] = \mathbb{E}[YX] = \langle Y, X \rangle.
\]
\item \emph{Linearity (in the first argument):} Note that
\begin{align}
\langle aX + bZ, Y \rangle &= \mathbb{E}[(aX + bZ)Y]\nonumber\\
&= \mathbb{E}[aXY + bZY]\nonumber\\
&= a\mathbb{E}[XY] + b\mathbb{E}[ZY]\nonumber\\
&= a\langle X, Y \rangle + b\langle Z, Y \rangle.\nonumber
\end{align}
Thus, linearity holds.
\item \emph{Positive-Definiteness:}
Evaluating the inner product of a random variable with itself gives
\[
\langle X, X \rangle = \mathbb{E}[XX] = \mathbb{E}[X^2].
\]
For any real-valued random variable, $X^2(s) \ge 0$ $\forall s \in \mathcal{S}$. Since the argument inside the expectation is non-negative everywhere, its average value must also be non-negative:
\[
\mathbb{E}[X^2] \ge 0.
\]
\item \emph{Strict Definiteness (Identity Property):}
\[
\langle X, X \rangle = 0 \iff X = 0 \quad (\text{almost surely}).
\]
We prove this logical equivalence in both directions. First, the forward direction: If $X = 0$ almost surely (a.s.), then $X^2 = 0$ a.s. The expectation of a random variable that is zero almost surely is exactly zero: $\mathbb{E}[X^2] = 0 \implies \langle X, X \rangle = 0$. 

For the reverse direction, suppose $\langle X, X \rangle = 0$, which implies $\mathbb{E}[X^2] = 0$. In probability theory, if the expectation of a non-negative random variable ($X^2 \ge 0$) is exactly zero, then that random variable must equal zero everywhere except possibly on a set of probability zero:
    \[
    \mathbb{E}[X^2] = 0 \implies X^2 = 0 \ \ \text{a.s.} \implies X = 0 \ \ \text{a.s.}
    \]
 \end{enumerate}
The core consequence of the fact that the expectation of the product of two random variables is an inner product is that probability theory inherits the frameworks and properties of Euclidean and Hilbert space geometry. By treating random variables as geometric vectors, we can solve complex statistical and financial problems using simple spatial concepts like distance, angles, and projections.

\subsection{Stochastic Process}
A (real) \emph{stochastic or random process} is a single valued function $X(s,t)$ that maps each outcome $s$ of an experiment, with a sample space $\mathcal{S}$, at each parameter (usually, time) value $t$ of a parameter set $\mathcal{T}\subseteq\mathbb{R}$ to a real number: $X:\mathcal{S}\times\mathcal{T}\rightarrow\mathbb{R}$. If $\mathcal{T}$ is a countable set, then the process is \emph{discrete-parameter process}; if $\mathcal{T}$ is an interval of $\mathbb{R}$, then the process is \emph{continuous-parameter process}. A discrete-parameter process is more conspicuously denoted by $X_n(s)$  ($n\in\mathbb{Z}$) rather than $X(s,t)$. It is customary to omit the reference to the outcome $s$ in the random process and denote it simply as $X(t)$ (or $X_n$). This is rather a relief as  specifying the outcome a priori is explicitly neither easy nor practically useful: An $s\in\mathcal{S}$ for practical purpose merely symbolizes a sequence of experimental realization (outcome),  say, $(X_0(s),X_1(s),X_2(s),X_3(s))$.  The elements or outcomes of sample space $\mathcal{S}$, thus, are able to specify abstractly every possible experimental realization or outcome. E.g., in an experiment of tossing a unbiased coin two times, $\mathcal{S}$ is not $\{H,T\}$ but set of all possible sequences of heads and tail---$\{(H,H), (H,T),(T,H),(T,T)\}$; of course, any of the two steps of the experiment is modelled by a random variable with sample space $\{H,T\}$.

At a fixed $t$, $X(t)$ is a random variable that can take all possible allowed values to which at least one of the experimental outcomes (not to be confused with $s$ mentioned above) at that $t$ is associated; every such value is called a \emph{state}. The set of all possible states across $\mathcal{T}$ is called \emph{state space} $\mathcal{X}$ of the process. If the state space is countable, then the process is \emph{discrete-state process} (commonly referred as \emph{chain}); otherwise, the process is \emph{continuous-state process}. For a fixed $s$, $X(t)$ is a \emph{sample/sample-path/sample-function/realization} of the process. Finally, with neither $s$ nor $t$ held fixed, $X(s,t)$ can be seen as specifying an \emph{ensemble} of time functions depending on outcome: Ensemble is the collection of all possible sample paths.

For the probabilistic description of a stochastic process, \emph{$n$th-order distribution}---defined as the joint distribution cdf
\begin{equation}
F(\{x_i\}_{i=1}^{i=n};\{t_i\}_{i=1}^{i=n})\equiv P(\{X(t_i)\le x_i\}_{i=1}^{i=n})
\end{equation}
---is a useful theoretical quantity. In principle, for the complete knowledge of $X(t)$ is equivalent to knowing all the $n$th-order distributions formed using all possible $t_i$'s. In place of such an impractical specification, following few quantities are commonly used to (partically) specify $X(t)$:
\begin{itemize}
\item \emph{Ensemble average} or \emph{mean}: $\mu_X(t)\equiv \mathbb{E}[X(t)]$.
\item \emph{Autocorrelation}: $R_X(t,t')\equiv \mathbb{E}[X(t)X(t')]$.
\item \emph{Autocovariance}: $C_X(t,t')\equiv\sigma_{X(t)X(t')}=R_X(t,t')-\mu_{X}(t)\mu_{X}(t')$.
\end{itemize}
An ideal but widely used important process is that of the \emph{white noise}: It is a stochastic process such that its autocovariance is of the form $C(t,t')=q(t)\delta(t-t')$ (or $C(n,n')=q(n)\delta_{nn'}$ for discrete-time process in obvious notations); usually, the mean of the process is taken as identically zero. If $X(t)$ and $X(t')$ ($t'\ne t$) are statistically independent (thus, implying uncorrelatedness), then it is a stronger form of white noise which we may call independent white noise. Any noise that is not purely white is generally referred to as a \emph{coloured noise}.

Fortunately, for some random processes of interest, their complete specification requires one to know a much smaller set of distributions than all possible $n$-th order distributions. Some of the important ones are as follows:
\begin{itemize}
\item \emph{Independent process}: If $X(t_i)$ and $X(t_j)$ are independent random variables $\forall i\ne j$, then $X(t)$ is an independent process. In this case, obviously any $n$th-order distribution can be written as a product of the first-order distributions. If  an independent process is white noise, then it is also called \emph{strictly white noise}.
\item \emph{$k$th order stationary process}: A stochastic process $X(t)$ is $k$th order stationary if $\forall i\le k$ and for every possible values of the set of parameters $\{t_i\}_{i=1}^{i=k}$, $F(\{x_i\}_{i=1}^{i=k};\{t_i\}_{i=1}^{i=k})=F(\{x_i\}_{i=1}^{i=k};\{t_i+\tau\}_{i=1}^{i=k})$ $\forall\tau\in\mathbb{R}$. 
\item \emph{Strict-sense stationary process}: Sometimes simple called the \emph{stationary process}, it satisfies the condition for the $k$th-order stationarity for all possible values of $k$. Evidently, since the $1$st-order distribution must be time-independent, the ensemble average, the variance $C_X(t,t)$  and similar quantities derived from the distribution must be constants. The quantities, like autocorrelation, calculated using the $2$nd-order distribution must depend on the time-difference only, i.e., $R_X(t,t')=R_X(|t-t'|)$. For stationary processes, sometimes \emph{mean-ergodicity} is assumed, i.e., it is assumed that the time-average of a single sample path (as time goes to infinity) is equal to its corresponding ensemble average; if the assumption of equality is true for any measurable function for the random variable, then we can say that \emph{ergodicity} has been assumed or that the ergodicity with respect to the probability distribution function has been assumed---the most strict notion of ergodicity. It is possible to have non-stationary processes that are mean-ergodic (and even additionally \emph{autocorrelation-ergodic}).
\item\emph{Weak-} or \emph{wide-sense stationary process}: Also aptly known as \emph{covariance stationarity}, it is defined through the collective requirements of finite variance (at all times), time independent mean, and time-difference-dependent autocovariance. Every strict-sense stationary process with finite second moments is weak-sense stationary, but the converse is generally false..
\item\emph{Gaussian} or \emph{normal process}: A stochastic process $X(t)$ is a Gaussian process, if for any $n$ and any $\{t_i\}_{i=1}^{i=n}\subseteq \mathcal{T}$, the sequence of random variables $\{X(t_i)\}_{i=1}^{i=n}$ is jointly normal. It is completely specified by the $2$nd-order distributions. A wide-sense stationary Gaussian process is strict-sense stationary as well. A useful straightforward fact is: an independent process consisting of Gaussian random variables is a \emph{Gaussian white noise process}.
\item\emph{Markov process}: Another important process that is completely specified by the $2$nd-order distributions is Markov process defined as the stochastic process such that $P(X(t_{n+1})\le x_{n+1}|\{X(t_i)=x_i\}_{i=1}^{i=n})=P(X(t_{n+1})\le x_{n+1})|X(t_n)=x_n)$ $\forall n$ and $\forall x_i$'s. It is also called \emph{memoryless process}. In the light of the commonly found definition of \emph{$p$th-order Markov process} that obeys $P(X(t_{n+1})\le x_{n+1})|\{X(t_i)=x_i\}_{i=1}^{i=n})=P(X(t_{n+1})\le x_{n+1})|\{X(t_i)=x_i\}_{i=n-p+1}^{i=n})$ $\forall n$ and $\forall x_i$'s, the memoryless process is also called $1$st-order Markov process and the $0$th-order Markov process is actually independent process.

\item \emph{Process with stationary independent increments}: It is a special class of \emph{non-stationary} Markov process $X(t)$ such that:
\begin{itemize}
    \item It has \emph{stationary increments}: For all $t,t'\in\mathcal{T}$ with $t>t'$ and all $\epsilon\ge0$, $X(t)-X(t')$ and $X(t+\epsilon)-X(t'+\epsilon)$ have the same distribution.
    \item It has \emph{independent increments}: For all $t_i\in\mathcal{T}$ (typically $\mathcal{T}=\mathbb{R}_{\ge0}$) with $0<t_1<t_2<\cdots<t_n$, the random variables $X(0),\,X(t_1)-X(0),\,X(t_2)-X(t_1),\dots,\,X(t_n)-X(t_{n-1})$ are mutually (jointly) independent.
\end{itemize}
\end{itemize}

To see how such a process is Markovian, fix $0 = t_0 < t_1 < \cdots < t_n < t$ and real numbers $x, x_0, \ldots, x_n$. Decomposing $X(t) = X(t_n) + [X(t) - X(t_n)]$ and substituting the conditioning value $X(t_n) = x_n$ gives
\begin{align}
& P\!\left(X(t) \leq x \,\big|\, \{X(t_i) = x_i\}_{i=0}^{n}\right) \nonumber \\
&\quad = P\!\left(X(t_n) + [X(t) - X(t_n)] \leq x \,\big|\, \{X(t_i) = x_i\}_{i=0}^{n}\right) \nonumber \\
&\quad = P\!\left(x_n + [X(t) - X(t_n)] \leq x \,\big|\, \{X(t_i) = x_i\}_{i=0}^{n}\right) \nonumber \\
&\quad = P\!\left([X(t) - X(t_n)] \leq x - x_n \,\big|\, \{X(t_i) = x_i\}_{i=0}^{n}\right).
\label{eq:midstep}
\end{align}
By the independent-increments assumption, $X(t) - X(t_n)$ is independent of the collection
\begin{equation}
\bigl\{ X(0),\; X(t_1) - X(t_0),\; \ldots,\; X(t_n) - X(t_{n-1}) \bigr\}.
\end{equation}
For each $k = 1, \ldots, n$, we can write
\begin{equation}
X(t_k) = X(0) + \sum_{i=0}^{k-1} \bigl[ X(t_{i+1}) - X(t_i) \bigr],
\end{equation}
so the random vector $(X(t_0), \ldots, X(t_n))$ is a measurable function of $X(0)$ and the earlier increments. Since measurable functions of random variables independent of $Z$ remain independent of $Z$, it follows that
\begin{equation}
X(t) - X(t_n) \;\text{is independent of}\; (X(t_0), \ldots, X(t_n)).
\end{equation}
Hence the conditioning in Eq.~\eqref{eq:midstep} can be dropped:
\begin{align}
& P\!\left([X(t) - X(t_n)] \leq x - x_n \,\big|\, \{X(t_i) = x_i\}_{i=0}^{n}\right) \nonumber \\
&\quad = P\!\left([X(t) - X(t_n)] \leq x - x_n\right).
\label{eq:drop1}
\end{align}
Applying the same independence argument to the singleton conditioning event $\{X(t_n) = x_n\}$ yields
\begin{align}
& P\!\left(X(t) \leq x \,\big|\, X(t_n) = x_n\right) \nonumber \\
&\quad = P\!\left([X(t) - X(t_n)] \leq x - x_n \,\big|\, X(t_n) = x_n\right) \nonumber \\
&\quad = P\!\left([X(t) - X(t_n)] \leq x - x_n\right).
\label{eq:drop2}
\end{align}
Combining Eqs.~\eqref{eq:midstep}, \eqref{eq:drop1}, and \eqref{eq:drop2}, we obtain
\begin{align}
& P\!\left(X(t) \leq x \,\big|\, X(t_0) = x_0, \ldots, X(t_n) = x_n\right) \nonumber \\
&\quad\quad\quad\quad \quad = P\!\left(X(t) \leq x \,\big|\, X(t_n) = x_n\right),
\end{align}
which is precisely the Markov property, thereby proving that a process with independent increments is Markovian. Note that the stationary increments property is not used in proving Markovianity; only independent increments are required.

\subsection{Markov Chain}
Let us now exclusively concentrate on the concepts related to discrete-time Markov chain which for brevity will be called \emph{Markov chain} henceforth. Without any loss of generality and for the simplification of notations, the state space $\mathcal{X}$ of a Markov chain is taken as $\mathbb{Z}$ or $\mathbb{Z}_{\ge0}$ or $\mathbb{N}$ (or a finite subset of these, in which case the chain is termed \emph{finite}) and the specific $N$ possible states are denoted by $i$, $j$, $k$, $l$ etc.; consequently, $P(X_n=i)$ is to be read as the probability of the system to be in state $i$ at time-step $n$. Thus, the fact that in a Markov chain the probability of the system to be in a future state, conditioned on the present state, is independent of the past states is encapsulated in the fact that the \emph{one-step transition probabilities} $p_{ij}^{(1)}(n)=p_{ij}(n)\equiv P(X_{n+1}=j|X_n=i)=P(X_{n+1}=j|X_n=i,X_{n-1}=k,\cdots,X_0=l)$ $\forall n,i,j,k,\cdots,l$. $p_{ij}$ can be thought of as the $(i,j)$-th element of an $N$-dimensional (possibly infinite) square matrix ${\sf P}(n)$, called the \emph{transition probability matrix} (also \emph{Markov matrix, stochastic matrix, transition matrix, and probability matrix}). Obviously, $\sum_{j\in\mathcal{X}}p_{ij}=1\,\forall i\in\mathcal{X}$---a property termed as \emph{row stochasticity}. One furthermore defines \emph{state probability (row) vector} at time $n$ as ${\boldsymbol{\pi}(n)}=[\pi_0(n)\,\pi_1(n)\,\cdots\,\pi_{N-1}(n)]\equiv[P(X_n=0)\,P(X_n=1)\,\cdots\,P(X_n=N-1)]$  that specifies the probability distribution of the chain at time $n$.  Naturally, the sum of all the elements of a state vector at any time step must be one.

If ${\sf P}(n)$ is independent of $n$, i.e., \emph{stationary} (not to be confused with stationarity of a chain), then the Markov chain is \emph{time-homogeneous}; the chain is \emph{nonhomogeneous} otherwise. However, this confusing dual meaning of the term---`stationary'---is somewhat unified on noting that applying Bayes' rule on $n$th order joint pmf, any stationary Markov chain can easily be shown to be time-homogeneous (but not vice versa).  Henceforth, we consider only (time-)homogeneous Markov chain unless specified otherwise. We note that
\begin{eqnarray}
&&P(X_{n+1}=j)=\sum_{i\in\mathcal{X}} P(X_{n+1}=j|X_n=i)P(X_n=i),\,\,\quad\nonumber\\
&&\implies\pi_j(n+1)=\sum_{i\in\mathcal{X}} p_{ij}\pi_i(n),\nonumber\\
&&\implies\boldsymbol{\pi}(n+1)=\boldsymbol{\pi}(n){\sf P},\label{eq:dtme}\\
&&\implies\boldsymbol{\pi}(n)=\boldsymbol{\pi}(0){\sf P}^n,
\end{eqnarray}
meaning that any time-homogeneous Markov chain is completely specified by initial probability vector $\boldsymbol{\pi}(0)$ and the transition matrix ${\sf P}$, i.e., equivalently, by the 2nd order distributions. For example, the $3$rd order joint pmf $P(X_2=i,X_1=j,X_0=k)=P(X_2=i|X_1=j,X_0=k)P(X_1=j,X_0=k)=P(X_2=i|X_1=j)P(X_1=j|X_0=k)P(X_0)=\pi_k(0)p_{kj}p_{ji}$; and so on for any higher order joint pmf. In literature, sometimes Eq.~\eqref{eq:dtme} is called \emph{discrete-time master equation}.

Similarly, the \emph{$n$-step transition probabilities}, $p_{ij}^{(n)}\equiv P(X_n=j|X_0=i)$, are merely the elements of ${\sf P}^n$ as can be seen from the following. Suppose the system reaches state $k$ from state $i$ after $n$ steps and then reaches state $j$ after $n'$ more steps, then because of the Markov property, $p_{ij}^{(n+n')}$ can be obtained by summing over all possible intermediate states $k$; therefore,
\begin{eqnarray}
p_{ij}^{(n+n')}=\sum_{k\in\mathcal{X}}p_{ik}^{(n)}p_{kj}^{(n')},\label{eq:dtcke}
\end{eqnarray}
known as the \emph{Chapman--Kolmogorov equation}. For $n=n'=1$, we note $p_{ij}^{(2)}$'s are merely the elements of ${\sf P}^2$. Extending the argument inductively, it is obvious that $p_{ij}^{(n)}$ are the elements of ${\sf P}^n$ and the Chapman--Kolmogorov equation is essentially the matrix identity, ${\sf P}^{n+n'}={\sf P}^{n}{\sf P}^{n'}$ $\forall n,n'\in\mathbb{Z}_{\ge0}$ (${\sf P}^0$ is taken to be identity matrix ${\sf I}$). 

One of the most interesting question in the theory of Markov chain is that of the \emph{limiting distribution}, $\boldsymbol{\pi}^\infty$ that is such a distribution that $\lim_{n\rightarrow\infty}\boldsymbol{\pi}(n)=\boldsymbol{\pi}^\infty$ $\forall \boldsymbol{\pi}(0)$.  The limit distribution is usually found by looking for \emph{stationary distribution} (or \emph{steady-state distribution}) $\boldsymbol{\pi}^*$ that is defined as the one that satisfies the following relation: $\boldsymbol{\pi}^*=\boldsymbol{\pi}^*{\sf P}$. Note that a time-homogeneous Markov chain is stationary iff the distribution $\boldsymbol{\pi}(0)$ is a stationary distribution of the Markov chain. $\boldsymbol{\pi}^\infty$ is a  $\boldsymbol{\pi}^*$ (but the converse need not be true) since
\begin{eqnarray}
&&\boldsymbol{\pi}^\infty=\lim_{n+1\rightarrow\infty}\boldsymbol{\pi}(0){\sf P}^{n+1},\nonumber\\
\implies&&\boldsymbol{\pi}^\infty=\left[\lim_{n\rightarrow\infty}\boldsymbol{\pi}(0){\sf P}^n\right]{\sf P},\nonumber\\
\implies&&\boldsymbol{\pi}^\infty=\boldsymbol{\pi}^\infty{\sf P}.
\end{eqnarray}
Obviously, by definition, $\boldsymbol{\pi}^\infty$ is insensitive to $\boldsymbol{\pi}(0)$ and hence it can also be equivalently defined in component form as $\pi^\infty_j=\lim_{n\rightarrow\infty}p^{(n)}_{ij}$ $\forall i,j\in\mathcal{X}$. It means that the rows of ${\sf P}^n$ are more and more similar to the limiting state probability vector as $n$ becomes larger and larger; consequently $\lim_{n\rightarrow\infty}\boldsymbol{\pi}(0){\sf P}^n=\boldsymbol{\pi}^\infty$ $\forall \boldsymbol{\pi}(0)$: Only when each column of ${\sf P}^n$, as $n\to\infty$, is a constant vector, can the multiplication by $\boldsymbol{\pi}(0)$ render the result independent of initial state vector because $\sum_{i\in\mathcal{X}} {\pi}_i(0)=1$.  To see this clearly, consider the example of transition matrix:
\[
\mathsf{P} = \begin{bmatrix} 1-\alpha & \alpha \\ \beta & 1-\beta \end{bmatrix}, \quad 0 < \alpha, \beta < 1.
\]
For stationary distribution, $\boldsymbol{\pi}^*=\begin{bmatrix} a & b \end{bmatrix}$, we have
\[
{\begin{bmatrix} a & b \end{bmatrix}} \mathsf{P} = \begin{bmatrix} a & b \end{bmatrix} \implies a = \frac{\beta}{\alpha+\beta}, \quad b = \frac{\alpha}{\alpha+\beta}.
\]
Higher-order transition matrix is then given as
\[
\mathsf{P}^n = \frac{1}{\alpha+\beta} \left\{ \begin{bmatrix} \beta & \alpha \\ \beta & \alpha \end{bmatrix} + (1-\alpha-\beta)^n \begin{bmatrix} \alpha & -\alpha \\ -\beta & \beta \end{bmatrix} \right\},
\]
whose limiting behaviour is:
\[
\lim_{n \to \infty} \mathsf{P}^n = \frac{1}{\alpha+\beta} \begin{bmatrix} \beta & \alpha \\ \beta & \alpha \end{bmatrix} = \begin{bmatrix} a & b \\ a & b \end{bmatrix}= \begin{bmatrix} \boldsymbol{\pi}^* \\ \boldsymbol{\pi}^*\end{bmatrix}= \begin{bmatrix} \boldsymbol{\pi}^\infty \\ \boldsymbol{\pi}^\infty\end{bmatrix}.
\]
Here note that all the rows are identical.

We naturally intuit that the answer to the question of the existence of the limiting distribution in a Markov chain should depend on the nature of the states in the Markov chain. Some important definitions classifying the states are as follows:
\begin{itemize}
\item \emph{Accessible state}: A state $j$ is accessible from a state $i$ (denoted by $i\rightarrow j$) if $p^{(n)}_{ij}>0$ for some $n\in\mathbb{Z}_{\ge0}$. It follows that any state is accessible from itself because $p^{(0)}_{ii}=1$. If $i\rightarrow j$ and $j\rightarrow i$, i.e., notationally $i\leftrightarrow j$, the two states are said to \emph{communicate}. 
\item \emph{Recurrent} or \emph{persistent state}: State $i$ is recurrent if the
probability $f_{ii}$, that the system starting from state $i$ comes back to $i$, is
unity. General $f_{ij}$ is defined as follows. Let the random variable \emph{first
passage time}, $\tau_j=\inf\{n\ge1: X_n=j\}$, with the convention $\inf\emptyset=\infty$,
be the number of steps required to visit state $j$ for the first time after the start
($\tau_j=\infty$ means that state $j$ is never visited). Define the \emph{first passage
probability in $n$ steps}, $f^{(n)}_{ij}\equiv P(\tau_j=n\mid X_0=i)$. Consequently, the
probability of eventually visiting state $j$ starting from state $i$ is the
\emph{eventual first passage probability}
$f_{ij}=\sum_{n=1}^{\infty}f_{ij}^{(n)}=P(\tau_j<\infty\mid X_0=i)$. For $j=i$, the
sequence $\{f^{(n)}_{ii}\}$ is aptly said to give the distribution of \emph{recurrence
times}. A recurrent state is further classified as follows: if the \emph{mean recurrence
time} of state $i$, $\mathbb{E}[\tau_i\mid X_0=i]$, is finite, then the recurrent state
$i$ is \emph{positive} or \emph{nonnull recurrent}; otherwise it is \emph{null
recurrent}. This finer classification makes sense only for infinite Markov chains; a
null recurrent state in a finite Markov chain is obviously an impossibility.
\item \emph{Transient state}: State $i$ is transient if $f_{ii}<1$; i.e., there is a finite probability that it will not be visited ever again.
\item \emph{Periodic state}: State $i$ is said to be periodic if the period of the state, $d(i)\equiv{\rm gcd}\{n\ge 1: p_{ii}^{(n)}>0\}>1$. Thus, it is a recurrent state such that only certain recurrence times (with $d(i)$ as their greatest common divisor) are allowed. $d(i)$ itself may not be a recurrence time.
\item \emph{Aperiodic state}: State $i$ is said to be aperiodic if $d(i)=1$. Thus, it is a recurrent state such that all recurrence times are allowed. It is easily argued that any state $i$ with $p_{ii}\ne0$ is aperiodic.
\item \emph{Absorbing state}: State $i$ is termed absorbing if $p_{ii}=1$; the state is \emph{nonabsorbing} otherwise. By definition, an absorbing state must be recurrent state (not vice versa) and aperiodic state (not vice versa).
\item \emph{Ergodic state}: State $i$ is called ergodic if it is aperiodic and positive recurrent. 
\end{itemize}

We realize that every state in a Markov chain must either be transient or recurrent. Think about a recurrent state: It is such that with full certainty it will be revisited; by Markov property, once revisited, it will be re-revisited with probability one, and so on. This means that the state will be visited infinitely many times with unit probability and hence, the mean of number of returns to the state must be infinity. Now think about a transient state: It is such a state that with a finite probability $1-f_{ii}$ it will not be revisited. If we refer `not returning' to the state as `success', then sequence of $k$ `failures' followed by a success is equivalent to a sequence independent Bernoulli trials with probability $f_{ii}^{k}(1-f_{ii})$.Therefore, the mean number of failures or returns is $\sum_{k=0}^{\infty}kf_{ii}^{k}(1-f_{ii})=f_{ii}/(1-f_{ii})$---a finite number. From a different angle, mean number of returns to any arbitrary state $i$ is nothing but $\sum_{n=1}^{\infty}p^{(n)}_{ii}$ (summation starts from $1$ as $n=0$ does not correspond to ``return'') because at any step $n$, one is either at $i$ (count 1) or not at $i$ (count 0), thereby expectation at step $n$ of being at $i$ is $1\times p^{(n)}_{ii}+0\times (1-p^{(n)}_{ii})$. Therefore, recurrent state and transient state are, respectively, equivalent to the states satisfying $\sum_{n=0}^{\infty}p^{(n)}_{ii}=\infty$ and $\sum_{n=0}^{\infty}p^{(n)}_{ii}<\infty$. Along similar line of arguments, it is obvious that given $j\rightarrow i$, the mean number of times one reaches $i$, starting from $j$ should be given by $\sum_{n=0}^{\infty}p^{(n)}_{ji}$ (summation starts from $0$ as $n=0$ does not contribute anything) and this should be finite if $i$ is transient. But for finiteness (i.e., convergence) of this sum, it must be that ${\lim_{n\rightarrow\infty}}p^{(n)}_{ji}=0$. Note that trivially $p^{(n)}_{ji}=0$ $\forall n$ if $j\nrightarrow i$. Summing up, if $i$ is transient, then ${\lim_{n\rightarrow\infty}}p^{(n)}_{ji}=0$ $\forall j\in\mathcal{X}$ necessarily. We mention, without proof, that this limit is also satisfied by any null recurrent state which is, thus, such a state that it is returned infinitely often with probability one but the probability that it will be occupied as $n\rightarrow\infty$ is zero.

Having formed some intuition about the recurrent and the transient states, we can now present a rather simple proof of an important proposition that \emph{any finite Markov chain has at least one positive recurrent state}: Observe that $\sum_{j\in\mathcal{X}}p^{(n)}_{ij}=1$ $\forall i,n$ because starting from a state, one must be somewhere in the state space at any $n$. Taking the limit $n\rightarrow\infty$ on both sides and letting the limit enter the summation (possible under the assumption of finite Markov chain), we get $\sum_{j\in\mathcal{X}}\lim_{n\rightarrow\infty}p^{(n)}_{ij}=1$. If every $j$ is either transient or null recurrent state, then the L.H.S. must be zero which is an impossibility. Hence, at least one $j$ must be a positive recurrent state. 

The relation `$\leftrightarrow$' (being reflexive, symmetric, and transitive) is an \emph{equivalence relation} and hence all the states of the Markov chain can be partitioned into \emph{communicating classes}. Note that if two states are in the same class, then either both of them are recurrent or both of them are transient. If two recurrent states are in the same class, then either both of them are null or both of them are nonnull. On a similar note, all the states of the same class have same period. Thus, it makes sense to talk about null recurrent class, nonnull/positive recurrent class, transient class, periodic class of a particular period, and aperiodic class; the definitions are obvious. 

Recurrent (null or nonnull) class is special in the sense that it is \emph{closed}, i.e., in case $i$ is recurrent then $p_{ij}=0$ where $j$ is any state not the equivalent class of $i$. This is because if it is so that $p_{ij}\ne0$ and ergo $i\rightarrow j$, then we must have $j\rightarrow i$ since $i$ is recurrent; but this is not possible since, by assumption, $j$ is not in the equivalent class of $i$. Summarizing, $p_{ij}\ne0$ is an impossibility. Thus, if there are more than one equivalent recurrent classes, one could just practically assume that the corresponding Markov chain has only that equivalent class where the chain starts initially. 

A Markov chain with one and only one communicating class is known as \emph{irreducible} or \emph{indecomposable}. An irreducible Markov chain consisting of only ergodic states is called \emph{ergodic Markov chain}. Thus, irreducible, aperiodic, and positive recurrent Markov chain is ergodic Markov chain. Note again that we can use the adjectives---aperiodic, nonnull recurrent, etc.---without any ambiguity, for the chain itself that is irreducible as the adjectives are class properties and we have a single class.

Now coming to the issue of existence and uniqueness of stationary distribution, it is known that \emph{in an irreducible Markov chain, a stationary distribution exists iff all the states are positive recurrent; furthermore, the $\boldsymbol{\pi}^*$ is unique with ${\pi}^*_i=(\mathbb{E}[\tau_i|X_0=i])^{-1}$}. (For comparison, recall that for null recurrent states, $\mathbb{E}[\tau_i|X_0=i]=\infty$.) It should be noted that the condition of irreducibility is crucial for the uniqueness. For example, recall that otherwise one could have the possibility of many positive recurrent classes; any such class (being closed) could lead to a stationary distribution formed by finding the appropriate values of the probability vector's components corresponding to the class and associating zeros to the other components' values for the states outside the class.  It is furthermore interesting to observe that there is no condition on the periodicity of the chain. Note that for finite chain, irreducible condition is enough to guarantee the existence and the uniqueness of the stationary distribution. This is so because, in the light of the fact that any finite Markov chain has at least one positive recurrent state, a finite irreducible Markov chain must have exclusively positive recurrent states. 

Actually, any (time-homogenous) finite Markov chain has at least one stationary state. This may be argued as follows: The row transition matrix has a trivial right-eigenvector---one with unity as all components---with right-eigenvalue equal to one. Since the right and the left eigenvalues of any square matrix are same, there must be a left-eigenvector, corresponding to the unity eigenvalue, that serves as the stationary distribution. It is in line with Brouwer's fixed-point theorem (stating that for any continuous function mapping a nonempty compact convex set to itself, there is a point in the set that is mapped to itself) applied to compact convex set of all probability distributions of the finite set, $\{1,2,\cdots, n\}$ ($n$ being the order of the transition matrix).  As an aside, it is of interest to note that the spectral radius of any transition matrix is one; and hence, in accordance with the Gershgorin circle theorem on the spectrum of complex square matrices, no eigenvalue has absolute value greater than one.

Whether $\boldsymbol{\pi}^*$ is reached asymptotically, i.e., whether $\boldsymbol{\pi}^\infty=\boldsymbol{\pi}^*$, is a question that requires constraint on the periodicity of the Markov chain for an affirmative answer. An intuitive idea behind this constraint can be seen as follows: Recall that $\pi^\infty_j=\lim_{n\rightarrow\infty}p^{(n)}_{ij}$ $\forall i,j$. It may be argued that a unique non-zero value of $p^{(n)}_{ii}$ in the limit of large $n$ (and hence $\pi^\infty_i$) does not make much sense if $i$ is a periodic state with period $d(i)$, since $p_{ii}^{(n)}=0$ for any $n$ not a multiple of $d(i)$. What can be proven is the following: For an aperiodic irreducible Markov chain with exclusively positive recurrent states, a unique $\boldsymbol{\pi}^*$, which is $\boldsymbol{\pi}^\infty$ as well, exists. In other words, \emph{a unique limiting invariant probability distribution exists for any ergodic Markov chain}. As an aside, we remark that the existence of a limiting distribution does not mean that the corresponding Markov chain must be ergodic, because limiting distribution can exist for reducible Markov chains: e.g., consider a Markov chain with an upper triangular transition matrix.

We now conclude the theoretical discussion on Markov chains with the important concept of (time-)reversibility. A general stochastic process $X(t)$ is called \emph{reversible} if any sequence ${X_{t_n}}$ has the same probability distribution as its \emph{reversed process} ${X_{\tau-t_n}}$, for all $n\in\mathbb{N}$, all $\tau\in\mathbb{R}$, and all $t_i$'s satisfying $t_1<t_2<\cdots<t_n$. From this definition, it follows that for a reversible process, both ${X_{t_n}}$ and ${X_{\tau'+t_{n}}}$ have the same distribution as ${X_{-t_n}}$ upon choosing $\tau=0$ and $\tau=\tau'$, respectively; hence, we conclude that a reversible process is stationary (though not necessarily vice versa). Therefore, when discussing reversible Markov chains, one need only consider stationary Markov chains. It is easy to see that a \emph{reversed chain} ${X_{\nu-n}}$ (for some arbitrary $\nu\in\mathbb{Z}$) corresponding to a Markov chain ${X_n}$ must itself be Markov, since the independence of events is a symmetric property. Note that the reversed chain can always be constructed, regardless of whether the process is reversible.

Now, for a reversible Markov chain, consider any two time indices $n$ and $n'$ with $n<n'$. By the definition of reversibility, the joint distribution of $(X_n, X_{n'})$ must coincide with that of the reversed pair $(X_{\nu-n}, X_{\nu-n'})$ for any $\nu\in\mathbb{Z}$. Choosing $\nu=n+n'$, the reversed indices become $\nu-n=n'$ and $\nu-n'=n$, so that the pair $(X_{\nu-n}, X_{\nu-n'})$ is simply $(X_{n'}, X_n)$. Reversibility therefore requires
\begin{eqnarray}
&&P(X_{n}=i,X_{n'}=j)=P(X_{n'}=i,X_{n}=j)\nonumber\\
\implies&&P(X_n=i)P(X_{n'}=j\mid X_n=i)=\nonumber\\
&&\qquad\qquad\qquad P(X_n=j)P(X_{n'}=i\mid X_n=j)\nonumber\\
\implies&&\pi_i p_{ij}=\pi_j p_{ji},
\end{eqnarray}
for all $i,j\in\mathcal{X}$. This is known as the \emph{local balance} or \emph{detailed balance} condition. (We have omitted the argument $n$ from the probability vector because it must necessarily be a stationary distribution.) Since the above steps can be reversed, we arrive at the following fact: detailed balance implies, and is implied by, reversibility in a stationary Markov chain. The stationary distribution of a reversible Markov chain is also called the \emph{equilibrium distribution}.

Summing the detailed balance condition over states $i$, we get $\boldsymbol{\pi}=\boldsymbol{\pi}{\sf P}$---the stationarity condition---that can now be interpreted as the \emph{global balance} condition: The rate of flow $\sum_{i\in\mathcal{X}}\pi_jp_{ji}$ (=$\pi_j$) out of state $j$ is equal to the rate of flow $\sum_{i\in\mathcal{X}}\pi_ip_{ij}$ into the state $j$. Of course, local balance implies global balance but not vice versa. By definition, global balance is required for stationarity of the Markov chain. So, there is no reason that a stationary Markov chain must satisfy local balance condition and hence be reversible. Therefore, even an ergodic Markov chain need not be reversible and hence, the corresponding limiting distribution need not be an equilibrium distribution. In contrast to being in equilibrium distribution, a Markov process is said to be in a \emph{non-equilibrium steady state (NESS)} if its probability distribution is stationary---satisfying global balance---but detailed balance is violated, so that nonzero stationary probability currents persist. In passing, we mention that \emph{Kolmogorov's criterion} for ergodic Markov chain states that the necessary and sufficient condition for the chain to be reversible is that starting from any state, the probability of coming back to the state along any sequence of states should be unchanged for the reversed sequence.

\subsection{Markov Chain: Examples}
\begin{figure}[t]
  \centering
  \begin{tikzpicture}[
    state/.style={circle, draw, thick, minimum size=7mm, inner sep=0pt, font=\small},
    edge/.style={->, >=stealth, semithick},
    cls/.style={draw=gray, dashed, thick, rounded corners=3pt},
    lbl/.style={font=\scriptsize, align=center}
  ]
    \node[state] (1)  at (0,0)      {1};
    \node[state] (2)  at (1.5,0)    {2};
    \node[state] (3)  at (1.5,-1.8) {3};
    \node[state] (4)  at (3.0,-1.8) {4};
    \node[state] (5)  at (4.5,-1.8) {5};
    \node[state] (6)  at (4.5,-3.6) {6};
    \node[state] (7)  at (6.0,-3.6) {7};
    \node[state] (8)  at (6.0,-5.4) {8};
    \node[state] (12) at (1.5,-5.4) {12};
    \node[state] (9)  at (4.0,-5.4) {9};
    \node[state] (13) at (1.5,-6.9) {13};
    \node[state] (11) at (2.6,-8.2) {11};
    \node[state] (10) at (4.0,-7.8) {10};

    \draw[edge] (1) to[bend left=25] (2);
    \draw[edge] (2) to[bend left=25] (1);
    \draw[edge] (1) to[out=150, in=210, looseness=6] (1);
    \draw[edge] (2) -- (3);
    \draw[edge] (3) to[bend left=25] (4);
    \draw[edge] (4) to[bend left=25] (3);
    \draw[edge] (4) to[bend left=25] (5);
    \draw[edge] (5) to[bend left=25] (4);
    \draw[edge] (5) -- (6);
    \draw[edge] (3) -- (12);
    \draw[edge] (6) to[bend left=25] (7);
    \draw[edge] (7) to[bend left=25] (6);
    \draw[edge] (7) -- (8);
    \draw[edge] (6) -- (9);
    \draw[edge] (8) to[out=30, in=-30, looseness=6] (8);
    \draw[edge] (12) -- (9);
    \draw[edge] (9) to[bend left=15] (10);
    \draw[edge] (10) to[bend left=15] (9);
    \draw[edge] (10) to[bend left=12] (11);
    \draw[edge] (11) to[bend left=12] (10);
    \draw[edge] (11) -- (13);
    \draw[edge] (13) -- (12);

    \draw[cls] (-1.1,-0.6) rectangle (2.0,0.6);
    \node[lbl, anchor=west] at (2.1,0) {transient\\aperiodic, $d=1$};

    \draw[cls] (1.0,-2.4) rectangle (5.0,-1.2);
    \node[lbl, anchor=west] at (5.1,-1.8) {transient\\periodic, $d=2$};

    \draw[cls] (4.0,-4.2) rectangle (6.5,-3.0);
    \node[lbl, anchor=east] at (3.9,-3.6) {transient\\periodic, $d=2$};

    \draw[cls] (5.5,-5.95) rectangle (7.15,-4.85);
    \node[lbl, anchor=north] at (6.3,-6.05) {absorbing\\(recurrent,\\aperiodic, d=1,\\ergodic)};

    \draw[cls] (1.0,-8.7) rectangle (4.7,-4.9);
    \node[lbl, anchor=north] at (2.85,-8.8) {recurrent, aperiodic, $d=1$,\\ergodic};
  \end{tikzpicture}
  \caption{\small A reducible Markov chain on 13 states.}
  \label{fig:markov}
\end{figure}
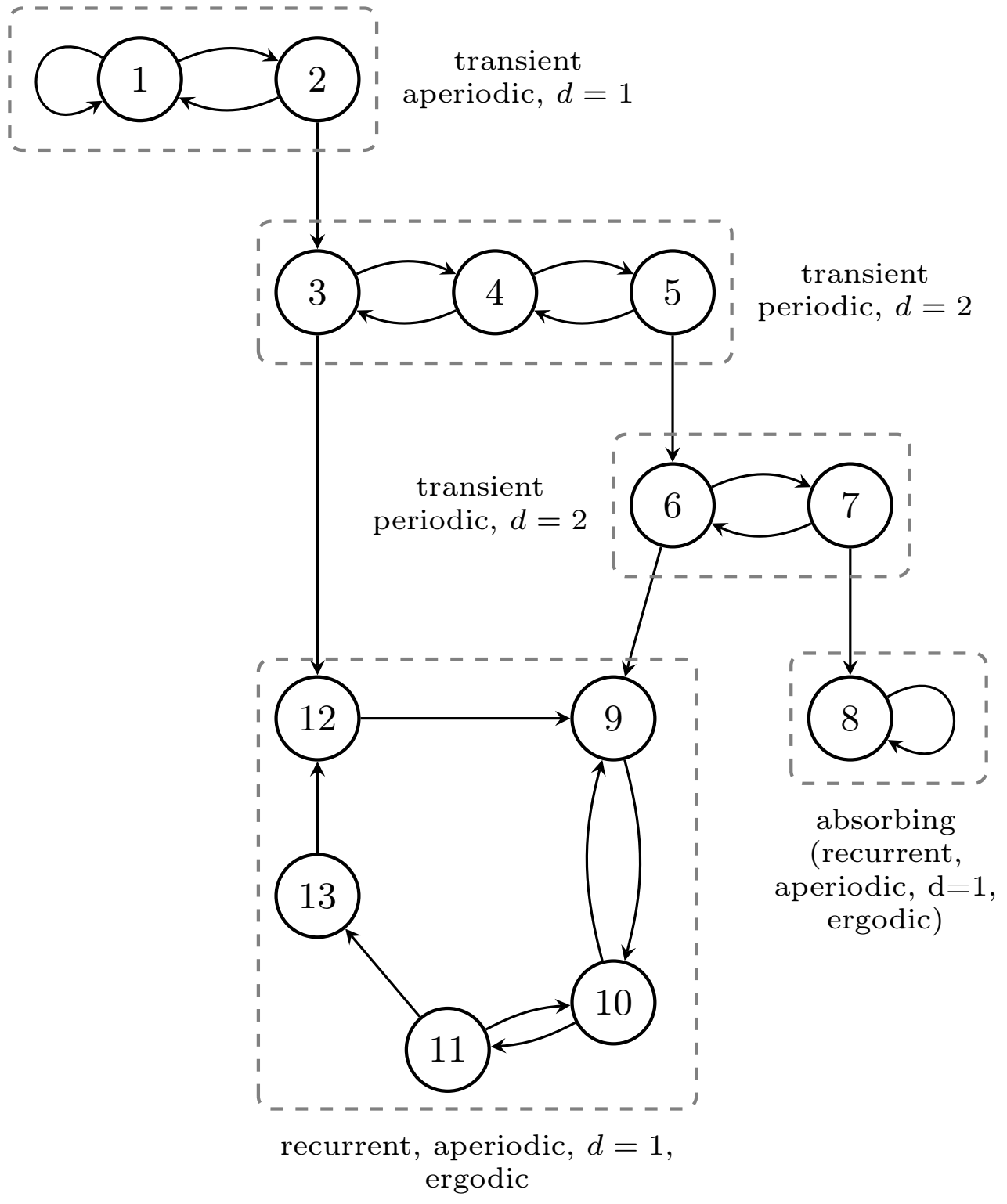

As an example, consider the Markov chain on the state space
$S=\{1,2,\dots,13\}$ shown in Fig.~\ref{fig:markov}. Each circle represents a state, an
arrow from $i$ to $j$ indicates $p_{ij}>0$, and a self-loop indicates $p_{ii}>0$. The
numerical values of the transition probabilities are not shown, since the classification
of states depends only on which transitions are possible. The dashed boxes enclose the
communicating classes, and each box is labeled with the properties of its states. We now
verify these labels.

The chain in Fig.~\ref{fig:markov} has five communicating classes, namely $\{1,2\}$,
$\{3,4,5\}$, $\{6,7\}$, $\{8\}$ and $\{9,\dots,13\}$, so it is reducible. Since
transience, recurrence and the period are class properties, it suffices to examine one
representative state in each class.

\emph{Class $\{1,2\}$:} Because $p_{11}>0$, the chain can return to state 1 in a single
step, so $d=1$ and the class is aperiodic. However, the transition $2\to3$ leads out of
the class and there is no path back. With positive probability the chain never returns, so
$f_{ii}<1$ and the class is transient. This shows that an aperiodic class need not be
recurrent.

\emph{Class $\{3,4,5\}$:} Inside this class the chain can only move back and forth along
$3\leftrightarrow4\leftrightarrow5$. Every return to a given state therefore takes an even
number of steps, and $d=2$. The class is periodic. It is also transient, because the
transitions $5\to6$ and $3\to12$ leave the class for good.

\emph{Class $\{6,7\}$:} The same reasoning applies: returns occur only at even times, so
$d=2$. The transitions $6\to9$ and $7\to8$ allow escape, so the class is transient.

\emph{Class $\{8\}$:} Here $p_{88}=1$, so state 8 is absorbing. Once the chain enters it,
it stays there forever. An absorbing state is automatically recurrent, and since it can
return to itself in one step, it is aperiodic with $d=1$.

\emph{Class $\{9,\dots,13\}$:} No arrow leaves this class, so it is closed. A finite
closed class is always recurrent, and in fact positive recurrent. To find its period, look
at two cycles in the class. The first is $12\to9\to10\to11\to13\to12$, which has length 5.
The second is $9\to10\to9$, which has length 2. Since $\gcd(5,2)=1$, we get $d=1$ and the
class is aperiodic. Note that this class has no self-loop, so it is aperiodic without one.
Being both positive recurrent and aperiodic, the class is ergodic.

Putting these results together, the transient states $1$ to $7$ are visited only finitely
many times. Wherever the chain starts, it is eventually captured by one of the two
recurrent classes, $\{8\}$ or $C\equiv\{9,\dots,13\}$. Each of these classes supports its
own stationary distribution: $\boldsymbol{\pi}^*_8$, which assigns probability $1$ to
state $8$, and $\boldsymbol{\pi}^*_C$, which is supported on $C$. Any convex combination of
the two is stationary, so the chain as a whole does not have a unique stationary
distribution. Since both recurrent classes are aperiodic, the limit
$\lim_{n\to\infty}p^{(n)}_{ij}=a_i\,\delta_{8j}+(1-a_i)\,\pi^*_{C,j}$ exists for every $i$
and $j$, where $a_i$ is the probability of eventual absorption in state $8$ starting from
$i$ ($a_8=1$, $a_i=0$ for $i\in C$, and $0<a_i<1$ for $i\le7$). The limit depends on the
initial state, so there is no limiting distribution independent of where the chain starts.
Finally, $\boldsymbol{\pi}^*_8$ trivially satisfies detailed balance, but
$\boldsymbol{\pi}^*_C$ cannot: since $\pi^*_{C,i}>0$ for all $i\in C$, detailed balance
would require $p_{ji}>0$ whenever $p_{ij}>0$, whereas, e.g., $p_{12,9}>0$ but
$p_{9,12}=0$. A net probability current circulates around the cycle
$12\to9\to10\to11\to13\to12$, so $\boldsymbol{\pi}^*_C$ is a nonequilibrium steady state.
Hence, apart from the trivial $\boldsymbol{\pi}^*_8$, the chain admits no equilibrium
distribution.

Next, let us illustrate some of the aforementioned concepts in some variants of random walks. The \emph{(simple) random walk}, $X_n$ with $n\in\mathcal{T}=\mathbb{Z}_{\ge0}$, is a discrete-state, discrete-time stochastic process defined as $X_0\equiv0$ and $X_n\equiv\sum_{i=1}^nY_i$ where $\{Y_i\}$ is a sequence of i.i.d. random variables with $P(Y_i=+1)=p$ and $P(Y_i=-1)=1-p$ $\forall i\in\mathbb{N}$. Note it is a Markov process since each $X_{i}-X_{i-1}$ is nothing but $Y_i$ and all $Y_i$'s (and $X_0$) are mutually independent by definition. Thus, random walk is a discrete-time \emph{Markov chain}. Note furthermore that the process $Y_n$ with $n\in\mathcal{T}=\mathbb{Z}_{\ge0}$ is stationary: all moments are independent of time. But the random walk itself is non-stationary as easily concluded by calculating mean ($\mathbb{E}[X_n]=n(2p-1)$) and variance ($\mathbb{E}[(X_n-\mathbb{E}[X_n])^2]=4np(1-p)$), which depend explicitly on time.

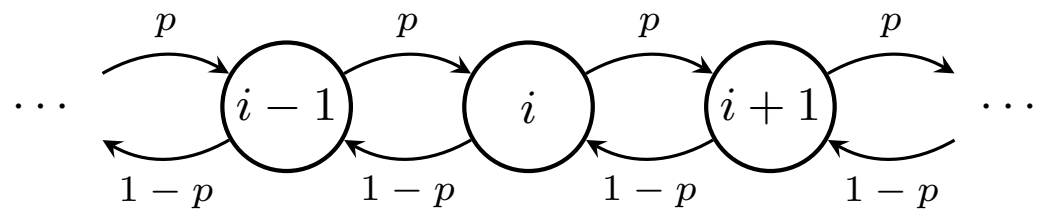
\begin{figure}[htbp]
  \small
  \centering
  \begin{tikzpicture}[
    state/.style={circle, draw, thick, minimum size=8.5mm, inner sep=0pt, font=\small},
    dots/.style={minimum size=7.5mm, inner sep=0pt},
    edge/.style={->, >=stealth, semithick},
    lab/.style={font=\scriptsize}
  ]
    \node[dots] (L) at (0.00,0) {$\cdots$};
    \node[state] (m1) at (1.60,0) {$i-1$};
    \node[state] (z) at (3.20,0) {$i$};
    \node[state] (p1) at (4.80,0) {$i+1$};
    \node[dots] (R) at (6.40,0) {$\cdots$};
    \draw[edge] (L) to[bend left=30] node[lab, auto] {$p$} (m1);
    \draw[edge] (m1) to[bend left=30] node[lab, auto] {$1-p$} (L);
    \draw[edge] (m1) to[bend left=30] node[lab, auto] {$p$} (z);
    \draw[edge] (z) to[bend left=30] node[lab, auto] {$1-p$} (m1);
    \draw[edge] (z) to[bend left=30] node[lab, auto] {$p$} (p1);
    \draw[edge] (p1) to[bend left=30] node[lab, auto] {$1-p$} (z);
    \draw[edge] (p1) to[bend left=30] node[lab, auto] {$p$} (R);
    \draw[edge] (R) to[bend left=30] node[lab, auto] {$1-p$} (p1);
  \end{tikzpicture}
  \caption{\small Simple random walk on the integers.}
  \label{fig:rw-simple}
\end{figure}
Firstly, consider the simple random walk (Fig.~\ref{fig:rw-simple}). Its state space is $\mathcal{X}=\mathbb{Z}$. The transition matrix ${\sf P}$ is given by
\[
p_{ij}=\left\{
\begin{array}{lc}
 p & {\rm if}\, j=i+1,   \\
1-p  & {\rm if}\, j=i-1,    \\
 0 & {\rm otherwise}.   
\end{array}\right.
\]
We note $\forall i,j$, $i\leftrightarrow j$; i.e., the chain is irreducible. For $p\ne0.5$, every state is transient; while for $p=0.5$, every state is null recurrent. This can be shown through the explicit calculation of $\sum_{n=0}^{\infty}p^{(n)}_{ii}$ and finding that it is finite in the former case, while it is infinite in the latter case. Furthermore, the chain is periodic with period $2$. Naturally, there cannot be any stationary or limiting or equilibrium distribution.

\begin{figure}[htbp]
  \small
  \centering
  \begin{tikzpicture}[
    state/.style={circle, draw, thick, minimum size=8.5mm, inner sep=0pt, font=\small},
    dots/.style={minimum size=7.5mm, inner sep=0pt},
    edge/.style={->, >=stealth, semithick},
    lab/.style={font=\scriptsize}
  ]
    \node[state] (s0) at (0.00,0) {$0$};
    \node[state] (s1) at (1.40,0) {$1$};
    \node[state] (s2) at (2.80,0) {$2$};
    \node[state] (s3) at (4.20,0) {$3$};
    \node[dots] (R) at (5.60,0) {$\cdots$};
    \draw[edge] (s0) to[bend left=30] node[lab, auto] {$p$} (s1);
    \draw[edge] (s1) to[bend left=30] node[lab, auto] {$1-p$} (s0);
    \draw[edge] (s1) to[bend left=30] node[lab, auto] {$p$} (s2);
    \draw[edge] (s2) to[bend left=30] node[lab, auto] {$1-p$} (s1);
    \draw[edge] (s2) to[bend left=30] node[lab, auto] {$p$} (s3);
    \draw[edge] (s3) to[bend left=30] node[lab, auto] {$1-p$} (s2);
    \draw[edge] (s3) to[bend left=30] node[lab, auto] {$p$} (R);
    \draw[edge] (R) to[bend left=30] node[lab, auto] {$1-p$} (s3);
    \draw[edge] (s0) to[out=150, in=210, looseness=6] node[lab, left] {$1-p$} (s0);
  \end{tikzpicture}
  \caption{\small Random walk on the non-negative integers with a self-transition at the origin.}
  \label{fig:rw-reflecting}
\end{figure}
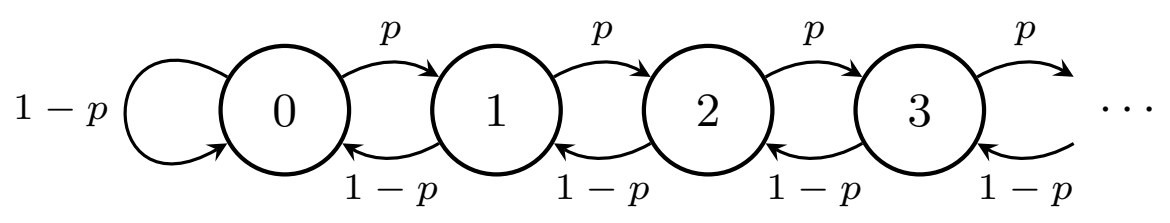
Secondly, consider random walk with one end unbounded and the other with self-transition $p_{00}\in(0,1)$ (Fig.~\ref{fig:rw-reflecting}), i.e., its state space is $\mathcal{X}=\mathbb{Z}_{\ge0}$. The corresponding transition matrix ${\sf P}$ is
\[
p_{ij}=\left\{
\begin{array}{lc}
1-p  & {\rm if}\, j=i=0, \\
 p & {\rm if}\, j=i+1,   \\
1-p  & \phantom{xxxxxii}{\rm if}\,i>0\,\&\, j=i-1,    \\
 0 & {\rm otherwise}.   
\end{array}\right.
\]
First we observe that the chain is irreducible. State $0$ is aperiodic, and hence so is every other state in the chain. Rather than trying to find whether the chain is recurrent or transient using their definitions, let us use the concept that if a stationary distribution exists in an irreducible Markov chain, then the chain is positive recurrent. To this end, we assume that $\boldsymbol{\pi}^*$ exists. Consequently, it is easy to see that $\pi^*_i=r\pi^*_{i-1}$ for $i\in\mathbb{N}$ and $r\equiv p/(1-p)$; and using this in the normalization condition $\sum_{i\in\mathcal{X}}\pi^*_i=1$, we get $\pi^*_0=1-r$ if $r<1$, i.e., if $0<p<1/2$. Concluding, for $p<1/2$, a stationary distribution [$\pi_i^*=r^i(1-r)$] exists, and hence the chain is positive recurrent. Due to aperiodicity, we can also claim that the distribution is limiting distribution as well. Furthermore, we observe that the condition of detailed balance, $\pi^*_{i-1}p=\pi^*_i(1-p)$ for $i\in\mathbb{N}$, is satisfied by $\boldsymbol{\pi}^*$ and so the limiting distribution is an equilibrium distribution as well. Similar arguments for the case $p\ge1/2$ tells us that no nonzero value of $\pi^*_0$ satisfies the normalization condition. Therefore, in this case, no stationary or limiting or equilibrium distribution exists. Ergo, the chain is either transient or null recurrent. 

\begin{figure}[htbp]
  \small
  \centering
  \begin{tikzpicture}[
    state/.style={circle, draw, thick, minimum size=8.5mm, inner sep=0pt, font=\small},
    dots/.style={minimum size=7.5mm, inner sep=0pt},
    edge/.style={->, >=stealth, semithick},
    lab/.style={font=\scriptsize}
  ]
    \node[state] (a1) at (0.00,0) {$1$};
    \node[state] (a2) at (1.40,0) {$2$};
    \node[dots] (D) at (2.80,0) {$\cdots$};
    \node[state] (aN1) at (4.20,0) {$N\!-\!1$};
    \node[state] (aN) at (5.60,0) {$N$};
    \draw[edge] (a2) to[bend left=30] node[lab, auto] {$1-p$} (a1);
    \draw[edge] (a2) to[bend left=30] node[lab, auto] {$p$} (D);
    \draw[edge] (D) to[bend left=30] node[lab, auto] {$1-p$} (a2);
    \draw[edge] (D) to[bend left=30] node[lab, auto] {$p$} (aN1);
    \draw[edge] (aN1) to[bend left=30] node[lab, auto] {$1-p$} (D);
    \draw[edge] (aN1) to[bend left=30] node[lab, auto] {$p$} (aN);
    \draw[edge] (a1) to[out=150, in=210, looseness=6] node[lab, left] {$1$} (a1);
    \draw[edge] (aN) to[out=30, in=-30, looseness=6] node[lab, right] {$1$} (aN);
  \end{tikzpicture}
  \caption{\small Random walk with absorbing boundaries at both ends.}
  \label{fig:rw-absorbing}
\end{figure}
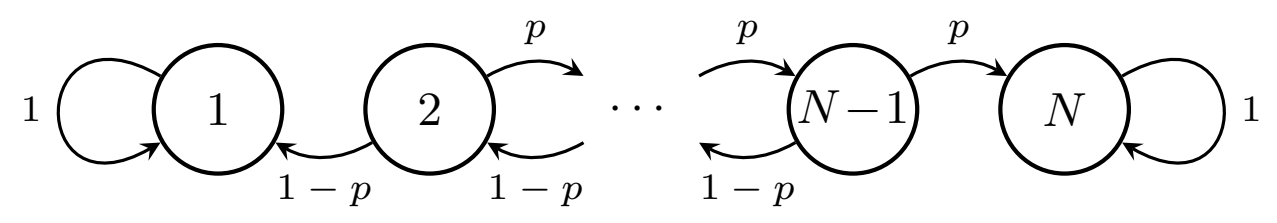
Thirdly, consider random walk with absorbing boundaries (Fig.~\ref{fig:rw-absorbing}). Its state space can be chosen as $\mathcal{X}=\{1,2,3,\cdots,N\}$. The transition matrix ${\sf P}$ is given by
\[
p_{ij}=\left\{
\begin{array}{ll}
 p & {\rm if}\, 1<i<N\,\&\, j=i+1,   \\
1-p  & {\rm if}\, 1<i<N\,\&\, j=i-1,    \\
1  & {\rm if}\, i=j=1\,{\rm or}\,N,   \\
 0 & {\rm otherwise}.   
\end{array}\right.
\]
Since $1\nleftrightarrow N$, it is not irreducible. In fact there are three equivalent classes: $\{1\}$, $\{N\}$, and $\{2,3,\cdots,N-1\}$---respectively absorbing (positive recurrent and aperiodic), absorbing (positive recurrent and aperiodic), and transient. Note that the absorbing states are trivially ergodic states but the chain is not ergodic. Furthermore, note that there are non-unique stationary distributions: e.g., $\boldsymbol{\pi}^*=[1\,0\,\cdots\,0\,0]$ and $\boldsymbol{\pi}^*=[0\,0\,\cdots\,0\,1]$ (in fact, any convex combination of them); starting from a transient state $k$, the chain converges to the mixture $h_k[1\,0\,\cdots\,0]+(1-h_k)[0\,\cdots\,0\,1]$, where $h_k$ is the probability of absorption at $1$, so which stationary distribution is reached depends on the initial state. We are witnessing a trivial illustration of a recurrent class's being closed.

\begin{figure}[htbp]
  \small
  \centering
  \begin{tikzpicture}[
    state/.style={circle, draw, thick, minimum size=8.5mm, inner sep=0pt, font=\small},
    dots/.style={minimum size=7.5mm, inner sep=0pt},
    edge/.style={->, >=stealth, semithick},
    lab/.style={font=\scriptsize}
  ]
    \node[state] (r1) at (90:2.1) {$1$};
    \node[state] (r2) at (30:2.1) {$2$};
    \node[state] (r3) at (-30:2.1) {$3$};
    \node[dots] (rD) at (-90:2.1) {$\cdots$};
    \node[state] (rN1) at (-150:2.1) {$N\!-\!1$};
    \node[state] (rN) at (150:2.1) {$N$};
    \draw[edge] (r1) to[bend left=20] node[lab, auto] {$p$} (r2);
    \draw[edge] (r2) to[bend left=20] node[lab, auto, inner sep=1pt] {$1-p$} (r1);
    \draw[edge] (r2) to[bend left=20] node[lab, auto] {$p$} (r3);
    \draw[edge] (r3) to[bend left=20] node[lab, auto, inner sep=1pt] {$1-p$} (r2);
    \draw[edge] (r3) to[bend left=20] node[lab, auto] {$p$} (rD);
    \draw[edge] (rD) to[bend left=20] node[lab, auto, inner sep=1pt] {$1-p$} (r3);
    \draw[edge] (rD) to[bend left=20] node[lab, auto] {$p$} (rN1);
    \draw[edge] (rN1) to[bend left=20] node[lab, auto, inner sep=1pt] {$1-p$} (rD);
    \draw[edge] (rN1) to[bend left=20] node[lab, auto] {$p$} (rN);
    \draw[edge] (rN) to[bend left=20] node[lab, auto, inner sep=1pt] {$1-p$} (rN1);
    \draw[edge] (rN) to[bend left=20] node[lab, auto] {$p$} (r1);
    \draw[edge] (r1) to[bend left=20] node[lab, auto, inner sep=1pt] {$1-p$} (rN);
  \end{tikzpicture}
  \caption{\small Random walk on a ring.}
  \label{fig:rw-ring}
\end{figure}
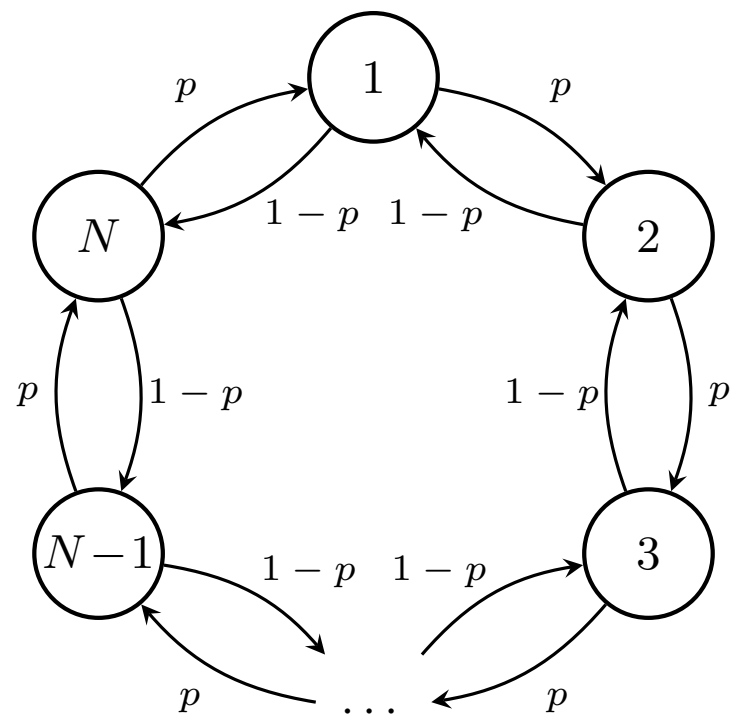
Finally, consider random walk on a ring (Fig.~\ref{fig:rw-ring}). Its state space can be chosen as $\mathcal{X}=\{1,2,3,\cdots,N\}$. The transition matrix ${\sf P}$ is given by
\[
p_{ij}=\left\{
\begin{array}{ll}
 p & {\rm if}\, j=i+1,   \\
  p & {\rm if}\, i=N\,\&\,j=1,   \\
1-p  & {\rm if}\, j=i-1,    \\
1-p  & {\rm if}\, i=1\,\&\,j=N,   \\
 0 & {\rm otherwise}.   
\end{array}\right.
\]
We note $\forall i,j$, $i\leftrightarrow j$; consequently, the chain is irreducible. For even $N$, every state is positive recurrent (the chain is finite) and periodic with period $2$. For odd $N$, every state is positive recurrent and aperiodic, and hence the chain is ergodic. Hence, unique stationary distribution, $\pi^*_i=1/N$ $\forall i$ exists for any $N$, but it is the limiting distribution only when $N$ is odd. For a stationary distribution, the detailed balance is satisfied only when $p=1/2$ because only in that case, $p_{ij}/N=p_{ji}/N$. Hence the ergodic random walk on a ring is reversible only if $p=1/2$ and hence, by definition, the equilibrium distribution exists in this case only.
%
\begin{center}
	$\symking\symqueen\symrook\symknight\symbishop\sympawn\epsdice{1}\,\epsdice{2}\,\epsdice{3}\,\epsdice{4}\,\epsdice{5}\,\epsdice{6}\,\spadesuit\heartsuit\clubsuit\diamondsuit$
\end{center}
\newpage
\newpage
\clearpage
\onecolumngrid
\section{Stochastic Calculus: A Co-evolution}
\epigraph{\textsf{The mathematics is not there till we put it there.}}{Arthur Eddington}
\twocolumngrid

\indent {\color{deepsea}{\LARGE \textgoth{T}}}he history of Brownian motion represents a seminal case of interdisciplinary co-evolution, where empirical anomalies in physics necessitated the development of entirely new mathematical paradigms. This intellectual trajectory was initiated by Einstein and Smoluchowski (1905--1906), who utilized statistical mechanics to transform the erratic kinematics of pollen grains into the first definitive proof of the atomic hypothesis. The subsequent experimental substantiation by Jean Perrin did more than validate the atomic theory; it exposed a fundamental mathematical pathology: the existence of continuous trajectories that were non-differentiable at every point. These jagged, self-similar paths represented a crisis for classical analysis, as they possessed infinite instantaneous velocity, rendering Newtonian heuristics insufficient for a rigorous description of microscopic motion.

The transition from ensemble probability densities to individual particle dynamics was pioneered by Paul Langevin (1908), who introduced the first stochastic differential equation (SDE) by augmenting Newton's Second Law with a fluctuating force term. While Langevin's trajectory-based approach provided a powerful physical engine, it operated in a mathematical vacuum, as the calculus required to handle such singular processes did not yet exist. This conceptual void was filled by the foundational axioms of Andrey Kolmogorov and the measure-theoretic framework of Norbert Wiener, which provided the necessary architecture to treat random paths as rigorous mathematical objects. The synthesis reached its analytical zenith with the stochastic calculus of Kiyosi It\^{o}. By internalizing the physical second-order effects---encapsulated in the identity $dW(t)^2 = dt$ ($W(t)$ being the Brownian--Wiener Process)---It\^{o} developed a specialized logic that successfully reconciled the ``pathological'' jaggedness of the physical world with the requirements of logical consistency.

Complementing this trajectory-based view, Adriaan Fokker and Max Planck developed the analytic counterpart to these fluctuations, demonstrating that the evolution of probability densities could be modelled through parabolic partial differential equations. This dualism between the Langevin SDE and the Fokker-Planck equation remains a cornerstone of modern statistical mechanics. Remarkably, the universality of these principles was prefigured in 1900 by Louis Bachelier, who identified these same dynamics within the {Paris Bourse}---the French stock exchange. Bachelier's treatment of financial `speculation' as a continuous-time random walk was eventually operationalized by Black, Scholes, and Merton, whose application of It\^{o} calculus to the pricing of {complex financial instruments} proved that the mathematical architecture of randomness is truly invariant. Ultimately, this convergence demonstrates that the same stochastic principles governing molecular entropy also dictate the systemic uncertainty of global economic systems.

Finally, it  is fascinating to realize that this co-evolution of physics and mathematics owes its origin to biology: While examining pollen grains of the wildflower \emph{Clarkia pulchella} suspended in water under a microscope, botanist Robert Brown in 1827 noted tiny sub-particles ($\sim$$1$--$2$ $\mu$m in size) inside the pollen performing continuous, irregular, jittery motion. While the initial hypotheses assumed this motion was a vital force unique to living matter, Brown systematically tested non-living specimens---including dried plant material, powdered glass, granite, coal dust, and soot---and observed the exact same random jittering across all materials. Although Brown could not explain the origin of the motion himself, his rigorous control experiments proved that the phenomenon was physical rather than biological.

\subsection{Brownian Motion \& Langevin Equation}
Brownian particle is modelled as a mesoscopic particle suspended in a fluid of significantly lighter, rapidly moving solvent molecules  which constitute a thermal bath at uniform temperature $T$. Assuming a Brownian particle of size $\sim 10^{-6}$ m with density of water molecule at room temperature, its velocity relaxation timescale---the characteristic duration over which a Brownian particle's initial momentum decays exponentially due to viscous drag---is  $\tau_d\sim 10^{-6}$ s. The individual molecular collisions occur on timescale $\tau_{\rm coll}\sim10^{-10}$.

To model the Brownian particle's motion, we assume that it is kicked by solvent particles by a time-dependent force, $\boldsymbol{\xi}(t)$ which is isotropic, with independent components ($\{\xi_i\}_{i=1,2,3}$) that are identically distributed and temporally uncorrelated for times much greater than $\tau_{\rm coll}$. So, we have
\begin{eqnarray}
\langle \xi_i(t)\rangle=0,\,\,\&\,\,\langle \xi_i(t)\xi_j(t')\rangle=\Gamma \delta_{ij}\delta(t-t'),\label{eq:mvarxi}
\end{eqnarray}
where $\Gamma$ is a constant. Note that over such a large timescale, on average there are many i.i.d. kicks, so it is reasonable to consider $\xi_i$ as Gaussian in line with the central limit theorem. Therefore, if $\xi_i(t)$ is considered an independent stochastic process, then it follows that $\xi_i(t)$ is a Gaussian white noise as is clear from above expression~(\ref{eq:mvarxi}). 

A remark is in order here: The \emph{Wiener--Khinchin theorem} proves that the \emph{power spectral density (PSD)}, which shows how the average power of a signal is distributed across different frequencies in the frequency domain, shares a direct Fourier transform relationship with the autocorrelation function. In the case of theoretical continuous white noise, the signal is completely uncorrelated with itself at any time shift other than zero, which collapses its autocorrelation into an infinitely tall, infinitely thin Dirac delta impulse. The Fourier transform of this delta impulse results in a perfectly flat PSD that stretches across all frequencies from negative infinity to positive infinity with constant amplitude. Because the total variance (or total average power) of a signal equals the mathematical integration of its PSD across the entire frequency spectrum, an ideal white noise signal cannot have a finite variance. Integrating a constant value over an infinite range yields infinite power, making ideal continuous white noise a useful mathematical abstraction rather than a physically realizable signal. Just to repeat from a different angle, we stress that if in white noise we enforce a finite variance or finite amplitude at the origin in the time domain, the area under its infinitely thin spike would shrink to zero, leading to a zero integral across the spectrum and resulting in a signal with absolutely zero power---effectively implying that there is no noise!

Finally, the \emph{Langevin equation} for the Brownian motion of particle of mass $m$ may be written as:
\begin{eqnarray}
m\dot{v}_i=-\gamma v_i+\xi_i(t)\label{eq:lle}.
\end{eqnarray}
This clearly describes a Markov process as noise term, $\xi_i(t)$, does not depend on past. Here, $\tau_d$ should be recognized as $m/\gamma$.

Motivated by this physical example, in more generality, a \emph{nonlinear Langevin equation} in a variable $x$ (need not mean `position'; it is abstract) is defined as the following form:
\begin{eqnarray}
\dot{x}=a(x,t)+b(x,t)\xi(t),\label{eq:nonlle}\qquad\qquad\\
\phantom{\dot{x}=}\left[\langle \xi(t)\rangle=0\,\,\&\,\,\langle \xi(t)\xi(t')\rangle=\delta(t-t')\right]\nonumber
\end{eqnarray}
where typically $a(x,t)$ and $b(x,t)$ are deterministic functions such that $a(x,t)$ is allowed to be nonlinear in $x$; $\xi(t)$ is Gaussian white noise. When $b$ is constant, the noise in the system is called \emph{additive}, otherwise the noise is deemed \emph{multiplicative}. If $a$ and $b$ both are time independent, then the equation with multiplicative noise can be transformed to one with additive noise by dividing the equation by $b$ and redefining variable and function. 

An intriguing conundrum presents itself: For $x$-dependent $b(x,t)$, which value of $x$ should be used to evaluate the noise amplitude, given that the stochastic kick due to $\xi(t)$ at $t=\tau$, say, will cause $x(t)$ just before and just after $\tau$ to be ``drastically different''? In a more visual but simpler delta-kick-train model of noise, a kick at \(t=\tau\) may produce a finite jump from $x_{\tau^-}$ to \(x_{\tau^+}\). For Gaussian white noise, by contrast, the trajectory $x(t)$ remains continuous, but its stochastic increment over $dt$ is known to be $\mathcal{O}(\sqrt{dt})$---vanishing as $dt\to0$, yet asymptotically larger than the deterministic $\mathcal{O}(dt)$ increment. This problem is not there in usual ordinary differential equations (consider, e.g., Eq.~(\ref{eq:nonlle}) with no noise term) because there variable $x$ changes smoothly and any value of $x$ chosen between $x$ and $x+\Delta x$ ($\Delta x\approx x+adt$) lead to unique determination of $a(x,t)$ in the equation (as second order in $dt$ terms vanish as $dt\rightarrow 0$). Since, in principle, the unbounded $\xi$ is uncorrelated with infinite variance, our desire to have $x$ as a continuous function seems to make no sense here. In other words, if we want both, then the nonlinear Langevin equation appears to be mathematically meaningless!

So our task is now to give the nonlinear Langevin equation a more acceptable mathematical form. To this end, let us focus on the case \emph{diffusive limit}---$t\gg \tau_d$---of the Brownian motion.This limit can be captured by ignoring $\dot{v}_i$ in comparison to $\gamma{v_i}$ in Eq.~\eqref{eq:lle}; setting all constants to unity for convenience, the position $x$ of the 1-dimensional Brownian particle is seen to satisfy:
\begin{eqnarray}
\dot{x}=\xi(t)\label{eq::dlbm}.
\end{eqnarray}
Let us formally write its solution as a stochastic process:
\begin{eqnarray}
{X}(t)=\int_0^t\xi(t')dt',\label{eq:dlbm}
\end{eqnarray}
assuming that the particle is invariable initiated at origin---$X(0)=0$.
We recall that $\xi$ is neither bounded nor almost everywhere continuous, and hence not Riemann-integrable---meaning the integral in Eq.~(\ref{eq:dlbm}) does not exist in usual common sense of integrals (area under the curve). So to give meaning to $X$ we need to give meaning to the integral.

Let us assume that the integral exists in some sense and $X(t)$ is continuous function of $t$. Therefore,
\begin{eqnarray}
&&{X}(t+\Delta t)=\lim_{\varepsilon\rightarrow0}\int_0^{t-\varepsilon}\xi(t')dt'+\int_t^{t+\Delta t}\xi(t')dt'\quad\nonumber\\
\implies&&dX(t)\equiv{X}(t+\Delta t)-{X}(t)=\int_t^{t+\Delta t}\xi(t')dt'.
\end{eqnarray}
It is immediately clear that $dX(t)$ should be Gaussian (if so is $\xi(t)$) with mean $\langle dX(t)\rangle=0$ and variance being
\begin{eqnarray}
&&\langle [dX(t)]^2\rangle=\int_t^{t+\Delta t}\int_t^{t+\Delta t}\langle\xi(t')\xi(t'')\rangle dt'dt''=\Delta t.\qquad
\end{eqnarray}
Also, for $t_1<t_2<t_3$, since $\xi(t)$ is uncorrelated:
\begin{eqnarray}
&&\langle[{X}(t_2)-{X}(t_1)][{X}(t_3)-{X}(t_2)]\rangle\nonumber\\
&&\phantom{\langle[{X}(t_2)-{X}(t_1)]}=\left\langle\left[\int_{t_1}^{t_2}\xi(t')dt'\right]\left[\int_{t_2}^{t_3}\xi(t')dt'\right]\right\rangle,\nonumber\\
&&\phantom{\langle[{X}(t_2)-{X}(t_1)]}=\left\langle\left[\int_{t_1}^{t_2}\xi(t')dt'\right]\right\rangle\left\langle\left[\int_{t_2}^{t_3}\xi(t')dt'\right]\right\rangle,\nonumber\\
&&\phantom{\langle[{X}(t_2)-{X}(t_1)]}=0.
\end{eqnarray}

Motivated by the above desiderata about $X(t)$, we introduce (one-dimensional) \emph{Brownian Process} or \emph{Wiener Process}, $W(t)$ ($t\in\mathcal{T}=\mathbb{R}_{\ge0}$; state space $\mathcal{X}=\mathbb{R}$) with following defining properties:
\begin{itemize}
\item $W(0) = 0$.
\item For $t < t'$: $W(t') - W(t) \sim \mathcal{N}(0, c(t'-t))$, $c$ being a proportionality constant.
\item Increments are independent.
\end{itemize}
When proportionality constant $c$ is unity, $W(t)$ is called \emph{standard} Wiener/Brownian process; unless otherwise specified, we shall assume $W(t)$ to be standard henceforth. Note that increments are, by construction, stationary. One can show that $W(t)$ exists, has continuous sample paths, and is non-differentiable almost everywhere. Furthermore, $W(t)$ is {continuous in probability}, i.e., $\forall\epsilon>0$ and $\forall t\in\mathcal{T}$, $P(|W(t+h)-W(t)|>\epsilon)\rightarrow0$ when $h\to0$. In fact, the Wiener process is a non-stationary Markov process. Markovianity is quickly seen as follow: For $t < t'$, the future increment $W(t') - W(t)$ is independent of the past history $\{W(u) : u \leq t\}$ by the independent increments property. Since $W(t') = W(t) + [W(t') - W(t)]$, the conditional distribution of $W(t')$ given the past depends only on $W(t)$, not on earlier values. For standard $W(t)$, the covariance is $\min(t,t')$ $\forall t,t'\in\mathcal{T}$---a signature of non-stationarity as made clear by the steps below:
\begin{eqnarray}
\langle W(t)W(t')\rangle=&&\langle W(t)[W(t)+W(t')-W(t)]\rangle\,\,(\textrm{say}, t'>t)\nonumber\\
=&&\langle W(t)W(t)\rangle+\langle W(t)[W(t')-W(t)]\rangle\nonumber\\
=&&t+\langle W(t)\rangle\langle[W(t')-W(t)]\rangle\nonumber\\
=&&t+0=\min(t,t')
\end{eqnarray}
[$\langle W(t)W(t')\rangle$ is, thus, not proportional to $t'-t$.] One can formally show that the Wiener process $W(t)$ is a Gaussian process because every finite collection of its random variables follows a multivariate Gaussian distribution which is completely specified by mean and covariance.

This mathematical construction of $W(t)$ and its properties appear rather 
intuitive when $W(t)$ is seen as a limit of a random walk which we now 
proceed to learn. A \emph{symmetric} 1D random walk is the random walk 
with equal probability for taking right or left step. Let it be denoted 
by $R_n$ with $n\in\mathcal{T}=\mathbb{Z}_{\ge0}$ and state space 
$\mathcal{X}=\mathbb{Z}$. It can be shown to converge to the standard 
Wiener process in the limit of very large $n$ and infinitesimal step size 
as follows: Consider a stochastic jump process, $Z_{\delta,\varepsilon}(t) \equiv \left\{\varepsilon R_{\lfloor t/\delta \rfloor} : 
t\in\mathbb{R}_{\ge0}\right\},$ with jumps of size $\pm\varepsilon$ at every $k\delta$ ($k\in\mathbb{Z}_{\ge0}$). 
It is obvious that $\mathbb{E}[Z_{\delta,\varepsilon}(t)]=0$ and 
$\mathbb{E}[Z_{\delta,\varepsilon}(t)^2]=\varepsilon^2\lfloor t/\delta \rfloor$ 
because $\mathbb{E}[R_n]=0$ and $\mathbb{E}[R_n^2]=n$. In the light of the central limit theorem, the choice of $\varepsilon=\sqrt{\delta}$ 
ensures that $Z_{\delta,\sqrt{\delta}}(t)\to\mathcal{N}(0,t)$ as $\delta\to 0$. 
It follows from the nature of the normal distribution that the increments of 
$Z_{\delta,\sqrt{\delta}}(t)|_{\delta\to 0}$ should be normally distributed 
as well. We also hope (without rigorous proof) that the property of stationary 
independent increments that $R_n$ is endowed with, owing to being the partial 
sum of i.i.d.\ random variables, carries over to the continuous time process 
in the limiting process discussed above. Furthermore, by construction, 
$Z_{\delta,\sqrt{\delta}}(0)|_{\delta\to 0}$ is obviously zero. In conclusion, 
$Z_{\delta,\sqrt{\delta}}(t)|_{\delta\to 0}$ is the standard Wiener process.

Finally, we understand that with Wiener process in mind, Eq.~(\ref{eq:nonlle}) is better cast, with $dW\equiv W(t+dt)-W(t)=\xi dt$, as
\begin{equation}
dx(t)=a(x,t)dt+b(x,t)dW(t),\label{eq:sdeini}
\end{equation}
which can be formally solved to the form
\begin{equation}
x(t)-x(0)=\int_{0}^ta[x(t'),t']dt'+\int_{0}^tb[x(t'),t')dW(t').
\end{equation}
The remaining task is now to assign a consistent mathematical meaning to the integral $\int_0^tb[x(t'),t')dW(t')$ which is a kind of stochastic version of Riemann--Stieltjes integral.
\subsection{Stochastic Integrals}
Consider a more general function $F(x,t)$ which need not be only deterministic. Let us define the stochastic integral 
\begin{equation}
I(t)\equiv\int_{t_0}^t F[x(t'),t'] dW(t')\label{eq:stoi}
\end{equation}
 as a limit (to be precisely defined later) of a Riemann--Stieltjes sum over a partition $t_0 < t_1 < \dots < t_n = t$:
\begin{equation}
I_n = \sum_{i=1}^n F[x(\tau_i), \tau_i] \Delta W_i,
\end{equation}
where $\Delta W_i\equiv W(t_i)-W(t_{i-1})$ and the sampling point $\tau_i$ is chosen such that $t_{i-1} \le \tau_i \le t_i$. In classical calculus, for continuous integrands, the choice of $\tau_i$ is irrelevant. However, the non-differentiable nature of Brownian motion introduces a fundamental pathology. Note, additionally, that $W(t)$ being random, $I_n$ is random as well and hence it makes practical sense only when thought of as some sort of average. 

Let us highlight the above issues through an illustrative choice of $F = W$. The expectation of the approximating sum is:
\begin{align}
\langle I_n \rangle &= \left\langle \sum_{i=1}^n W(\tau_i) [W(t_i) - W(t_{i-1})] \right\rangle \nonumber \\
&= \sum_{i=1}^n \left[ \langle W(\tau_i)W(t_i) \rangle - \langle W(\tau_i)W(t_{i-1}) \rangle \right].
\end{align}
Recalling the covariance $\langle W(t')W(t) \rangle = \min(t', t)$, and given $t_{i-1} \le \tau_i \le t_i$, we have:
\begin{equation}
\langle I_n \rangle = \sum_{i=1}^n [\min(\tau_i, t_i) - \min(\tau_i, t_{i-1})] = \sum_{i=1}^n [\tau_i - t_{i-1}].
\end{equation}
Parameterizing the evaluation point as $\tau_i = c t_i + (1-c)t_{i-1}$ for $c \in [0, 1]$, we find:
\begin{eqnarray}
&&\langle I_n \rangle = \sum_{i=1}^n [c t_i + (1-c)t_{i-1} - t_{i-1}] = c \sum_{i=1}^n (t_i - t_{i-1})\nonumber\\
\implies&&\langle I_n \rangle = c(t - t_0).
\end{eqnarray}
The result clearly depends explicitly on $c$, implying that the stochastic integral is ill-defined without a convention.

Finally, $I(t)$ has to be defined as a limit when $n\rightarrow\infty$. This has to be done carefully as now we are dealing with random functions. In fact, note that $I(t)$ is, technically, a stochastic process. In the theory of stochastic calculus, the limit $I_n \to I$ is mostly taken in the \emph{mean-square (ms)} sense:
\begin{eqnarray}
&&\underset{n \to \infty}{\text{ms-lim}} I_n(t)= I(t)\implies\nonumber\\
&&\lim_{n \to \infty} \langle [I_n(t) - I(t)]^2 \rangle =\lim_{n \to \infty} \int_{s\in\mathcal{S}}[I_n(s,t) - I(s,t)]^2 dP(s)\nonumber\\
&&\phantom{\lim_{n \to \infty} \langle [I_n(t) - I(t)]^2 \rangle }= 0,
\end{eqnarray}
where \(\mathcal{S}\) is the sample space and the integral is taken with respect to the probability measure \(P\).  Note an individual sample point \(s\) is not a number here; it is an entire, complete, continuous timeline (sample path) of a Wiener process from starting time \(t_{0}\) to the final time \(t\). The mean-square convergence is a strong convergence requirement as it implies both stochastic convergence [${\rm Prob}(|I_n(t) - I(t)| > \epsilon) \to 0$] and convergence in distribution (cdf's of $I_n$ and $I$ become exactly the same as $n\rightarrow \infty$).

So, in general, it appears that we should define:
\begin{equation}
I(t) \equiv \underset{n \to \infty}{\text{ms-lim}} \sum_{i=1}^n F[x(c t_i + (1-c)t_{i-1}), c t_i + (1-c)t_{i-1}] \Delta W_i.\label{eq:sdec}
\end{equation}
Three primary conventions to interpret the stochastic integral are used to resolve the rather general ambiguities showcased by the example above:
\begin{enumerate}
\item \emph{It\^{o}} ($c=0$): Evaluated at the pre-point, $\tau_i = t_{i-1}$. 
\begin{equation}
I(t) \equiv \underset{n \to \infty}{\text{ms-lim}} \sum_{i=1}^n F[x(t_{i-1}), t_{i-1}] \Delta W_i.
\end{equation}
The {It\^o convention} is foundational in pure mathematics and theoretical probability due to its martingale properties (the future value of $I(t) $ on average equals its exact present value), which simplify many proofs and enable a rigorous theory of stochastic calculus. It is standard in mathematical finance for modelling asset prices, where the non-anticipative nature of the integral---using information up to time $t_{i-1}$---aligns with the requirement that trading strategies cannot see into the future. It\^o 's choice also yields the zero expectation property for It\^o  integrals with respect to Brownian motion, a clean feature for deriving stochastic differential equations and Fokker--Planck equations, though the latter acquire an additional spurious drift term when the noise is multiplicative.
\item \emph{Stratonovich} ($c=1/2$): Evaluated at the mid-point.
\begin{align}
I(t) \equiv \underset{n \to \infty}{\text{ms-lim}} \sum_{i=1}^n F\left[x\left(\frac{t_i+t_{i-1}}{2}\right), \frac{t_i+t_{i-1}}{2}\right] \Delta W_i.\label{eq:sic1}
\end{align}
The {Stratonovich convention} emerges naturally in physical systems with smooth, real noise limits, such as diffusion processes derived from Hamiltonian dynamics with fast variables or in the modelling of inertial particles in turbulence. Its chain rule matches that of ordinary calculus, making it convenient for coordinate transformations and for systems where noise approximates a physical process with finite correlation time. Stratonovich integrals are widely used in engineering, control theory, and nonequilibrium statistical mechanics, particularly when analyzing stochastic differential equations on manifolds or in the presence of conserved quantities, because they preserve geometric structure and obey the classical chain rule without extra correction terms.

It should be pointed out that Stratonovich, however, defined $I(t)$ as
\begin{equation}
\quad I(t) \equiv \underset{n \to \infty}{\text{ms-lim}}\sum_{i=1}^n F\left[\frac{x(t_i)+x(t_{i-1})}{2}, t_{i-1}\right] \Delta W_i,\label{eq:sic2}
\end{equation}
where if $F(x, t)$ is differentiable in $t$, then $I(t) $ is known to be independent of the specific choice of value for $\tau_i\in[t_{i-1},t_i]$. More importantly, definition in  Eq.~(\ref{eq:sic2}) need not be same as Eq.~(\ref{eq:sic1}); they, however, would be same if $F(x,t)$ is additionally smooth in $x$.
\item \emph{H\"{a}nggi--Klimontovich} ($c=1$): Evaluated at the post-point, $\tau_i = t_{i}$.
\begin{equation}
I(t) \equiv \underset{n \to \infty}{\text{ms-lim}} \sum_{i=1}^n F[x(t_{i}), t_{i}] \Delta W_i.
\end{equation}
The {H\"anggi--Klimontovich convention}---also known a isothermal convention---is prominent in kinetic theory and statistical physics for ensuring detailed balance in thermal equilibrium. By evaluating the noise term at the post-point, this convention guarantees that the steady-state distribution of a Langevin system with multiplicative noise is the Boltzmann distribution, provided the fluctuation--dissipation relation holds. It is essential in modelling thermal baths and in simulations of molecular dynamics where thermodynamic consistency is paramount. The convention also appears in certain field-theoretic approaches to stochastic quantization and in the study of active matter, where it helps maintain a consistent entropy production framework under coordinate changes.
\end{enumerate}


\subsection{It\^o Integral}
Now let us elaborate on It\^o integrals---the stochastic integrals with It\^o convention imposed and know some of its useful core results. In such a case, it can be proven that Eq.~(\ref{eq:stoi})  exists if $F(t')$ is continuous, \emph{nonanticipating}, and bounded in $L^2$ on the interval $[t_0,t]$. Nonanticipating property of a function is a rather intuitive condition imposed to make sure that future does not affect the past: $F(t)$ is said to be nonanticipating if $F(t)$ is statistically independent of $W(t') - W(t)$ $\forall t', t$ such that $t' > t$. By definition, $W(t)$ and any deterministic function are such functions. With nonanticipating $F(t)$ and deterministic $f$,  $\int_{t_0}^t f(F(t')) dt'$ and $\int_{t_0}^t f(F(t')) dW(t')$ are obviously nonanticipating too---the last one is so because the It\^o integral only involves integrand for $t'<t$ and $W(t')$ for $t'\le t$. Henceforth, unless otherwise specified, we shall work only with nonanticipating functions  that are continuous and bounded in $L^2$.

The very first important result follows:
\begin{align}
\left\langle\int_{t_0}^t F[x(t'),t'] dW(t')\right\rangle=0,\label{eq:stoia}
\end{align}
since ms-lim and expectation operations can be swapped in the definition of the integral, and $F(t_{i-1})$, by the nonanticipatory property, is independent of $\Delta W_i$. 

Next, one can deduce the so-called \emph{generalized It\^o isometry} or \emph{bilinear It\^o isometry}:
\begin{align}
\left\langle \int_{t_0}^t F(t') dW(t') \int_{t_0}^t G(t') dW(t') \right\rangle = \int_{t_0}^t \langle F(t') G(t') \rangle dt',\label{eq:ii}
\end{align}
which is called \emph{It\^o isometry} when $F=G$. The proof of Eq.~(\ref{eq:ii}) follows below.

Consider a partition $t_0<t_1<\dots<t_n=t$ with $\Delta W_i=W(t_i)-W(t_{i-1})$. We introduce shorthand notations: $F_{i-1}=F(t_{i-1})$, $G_{i-1}=G(t_{i-1})$. We have
\begin{align}
\Big\langle \sum_{i=1}^n F_{i-1} \Delta W_i &\cdot \sum_{j=1}^n G_{j-1} \Delta W_j \Big\rangle \nonumber \\
&= \Big\langle \sum_{i,j=1}^n F_{i-1} G_{j-1} \Delta W_i \Delta W_j \Big\rangle\nonumber\\
&= \Big\langle \sum_{i=1}^n F_{i-1} G_{i-1} (\Delta W_i)^2 \Big\rangle \nonumber \\
&\quad + \Big\langle \sum_{i\neq j} F_{i-1} G_{j-1} \Delta W_i \Delta W_j \Big\rangle,\label{eq:ii1}
\end{align}
where we separate diagonal ($i=j$) and off-diagonal ($i\neq j$) terms. The off-diagonal term can be evaluated as follows:
\begin{align}
&\Big\langle \sum_{i\neq j} F_{i-1} G_{j-1} \Delta W_j \Delta W_i \Big\rangle\nonumber\\
&\phantom{xxxxxx}=\Big\langle \sum_{i> j} (F_{i-1} G_{j-1}+F_{j-1} G_{i-1}) \Delta W_i \Delta W_j \Big\rangle\nonumber\\
&\phantom{xxxxxx}=\sum_{i> j}\Big\langle (F_{i-1} G_{j-1}+F_{j-1} G_{i-1}) \Delta W_j\Big\rangle \Big\langle  \Delta W_i \Big\rangle\nonumber\\
&\phantom{xxxxxx}= 0.\label{eq:ii2}
\end{align}

For $i=j$, using $\mathbb{E}[(\Delta W_i)^2] = \Delta t_i$,
\begin{align}
\langle F_{i-1} G_{i-1} (\Delta W_i)^2 \rangle 
&= \langle F_{i-1} G_{i-1} \rangle \Delta t_i.\label{eq:ii3}
\end{align}
As $\max_i \Delta t_i \to 0$, the sum converges:
\begin{align}
\sum_{i=1}^n \langle F_{i-1} G_{i-1} \rangle \Delta t_i 
&\to \int_{t_0}^t \langle F(t') G(t') \rangle \, dt'. \label{eq:ii4}
\end{align}

In conclusion, Eq.~(\ref{eq:ii1}), Eq.~(\ref{eq:ii2}), Eq.~(\ref{eq:ii3}), and Eq.~(\ref{eq:ii4}), together combine to yield the aforementioned (generalized) It\^o isometry [Eq.~(\ref{eq:ii})] because additionally mean-square convergence preserves inner products (recall, expectation defines inner product), i.e., 
$$\underset{n \to \infty}{\text{ms-lim}}\mathbb{E}[X_n Y_n] = \mathbb{E}\left[ \left(\underset{n \to \infty}{\text{ms-lim}} X_n\right) \left(\underset{n \to \infty}{\text{ms-lim}} Y_n\right) \right]
$$
for any from sequence of random variables---$\{X_n\}$ and $\{Y_n\}$.

It\^o isometry fixes a convention for $\delta$-function integration where argument of the function is either the lower or the upper limit of the integration. To this end, recall in the discussion of Zwanzig harmonic bath model and Langevin equation where rather than dealing with the more technical $dW(t)$, we just use heuristic $\xi(t) dt$ (with $\langle\xi(t) \xi(t') \rangle=\delta(t - t')$). Restricting $F$ and $G$ to be  deterministic, consider
\begin{align}
&\left\langle \int_{t_0}^t F(t') dW(t') \int_{t_0}^{t'} G(t'') dW(t'') \right\rangle\nonumber\\
=&\int_{t'=t_0}^t \int_{t''=t_0}^{t'} \left\langle F(t') G(t'')  \right\rangle \delta(t' - t'')dt'' dt'.
\end{align}
The L.H.S. of this equation is zero because of Eq.~(\ref{eq:stoia})---note that $F(t') \int_{t_0}^{t'} G(t'') dW(t'')$ is nonanticipating function. Therefore, the R.H.S.---which must be zero---forces to adopt the convention that for arbitrary $H$, 
\begin{align}
\int_{t_0}^{t_{\rm upper}} H(t) \delta(t - t_{\rm upper})dt=0.\label{eq:upperl}
\end{align}
For the convention corresponding to the lower limit, let us note that from Eq.~(\ref{eq:ii}):
\begin{align}
&\int_{t_0}^t \langle F(t'') G(t'') \rangle dt''\nonumber\\
=&\left\langle \int_{t_0}^t F(t') dW(t') \int_{t_0}^t G(t') dW(t') \right\rangle \nonumber\\
=&\left\langle\int_{t_0}^t \int_{t_0}^t F(t') G(t'') \xi(t') \xi(t'') \right\rangle dt' dt'' \nonumber\\
=&\int_{t_0}^t \int_{t_0}^t \left\langle F(t') G(t'')  \right\rangle \delta(t' - t'')dt' dt''  \nonumber\\
=& \left[\int_{t''=t_0}^t \int_{t_0}^{t''}+\int_{t''=t_0}^t \int_{t''}^t\right]\left\langle F(t') G(t'')  \right\rangle \delta(t' - t'')dt' dt''\nonumber\\
=& \int_{t''=t_0}^t \int_{t''}^t\left\langle F(t') G(t'')  \right\rangle \delta(t' - t'')dt' dt''.
\end{align}
because due to Eq.~(\ref{eq:upperl}), the integral corresponding to $\int_{t''=t_0}^t \int_{t_0}^{t''}$ vanish. Of course, then we must conclude that if Eq.~(\ref{eq:upperl}) were to hold, the following complimentary convention that for arbitrary $H$ must also be adopted:
\begin{align}
\int_{t_{\rm lower}}^{t} H(t') \delta(t' - t_{\rm lower})dt'=H(t_{\rm lower}).\label{eq:lowerl}
\end{align}

In contrast, in the Stratonovich interpretation, the Dirac delta at an integration endpoint is treated symmetrically by replacing it with a symmetric, sharply peaked regularized distribution. In the limit where the peak becomes infinitely narrow, exactly half of its total weight lies inside the integration interval and half lies outside. Therefore, whether the endpoint is the upper or lower limit, the delta-function integral contributes one-half of the function evaluated at that endpoint: 
\begin{align}
\int_{t_0}^{t_*} H(t) \delta(t - t_*)dt=\int_{t_*}^{t} H(t') \delta(t' - t_*)dt'=\frac{1}{2}H\left(t_*\right).\label{eq:mid1/2}
\end{align}
This endpoint prescription reflects the midpoint rule underlying Stratonovich stochastic integration.

An extremely useful result in It\^o calculus is as follows: For an arbitrary nonanticipating function $F(t')$, the mean-square limit results for It\^o integrals involving higher integral powers of $[dW(t')]$ and the product $dW(t')dt'$ are defined as follows:
\begin{align}
&\int_{t_0}^t F(t') [dW(t')]^m = \begin{cases}\int_{t_0}^t F(t') \, dt', & \text{for } m = 2, \\ 0, & \text{for } m > 2, \end{cases}\\
&\int_{t_0}^t F(t') \, dW(t') dt' = 0.
\end{align}
So essentially, we are arming ourselves with the heuristics: 
\begin{eqnarray}
[dW(t)]^{2} \equiv dt;\, [dW(t)]^{m>2} \equiv 0\,;\, dW(t)dt \equiv 0.\qquad\label{eq:heuristics}
 \end{eqnarray}
Basically, we are observing that $dW(t)\sim\sqrt{dt}$; hence, any integral with power of $dt$ more than unity vanishes. As an example, let us use it to calculate $\int_{t_0}^{t}W(t')^ndW(t')$. Let's expand $d[W(t)^{n+1}]$ using the binomial theorem:
\begin{align}
d[W(t)]^{n+1} &= [W(t) + dW(t)]^{n+1} - W(t)^{n+1}\nonumber\\
 &= \sum_{m=1}^{n+1}{}^{n+1}C_mW(t)^{n+1-m} [dW(t)]^m\nonumber\\
&= (n+1) W(t)^n dW(t) + \frac{(n+1)n}{2} W(t)^{n-1} dt,
\end{align}
where to arrive at the last equality, we have used the above heuristics (\ref{eq:heuristics}). Finally, integrating both sides from $t_0$ to $t$ and rearranging  yields:
\begin{align}
&\int_{t_0}^t W(t')^n dW(t') \nonumber\\
=& \frac{1}{n+1} \left[ W(t)^{n+1} - W(t_0)^{n+1} \right] - \frac{n}{2} \int_{t_0}^t W(t')^{n-1} dt'.
\end{align}

\subsection{Stochastic Differential Equation}
\emph{Stochastic differential equation (SDE)} is a differential equation where one or more of the terms are driven by a stochastic (random) process, making the solution itself a stochastic process. Mathematically, a completely general SDE can be defined without restricting the noise to a Wiener process. However, here we shall concentrate only on the one-dimensional SDE that restricts the general noise term strictly to a Wiener process:
\begin{align}
dX(t)=a[X(t),t]dt+b[X(t),t]dW(t), 
\end{align}
where by using capital $X$ (and not small-case $x$), we emphasize that the equation looks for solution of the stochastic process $X(t)$; although it is common among non-mathematicians to use small-case $x$ (realization of $X$) in the SDE as we shall also do [see Eq.~(\ref{eq:sdeini})] whenever resorting to $X$ is not explicitly necessary. While solving the SDE, if we resort to It\^o interpretation, it is called an \emph{It\^o SDE}; if we resort to Stratonovich interpretation, it is called an \emph{Stratonovich SDE}. The \emph{drift coefficient} $a(x,t)$ encapsulates deterministic trend whereas $b(x,t)$ is \emph{diffusion coefficient} denoting the noise intensity. A note of caution: when reading biology-heavy literature, one may find the term `drift' associate to diffusion.

Setting aside the questions about existence and uniqueness of the solution of It\^o SDE, we shall assume the solution to be so henceforth. However, let us prove that the solution is Markov: For $t' < t$, the solution 
\[
X(t) = X(t') + \int_{t'}^t a(X(u),u)du + \int_{t'}^t b(X(u),u)dW(u).
\]
The evolution from time $t'$ to $t$ depends on: the current state $X(t')$ and future increments $\{dW(u) : t' < u \leq t\}$. Since the Wiener process has independent increments, $W(t) - W(t')$ is independent of all information available at time $t'$. Therefore, given $X(t')$, the future value $X(t)$ is independent of the past history $\{X(u) : u < t'\}$ because, for any interval---technically, any Borel set---$\mathcal{B}$ of the state space,
\begin{align*}
&P[X(t) \in \mathcal{B}\mid X(t'), \text{past}] = \\
&\phantom{xxxxxxx} P\left[\left(X(t')+ \int_{t'}^t a(X(t''),t'')dt''+\right.\right. \\
&\phantom{xxxxxxxxx} \left.\left.\int_{t'}^t  b(X(t''),t'')dW(t'')\right)\in \mathcal{B}\bigg| X(t'), \text{past}\right]\\
&\phantom{xxxxxxx}= P[X(t) \in \mathcal{B}\mid X(t')].
\end{align*}
The equality holds since the random integral depends only on independent future increments. Thus, we have arrived at the {Markov property}: the future is conditionally independent of the past given the present.

One of the most fundamental results in It\^o calculus is the \emph{It\^o formula/rule/lemma} which finds what It\^o SDE a smooth function of $x$ satisfies, when $x$ itself satisfies the It\^o SDE: $dx = a dt + b dW$, where we suppress the arguments for notational convenience. We apply Taylor expansion to $df(x(t))$:
\begin{align*}
df(x(t)) &= f(x + dx) - f(x) \\
&= f(x) + f'(x) \, dx + \frac{1}{2} f''(x) \, dx^2 + \mathcal{O}(dx^3) - f(x) \\
&= f'(x) \, dx + \frac{1}{2} f''(x) \, dx^2,
\end{align*}
where we substitute $dx = a \, dt + b \, dW$ and
\begin{align*}
dx^2 &= (a \, dt + b \, dW)^2= a^2 \, dt^2 + 2ab \, dt \, dW + b^2 \, dW^2,
\end{align*}
to finally arrive at [on rearranging and recalling Eq.~(\ref{eq:heuristics})]:
\begin{align}
df(x) = \left[af'(x) + \frac{1}{2}b^2 f''(x)\right] dt + f'(x)b \, dW.\label{eq:ifor}
\end{align}
This is the It\^o formula. Evidently, the normal chain rule of ordinary calculus does not work here. In summary, if $x(t)$ satisfies an It\^o SDE, then $f(x(t))$ satisfies an It\^o SDE with $af'(x) + \frac{1}{2}b^2 f''(x)$ as the drift component and $f'(x)b$ as the diffusion component. {The extra term} $\frac{1}{2}b^2 f''(x)$ is the {stochastic correction}---it arises because $dW^2 = dt$ in It\^o calculus, unlike classical analysis.

One of the most useful fact is that although Stratonovich and It\^o SDEs are not identical and they follow different rules, to every Stratonovich SDE, there correspond an It\^o SDE (and vice versa) such that both give formally the same solutions. To show this let us consider an {It\^o SDE}:
\begin{equation}
dx = a(x,t)\,dt + b(x,t)\,dW,\label{eq:eto}
\end{equation}
and seek an equivalent {Stratonovich SDE}:
\begin{equation}
dx = \alpha(x,t)\,dt + \beta(x,t)\circ dW,\label{eq:strateno}
\end{equation}
where $\circ$ denotes Stratonovich interpretation should be adopted.

Consider partition $t_0 < t_1 < \cdots < t_n = t$ along with the shorthand notations: $\Delta t_i \equiv t_i - t_{i-1}$, $\Delta W_i\equiv W(t_i) - W(t_{i-1})$, and $x_i\equiv x(t_i)$. The Stratonovich integral is:
\begin{equation}
\int_{t_0}^{t} \beta(x,t')\circ dW(t')\equiv \underset{n \to \infty}{\text{ms-lim}} \sum_{i=1}^n \beta\!\left(\frac{x_{i-1}+x_i}{2},t_{i-1}\right)\Delta W_i,\label{eq:betas}
\end{equation}
as per Eq.~(\ref{eq:sic2})

In the It\^o interpretation, by ignoring higher order terms, the increment $\Delta x_i(\equiv x_i-x_{i-1}$) is
\begin{equation}
\Delta x_i = a_{i-1}\Delta t_i + b_{i-1}\Delta W_i,\label{eq:delxito}
\end{equation}
with shorthand: $a_{i-1} \equiv a(x_{i-1}, t_{i-1})$ and $b_{i-1}\equiv b(x_{i-1}, t_{i-1})$. Using $x_i$ as $x_{i-1}+\Delta x_i$ in the argument of $\beta$ in Eq.~(\ref{eq:betas}), we arrive at 
\begin{equation}
\beta(x_{i-1/2},t_{i-1}) = \beta_{i-1} + \frac{1}{2}\beta'_{i-1}\Delta x_i + \frac{1}{8}\beta''_{i-1}(\Delta x_i)^2 + \cdots
\end{equation}
where we have set midpoint $x_{i-1/2} \equiv x_{i-1} + \frac{1}{2}\Delta x_i$ and adopted the obvious convention that a subscript $i-1$ means that the corresponding function (e.g., $\beta_{i-1}$) is being evaluated at $(x_{i-1},t_{i-1})$. Subsequently, putting Eq.~(\ref{eq:delxito}) in the above expansion and putting that in Eq.~(\ref{eq:betas}), we arrive at 
\begin{equation}
\int_{t_0}^{t} \beta(x,t')\circ dW(t')= \int_{t_0}^{t} \beta\,dW + \frac{1}{2}\int_{t_0}^{t} \beta' b\,dt,
\end{equation}
where we have made use of Eq.~(\ref{eq:heuristics}) and dropped all term that of higher order in $\Delta t_i$---we keep $dW_i$, $\Delta t_i$ and $\Delta W_i^2$ but dropped $dt_idW_i$ and $dW_i^2$. Note here the first integral is It\^o, and the second is a correction term---noise-induced drift term (the \emph{spurious drift}).

In the light of the above integral, Stratonovich SDE can be cast as:
\begin{equation}
dx = \alpha\,dt + \beta\circ dW = \left[\alpha + \frac{1}{2}\beta'b\right]dt + \beta\,dW.
\end{equation}
Comparing it with Eq.~(\ref{eq:eto}), we conclude that Stratonovich SDE (\ref{eq:strateno}) is equivalent to an It\^o SDE with
\begin{align}
 b=\beta\,\,\textrm {and}\,\,a= \alpha + \frac{1}{2}\beta'b= \alpha + \frac{1}{2}\beta'\beta.\label{eq:s2i}
\end{align}
Conversely, It\^o SDE (\ref{eq:eto}) is equivalent to a Stratonovich SDE (\ref{eq:strateno}) with
\begin{align*}
 \beta=b\,\,\textrm {and}\,\,\alpha= a-\frac{1}{2}b'b.
\end{align*}
Therefore, for our limited pedagogical purpose, whether it is Stratonovich or It\^o interpretation for an SDE, the mathematical tools learnt to deal with It\^o SDEs is enough to deal with both.

In fact, for different stochastic integral conventions [recall $c$ in Eq.~(\ref{eq:sdec})] that follow different calculus rules, there exists a conversion formula between them such that all give formally the same solutions. In the way similar to what led us at the correspondence between Stratonovich or It\^o interpretations above, one can show that if one considers a general SDE with parameter $c \in [0,1]$:
\begin{equation}
dx = a_c(x,t)\,dt + b_c(x,t)\,dW_c,\label{eq:general_sde}
\end{equation}
where $dW_c$ denotes integration over it should to be done with evaluation point parameterized by $c$, then for conversion between any two conventions---$c=c_1$ and $c=c_2$---is obtained through following relations:
\begin{equation}
b_{c_2}=b_{c_1}=b \,({\textrm {say}})\,\,\textrm {and}\,\,a_{c_2} = a_{c_1} + (c_1 - c_2)b'b.\label{eq:170}
\end{equation}

Of course, a Stratonovich SDE and an It\^o SDE with exactly same drift and diffusion terms will give, in general, completely different solutions. Let us illustrate this by using the example of  \emph{Geometric Brownian Motion} (GBM):
\begin{equation}
dx = \mu x\,dt + \sigma x\,dW(t), \qquad x(t_0)=x_0>0,
\label{eq:gbm_sde}
\end{equation}
where $\mu$ and $\sigma$ are constants (not to be confused with mean and standard deviation); the SDE is It\^o as notationally clear (absence of `$\circ$'). As It\^o SDE, it is also defined to describe \emph{Economic Brownian Motion}.

Let $y=\ln x$. By It\^o's lemma:
\begin{align}
dy = \frac{1}{x}dx - \frac{1}{2x^2}(dx)^2&= \mu dt + \sigma dW - \frac{\sigma^2}{2}dt \nonumber \\
    &= \left(\mu - \frac{\sigma^2}{2}\right)dt + \sigma dW.
\end{align}
Integrating from $t_0$ to $t$:
\begin{equation}
y(t) = y(t_0) + \left(\mu - \frac{\sigma^2}{2}\right)(t-t_0) + \sigma[W(t)-W(t_0)].
\end{equation}
Thus,
\begin{equation}
x(t) = x_0\exp\!\left[\left(\mu - \frac{\sigma^2}{2}\right)(t-t_0) + \sigma\Delta W\right],
\label{eq:ito_sol}
\end{equation}
where $\Delta W \equiv W(t)-W(t_0)$.

Now we can calculate some moments of this stochastic process. We shall find it handy to note that $\mathbb{E}[e^{\sigma\Delta W}] = e^{\sigma^2(t-t_0)/2}$ which follows from Eq.~(\ref{eq:mgfg0}) since $\Delta W \sim \mathcal{N}(0,t-t_0)$. Equivalently, the fact is that if $y$ is a Gaussian random variable, then $\mathbb{E}[\exp(y)]=\exp({\mathbb{E}[y^2]}/{2})$. Using this fact about the Gaussian variable, first, we find
\begin{equation}
\mathbb{E}[x(t)] = x_0 e^{\mu(t-t_0)}.\label{eq:mgbmi}
\end{equation}
For $t' \geq t \geq t_0$, without any loss of generality, the autocorrelation turns out to be
\begin{align}
\mathbb{E}[x(t)x(t' )] &= x_0^2 \exp\!\Big[2\mu(t-t_0) + \left(\mu - \tfrac{\sigma^2}{2}\right)(t' -t) \nonumber \\
&\quad + \tfrac{\sigma^2}{2}\big(2(t-t_0) + (t' -t) + 2(t-t_0)\big)\Big] \nonumber \\
&= x_0^2 e^{\mu[(t-t_0)+(t' -t_0)] + \sigma^2(t-t_0)}.\label{eq:ito_corr}
\end{align}
Using Eq.~(\ref{eq:mgbmi}) and Eq.~(\ref{eq:ito_corr}), variance is then found to be
\begin{align}
\mathrm{Var}[x(t)] &= x_0^2 e^{2\mu(t-t_0)}\left(e^{\sigma^2(t-t_0)} - 1\right).
\end{align}

Now consider the geometric Brownian motion under Stratonovich interpretation:
\begin{equation}
dx = \mu x\,dt + \sigma x\circ dW(t), \qquad x(t_0)=x_0>0.
\label{eq:gbm_sde}
\end{equation}
From  expressions in (\ref{eq:s2i}), the equivalent It\^o form is obviously:
\begin{equation}
dx = \left(\mu + \frac{\sigma^2}{2}\right)x\,dt + \sigma x\,dW(t).
\label{eq:strat_to_ito}
\end{equation}
Integrating leads to
\begin{equation}
x(t) = x_0\exp\!\left[\mu(t-t_0) + \sigma\Delta W\right].
\label{eq:strat_sol}
\end{equation}

Again, as before, the mean, the autocorrelation, and the variance are found, respectively, as follows:
\begin{align}
\mathbb{E}[x(t)] &= x_0 e^{(\mu + \sigma^2/2)(t-t_0)},\\
\mathbb{E}[x(t)x(t' )] &= x_0^2 e^{\mu[(t-t_0)+(t' -t_0)]} \nonumber \\
&\quad \times \mathbb{E}[e^{\sigma(W(t)+W(t' )-2W(t_0))}] \,\,(t'  \geq t \geq t_0)\nonumber \\
=& x_0^2 e^{\mu[(t-t_0)+(t' -t_0)] + \frac{\sigma^2}{2}[t+t' -2t_0 + 2(t-t_0)]},
\label{eq:strat_corr}\\
\mathrm{Var}[x(t)] &= x_0^2 e^{(2\mu+\sigma^2)(t-t_0)}\left(e^{\sigma^2(t-t_0)} - 1\right).
\end{align}
These are clearly different from what the It\^o interpretation of the SDE furnished earlier. 

Thus, the choice between It\^o and Stratonovich interpretations is not merely notational---it has real mathematical and physical consequences for the process dynamics and the resulting distributions. This distinction explains why in economics literature, we see the ($-\sigma^2/2$) term in the exponent of the GBM solution; while in some physics contexts using Stratonovich calculus, this term does not appear. They are different models unless explicitly reconciled through the drift correction. The aforementioned choice aligns perfectly with financial reality: An investor must make a trading decision ($x$ is, say, stock-price) at time \(t\) using only the historical data available up to that exact moment. They cannot look ahead to see the price at the end or middle of the interval. Whereas for a physicist,  if a physical system experiences a continuous, smoothly changing force (even a highly erratic one), Stratonovich SDE provides the natural limit: e.g., a tiny colloidal bead held in place by an optical tweezer (a laser trap) inside a fluid where instead of just being bumped by water molecules (which is standard additive Brownian motion), the intensity of the laser or the viscosity of the fluid is modulated by rapid, random external fluctuations.

According to the Wong--Zakai Theorem, when you take the mathematical limit of a real, continuous physical system as the noise correlation time goes to zero (\(\tau \to 0\)), the system automatically converges to the Stratonovich interpretation, not It\^o

\subsection{It\^o--Stratonovich Dichotomy}  
Understandably, a question comes up often: How should one interpret the following SDE:
\begin{equation}
dx(t) = a(x,t)\,dt + b(x,t)\,dW(t),\label{eq:stosy1}
\end{equation}
or for that matter, the nonlinear Langevin equation,
\begin{equation}
\dot{x}(t) = a(x,t)\,dt + b(x,t)\xi(t),\label{eq:stosy2}
\end{equation}
intended to imply similar meaning? Is there a prescription about when to adopt It\^o convention and when to adopt Stratonovich convention?

To discuss this matter, let us first delve deeper into the GBM in little more detail. To make things less abstract, consider a population, whose size or density [say, $x(t)$] is changing in a fluctuation environment. Over a short
time interval $[t,t+\Delta t]$, suppose that the population increment
is proportional to the current population:
\begin{equation}
\Delta x
=
x(t)\,r_{\Delta t},
\label{eq:population-proportional-increment}
\end{equation}
where $r_{\Delta t}$ is the random per-capita increment over the
interval. This proportionality describes density-independent growth:
each individual is exposed to the same average demographic conditions
and to the same environmental fluctuations. Assume that the per-capita increment has a constant deterministic
component and a random environmental component,
\begin{equation}
r_{\Delta t}
=
\mu\,\Delta t+\sigma\,\Delta W,
\label{eq:random-per-capita-increment}
\end{equation}
where $\mu$ is a constant per-capita drift coefficient, $\sigma\geq0$
is a constant environmental-noise amplitude, and
\begin{equation}
\Delta W
=
W(t+\Delta t)-W(t)
\sim \mathcal N(0,\Delta t).
\end{equation}
Substituting Eq.~\eqref{eq:random-per-capita-increment} into
Eq.~\eqref{eq:population-proportional-increment} gives
\begin{equation}
x(t+\Delta t)-x(t)
=
\mu x(t)\Delta t+\sigma x(t)\Delta W.
\label{eq:discrete-population-gbm}
\end{equation}
Because the coefficients on the right-hand side are evaluated at the beginning of the interval and therefore mathematically, it corresponds to the It\^o convention: As $\Delta t\to 0$, we get the form resembling Eq.~(\ref{eq:stosy1}) as It\^o SDE. This could be done because implicitly we have assumed the environmental fluctuations to be of a particular kind that describes fresh, unpredictable shocks---like pesticide applications, harvesting episodes, accidental toxic releases, sudden fires or floods, and abrupt disease introductions---on the current state ($x(t)$), i.e., immediately before an event, the population is $x(t)$ and the event removes or adds a random fraction based on this pre-event population. 

However, if the environmental fluctuation is of a different kind, then the situation is different; specifically,  suppose the environmental forcing and the population evolve together during each short fluctuation. As an example, consider an environment of rapidly fluctuating temperature and some microorganisms---whose per capita growth rate ($\dot{x}/x$) depends on temperature---in that environment, i.e.,  
\[ \dot{x}=\bigl[\mu+\sigma T_\tau(t)\bigr]x(t),
\]
where ($T_\tau(t)$) is a rapidly varying but continuous temperature fluctuation with correlation time ($\tau>0$).
During a warm fluctuation, the population starts growing immediately. Later portions of that same temperature fluctuation therefore act on a population that has already changed. If the correlation time becomes very short and the appropriately scaled integrated temperature converges to Brownian motion, then the limiting equation can be argued to be---in line with the \emph{Wong--Zakai Theorem}---of the form resembling Eq.~(\ref{eq:stosy1}) as Stratonovich SDE. We emphasize that it so happened  because the environmental forcing and the population evolved together during each short fluctuation.

Let's understand it better. Let $\xi_\tau(t)$ be a random environmental process with correlation
time $\tau>0$. Thus, different environmental realizations produce
different functions $t\mapsto\xi_\tau(t)$. We assume that, for every
fixed $\tau>0$, almost every sample path of $\xi_\tau$ is sufficiently
regular in time (for example, continuous and piecewise smooth) that
the population equation for the population state $x_\tau(t)$ can be interpreted pathwise as an ordinary
random differential equation:
\begin{equation}
\dot{x}_\tau
=
x_\tau(t)\left[\mu 
+
\sigma \xi_\tau(t)\right].
\label{eq:colored-noise-population}
\end{equation}
Note that the explicit appearance of $\tau$ in the population state identifies the environment it is in.
Although the coefficients are random, this is an ordinary differential
equation for each fixed realization of $\xi_\tau$. Ordinary calculus
therefore applies before the limit $\tau\downarrow0$ is taken. 

Integrating Eq.~\eqref{eq:colored-noise-population} over the interval $[t,t+\Delta t]$ gives
\begin{align}
x_\tau(t+\Delta t)-x_\tau(t)
=&
\mu
\int_t^{t+\Delta t}x_\tau(t')\,dt'\nonumber\\
&+
\sigma
\int_t^{t+\Delta t}x_\tau(t')\xi_\tau(t')\,dt'.
\label{eq:integrated-colored-noise-population}
\end{align}
The environmental forcing therefore acts not only on the population
present at time $t$, but on the population as it evolves during the
whole interval. 

For a sufficiently short interval, the population appearing in the
integrals may be represented by its midpoint value. Since
$x_\tau(t)$ is smooth for every fixed $\tau>0$, midpoint quadrature
gives
\begin{equation}
\int_t^{t+\Delta t}x_\tau(t')\,dt'
=
x_\tau\!\left(t+\frac{\Delta t}{2}\right)\Delta t
+
\mathcal{O}(\Delta t^3),
\label{eq:midpoint-deterministic-integral}
\end{equation}
and, in the same symmetric sense,
\begin{equation}
\int_t^{t+\Delta t}x_\tau(t')\xi_\tau(t')\,dt'
\approx
x_\tau\!\left(t+\frac{\Delta t}{2}\right)
\int_t^{t+\Delta t}\xi_\tau(t')\,dt'.
\label{eq:midpoint-random-integral}
\end{equation}
Moreover, over a short interval,
\begin{equation}
x_\tau\!\left(t+\frac{\Delta t}{2}\right)
=
\frac{x_\tau(t)+x_\tau(t+\Delta t)}{2}
+
\mathcal{O}(\Delta t^2).
\label{eq:midpoint-endpoint-average}
\end{equation}
Introducing the integrated environmental fluctuation
\begin{equation}
\Delta W_\tau
=
\int_t^{t+\Delta t}\xi_\tau(s)\,ds,
\label{eq:regularized-noise-increment}
\end{equation}
the short-time population increment becomes
\begin{equation}
\Delta x_\tau
\approx
\frac{x_\tau(t)+x_\tau(t+\Delta t)}{2}
\left(
\mu \Delta t+\sigma\Delta W_\tau
\right).
\label{eq:colored-noise-midpoint-increment}
\end{equation}

Now it is a rather technical fact but letting the environmental correlation time tend to zero while scaling
the amplitude of $\xi_\tau$ so that its integrated covariance remains
finite, it can be shown that under the standard functional central-limit assumptions,
\begin{equation}
\Delta W_\tau
=
\int_t^{t+\Delta t}\xi_\tau(t')\,dt'
\;\xrightarrow[]{\text{distribution}}\;
\Delta W,
\label{eq:colored-to-white-noise}
\end{equation}
where $W(t)$ is a Wiener process. The limiting short-time rule is
therefore
\begin{equation}
\Delta x
=
\frac{x(t)+x(t+\Delta t)}{2}
\left(
\mu\Delta t+\sigma\Delta W
\right).
\label{eq:stratonovich-midpoint-rule}
\end{equation}
This is precisely the midpoint discretization that defines the
Stratonovich stochastic integral.

The midpoint prescription should therefore not be viewed as an
arbitrary numerical choice. It preserves, in the idealized white-noise
model, the symmetric within-interval response that was present in the
original coloured-noise dynamics. When the correlation time is sent to zero, this
ordinary-calculus structure survives as the Stratonovich convention. Moreover, the fact that the average contains $(x(t+\Delta t))$ does not mean that the original coloured-noise model anticipates the future. For every $(\tau>0)$, the dynamics are an ordinary causal differential equation. The symmetric endpoint expression is merely a short-interval approximation to the population's actual continuous evolution within that interval.

The above discussion in the illustrative context of real-life GBM applies generally to the case of \emph{external noise}---the kinds we considered above and whose some more example could be: electromagnetic interference, mechanical vibrations (e.g., wind, seismic), external voltage/power supply noise, macroeconomic market shocks, and acoustic background noise. Here one knows what is the deterministic part of the stochastic equation because when the noise is switched off---which can practically or in principle be done---the deterministic part $(a(x,t))$ is what all remains; after all, the system is not supposed to be influencing the noise. In fact, if the noise is external, one may think of writing a Langevin equation corresponding to a stochastic system phenomenologically by just adding a noise term (like $\sigma dW$ to per capita growth rate in population following GBM) to the known deterministic term, and interpret the equation either via It\^o or Stratonovich convention depending on whether the noise fluctuations are discrete or coloured with short correlation time.

However, in the case of \emph{internal noise}, it becomes very hard to separate out deterministic and noisy part in such a manner that an ad hoc Langevin equation may be confidently prescribed for the system-dynamics. This is because noise is, in fact, generated by the system itself by virtue of being consisting of discrete particles about whose individual dynamics one is uncertain; one cannot just tune a parameter and shut off the noisy aspect without influencing the macroscopic deterministic manifestation of the system. Some examples of internal noise are: chemical noise in gene expression (due to low molecule counts), demographic noise in ecosystems and epidemics (from discrete birth/death/infection events), shot noise in photodetectors and nano-circuits (from discrete photons/electrons), and thermal noise in Brownian motion (from discrete molecular collisions). Naturally, here writing a Langevin equation in ad hoc manner and hoping to fix coefficients ($a$ and $b$) therein from data of the system under consideration is a dangerous approach. There is no other way out than to take the bull by the horns: One must embrace the inherent uncertainty and go for a coherent probabilistic description of the system and its dynamics. And, of course, such an approach will be applicable to systems with external noise as well, for probability theory does not discriminate uncertainties based on their origin.

\begin{center}
	$\symking\symqueen\symrook\symknight\symbishop\sympawn\epsdice{1}\,\epsdice{2}\,\epsdice{3}\,\epsdice{4}\,\epsdice{5}\,\epsdice{6}\,\spadesuit\heartsuit\clubsuit\diamondsuit$
\end{center}
\newpage
\newpage
\clearpage
\onecolumngrid
\section{Probability Evolution: An Alternative}
\epigraph{\textsf{The lamps are different, but the light is the same.}}{Rumi}
\twocolumngrid

\indent {\color{deepsea}{\LARGE \textgoth{F}}}ollowing temporal evolution of probability of the state of a stochastic system gives equivalent information that can be extracted from the corresponding SDE tracking the evolution of corresponding stochastic trajectories directly. To see this concretely, consider the example of a stochastic process $X(t)$ following It\^o SDE. For a smooth test function $f(x)$ that is non-anticipating, It\^o formula (\ref{eq:ifor}) on averaging yields:
\begin{equation}
    \frac{d}{dt}\langle f(x)\rangle
    = \Big\langle a\,f'
      + \tfrac{1}{2}b^{2}f''\Big\rangle,
    \label{eq:avg}
\end{equation}
since $\langle f'b\,dW\rangle = 0$. Let $p(x,t|x_0,t_0)$---written simply as $p$ below for brevity---be the conditional probability density of $x(t)$, so that with obvious definition of angular brackets: $\langle g(x)\rangle \equiv \int g(x)pdx$, Eq.~(\ref{eq:avg}) becomes
\begin{align}
    \int f\,\partial_t pdx = \int\Big[af' + \tfrac{1}{2}b^{2} f''\Big]\,pdx.
\end{align}
Assuming $p$ and its derivatives vanish as
$|x|\to\infty$, we get on integrating by parts and collecting terms:
\begin{align}
    \int f(x)\Big\{\partial_t p+ \partial_x(a p) - \partial_x^{2}\big(\tfrac{1}{2}b^{2} p\big)\Big\}dx = 0.
\end{align}
Here we have denoted partial derivatives rather explicitly. Since $f(x)$ is arbitrary, the term inside second brackets must vanish; therefore,
\begin{align}
    \partial_t p(x,t|x_0,t_0)=& -\partial_x\!\big[a(x,t)p(x,t|x_0,t_0)\big]\nonumber\\
    &+ \tfrac{1}{2}\partial_x^{2}\big[b^{2}(x,t)p(x,t|x_0,t_0)\big].
    \label{eq:FP}
\end{align}
This is known as \emph{forward Fokker--Planck equation} which describes how the probability density of the stochastic process evolves in time. This illustration motivates us to develop an alternative technique of analyzing stochastic processes using dynamics of their probabilities.
\subsection{Chapman--Kolmogorov Equation}
Let us consider general stochastic Markov process $X(t)$ with continuous one-dimensional state space and continuous time-parameter. The probability distributions functions must be able to completely characterize the stochastic process. As an example, we observe how Wiener process can be defined using distributions.

We claim that $W(t)$ can be defined by specifying the transition (conditional) probability density as
\begin{equation}
p(x_2,t_2\mid x_1,t_1)
=
\frac{1}{\sqrt{2\pi (t_2-t_1)}}
\exp\left[
-\frac{(x_2-x_1)^2}{2(t_2-t_1)}
\right],
\label{eq:3.11}
\end{equation}
for $t_0 \le t_1 < t_2$, and the initial distribution at time $t_0$ as
\begin{equation}
\rho_0(x_0)=\delta(x_0).
\end{equation}
We now prove that the three defining axioms of the one-dimensional standard Wiener process follow from here:

First, consider the axiom that \(W(t_0)=0\). The initial distribution is
\[
\rho_0(x_0)=\delta(x_0).
\]
Therefore,
\[
 P(W(t_0)=0)=1,
\]
which is equivalent to
\[
{W(t_0)=0\quad\text{almost surely}.}
\]

Next, let's focus on the axiom of {Gaussian increments}. Let $t_0 \le t_1 < t_2$. Conditional on $W(t_1)=x_1$, Eq.~\eqref{eq:3.11} states that
\[
[W(t_2)\mid W(t_1)=x_1]
\sim
\mathcal N\!\left(x_1,(t_2-t_1)\right).
\]
Defining the generic increment
\[
\Delta W=W(t_2)-W(t_1),
\]
gives conditional on $W(t_1)=x_1$ as
\[
[\Delta W\mid W(t_1)=x_1]
=
W(t_2)-x_1.
\]

Using Eq.~\eqref{eq:3.11}, its conditional density is
\begin{equation*}
\begin{split}
p_{\Delta W\mid W(t_1)=x_1}(u)
&=
p(x_1+u,t_2\mid x_1,t_1)\\
&=
\frac{1}{\sqrt{2\pi (t_2-t_1)}}
\exp\left[
-\frac{u^2}{2(t_2-t_1)}
\right].
\end{split}
\end{equation*}
This density does not depend on $x_1$. Hence the increment is independent of $W(t_1)$, and
\[
\Delta W\sim
\mathcal N\!\left(0,(t_2-t_1)\right).
\]
Thus,
\[
{
W(t_2)-W(t_1)\sim
\mathcal N\!\left(0,(t_2-t_1)\right)
}.
\]

Finally, we turn to the axiom of independence of increments. Consider arbitrary times
\[
t_0 < t_1 < \cdots < t_n
\]
and define the increments
\[
\Delta_k W=W(t_k)-W(t_{k-1}),
\qquad k=1,\ldots,n.
\]
The joint density of the process values is
\[
p_n(x_1,\ldots,x_n)
=
\prod_{k=1}^{n}
p(x_k,t_k\mid x_{k-1},t_{k-1}),
\]
where $x_0=0$ at $t_0$.

Using Eq.~\eqref{eq:3.11},
\begin{equation*}
\begin{split}
&p_n(x_1,\ldots,x_n) = \\
&\prod_{k=1}^{n}
\frac{1}{\sqrt{2\pi (t_k-t_{k-1})}}
\exp\left[
-\frac{(x_k-x_{k-1})^2}
{2(t_k-t_{k-1})}
\right].
\end{split}
\end{equation*}

Now make the change of variables
\[
u_k=x_k-x_{k-1},
\qquad k=1,\ldots,n.
\]
Since this transformation is triangular, its Jacobian,
\[
\left|
\frac{\partial(x_1,\ldots,x_n)}
{\partial(u_1,\ldots,u_n)}
\right|=1.
\]
Therefore, the joint density of the increments is
\begin{equation*}
\begin{split}
&p_{\Delta_1W,\ldots,\Delta_nW}(u_1,\ldots,u_n)\\
&=
\prod_{k=1}^{n}
\frac{1}{\sqrt{2\pi (t_k-t_{k-1})}}
\exp\left[
-\frac{u_k^2}
{2(t_k-t_{k-1})}
\right]\\
&=
\prod_{k=1}^{n}
p_{\Delta_kW}(u_k).
\end{split}
\end{equation*}
The joint density factorizes into the product of the marginal densities. Consequently, the increments are mutually independent. Since the choice of the times $t_0,\ldots,t_n$ was arbitrary, all increments over disjoint time intervals are independent. This establishes the claim we set out to establish about Wiener process.

Now, we want to present a fundamental identity relating probability distributions in any Markov process. Consider $t_1 \le t_2 \le t_3$ and let a joint probability density function of $X(t)$ be denoted by $p(x_1, t_1; x_2, t_2; x_3, t_3)$. We have
\begin{align}
    &p(x_1, t_1;\, x_2, t_2; \,x_3, t_3) \nonumber \\
    &\quad = p(x_3, t_3 | x_2, t_2;\, x_1, t_1) p(x_2, t_2;\,x_1, t_1) \nonumber \\
    &\quad = p(x_3, t_3 | x_2, t_2) p(x_2, t_2|x_1, t_1) p(x_1, t_1) 
\end{align}
where we have made use of the Markovian nature of the process and the definition of the conditional probability. Now integrating this relation over $x_2$ on both sides, we obtain
\begin{align}
   &p(x_1, t_1;\,x_3, t_3)= \int dx_2\,p(x_1, t_1;\, x_2, t_2; \,x_3, t_3) \quad\nonumber \\
    & \phantom{p(x_1)}=  p(x_1, t_1) \int dx_2\,p(x_3, t_3 | x_2, t_2) p(x_2, t_2|x_1, t_1),\nonumber \\
    \implies& p(x_3, t_3 | x_1, t_1) = \int dx_2 \, p(x_3, t_3 | x_2, t_2) p(x_2, t_2 | x_1, t_1).\label{eq:ckec}
\end{align}
This relation is known as the \emph{Chapman--Kolmogorov equation} obeyed by Markov processes. 

A converse result also holds true: Any normalized non-negative function that obeys Eq.~(\ref{eq:ckec}) and the law of total probability---
\begin{align}
   &p(x_1, t_1)= \int dx_2\,p(x_1, t_1|\, x_2, t_2) p(x_2, t_2)\label{eq:lotp}
\end{align}
---uniquely defines a Markov process. This can be shown as follows:

Let \(p(x,t| x',t')\) be non-negative functions satisfying
\begin{align}
\int dx\,p(x,t| x',t')=1
\label{eq:naii}
\end{align}
and the Chapman--Kolmogorov equation [Eq.~(\ref{eq:ckec})]. Let \(\rho_0(x_0)\) be an initial probability density. For \(t_0<t_1<\cdots<t_n\), define
\begin{align}
p_n(x_0,\ldots,x_n)
\equiv
\rho_0(x_0)
\prod_{k=1}^{n}
p(x_k,t_k\mid x_{k-1},t_{k-1}).\label{eq:Pndef}
\end{align}
Since \(\rho_0\geq 0\) and \(p\geq 0\), we have \(p_n\geq 0\). Moreover,
\[
\int dx_0\cdots dx_n\,p_n(x_0,\ldots,x_n)=1
\]
by repeated use of Eq.~(\ref{eq:naii}).

The distributions are consistent. Indeed, integrating over an intermediate
variable \(x_j\) gives
\[
\begin{aligned}
\int dx_j\,p_n
&=
\rho_0(x_0)
\prod_{k\neq j,j+1}
p(x_k,t_k\mid x_{k-1},t_{k-1})\quad\\
\quad\times
\int dx_j\,
&p(x_j,t_j\mid x_{j-1},t_{j-1})
p(x_{j+1},t_{j+1}\mid x_j,t_j)\\
&=
\rho_0(x_0)
\prod_{k\neq j}
p(x_k,t_k\mid x_{k-1},t_{k-1}),
\end{aligned}
\]
where Eq.~(\ref{eq:ckec}) has been used. Thus, the \(p_n\) define consistent
finite-dimensional probability distributions and therefore a stochastic
process.

Finally,
\[
\begin{aligned}
&p(X({t_n})=x_n\mid X({t_0})=x_0,\cdots,X({t_{n-1}})=x_{n-1})\\
&\qquad =
\frac{p_n(x_0,\ldots,x_n)}
     {p_{n-1}(x_0,\ldots,x_{n-1})}
=
p(x_n,t_n\mid x_{n-1},t_{n-1}),
\end{aligned}
\]
which is clearly seen to depend---using definition of $p_n$ from Eq.~(\ref{eq:Pndef})--- only on the immediately preceding state. Hence the process is Markovian.

Conversely, any Markov process with initial density \(\rho_0\) and transition
probability \(p\) must have the finite-dimensional distributions
\[
p_n(x_0,\ldots,x_n)
=
\rho_0(x_0)
\prod_{k=1}^{n}
p(x_k,t_k\mid x_{k-1},t_{k-1}).
\]
Therefore these functions determine the Markov process uniquely in
distribution.

\subsection{Master Equation}
The Chapman--Kolmogorov equation is an exact, nonlinear integral consistency condition satisfied by the transition probabilities of any Markov process. In its integral form it relates transition probabilities across three time points and is generally difficult to solve directly. The \emph{master equation} is the equivalent \emph{integro-differential form} of the Chapman--Kolmogorov equation: it is obtained by taking the short-time limit of the Chapman--Kolmogorov relation together with a short-time ansatz for the transition probability that separates a ``no-jump'' contribution from a ``jump'' contribution characterized by transition rates. The result is a linear, first-order-in-time equation for the probability distribution with a transparent gain--loss structure, which forms the starting point for most kinetic and stochastic descriptions in statistical physics.

We begin from the Chapman--Kolmogorov equation,~Eq.~(\ref{eq:ckec}), choosing three time points $t_1 = t$, $t_2 = t$, and $t_3 = t + \Delta t$, so that the intermediate step is separated from the initial one by an infinitesimal interval:
\begin{align}
&p(x_3, t+\Delta t \mid x_1, t) \nonumber \\
&\quad = \int dx_2 \, p(x_3, t+\Delta t \mid x_2, t)\, p(x_2, t \mid x_1, t).\label{eq:ckemod}
\end{align}
At equal times, the conditional probability collapses to a delta function,
\begin{equation}
p(x_2, t \mid x_1, t) = \delta(x_2 - x_1),
\end{equation}
which reflects the fact that no evolution has occurred yet.

Let us use short-time ansatz: For a small interval $\Delta t$, we expand the transition probability into a ``no-jump'' piece and a ``jump'' piece:
\begin{align}
p(x_2, t+\Delta t \mid x_1, t) =& \delta(x_2 - x_1)\bigl[1 - a^{(0)}(x_1, t)\,\Delta t\bigr] \nonumber \\
&\quad + R_t(x_2 \mid x_1)\,\Delta t + \text{h.o.t}.\label{eq:ansatzst}
\end{align}
(`h.o.t.' means higher order terms.) Here $a^{(0)}(x_1, t)\,\Delta t$ is the probability that a transition occurs out of $x_1$ during $\Delta t$, $R_t(x_2 \mid x_1)$ is the \emph{transition rate} from $x_1$ to $x_2$, and the delta term encodes the ``survival'' probability of remaining at $x_1$. Integrating the ansatz over $dx_2$ and using $\int dx_2\, p = 1$:
\begin{equation}
1 \simeq 1 - a^{(0)}(x_1, t)\,\Delta t + \int dx_2 \, R_t(x_2 \mid x_1)\,\Delta t,
\end{equation}
which immediately yields the \emph{total exit rate},
\begin{equation}
{\; a^{(0)}(x_1, t) = \int dx_2 \, R_t(x_2 \mid x_1). \;}\label{eq:totalexitrate}
\end{equation}
The factor $1 - a^{(0)}(x_1, t)\,\Delta t$ is therefore the probability that no transition occurs in the interval $\Delta t$.

Substituting the short-time ansatz,~Eq.~(\ref{eq:ansatzst}), back into~Eq.~(\ref{eq:ckemod}) gives
\begin{align}
p(x_3, t+\Delta t \mid x_1, t) =& \int dx_2 \Bigl[\delta(x_3 - x_2)\bigl(1 - a^{(0)}(x_2, t)\Delta t\bigr) \nonumber \\
+& R_t(x_3 \mid x_2)\Delta t \Bigr] p(x_2, t \mid x_1, t) \nonumber \\
=& p(x_3, t \mid x_1, t)\bigl[1 - a^{(0)}(x_3, t)\Delta t\bigr] \nonumber \\
+& \int dx_2 \, R_t(x_3 \mid x_2)\,\Delta t\, p(x_2, t \mid x_1, t).
\label{eq:assembled}
\end{align}
To obtain a differential equation, we rearrange Eq.~\eqref{eq:assembled}, moving the survival contribution to the left-hand side:
\begin{align}
&p(x_3, t+\Delta t \mid x_1, t) - p(x_3, t \mid x_1, t) \nonumber \\
&\quad = -a^{(0)}(x_3, t)\,\Delta t\, p(x_3, t \mid x_1, t) \nonumber \\
&\qquad + \int dx_2 \, R_t(x_3 \mid x_2)\,\Delta t\, p(x_2, t \mid x_1, t).
\end{align}
Dividing by $\Delta t$ and taking $\Delta t \to 0$ produces the time derivative on the left. At this point we {relabel the variables}---$x_3 \to x$, $(x_1, t) \to (x_0, t_0)$, and $x_2 \to x'$--- to expose the standard form of the \emph{master equation}:
\begin{align}
\frac{\partial p(x, t \mid x_0, t_0)}{\partial t}
&= \int dx' \, R_t(x \mid x')\, p(x', t \mid x_0, t_0) \nonumber \\
&\quad - \int dx' \, R_t(x' \mid x)\, p(x, t \mid x_0, t_0),
\label{eq:master}
\end{align}
where we can made use of Eq.~\eqref{eq:totalexitrate}. Eq.~\eqref{eq:master} is the integro-differential form of the Chapman--Kolmogorov equation.

The two terms on the R.H.S. have a clear physical meaning:
\begin{itemize}
\item {Gain term:} $\int dx' \, R_t(x \mid x')\, p(x', t \mid x_0, t_0)$, the probability flowing \emph{into} state $x$ from all other states $x'$.
\item {Loss term:} $\int dx' \, R_t(x' \mid x)\, p(x, t \mid x_0, t_0)$, the probability flowing \emph{out} of state $x$ into all other states $x'$, with total exit rate $a^{(0)}(x, t)$.
\end{itemize}

We note that since the unconditional distribution $p(x,t)$ can be obtained by averaging the conditional distribution over the initial state weighted by its distribution $p(x_0, t_0)$:
\begin{equation}
p(x, t) = \int dx_0 \, p(x, t \mid x_0, t_0)\, p(x_0, t_0),
\end{equation}
multiplying both sides of the master equation~\eqref{eq:master} by $p(x_0, t_0)$ and integrating over $x_0$ yields:
\begin{align}
\frac{\partial p(x, t)}{\partial t}
&= \int dx' \, R_t(x | x')\, p(x', t)- \int dx' \, R_t(x' | x)\, p(x, t),
\end{align}
which has exactly the same form as Eq.~\eqref{eq:master}.

Finally we remark, that for a system with a discrete set of states $\{n\}$, integrals are replaced by sums, $\int dx' \to \sum_{n'}$, giving the master equation:
\begin{align}
\frac{\partial p(n, t \mid n_0, t_0)}{\partial t}
&= \sum_{n'} R_t(n \mid n')\, p(n', t \mid n_0, t_0) \nonumber \\
&\quad - \sum_{n'} R_t(n' \mid n)\, p(n, t \mid n_0, t_0).\label{eq:masterdis}
\end{align}

\subsection{Forward Fokker--Planck Equation}
For notational convenience in the intermediate steps, we will temporarily suppress the initial conditions $(x_0, t_0)$ and simply write $p(x,t) \equiv p(x,t|x_0,t_0)$. So consider the master equation~(\ref{eq:master}).

Define the jump size
\begin{equation}
    \Delta \equiv x-x',
    \qquad\text{so that}\qquad
    x' = x-\Delta,
    \qquad dx' = -d\Delta .
    \label{eq:Delta_def}
\end{equation}
It is convenient to express the rate in terms of the \emph{initial} state
$x'$ and the jump $\Delta$ at time $t$:
\begin{equation}
    R_t(x|x') \;=\; R_t(x',\Delta)
    \qquad\bigl(\text{rate from }x'\text{ to }x'+\Delta\bigr).
    \label{eq:R_initial}
\end{equation}
We assume that 
\begin{itemize}
\item only \emph{small} jumps occur, so $R_t(x',\Delta)$ is
sharply peaked around $\Delta=0$; 
\item $R_t$ varies slowly with its first
argument $x'$; and 
\item the solution $p(x,t)$ also varies slowly with $x$.  
\end{itemize}
These conditions justify a Taylor expansion of the gain
integrand about the point $x$.

We use $\Delta$ to change of variables in the master equation. In the gain term, substituting $x'=x-\Delta$ and $dx'=-d\Delta$:
\begin{align}
    \int dx'\,R_t(x|x')\,p(x',t)
    &=
    \int d\Delta\,
    R_t(x-\Delta,\Delta)\,
    p(x-\Delta,t);
    \label{eq:gain_delta}
\end{align}
Similarly, in the loss term, where $x$ is the initial state and $x'$ the final state, using the {same} $\Delta=x-x'$ we have $x'=x-\Delta$, and the jump from $x$ to $x-\Delta$ is $-\Delta$.  Hence $R_t(x'|x)=R_t(x,-\Delta)$, giving
\begin{align}
    \int dx'\,R_t(x'|x)\,p(x,t)
    &=
    p(x,t)\int d\Delta\,
    R_t(x,-\Delta).
    \label{eq:loss_delta}
\end{align}
Putting both pieces together,
\begin{align}
    \frac{\partial p(x,t)}{\partial t}
    &=
    \int d\Delta\,R_t(x-\Delta,\Delta)\,p(x-\Delta,t)
    \nonumber\\
    &\quad
    -\,p(x,t)\int d\Delta\,R_t(x,-\Delta).
    \label{eq:master_delta}
\end{align}
Note that the variable change $x'\to\Delta$ essentially means
$\int dx'f(x')=-\!\int d(x-x')f(x-\Delta)=\!\int d\Delta\,f(x-\Delta)$,
which is correct because $d\Delta=-dx'$ and the limits reverse, absorbing
the sign.

Next, because $\Delta$ is small, we expand $R_t(x-\Delta,\Delta)\,p(x-\Delta,t)$
about the point $x$ (treating $\Delta$ as a fixed parameter in the first
argument of $R_t$) in the Talyor series:
\begin{align}
    &R_t(x-\Delta,\Delta)\,p(x-\Delta,t)
    \nonumber\\
    &\qquad\qquad\qquad=
    \sum_{k=0}^{\infty}
    \frac{(-\Delta)^k}{k!}\,
    \frac{\partial^k}{\partial x^k}
    \Bigl[R_t(x,\Delta)\,p(x,t)\Bigr].
    \label{eq:taylor}
\end{align}
Inserting Eq.~\eqref{eq:taylor} into the gain integral of
Eq.~\eqref{eq:master_delta} yields
\begin{align}
    \frac{\partial p(x,t)}{\partial t}
    &=
    \int d\Delta\,
    R_t(x,\Delta)\,p(x,t)
    \nonumber\\
    &\quad+
    \sum_{k=1}^{\infty}
    \frac{(-1)^k}{k!}
    \int d\Delta\,
    \Delta^k\,
    \frac{\partial^k}{\partial x^k}
    \Bigl[R_t(x,\Delta)\,p(x,t)\Bigr]
    \nonumber\\
    &\quad-
    p(x,t)\int d\Delta\,
    R_t(x,-\Delta).
    \label{eq:expanded}
\end{align}

Here, the two zeroth-order terms together
\begin{equation}
    p(x,t)\!\int\!d\Delta\,R_t(x,\Delta)
    \;-\;
    p(x,t)\!\int\!d\Delta\,R_t(x,-\Delta)=0.
    \label{eq:zeroth}
\end{equation}
This is because changing the integration variable $\Delta\to-\Delta$ in the second
integral shows the two integrals are identical. Therefore the zeroth-order
contributions cancel exactly, leaving only $k\ge1$ terms.

Finally, we define the $k$th Kramers--Moyal coefficients:
\begin{equation}
    {
    a^{(k)}(x,t)
    \equiv
    \int d\Delta\,
    \Delta^k\,
    R_t(x,\Delta)
    }.
    \label{eq:KM_def}
\end{equation}
Since the derivative $\partial^k/\partial x^k$ acts on $x$ and the
integral is over $\Delta$, we may interchange them. Substituting the definition~\eqref{eq:KM_def} into Eq.~\eqref{eq:expanded}, interchanging the derivative and integral, and restoring the full conditional probability $p(x,t|x_0,t_0)$, we obtain the \emph{Kramers--Moyal expansion}:
\begin{align}
    \frac{\partial p(x,t|x_0,t_0)}{\partial t}=
    \sum_{k=1}^{\infty}
    \frac{(-1)^k}{k!}\,
    \frac{\partial^k}{\partial x^k}
    \Bigl[ a^{(k)}(x,t)\,p(x,t|x_0,t_0) \Bigr].
    \label{eq:KM}
\end{align}

In a situation where $a^{(k)}(x,t)\approx0$ for $k>2$,
the infinite sum in Eq.~\eqref{eq:KM} truncates at
$k=2$, giving the one-dimensional \emph{(forward) Fokker--Planck equation} for the full conditional probability:
\begin{align}
    \frac{\partial p(x,t|x_0,t_0)}{\partial t}
    &=
    -\frac{\partial}{\partial x}
    \Bigl[a^{(1)}(x,t)\,p(x,t|x_0,t_0)\Bigr]
    \nonumber\\
    &\quad+\;
    \frac{1}{2}
    \frac{\partial^2}{\partial x^2}
    \Bigl[a^{(2)}(x,t)\,p(x,t|x_0,t_0)\Bigr].
    \label{eq:FPE}
\end{align}
The first term describes deterministic \emph{drift}; the second term describes
\emph{diffusion}. In this context, a useful theorem---Pawula's theorem---states that the Kramers--Moyal expansion
{either} terminates after the second term {or} must contain
infinitely many non-zero terms. Any finite truncation at order $k>2$
would produce a solution that can become negative, violating the axioms of probability.

Eq.~(\ref{eq:FPE}) can be written as a \emph{conservation law}:
\begin{align}
\frac{\partial p(x,t|x_0,t_0)}{\partial t}
&+
\frac{\partial J(x,t|x_0,t_0)}{\partial x}
=0,
\label{eq:probability-conservation}
\end{align}
where the \emph{probability current} is
\begin{align}
J(x,t|x_0,t_0)
=&
a^{(1)}(x,t)p(x,t|x_0,t_0)\nonumber\\
&-\frac{1}{2}
\frac{\partial}{\partial x}
\left[a^{(2)}(x,t)p(x,t|x_0,t_0)\right].
\label{eq:probability-current}
\end{align}
The drift transports
probability, while diffusion spreads it. In passing we remark that since the unconditional distribution $p(x,t)$ can be obtained by averaging the conditional distribution over the initial state weighted by its distribution $p(x_0, t_0)$, multiplying both sides of the forward Fokker--Planck equation~\eqref{eq:master} by $p(x_0, t_0)$ and integrating over $x_0$ yields the fact that $p(x,t)$ obey the same equation.

Next we want to see connection of Eq.~(\ref{eq:FPE}) with the Langevin equation. We begin by noting that Eq.~\eqref{eq:KM_def} implies that the Kramers--Moyal coefficients can be defined as:
 \begin{align}
    a^{(k)}(x,t)
    \equiv
    \int d\Delta\,
    \Delta^k\,
    R_t(x+\Delta|x).
    \end{align}
We observe that $\Delta$ is a dummy variable here and hence to remove any confusion, we introduce $\Delta'$ to recast the above equation as follows:
 \begin{align}
    a^{(k)}(x,t)
=
    \int d\Delta'\,
    {\Delta'}^k\,
    R_t(x'|x),
    \end{align}
    where we realize that $\Delta'$ should be recognized as $x' - x$; consequently, this equation can equivalently be seen as:
  \begin{align}
    a^{(k)}(x,t)
    =
    \int dx'\,
    {(x'-x)}^k\,
    R_t(x'|x).\label{eq:akx'x}
    \end{align}   
In terms of the continuous stochastic process $X(t)$, these Kramers--Moyal coefficients---defined via the transition moments---represent the short-time conditional moments of the state increments. This is clear from the following sequence of steps:
\begin{align}
 &\lim_{\Delta t \to 0} \frac{1}{\Delta t} \langle [X(t+\Delta t) - X(t)]^k \rangle_{X(t)=x}\nonumber\\
 =&\lim_{\Delta t \to 0}\frac{1}{\Delta t}\int dx'(x'-x)^kp(x',t+\Delta t\mid x,t)\nonumber\\
 =&\lim_{\Delta t \to 0}\frac{1}{\Delta t}\int dx'(x'-x)^k\left[\delta(x'-x)\{1-a^{(0)}(x,t)\Delta t\}\right.\nonumber\\
 &\phantom{\lim_{\Delta t \to 0}\frac{1}{\Delta t}\int dx'(x'-x)^k[}\left.+R_t(x'|x)\Delta t\right]\nonumber\\
     =&  a^{(k)}(x,t)
     \label{eq:ak_def}
\end{align}
due to Eq.~\eqref{eq:akx'x}.

Now consider the one-dimensional nonlinear Langevin equation
\begin{equation}
    \frac{dx}{dt} = a(x,t) + b(x,t)\xi(t), \quad \langle \xi(t)\xi(t') \rangle = \delta(t-t'),
    \label{eq:langevin}
\end{equation}
where $\xi(t)$ is Gaussian white noise. Integrating Eq.~\eqref{eq:langevin} over a small time interval $[t, t+\Delta t]$ gives the finite state increment:
\begin{align}
    x(t+\Delta t) - x(t) &= \int_t^{t+\Delta t} dt' \, a(x(t'), t') \nonumber \\
    &\quad + \int_t^{t+\Delta t} dt' \, b(x(t'), t')\xi(t').
\end{align}

Because the arguments $x(t')$ evolve during the time step, we must expand the drift and the diffusion functions about the initial state $x(t) \equiv x$. For example:
\begin{equation}
    a(x(t'),t') = a(x,t') + a'(x,t')\bigl(x(t')-x\bigr) + \dots
\end{equation}
Expanding $b(x(t'),t')$ similarly, and iteratively substituting
\[
x(t')-x \approx \int_t^{t'} dr \, b(x,r)\xi(r)
\]
for the lowest-order noise contribution, we obtain the expanded state increment:
\begin{align}
    x(t+\Delta t) - x &= \int_t^{t+\Delta t} dt' \, a(x, t') + \dots \nonumber \\
    &\quad + \int_t^{t+\Delta t} dt' \, b(x, t')\xi(t') \nonumber \\
    &\quad + \int_t^{t+\Delta t} dt' \, b'(x, t')\bigl(x(t')-x\bigr)\xi(t') + \dots
    \label{eq:expansion}
\end{align}

To find the first Kramers--Moyal coefficient, or effective drift, we evaluate the ensemble average
\[
a^{(1)}\Delta t = \langle x(t+\Delta t) - x(t) \rangle;
\]
recall Eq.~\eqref{eq:ak_def}. Taking the expectation value of Eq.~\eqref{eq:expansion}, the linear noise term $\langle \xi(t') \rangle$ vanishes, and we are left with the correlation in the third term:
\begin{align}
    a^{(1)} \Delta t &= \int_t^{t+\Delta t} dt' \, a(x, t') \nonumber \\
    &\quad + \int_t^{t+\Delta t} dt' \, b'(x, t')
    \int_t^{t'} dr \, b(x, r)\delta(r-t').
\end{align}

The evaluation of the integral over the Dirac delta function,
\[
\int_t^{t'} \delta(r-t')\,dr,
\]
depends strictly on the chosen stochastic-calculus convention: Recall [Eq.~(\ref{eq:upperl}) and Eq.~(\ref{eq:mid1/2})] that under {It\^o convention}, the integral gives $0$; whereas under {Stratonovich convention}, the integral yields a value of $1/2$. Finally, dividing by $\Delta t$ and taking the limit $\Delta t \to 0$, we find the first Kramers--Moyal coefficient:
\begin{align}
    \implies a^{(1)} &= a(x,t) \quad \text{(It\^o)}, \\
    \implies a^{(1)} &= a(x,t) + \frac{1}{2}b(x,t)b'(x,t)
    \quad \text{(Stratonovich)}.
\end{align}
The extra term in the Stratonovich interpretation is often referred to as the noise-induced, or spurious, drift.

To determine the higher-order Kramers--Moyal coefficients, we calculate $\langle (\Delta x)^k \rangle$, where $\Delta x \equiv x(t+\Delta t) - x(t)$. For the leading-order behaviour, it suffices to retain only the most dominant terms in the increment:
\begin{equation}
    \Delta x \approx a(x,t)\Delta t
    + \int_t^{t+\Delta t} dt' \, b(x,t')\xi(t').
\end{equation}

For the second order moments, squaring the increment yields three terms:
\begin{align}
    (\Delta x)^2 &\approx a^2 (\Delta t)^2
    + 2a\Delta t \int_t^{t+\Delta t} dt' \, b(x,t')\xi(t') \nonumber \\
    &\quad + \int_t^{t+\Delta t}\int_t^{t+\Delta t}
    dt'\,dr \, b(x,t')b(x,r)\xi(t')\xi(r).
\end{align}
Taking the expectation value, the linear noise term vanishes, $\langle \xi(t') \rangle = 0$; the deterministic term is of order $\mathcal{O}(\Delta t^2)$; and the final term involves the noise covariance $\langle \xi(t')\xi(r) \rangle = \delta(t'-r)$:
\begin{align}
    \langle (\Delta x)^2 \rangle
    &= \mathcal{O}(\Delta t^2)
    + \int_t^{t+\Delta t} dt' \, b^2(x,t') \nonumber \\
    &\approx b^2(x,t)\Delta t + \mathcal{O}(\Delta t^2).
\end{align}
Applying the definition from Eq.~\eqref{eq:ak_def}, we divide by $\Delta t$ and take the limit:
\begin{equation}
    a^{(2)}(x,t)
    = \lim_{\Delta t \to 0}
    \frac{b^2\Delta t+\mathcal{O}(\Delta t^2)}{\Delta t}
    = b^2(x,t).
\end{equation}
This result is independent of the chosen stochastic-calculus convention, It\^o or Stratonovich.

For higher-order moments ($k > 2$), the expansion of $\langle (\Delta x)^k \rangle$ is dominated by the expectation values of products of the noise term:
\begin{equation}
    \langle (\Delta x)^k \rangle
    \sim
    \left\langle
    \left(
    \int_t^{t+\Delta t} dt'\, b(x,t')\xi(t')
    \right)^k
    \right\rangle.
\end{equation}

For Gaussian white noise, expectation values of odd powers of $\xi$ vanish exactly by \emph{Isserlis' theorem}, also known as \emph{Wick's probability theorem}. Thus,
\begin{equation}a^{(k)}=0\quad\,\textrm{for all odd}\,\, k\geq 3.\end{equation}
For even $k$, Isserlis' theorem dictates that the expectation value of $k$ noise variables splits into a sum of products of pairwise covariances, represented by delta functions. Integrating over $k/2$ delta functions within a $k$-dimensional time integral over a domain of length $\Delta t$ yields a result proportional to $(\Delta t)^{k/2}$. Therefore, for even $k$:
\begin{equation}
    \langle (\Delta x)^k \rangle
    \sim \mathcal{O}\left(\Delta t^{k/2}\right).
\end{equation}
Since $k \geq 4$ for higher-order even moments, $k/2 \geq 2$. Consequently:
\begin{equation}
    a^{(k)}(x,t)
    = \lim_{\Delta t \to 0}
    \frac{\mathcal{O}(\Delta t^{\geq 2})}{\Delta t}
    = 0.
\end{equation}
This explicitly demonstrates that, for processes driven by Gaussian white noise, all Kramers--Moyal coefficients above the second order vanish identically:
\begin{equation}
    a^{(k)}=0 \quad \text{for } k>2.
\end{equation}

By substituting the derived coefficients into the general Kramers--Moyal expansion [Eq.~(\ref{eq:KM})], the series truncates exactly at the second term, yielding the forward Fokker--Planck equation.
For the {It\^o} case, the equation is:
\begin{equation}
    \frac{\partial p}{\partial t}
    = -\frac{\partial}{\partial x}[a(x,t)p]
    + \frac{1}{2}\frac{\partial^2}{\partial x^2}[b^2(x,t)p].\label{eq:253}
\end{equation}
For the {Stratonovich} case, incorporating the noise-induced drift, the forward Fokker--Planck equation takes the form:
\begin{equation}
    \frac{\partial p}{\partial t}
    = -\frac{\partial}{\partial x}
    \left[
    \left(a+\frac{1}{2}bb'\right)p
    \right]
    + \frac{1}{2}\frac{\partial^2}{\partial x^2}[b^2p],
\end{equation}
which can be equivalently rewritten in the standard Stratonovich differential form:
\begin{equation}
    \frac{\partial p}{\partial t}
    = -\frac{\partial}{\partial x}[a(x,t)p]
    + \frac{1}{2}\frac{\partial}{\partial x}
    \left[
    b(x,t)\frac{\partial}{\partial x}(b(x,t)p)
    \right].\label{eq:255}
\end{equation}
Finally, for the anti-It\^o (H\"anggi--Klimontovich) convention, $c=1$,
Eq.~(\ref{eq:170}) with $c_1=1$ and $c_2=0$ shows that the equivalent It\^o SDE
has drift $a+bb'$. Substituting this into Eq.~(\ref{eq:253}) gives
\begin{align}
    \frac{\partial p}{\partial t}
    = -\frac{\partial}{\partial x}\big[a(x,t)\,p\big]
      + \frac{1}{2}\frac{\partial}{\partial x}
        \left[b^2(x,t)\,\frac{\partial p}{\partial x}\right],
    \label{eq:FP-antiIto}
\end{align}
since $\tfrac12\partial_x^2(b^2p)=\tfrac12\partial_x(b^2\partial_x p)+\partial_x(bb'p)$.
Eqs.~(\ref{eq:253}), (\ref{eq:255}) and~\eqref{eq:FP-antiIto} differ only in where
$b$ sits relative to the derivatives, and all three are special cases of
\begin{eqnarray}
    \frac{\partial p}{\partial t}
    = -\frac{\partial}{\partial x}\big[a\,p\big]
      + \frac{1}{2}\frac{\partial}{\partial x}
        \left[b^{2c}\,\frac{\partial}{\partial x}\big(b^{2(1-c)}\,p\big)\right],
\end{eqnarray}
where $c\in[0,1]$.
\subsection{Boundary Conditions}

The forward Fokker--Planck equation is a partial differential equation and hence its solutions depend crucially on the boundary conditions. This means that even stationary solutions which may correspond to either an equilibrium state or a nonequilibrium steady state must depend on boundary conditions. Understanding boundary conditions is, thus, essential because they encode physical constraints imposed by the system's environment or geometry, thereby influencing the behaviour and long-term states of the probability distribution. Let us try to get a deeper understanding of this fact through the theoretical foundations and practical implications of imposing boundary conditions on the 1D forward Fokker--Planck equation equation.

Consider the probability density \( p(x,t) \) evolving according to:
\begin{equation}
\partial_t p(x,t) = -\partial_x J(x,t),
\label{eq:fp_continuity}
\end{equation}
where the probability flux \( J(x,t) \) is defined---with $A$ and ${B}/2$, respectively, as the drift and the diffusion coefficients, for the notational convenience---as
\begin{equation}
J(x,t) = A(x,t) p(x,t) -\frac{1}{2} \partial_x \big[ B(x,t) p(x,t) \big].
\label{eq:flux}
\end{equation}
Eq.~\eqref{eq:fp_continuity} can be interpreted as a continuity equation for probability: the time rate of change of probability at point \(x\) equals the net inflow or outflow of probability current through \(x\). Integrating over a spatial domain and applying boundary conditions determines the total probability flow and its conservation.

 For convenience let the Fokker--Planck process be occurring between two boundaries at $x_a$ and $x_b$. We consider five important types of boundary conditions:
\begin{enumerate}
\item \emph{Dirichlet (Absorbing) Boundary Condition:}  
\[
p(x_b, t) = 0.
\] 
The probability density vanishes at the boundary \(x_b\), modeling absorption or removal of particles at the boundary.

{Physical interpretation:} Probability is lost through the boundary, so the flux \(J(x_b,t)\) can be nonzero. This boundary models situations such as a particle escaping a confined domain or being absorbed at the system edge.

{Example:} Consider diffusion of molecules in a container with a small hole that allows particles to escape; the boundary at the hole acts as an absorbing boundary where \(p=0\).
\item \emph{Reflecting (Hard Wall) Boundary Condition:}  
\[
J(x_b, t) = 0.
\] 
No net probability current crosses the boundary, implying the system confines particles within the domain.

{Physical interpretation:} Particles are reflected back into the domain; probability flux at the boundary is zero.

{Example:} A particle confined in a box with perfectly reflecting walls, such as a Brownian particle trapped between hard barriers, cannot leave the domain.

\item \emph{Neumann (Flat Profile) Boundary Condition:}  
\[
\partial_x p(x_b, t) = 0.
\] 
The spatial derivative of the probability density is zero at the boundary (flat profile).

{Physical interpretation:} The probability distribution is locally flat at the boundary. Unlike reflecting conditions, Neumann conditions do not guarantee zero flux; the flux may be nonzero depending on drift and diffusion terms. 

{Example:} Consider a channel connecting to a large reservoir, where the particle density gradient is zero at the boundary due to reservoir buffering, but net drift can induce a flux.

A caveat: Both Neumann and Reflecting boundary conditions involve spatial derivatives or fluxes at the boundary, but they impose fundamentally different constraints: They coincide only when special conditions are met; e.g., in case the diffusion is not $x$-dependent, the vanishing of drift term at the boundary [ \( A(x_b) p(x_b) = 0 \)] makes Neumann and Reflecting conditions equivalent.

\item \emph{Periodic (Loop) Boundary Condition:}  
\[
p(x_a, t) = p(x_b, t), \quad J(x_a, t) = J(x_b, t).
\] 
The domain is topologically a loop, where probability density and flux match at both ends.

{Physical interpretation:} The system forms a closed ring; probability and flux are continuous through the boundaries.

{Example:} Diffusion of ions around a circular ring-shaped conductor or molecular motors moving along a circular track.

\item \emph{Natural (Infinite Domain) Boundary Condition:}  
\[
\lim_{x \to \pm \infty} p(x,t) = 0, \quad \lim_{x \to \pm \infty} J(x,t) = 0.
\] 
Probability density and flux vanish at infinity.

{Physical interpretation:} The system extends infinitely, and particles are localized so the probability vanishes far away.

{Example:} Free Brownian motion in the whole real line where particles spread but become very unlikely far from the origin.

\end{enumerate}

Now how these five boundary conditions influence the stationary solutions. Assuming a stationary distribution ($p_s(x)$ such that $\partial_t p_s = 0$), implies:
\[
\partial_x J(x) = 0 \implies J(x) = C,
\]
where \(C\) is a constant flux throughout the domain. If \(C=0\), the system is in \emph{equilibrium}, with no net probability flow; if \(C \neq 0\), a \emph{non-equilibrium steady state} (NESS) exists with persistent probability current. Different boundary conditions constrain the value of \(C\), determining whether the system can realize equilibrium or a NESS. The five boundary conditions introduced earlier imply the following cases:
\begin{itemize}
\item \emph{Dirichlet (Absorbing) Boundaries:}  
Since \(p(x_b) = 0\) at the boundaries, the stationary density typically must vanish at the edges. Because probability escapes the domain, the only stationary solution consistent with normalization is the trivial one \(p_s=0\), corresponding to decay over time rather than a true steady state. Thus, no equilibrium or steady flux \(C\) exists; the system relaxes to zero probability. In short, Dirichlet boundaries imply no stationary equilibrium or NESS.

\item \emph{Reflecting (Hard Wall) Boundaries:}  
The zero flux condition \(J(x_b) = 0\) at boundaries enforces \(C=0\). Hence, the system can only be in equilibrium with no net probability current. Reflecting boundaries naturally support equilibrium stationary distributions confined within finite domains and preclude NESS.

\item \emph{Neumann (Flat Profile) Boundaries:}  
Here \(\partial_x p(x_b) = 0\) fixes the gradient, but not flux, at the boundaries. Because flux depends on drift and diffusion, \(J(x_b)\) can be zero or nonzero. Thus, both equilibrium (\(C=0\)) and non-equilibrium steady states (\(C \neq 0\)) can arise, depending on system parameters. Neumann conditions allow for a richer set of steady states, including persistent currents when drift drives flow at the boundaries.

\item \emph{Periodic (Loop) Boundaries:}  
The periodicity \(p(x_a) = p(x_b)\) and \(J(x_a) = J(x_a)\) naturally admits constant nonzero flux \(C\). Such systems commonly support NESS with sustained circulation of probability around the loop, corresponding to persistent currents and nonzero \(C\). Equilibrium (\(C=0\)) is also possible but less generic.

\item \emph{Natural (Infinite Domain) Boundaries:}  
The decay conditions \(p(x) \to 0\), \(J(x) \to 0\) as \(x \to \pm \infty\) imply zero flux at infinity, so \(C=0\). Hence, stationary solutions represent equilibrium localized distributions that vanish at infinity. NESS does not arise in this setting due to boundary conditions forbidding net flux escape or circulation. Importantly, the stationary distribution is normalizable only if the drift provides sufficient confinement; otherwise, probability spreads indefinitely and no true stationary state exists.
\end{itemize}

Finally for completeness, since there are two boundaries in 1D problems, we consider all 15 possible pairs of boundary conditions, chosen from \{Dirichlet (D), Reflecting (R), Neumann (N), Periodic (P), Natural (Nat)\}. Table~\ref{tab:BCs} summarizes the conclusions. Of course, having periodic BC on one boundary and some other BC on the other boundary makes no sense because, by definition, defining periodic BC requires its specifying on both the boundaries simultaneously. The Dirichlet--Natural boundary pair is mathematically incompatible too. To make it  crystal clear, consider the Fokker--Planck equation on the semi-infinite domain \([a, \infty)\) with boundary conditions
\[
p(x_a,t) = 0, \quad \lim_{x \to \infty} p(x,t) = 0, \quad \lim_{x \to \infty} J(x,t) = 0.
\]

Integrating the continuity equation over the domain,
\[
\frac{d}{dt} \int_{x_a}^\infty p(x,t) \, dx = -J({x_a},t) + \lim_{x \to \infty} J(x,t) = -J({x_a},t).
\]

Since \(p({x_a},t) = 0\), the flux \(J({x_a},t) \geq 0\) represents an absorbing boundary where probability constantly escapes, causing
\[
\frac{d}{dt} \int_{x_a}^\infty p(x,t) \, dx < 0.
\]

With no inflow from infinity (\(\lim_{x\to \infty} J(x,t) = 0\)), the total probability strictly decreases over time, violating normalization:
\[
\int_{x_a}^\infty p(x,t) \, dx < 1 \quad \forall t > 0.
\]
Therefore, no normalized stationary distribution \(p_s(x)\) exists. The boundary value problem is ill-posed for a probability density and no physically meaningful stationary or time-dependent solution with conserved total probability exists.

\begin{table}[h]
\label{tab:BCs}
\caption{Admissibility and Stationary State Classification for Boundary Condition (BC) Pairs.}
\begin{tabular}{cccc}
\hline
\,BC Pair\, & \,Admissible?\, & \,\(p_s = 0\)?\, & \,Equilibrium/NESS?\, \\
\hline
D--D     & Yes & Yes & None \\
D--R     & Yes & Yes & None \\
D--N     & Yes & Yes & None \\
D--P     & No  & --  & -- \\
D--Nat   & No  & --  & -- \\
R--R     & Yes & No  & Equilibrium \\
R--N     & Yes & No  & Equilibrium \\
R--P     & No  & --  & -- \\
R--Nat   & Yes & No  & Equilibrium \\
N--N     & Yes & No  & Both \\
N--P     & No  & --  & -- \\
N--Nat   & Yes & No  & Both \\
P--P     & Yes & No  & NESS \\
P--Nat   & No  & --  & -- \\
Nat--Nat & Yes & No  & Equilibrium \\
\hline
\end{tabular}
\end{table}

Conclusion is that the existence and character of a stationary state for a Fokker--Planck process are intimately tied to the boundary conditions. Reflecting walls allow equilibrium Boltzmann-like distributions, absorbing boundaries lead to depletion, periodic domains may support non-equilibrium steady states, and natural boundaries on an infinite line often forbid stationary solutions when a constant drift is present. As to be discussed now, the simple example of sedimentation under gravity illustrates how the geometry of confinement dictates the long-time statistical behaviour.

Consider a particle sedimenting in a viscous medium. Choose the vertical coordinate \(z\) {positive upward}; then gravity points downward, i.e., the constant force is \(-mg\) with \(g>0\). The overdamped It\^o equation reads
\begin{equation}
dz = -g\,dt + \sqrt{2D}\,dW(t),
\label{eq:SDE}
\end{equation}
so the drift is \(A(z) = -g\) and the diffusion coefficient is \({B(z)}/2 = D\). The Fokker--Planck flux becomes
\begin{equation}
J(z,t) = -g\,p - D\frac{\partial p}{\partial z}.
\label{eq:flux_grav}
\end{equation}

{Absorbing Boundaries:} If the boundaries are absorbing (\(p=0\) at \(z=0,L\)), the only stationary solution that satisfies both conditions is \(p_s(z) = 0\). Probability mass decays exponentially with time as particles are trapped at the boundaries. Essentially, absorbing boundaries model sinks or traps. A particle that reaches the bottom \((z=0)\) or top \((z=L)\) is permanently removed. Consequently, the probability density inside the domain eventually vanishes; all particles sediment to the bottom or escape through the top. This describes, for example, sedimentation into a collecting plate.

{Reflecting Boundaries:} Let the particle be confined between two reflecting walls at \(z=0\) and \(z=L\). Setting \(J=0\)  (because Reflecting BC forces so) in~\eqref{eq:flux_grav} gives
\begin{equation}
\frac{dp_s}{dz} = -\frac{g}{D} p_s,
\end{equation}
whose normalized solution is the exponential (barometric) distribution
\begin{equation}
p_s(z) = \frac{g/D}{1 - e^{-gL/D}} e^{-gz/D}, \qquad 0 \le z \le L.
\end{equation}
Thus, the reflecting walls trap the particle in a box. Gravity pulls it downward, but diffusion spreads it upward. The competition produces an exponential density profile, exactly analogous to the barometric law for an isothermal atmosphere: the probability of finding the particle at height \(z\) decays as \(\exp(-gz/D)\), where \(D\) plays the role of \(k_B T / (mg)\).

{Periodic Domain:} Take a ring of circumference \(L\) (periodic coordinate \(z\)). The constant flux solution of Eq.~\eqref{eq:flux_grav} with \(J = C\) is
\begin{equation}
p_s(z) = -\frac{C}{g} + K e^{-gz/D}.
\end{equation}
Periodicity \(p_s(0) = p_s(L)\) and \(J(0) = J(L)\) force \(C = -g / L\) (a non-zero current) and \(K = 0\), yielding the uniform density \(p_s(z) = 1 / L\). Thus, with a constant drift, the periodic system cannot sustain a stationary density unless the drift vanishes; instead, probability circulates with a constant current \(J = -g / L\).

Therefore, on a periodic ring, gravity drives a persistent probability current. Particles constantly drift downward (in the negative-\(z\) direction), but because the domain is closed, they re-enter at the top after completing a loop. The result is a uniform density (no height-dependent gradient) accompanied by a steady circulation of probability--a clear example of a nonequilibrium steady state.

The periodic boundary applied to a purely gravitational potential appears unphysical---it would require particles to be teleported from the top to the bottom without conserving potential energy. This mathematical idealization, however, is physically realized in systems driven by non-conservative forces on a closed loop, such as a colloidal particle on a rapidly rotating ring (where the centrifugal force provides a constant tangential drive) or a charged particle driven by a constant electric field around a circular electrode. In those real systems the same overdamped SDE structure emerges, and the periodic-sedimentation model therefore captures the essential features of a genuine nonequilibrium steady state sustained by external energy injection.

{Neumann Boundaries:} Consider Neumann (flat profile) boundary conditions at \(z=0\) and \(z=L\),
\begin{equation}
\frac{\partial p}{\partial z}(0,t) = 0, \quad \frac{\partial p}{\partial z}(L,t) = 0.
\end{equation}
Using the flux form~\eqref{eq:flux_grav}, the Neumann condition does not enforce zero flux; instead,
\[
\frac{\partial p}{\partial z} = 0 \implies \quad J(z_b, t) = -g \, p(z_b,t),
\]
which can be nonzero if \(g \neq 0\) and \(p(z_b,t) \neq 0\).

The stationary solution \(p_s(z)\) satisfies
\begin{equation}
\frac{dp_s}{dz} = -\frac{g}{D} p_s + \frac{C}{D},
\end{equation}
where \(C\) is the constant stationary flux, \(J = C\). Solving this with Neumann conditions leads to
\begin{equation}
p_s(z) = \frac{C}{g} + K e^{-gz/D},
\end{equation}
where constants \(C, K\) are fixed by the Neumann conditions:
\[
\left.\frac{dp_s}{dz}\right|_{z=0} = 0, \quad \left.\frac{dp_s}{dz}\right|_{z = L} = 0.
\]

This system admits two distinct cases:
\begin{itemize}
    \item {Equilibrium (Zero Flux):} When the stationary flux is zero (\(C=0\)), the solution reduces to the reflecting boundary case,
    \[
    p_s(z) = K e^{-gz/D},
    \]
    with no net probability current through the boundaries. The system reaches thermodynamic equilibrium with a detailed balance distribution and zero steady-state flux.
    
    \item {Non-Equilibrium Steady State (NESS, Nonzero Flux):} When the stationary flux is nonzero (\(C \neq 0\)), the solution becomes
    \[
    p_s(z) = \frac{C}{g} + K e^{-gz/D},
    \]
    allowing a persistent, nonzero probability current \(J=C\) flowing through the system. This models a system connected to reservoirs or environments at the boundaries that maintain a steady flux, breaking detailed balance and producing a genuine non-equilibrium steady state.
\end{itemize}

 In this sedimentation scenario, Neumann boundaries correspond to impermeable walls where the probability density gradient vanishes but a net flux of particles can still exist if maintained by external sources or sinks. When the flux \(C=0\), particles accumulate in an equilibrium barometric distribution as in the reflecting case. However, if \(C \neq 0\), the system sustains a steady downward or upward probability current, modeling situations where particles are continuously supplied or removed at the boundaries, such as injection or extraction processes in a confined sedimentation column. This leads to a non-equilibrium steady state with a persistent sedimentation flow through the domain despite the walls being impermeable to gradients.

Neumann boundary conditions can be viewed as a generalization that bridges reflecting and periodic cases. When the flux \(C=0\), Neumann BCs coincide with reflecting boundaries, both enforcing no net particle flow across boundaries and resulting in equilibrium distributions confined within the domain. In contrast, allowing nonzero flux (\(C \neq 0\)) in Neumann BCs enables a non-equilibrium steady state resembling that of periodic boundaries, where persistent currents circulate indefinitely. This correspondence makes physical sense, as Neumann boundaries control gradient flatness rather than flux directly, thereby accommodating both impermeable equilibrium confinement and steady flux scenarios, linking smoothly the concepts of reflecting and periodic boundary behaviours.

{Natural Boundaries:} On the infinite line \(z \in (-\infty, \infty)\), the normalizability condition \(\int_{-\infty}^{\infty} p_s(z) \, dz = 1\) cannot be satisfied by the exponential solution \(p_s(z) \propto e^{-gz/D}\) when \(g > 0\), because the density grows without bound as \(z \to -\infty\). Hence no normalizable stationary state exists: the particle sediments indefinitely downward. What is physically happening is that with no boundaries to reflect or absorb the particle, gravity pulls it ever downward. Diffusion cannot counteract the constant drift on an infinite domain; the probability density simply spreads and drifts downward forever, never reaching a time-independent shape. This corresponds to unbounded sedimentation in an infinitely deep column.
\subsection{It\^o--Stratonovich Dichotomy Revisited}  
Now we are well-equipped to further comprehend the It\^o--Stratonovich dichotomy. We observe that the master equation involves no approximation beyond the assumption of Markovianity. So validating appropriateness of an ad hoc SDE in a system of interest is now possible. For illustrative purpose, we consider a system with internal noise---a \emph{demographic noise} to be precise: Demographic noise is defined as the random fluctuations that occur simply because the system is made up of discrete, individual units (like atoms, molecules, or animals) that undergo probabilistic events one at a time. Specifically, consider the process of radioactive decay. Let us use $N(t)$ as the notation for this stochastic process. Suppose a radioactive material with $N(0)=n_0$ active nuclei decaying at a constant rate $\lambda$. The process $X(t)\equiv n_0-N(t)$ is known to be a \emph{Poisson process}. 

A Poisson process is a kind of \emph{counting process} [$X(t)\ge0$, $X(0)=0$, $X(t)\in \mathbb{Z}$, $X(t')\le X(t)$ if $t'<t$, and $X(t) -X(t')$ is equal to the number of events between $t'$ and $t$] with rate $\lambda$ if
\begin{enumerate}
\item X(t) has independent and stationary increments.
\item $P[X(t+\Delta t)-X(t)=1]=\lambda\Delta t +\mathcal{O}(\Delta t^{>1}).$
\item $P[X(t+\Delta t)-X(t)\ge2]=\mathcal{O}(\Delta t^{>1}).$
\end{enumerate}
Note $\lim_{\Delta t\to 0}\frac{\mathcal{O}(\Delta t^{>1})}{\Delta t}=0$. The set of three conditions above can be shown to be equivalent to the following set:
\begin{enumerate}
\item X(t) has independent increments.
\item The number of events in any interval of length $t$ is Poisson distributed with mean $\lambda t$, i.e., $\forall\,t,t' $  $$P[X(t+t')-X(t')=k]=e^{-\lambda t}\frac{(\lambda t)^k}{k!}.$$
Stationarity follows automatically as the distribution is not explicitly dependent on $t'$.
\end{enumerate}

Now, we want to guess the Langevin equation for nuclear decay using physical intuition. We start with the deterministic foundation. The standard macroscopic decay law states that the decay rate is proportional to the amount of substance present: Any stochastic version of this process must reduce to this deterministic drift when the noise is `turned off'. Therefore, the core structure of the desired equation must look like:
\begin{equation}
    \frac{dN(t)}{dt\phantom{(t)}} = -\lambda N(t) + \text{noise}.
\end{equation}
Now, the noise term must physically vanish when $N(t)$ is zero. If one were to try using simple additive noise (e.g., $+ \sigma \, dW_t$ with a constant amplitude $\sigma$), the model would fail physically. If the number of nuclei $N$ reaches zero, a constant noise term would continue fluctuating the value, leading to a physically impossible negative number of nuclei. Thus, the noise amplitude $b(N)$ must be a function of $N$ such that $b(0) = 0$. As far as the noisy fluctuations in radioactive decays concerned, they are independent, discrete events; consequently, their statistics follow Poisson distributions. A Poisson distribution has unique property that all its cumulants are identical. Hence, since the mean rate of decay is proportional to $N$, so should be the variance; and hence, the noise-amplitude should go as $\sqrt{N}$ (proportional to the standard deviation). Moreover, since the exact same microscopic process (individual nuclei decaying at rate $\lambda$) governs both the predictable macroscopic drift and the random demographic fluctuations, a single parameter must rule both. The rate constant $\lambda$ should also scale the noise. To match dimensions and units properly, putting the aforementioned heuristic considerations together yields the following guess of a Langevin equation for nuclear decay:
\begin{equation} \label{eq:langevinai1}
      \frac{dN(t)}{dt\phantom{(t)}} = -\lambda N(t)+ \sqrt{\lambda N(t)}\,\xi,
\end{equation}
or in an SDE form:
\begin{equation} \label{eq:langevinai2}
dN(t) = -\lambda N(t) dt + \sqrt{\lambda N(t)} \,dW(t).
\end{equation}
This represents a process with a deterministic exponential decay and state-dependent (multiplicative) noise that naturally turns itself off as the radioactive sample depletes.

We know an SDE with multiplicative noise is mathematically ambiguous until an interpretation is chosen. We note that the guesswork SDE~\eqref{eq:langevinai2} is supposed to be an approximation of a discrete master Equation (a Markovian jump process). In such processes, the probability of a decay event happening in the next infinitesimal instant $dt$ depends strictly on the number of atoms present {right now} at the beginning of the interval. This perfectly aligns with It\^o calculus, which evaluates functions at the start of the time step (non-anticipating). Conversely, the Stratonovich interpretation evaluates at the midpoint, implying a correlation between the state and the noise during the interval. While appropriate for systems driven by continuous external coloured noise (e.g., a fluctuating environment) or by internal noise arising from integrating out fast continuous internal degrees of freedom (where the Stratonovich interpretation becomes the natural choice), it is entirely inappropriate for internal, discrete {demographic noise}. Just to make the argument more concrete: if this equation were interpreted in the Stratonovich sense, the multiplicative noise would introduce an additional `noise-induced drift' which would physically alter the average decay rate, blatantly violating the known fundamental physics of the macroscopic system.
 
Before we compare the heuristically derived equation with the full exact treatment of the master equation, we calculate the first three moments $\langle N^m(t)\rangle$ ($m=1,2,3$) under the It\^o interpretation. To this end, we observe that for a general It\^o SDE $dN=a(N)\,dt+b(N)\,dW$, It\^o's lemma [Eq.~(\ref{eq:ifor})] applied to $f(N)=N^m$ and taking expectations yields
\begin{equation}
\frac{d\langle N^m\rangle}{dt} 
= \Bigl\langle a(N)\,m N^{m-1} + \tfrac{1}{2}b^2(N)\,m(m-1)N^{m-2}\Bigr\rangle.\label{eq:sdeai}
\end{equation}
Substituting $a(N)=-\lambda N$ and $b^2(N)=\lambda N$ into 
Eq.~\eqref{eq:sdeai} gives the closed hierarchy:
\begin{equation}
\frac{d\langle N^m\rangle}{dt} 
= -\lambda m\,\langle N^m\rangle 
+ \tfrac{1}{2}\lambda m(m-1)\,\langle N^{m-1}\rangle.
\label{eq:hierarchy}
\end{equation}
Crucially, the moment of order $m$ depends only on moments of order $\leq m$, so the hierarchy can be solved recursively without closure 
approximations.

Setting $m=1$ in Eq.~\eqref{eq:hierarchy},
\begin{equation}
\frac{d\langle N\rangle}{dt} = -\lambda\langle N\rangle,
\end{equation}
which with $\langle N(0)\rangle = n_0$ gives
\begin{equation}
{\;\langle N(t)\rangle = n_0q,\;}
\end{equation}
where $q\equiv e^{-\lambda t}$. 
This is precisely the deterministic decay law. For the second moment ($m=2$), one has
\begin{equation}
\frac{d\langle N^2\rangle}{dt} 
= -2\lambda\langle N^2\rangle + \lambda\langle N\rangle 
= -2\lambda\langle N^2\rangle + \lambda n_0 e^{-\lambda t}.
\end{equation}
Using the integrating factor $e^{2\lambda t}$ and the fact that $\langle N^2(0)\rangle = n_0^2$ we find $C=n_0^2-n_0$, we arrive on integration from $0$ to $t$ at
\begin{equation}
{\;\langle N^2(t)\rangle 
= n_0^2\,q^2 + n_0\,q(1-q)=n_0 q + n_0(n_0-1)q^2.\;}
\end{equation}
The third moment can be calculated similarly to arrive at
\begin{align}
\langle N^3(t)\rangle 
&= n_0^3q^3
+3n_0(n_0-1)(q^2-q^3)
+\frac{3}{2} n_0(q-q^3)\nonumber\\
&=\frac{3}{2}n_0q
+3n_0(n_0-1)q^2
+\left(n_0^3-3n_0^2+\frac{3}{2}n_0\right)q^3.\label{eq:is3m}
\end{align}

How true are the results of the It\^o SDE? Well, the master equation formalism helps us determine that. Let $p_n(t)\equiv p(n,t\mid n_0,t_0=0)$ denote the probability that $n$ undecayed nuclei remain at time $t$. Therefore, we can now put Eq.~(\ref{eq:masterdis}) to use but first we need to ascertain the transition rates. During a small time interval $\Delta t$, each of the $n'$ nuclei present at time $t$ may decay with probability $\lambda \Delta t$, independently of the others. Therefore, the probability of exactly one decay is $n'\lambda \Delta t$ to first order in $\Delta t$, while the probability of two or more decays is $\mathcal{O}(\Delta t^2)$. Hence, the transition probability from state $n'$ to state $n$ is
\begin{equation}
    R_t(n|n')\Delta t =
    \begin{cases}
        0, & n > n'; \\[4pt]
        n'\lambda \Delta t, & n=n'-1; \\[4pt]
        \mathcal{O}(\Delta t^2), & n<n'-1.
    \end{cases}
\end{equation}
This reflects the fact that, to leading order in $\Delta t$, the number of active nuclei can decrease by at most one in a single time step.

Consequently, the master equation is:
\begin{equation}
    \dot{p}_n(t)= \sum_{n'} R_t(n|n')\,p_{n'}(t)- \sum_{n'} R_t(n'|n)\,p_n(t).
\end{equation}
The gain term receives contributions only from the state $n+1$, while the loss term comes from transitions out of state $n$. Thus,
\begin{equation}
    \dot{p}_n(t)
    = \lambda (n+1)p_{n+1}(t) - \lambda n\,p_n(t).
\end{equation}

We now derive the first three moments beginning with the first moment:
\begin{align}
    \frac{d}{dt}\langle N(t)\rangle
    &= \sum_{n=0}^{\infty} n\,\dot{p}_n(t) \nonumber \\
    &= \sum_{n=0}^{\infty} n\left[\lambda(n+1)p_{n+1}(t)-\lambda n\,p_n(t)\right] \nonumber \\
    &= \lambda \sum_{n=0}^{\infty} n(n+1)p_{n+1}(t)
       - \lambda \sum_{n=0}^{\infty} n^2 p_n(t) \nonumber \\
    &= \lambda \sum_{k=1}^{\infty} (k-1)k\,p_k(t)
       - \lambda \sum_{k=0}^{\infty} k^2 p_k(t) \nonumber \\
    &= -\lambda \sum_{k=1}^{\infty} k\,p_k(t)= -\lambda \langle N(t)\rangle.
\end{align}
Thus,
\begin{equation}
    \langle N(t)\rangle = n_0 q,
\end{equation}
with $q=e^{-\lambda t}$ like before. Similar manipulations give consider the second moment's and the third moment's time derivatives respectively as
\begin{align}
    \frac{d}{dt}\langle N^2(t)\rangle
    &= -2\lambda \langle N^2(t)\rangle + \lambda \langle N(t)\rangle, \\
    \frac{d}{dt}\langle N^3(t)\rangle
    &= -3\lambda \langle N^3(t)\rangle
       + 3\lambda \langle N^2(t)\rangle
       - \lambda \langle N(t)\rangle.
\end{align}
Their solutions respectively are:
\begin{align}
    \langle N^2(t)\rangle &= n_0 q + n_0(n_0-1)q^2, \\
    \langle N^3(t)\rangle &= n_0 q
    + 3n_0(n_0-1)q^2
    + n_0(n_0-1)(n_0-2)q^3.\label{eq:me3m}
\end{align}

Taken together, these results show that the It\^o Langevin equation
is a useful diffusion approximation to the exact master equation,
but it is not fully equivalent to the underlying discrete decay
process. The It\^o SDE and the master equation produce identical
first moments and identical second moments, and therefore the same variance.
They disagree, however, at the third moment---compare Eqs.~(\ref{eq:is3m}) and~(\ref{eq:me3m}).  The connection with the Fokker--Planck equation clarifies the origin of this mismatch. The It\^o SDE is equivalent to the
diffusion equation
\begin{equation}
{
\frac{\partial p}{\partial t}
=
\lambda\frac{\partial}{\partial N}(Np)
+\frac{\lambda}{2}
\frac{\partial^2}{\partial N^2}(Np).
}
\end{equation}
This is the second-order Kramers--Moyal
approximation to the exact master equation. It retains the first
two jump coefficients---the drift and diffusion---but neglects
the nonzero higher-order coefficients associated with the finite
jump \(N\to N-1\). Consequently, it reproduces the exact first
two moments but not the complete higher-order statistics.
Strictly speaking, the exact description is the discrete master
equation; no ordinary second-order Fokker--Planck equation can
represent the finite jump process exactly at all orders. The
Fokker--Planck equation above is therefore the natural and best
diffusion approximation when \(N\) is sufficiently large. It is important that this Fokker--Planck equation be associated
with the It\^o interpretation. In this formulation there is no
additional noise-induced drift---additional \(\lambda/4\) factor---which appears only if the
same SDE is instead interpreted in the Stratonovich sense and
then converted to It\^o form. Consequently, the Stratonovich SDE can be shown to yield: $\langle N(t)\rangle
=n_0q+\frac{1}{4}(1-q)$, which is already incorrect at first order. Message: for internal noise, both It\^o and Stratonovich interpretations could lead to incorrect results---one of them may be at best a better approximation than the other.

\begin{center}
	$\symking\symqueen\symrook\symknight\symbishop\sympawn\epsdice{1}\,\epsdice{2}\,\epsdice{3}\,\epsdice{4}\,\epsdice{5}\,\epsdice{6}\,\spadesuit\heartsuit\clubsuit\diamondsuit$
\end{center}
\newpage
\newpage
\clearpage
\onecolumngrid
\section{Ornstein--Uhlenbeck Process: Pause \& Ponder}
\epigraph{\textsf{Reflective thinking turns experience into insight.}}{John C. Maxwell}
\twocolumngrid

\indent {\color{deepsea}{\LARGE \textgoth{A}}}he \emph{Ornstein--Uhlenbeck (OU) process} is a fundamental stochastic process describing a system that exhibits mean reversion toward a central equilibrium
state while remaining subject to random fluctuations. Introduced as
a physically consistent modification of standard Brownian motion,
it is uniquely characterized, under suitable regularity
assumptions, as a stationary Gaussian Markov process. It therefore
occupies a central position in statistical mechanics and applied
probability. In physical systems, it describes the velocity of a
Brownian particle experiencing viscous friction, as well as the
position of an overdamped Brownian particle subject to a Hookean
restoring force. Beyond physics, variants of the process appear in
financial mathematics, evolutionary biology, neuroscience, and
other disciplines. Its theoretical importance lies in its
mathematical tractability and in its role as the canonical model of
a continuous-time mean-reverting system in which deterministic
relaxation and stochastic diffusion are balanced. 

Interestingly an instructive toy model---\emph{Ehrenfest urn model}---is intimately connected with the
OU process, which emerges naturally in
an appropriate large-system scaling limit. 
The Ehrenfest model consists of a closed system containing \(N\)
distinguishable balls, numbered \(1,\ldots,N\), distributed between
two identical containers denoted Urn~A and Urn~B. The system evolves
through a sequence of random transfer events. At each step, one of
the \(N\) balls is selected uniformly at random, located in its
current urn, and transferred to the opposite urn.

The goal of this section is to study {Ehrenfest urn model} and its limit, the OU process so that through them we can revisit several central ideas of nonequilibrium statistical mechanics, including relaxation toward equilibrium, detailed
balance, reversibility, recurrence, stochastic differential equation, and Fokker--Planck equation.

\subsection{Discrete-Time Master Equation}

The standard Ehrenfest Urn process is a discrete-time, finite-state
Markov chain. Its time-parameter space is
\begin{equation}
    \mathcal{T}=\mathbb{Z}_{\ge0}=\{0,1,2,\ldots\},
\end{equation}
where each increment of time corresponds to one ball-transfer
event.

Two distinct but compatible Markov descriptions are useful. The first
retains the identities of all balls and describes the system at the
microscopic level. The second records only the total occupancy of one
urn and is obtained by coarse-graining the microscopic process.

First, at the microscopic level, every ball is distinguishable. A complete
specification of the system records, for each ball
$i\in\{1,\ldots,N\}$, the urn---$A$ or $B$---in which it is located. A
microstate is therefore a vector
\begin{equation}
    \omega=(\sigma_1,\sigma_2,\ldots,\sigma_N),
    \qquad
    \sigma_i\in\{A,B\}.
\end{equation}
The microscopic state space is
\begin{equation}
    \Omega_{\mathrm{micro}}=\{A,B\}^{N},
    \qquad
    |\Omega_{\mathrm{micro}}|=2^{N}.
\end{equation}
Therefore, the state space, $\mathcal{X}$, is the set of $2^N$
vertices of an $N$-dimensional hypercube.

Let $\tilde{\sf P}$ denote the microscopic one-step transition
probability matrix; its elements are denoted as
$\tilde{p}_{\omega\omega'}$, the probability of transitioning from
state $\omega$ to $\omega'$. Since a single ball is selected uniformly
and moved to the opposite urn,
\begin{equation}
\tilde{p}_{\omega\omega'}
=
\begin{cases}
\dfrac{1}{N},
&
\text{if }d_H(\omega,\omega')=1,\\[6pt]
0,
&
\text{otherwise},
\end{cases}
\label{eq:micro-kernel}
\end{equation}
where $d_H$ is the \emph{Hamming distance}, namely the shortest-path
distance between two vertices along the edges of the hypercube.
Consequently,
\begin{equation}
    \tilde{p}_{\omega\omega'}
    =
    \tilde{p}_{\omega'\omega}.
\label{eq:micro-kernel-symmetry}
\end{equation}
The microscopic transition matrix is therefore symmetric.

The probability state vector is a $2^N$-dimensional row vector,
denoted by
\begin{equation}
\tilde{\boldsymbol{\pi}}(t)
=
\big[
\tilde{\pi}_{\omega_1}(t)\,
\tilde{\pi}_{\omega_2}(t)\,
\cdots\,
\tilde{\pi}_{\omega_{2^N}}(t)
\big],
\end{equation}
where $\omega_i\in\Omega_{\rm micro}$. Its evolution is
\begin{equation}
    \tilde{\boldsymbol{\pi}}(t+1)
    =
    \tilde{\boldsymbol{\pi}}(t)\tilde{\sf P}.
\label{eq:micro-evolution}
\end{equation}

A stationary distribution is a probability row vector
$\tilde{\boldsymbol{\pi}}^{*}$ satisfying
\begin{equation}
    \tilde{\boldsymbol{\pi}}^{*}
    =
    \tilde{\boldsymbol{\pi}}^{*}\tilde{\sf P}.
\label{eq:micro-stationarity}
\end{equation}
Because $\tilde{\sf P}$ is symmetric and each row sums to one, each column also sums to
one. The uniform distribution,
\begin{equation}
    \tilde{\pi}^{*}_{\omega}
    =
    \frac{1}{2^N},
    \qquad
    \omega\in\Omega_{\mathrm{micro}},
\label{eq:micro-stationary-distribution}
\end{equation}
is normalized, since $\Omega_{\mathrm{micro}}$ has $2^N$ elements, and it is
stationary. Indeed, for every $\omega'$,
\begin{equation}
    \sum_{\omega\in\Omega_{\mathrm{micro}}}
    \tilde{\pi}^{*}_{\omega}
    \tilde{p}_{\omega\omega'}
    =
    \frac{1}{2^N}
    \sum_{\omega\in\Omega_{\mathrm{micro}}}
    \tilde{p}_{\omega\omega'}
    =
    \frac{1}{2^N}
    =\tilde{\pi}^{*}_{\omega'}.
\end{equation}

The microscopic chain is irreducible for $N\geq1$, since any vertex of the hypercube
can be reached from any other by changing the positions of the balls one at a time.
Because the state space is finite, an irreducible chain is positive recurrent, so the
stationary distribution in Eq.~\eqref{eq:micro-stationary-distribution} exists and is
unique. However, the chain has period two: every transition changes the parity of the
Hamming weight, and the chain cannot return to its starting state in an odd number of
steps. Consequently, the stationary distribution is not a limiting distribution:
$\tilde{\sf P}^n$ applied to a fixed initial state oscillates with the parity of $n$, and
only time averages converge to it.

The microscopic chain satisfies detailed balance with respect to
$\tilde{\boldsymbol{\pi}}^{*}$. For all
$\omega,\omega'\in\Omega_{\mathrm{micro}}$,
\begin{equation}
    \tilde{\pi}^{*}_{\omega}
    \tilde{p}_{\omega\omega'}
    =
    \tilde{\pi}^{*}_{\omega'}
    \tilde{p}_{\omega'\omega}.
\label{eq:micro-detailed-balance}
\end{equation}
This follows immediately from the symmetry of $\tilde{\sf P}$ and the uniformity of
$\tilde{\boldsymbol{\pi}}^{*}$. Thus, the chain is reversible: there is no net
probability current between any pair of states, and the stationary distribution is an
equilibrium distribution.

Next, at the macroscopic level, the identities of the balls are discarded
and only the occupancy of Urn~$A$ is retained:
\begin{equation}
    m_t
    \equiv
    \#\{\text{balls in Urn~A at time }t\}.
\label{eq:macro-variable}
\end{equation}
The corresponding macrostate space is
\begin{equation}
    \mathcal{S}=\{0,1,2,\ldots,N\},
    \qquad
    |\mathcal{S}|=N+1.
\end{equation}

A macrostate $m$ corresponds to
\begin{equation}
    \Omega(m)={}^{N}C_{m}
\end{equation}
labelled microstates, because one must choose which $m$ of the $N$
balls occupy Urn~$A$. Thus, the mapping
\begin{equation}
    \omega\longmapsto m(\omega)
\end{equation}
is a many-to-one coarse-graining of the microscopic state space.

Because the microscopic transition rule is invariant under
permutations of the ball labels and depends only on whether the
selected ball lies in Urn~$A$ or Urn~$B$, this coarse-graining is
Markovian. Hence, $\{m_t\}_{t\in\mathcal{T}}$ is itself a Markov chain
on $\mathcal{S}$.

If the current occupancy is $m$, then $m$ balls are in Urn~$A$ and
$N-m$ are in Urn~$B$. Selecting a ball in Urn~$A$ produces the
transition $m\to m-1$, whereas selecting a ball in Urn~$B$ produces
the transition $m\to m+1$. Therefore,
\begin{align}
    {p}_{m,m-1}
    &=\frac{m}{N},
    \label{eq:macro-down}\\
    {p}_{m,m+1}
    &=\frac{N-m}{N},
    \label{eq:macro-up}\\
    {p}_{m,m}
    &=0.
\end{align}
All other transition probabilities vanish. These elements define the
macroscopic transition matrix ${\sf P}$. In the ordering
$\mathcal{S}=(0,1,\ldots,N)$, it has the tridiagonal form
\begin{equation}
{\sf P}
=
\begin{pmatrix}
0 & 1 & 0 & \cdots & 0\\
\frac{1}{N} & 0 & \frac{N-1}{N} & \cdots & 0\\
0 & \frac{2}{N} & 0 & \ddots & \vdots\\
\vdots & \ddots & \ddots & \ddots & 1\\
0 & \cdots & 0 & 1 & 0
\end{pmatrix},
\label{eq:macro-transition-matrix}
\end{equation}
where the displayed boundary entries are understood through Eqs.~\eqref{eq:macro-down}--\eqref{eq:macro-up}. In particular,
${\sf P}$ is not symmetric in general.

The probability state vector is an $(N+1)$-dimensional row vector,
denoted by
\begin{equation}
\boldsymbol{\pi}(t)
=
\big[
\pi_{m_1}(t)\,
\pi_{m_2}(t)\,
\cdots\,
\pi_{m_{N+1}}(t)
\big],
\end{equation}
where $m_i\in\mathcal{S}$. Its evolution is
\begin{equation}
    \boldsymbol{\pi}(t+1)
    =
    \boldsymbol{\pi}(t){\sf P}.
\label{eq:macro-evolution}
\end{equation}

The stationary distribution
$\boldsymbol{\pi}^{*}$ satisfies
\begin{equation}
    \boldsymbol{\pi}^{*}
    =
    \boldsymbol{\pi}^{*}{\sf P}.
\label{eq:macro-stationarity}
\end{equation}
It is given by the binomial distribution
\begin{equation}
    \pi_m^{*}
    =
    \frac{{}^{N}C_m}{2^N},
    \qquad
    m\in\mathcal{S}.
\label{eq:macro-stationary-distribution}
\end{equation}
This is precisely the distribution obtained by aggregating the
uniform microscopic stationary distribution:
\begin{equation}
    \pi_m^{*}
    =
    \sum_{\omega\in\Omega(m)}
    \tilde{\pi}_{\omega}^{*}
    =
    \frac{|\Omega(m)|}{2^N}
    =
    \frac{{}^{N}C_m}{2^N}.
\end{equation}

The macroscopic chain is irreducible for $N\geq1$, since every state in
$\mathcal{S}$ can be reached from every other state by successive
increments or decrements. It is also periodic with period two, because
each transition changes the parity of $m$. Therefore,
\eqref{eq:macro-stationary-distribution} is the unique stationary
distribution, but it is not generally a limiting distribution at
individual times. 

Although ${\sf P}$ is not symmetric, the macroscopic chain satisfies
detailed balance with respect to
$\boldsymbol{\pi}^{*}$. For neighbouring states,
\begin{align}
    \pi_m^{*}p_{m,m+1}
    &=
    \frac{{}^{N}C_m}{2^N}\frac{N-m}{N},\\
    \pi_{m+1}^{*}p_{m+1,m}
    &=
    \frac{{}^{N}C_{m+1}}{2^N}\frac{m+1}{N}.
\end{align}
Using
\begin{equation}
    {}^{N}C_m(N-m)
    =
    {}^{N}C_{m+1}(m+1),
\end{equation}
we obtain
\begin{equation}
    \pi_m^{*}p_{m,m+1}
    =
    \pi_{m+1}^{*}p_{m+1,m}.
\label{eq:macro-detailed-balance-neighbor}
\end{equation}
For non-neighbouring states, both transition probabilities are zero.
Hence, for all $m,m'\in\mathcal{S}$,
\begin{equation}
    \pi_m^{*}p_{m,m'}
    =
    \pi_{m'}^{*}p_{m',m}.
\label{eq:macro-detailed-balance}
\end{equation}

Thus, detailed balance is preserved by this coarse-graining, even
though the transition matrix itself ceases to be symmetric. The
asymmetry of ${\sf P}$ is exactly compensated by the unequal
multiplicities ${}^{N}C_m$ of the macrostates. Thus, the stationary distribution is an equilibrium distribution.

\subsection{Lumpability \& Coarse-Graining}
\label{sec:lumpability}

The macroscopic Ehrenfest chain is obtained by coarse-graining the microscopic chain.
More generally, coarse-graining a Markov chain does not automatically produce another
Markov chain. The relevant condition is \emph{strong lumpability}.

Let a finite microscopic Markov chain have state space $\Omega_{\mathrm{micro}}$ and
transition matrix $\tilde{\sf P}$. Suppose that $\Omega_{\mathrm{micro}}$ is
partitioned into nonempty, disjoint macrostates,
\begin{equation}
    \mathcal{A}=\{A_1,A_2,\ldots,A_r\},\,\,    \bigcup_{i=1}^{r}A_i=\Omega_{\mathrm{micro}},
    \,\,
    A_i\cap A_j=\emptyset\ \ (i\neq j).
\end{equation}
The \emph{coarse-grained} (also called \emph{aggregated} or \emph{lumped}) process
$\hat{X}_t$ records only the partition element occupied by the microscopic chain $X_t$:
\begin{equation}
    \hat{X}_t=A_i
    \quad\Longleftrightarrow\quad
    X_t\in A_i.
\end{equation}

The partition $\mathcal{A}$ is \emph{strongly lumpable} with respect to $\tilde{\sf P}$
if, for every pair of macrostates $A_i,A_j\in\mathcal{A}$ (including $i=j$), the total
probability of transitioning from a microstate in $A_i$ to the macrostate $A_j$ is
independent of the particular microstate in $A_i$:
\begin{equation}
    \sum_{\omega'\in A_j}\tilde{p}_{\omega\omega'}
    =
    \sum_{\omega'\in A_j}\tilde{p}_{\bar{\omega}\omega'},\,\,
    \forall\,\omega,\bar{\omega}\in A_i,\,\,
    \forall\,A_i,A_j\in\mathcal{A}.
\label{eq:strong-lumpability}
\end{equation}
When \eqref{eq:strong-lumpability} holds, the quantities
\begin{equation}
    \hat{p}_{ij}=\sum_{\omega'\in A_j}\tilde{p}_{\omega\omega'},
    \qquad
    \omega\in A_i,
\label{eq:lumped-transition-probability}
\end{equation}
are well defined. Each row sums to one, $\sum_j\hat p_{ij}=1$, so they form the
transition matrix
\begin{equation}
    \hat{\sf P}=\big(\hat{p}_{ij}\big)_{i,j=1}^{r}
\end{equation}
of a time-homogeneous coarse-grained Markov chain.

The \emph{Kemeny--Snell lumpability theorem} states that the coarse-grained process
$\hat{X}_t$ is Markov for every initial distribution of the microscopic chain if and only
if the partition is strongly lumpable. The future macrostate then depends only on the
current macrostate, and not on the unresolved microscopic state or on the preceding
macro-history. If strong lumpability fails, $\hat{X}_t$ is
non-Markovian for at least some initial distributions.

Next, suppose $\tilde{\boldsymbol{\pi}}^*$ is stationary for the microscopic chain.
Define the aggregate distribution by adding up the probability sitting in each block,
\begin{equation}
\hat{\pi}^*_i=\sum_{\omega\in A_i}\tilde{\pi}^*_{\omega},
\label{eq:slsd}
\end{equation}
which is a probability vector, $\sum_i\hat{\pi}^*_i=\sum_{\omega}\tilde{\pi}^*_{\omega}=1$.
Then, under strong lumpability, $\hat{\boldsymbol{\pi}}^*$ is a stationary distribution of
the lumped chain. Indeed, using \eqref{eq:lumped-transition-probability} for each
$\omega\in A_i$ and then the stationarity of $\tilde{\boldsymbol{\pi}}^*$,
\begin{align}
\sum_{i}\hat{\pi}^*_i\hat{p}_{ij}
&=\sum_{i}\sum_{\omega\in A_i}\tilde{\pi}^*_{\omega}
  \sum_{\omega'\in A_j}\tilde{p}_{\omega\omega'}
  \,\,\text{(by Eq.~\eqref{eq:lumped-transition-probability}, Eq.~\eqref{eq:slsd})}\nonumber\\
&=\sum_{\omega'\in A_j}\sum_{\omega\in\Omega_{\mathrm{micro}}}
  \tilde{\pi}^*_{\omega}\tilde{p}_{\omega\omega'}
  \,\,\,\,\text{(blocks partition }\Omega_{\mathrm{micro}})\nonumber\\
&=\sum_{\omega'\in A_j}\tilde{\pi}^*_{\omega'}
  \,\,\qquad\qquad\quad\,\,\text{(stationarity)}\nonumber\\
&=\hat{\pi}^*_j .
\end{align}

The intuition is a conservation-of-probability argument. If one starts the microscopic
chain in $\tilde{\boldsymbol{\pi}}^*$ and runs it one step, nothing changes; in
particular the total probability inside each block $A_i$ is unchanged. Strong
lumpability guarantees that the evolution of these block totals is governed by
$\hat{\sf P}$ alone, so the block totals are invariant under the lumped chain too, which
is the statement $\hat{\boldsymbol{\pi}}^*=\hat{\boldsymbol{\pi}}^*\hat{\sf P}$.

A caveat concerns uniqueness. Irreducibility passes from the microscopic to the lumped
chain, since a path of microstates from $A_i$ to $A_j$ gives a path of blocks. Hence if
the microscopic chain is irreducible, $\hat{\boldsymbol{\pi}}^*$ is the unique
stationary distribution of the lumped chain. The converse fails: the lumped chain may be
irreducible while the microscopic chain is not, in which case several microscopic
stationary distributions can project onto the same $\hat{\boldsymbol{\pi}}^*$.

Strong lumpability preserves not only the Markov property but also reversibility:

Let $\tilde{\sf P}$ be a finite-state Markov transition matrix on
$\Omega_{\mathrm{micro}}$, and let $\mathcal{A}=\{A_1,\ldots,A_r\}$ be a strongly
lumpable partition. Suppose that the microscopic chain is reversible with respect to a
stationary distribution $\tilde{\boldsymbol{\pi}}^*$, so that
\begin{equation}
    \tilde{\pi}^*_{\omega}\tilde{p}_{\omega\omega'}
    =
    \tilde{\pi}^*_{\omega'}\tilde{p}_{\omega'\omega},
    \quad
    \forall\,\omega,\omega'\in\Omega_{\mathrm{micro}}.
\label{eq:micro-reversibility-burke-rosenblatt}
\end{equation}
Then the lumped chain $\hat{\sf P}$ is reversible with respect to the aggregate
distribution $\hat{\pi}^*_i=\sum_{\omega\in A_i}\tilde{\pi}^*_{\omega}$:
\begin{equation}
    \hat{\pi}^*_i\hat{p}_{ij}
    =
    \hat{\pi}^*_j\hat{p}_{ji},
    \quad
    \forall\,i,j\in\{1,\ldots,r\}.
\label{eq:macro-reversibility-burke-rosenblatt}
\end{equation}

{Proof:} Using Eq.~\eqref{eq:slsd} and Eq.~\eqref{eq:lumped-transition-probability},
and then microscopic detailed balance Eq.~\eqref{eq:micro-reversibility-burke-rosenblatt},
\begin{align}
    \hat{\pi}^*_i\hat{p}_{ij}
    =
    \sum_{\omega\in A_i}\tilde{\pi}^*_{\omega}\hat{p}_{ij}
    =
    \sum_{\omega\in A_i}\sum_{\omega'\in A_j}
    \tilde{\pi}^*_{\omega}\tilde{p}_{\omega\omega'}
    =
    \sum_{\omega'\in A_j}\tilde{\pi}^*_{\omega'}
    \sum_{\omega\in A_i}\tilde{p}_{\omega'\omega}.
\end{align}
By Eq.~\eqref{eq:lumped-transition-probability}, the inner sum equals $\hat{p}_{ji}$, so
\begin{equation}
    \hat{\pi}^*_i\hat{p}_{ij}
    =
    \Bigg(\sum_{\omega'\in A_j}\tilde{\pi}^*_{\omega'}\Bigg)\hat{p}_{ji}
    =
    \hat{\pi}^*_j\hat{p}_{ji}.
\end{equation}
This is detailed balance for the lumped chain, proving its reversibility.

Both results were originally given for discrete time, but they carry over to
continuous-time finite Markov chains, with the transition matrix replaced by the
generator.

In view of the above general results, it is not a surprise that there are
correspondences, including for detailed balance, between the stationary distributions of
the microscopic and the macroscopic Ehrenfest urn models: the macroscopic sample space
$\mathcal{S}$ is a partition of the microscopic sample space $\Omega_{\mathrm{micro}}$.
In the notation of this section, an element of $\mathcal{S}$ is
$A_k=\{\omega:\text{exactly }k\text{ balls lie in Urn-A}\}$. Let's fix $\omega\in A_k$. A step
moves one uniformly chosen ball; since $k$ balls lie in Urn-A and $N-k$ in Urn-B,
\begin{equation}
    \sum_{\omega'\in A_{k-1}}\tilde{p}_{\omega\omega'}=\frac{k}{N},
    \qquad
    \sum_{\omega'\in A_{k+1}}\tilde{p}_{\omega\omega'}=\frac{N-k}{N},
\end{equation}
and the sum over every other block, including $A_k$, vanishes, because each step changes
the number of balls in Urn-A by exactly one. These sums depend only on $k$, not on which
balls sit where, so Eq.~\eqref{eq:strong-lumpability} holds: the Ehrenfest urn model is
\emph{strongly lumpable}. The lumped transition probabilities are the macroscopic ones,
$\hat p_{k,k-1}=k/N$ and $\hat p_{k,k+1}=(N-k)/N$. The Kemeny--Snell theorem then
guarantees that the macroscopic process is Markov, and reversibility is preserved. The
aggregate of the uniform stationary distribution is the binomial distribution,
$\hat\pi^*_k={}^{N}C_{k}\,2^{-N}$, which satisfies detailed balance for the macroscopic
chain. The underlying reason for all this is symmetry: the dynamics are invariant under
permutations of the ball labels, and the blocks $A_k$ are the orbits of these
permutations.

A coarse-graining of a Markov chain is Markov for every initial distribution only if
the lumpability condition holds, and for a generic partition it does not. The Ehrenfest
urn satisfies it because the dynamics are invariant under permutations of the ball
labels. Further insight from the Ehrenfest urn model follows: in passing to the
macroscopic description, statistical information is lost. The \emph{Shannon entropy} of
the uniform microscopic stationary distribution is $H_{\text{micro}}=\log_2(2^N)=N$
bits. For the lumped (binomial) stationary distribution
$\hat\pi^*_k={}^{N}C_{k}\,2^{-N}$,
\begin{equation}
H_{\text{macro}}
=-\sum_{k=0}^{N}\hat\pi^*_k\log_2\hat\pi^*_k
=N-\frac{1}{2^N}\sum_{k=0}^{N}{}^{N}C_{k}\,\log_2\!\big({}^{N}C_{k}\big).
\end{equation}
The information lost in the lumping step,
$\Delta H=H_{\text{micro}}-H_{\text{macro}}$, is the second term: the average of
$\log_2|A_k|$, i.e.\ the entropy of the microstate within its block. For large $N$,
$H_{\text{macro}}\simeq\tfrac12\log_2(\pi eN/2)$ grows only logarithmically, whereas
$H_{\text{micro}}=N$ grows linearly. Hence $\Delta H\to\infty$ and the retained fraction
$H_{\text{macro}}/H_{\text{micro}}\simeq(\log_2N)/N\to0$, even though $H_{\text{macro}}$
itself still diverges slowly.

Stretching the conclusion to real-world thermodynamics, the macroscopic description
retains a vanishing fraction of the microscopic information. Macroscopic variables, such
as temperature, pressure, or the number of balls in an urn, are insensitive to the
identities and exact locations of individual particles. Since $\Delta H$ grows linearly
with $N$, most of the microscopic information is discarded in coarse-graining. The
projection is many-to-one and hence non-invertible: the microstate cannot be
reconstructed from the macrostate. Nevertheless, for a strongly lumpable partition the
retained information, $H_{\text{macro}}$, suffices to form a closed Markov chain, so the
discarded information is not needed to predict the evolution of the macrostate.

A fundamental question arises: can every Markovian macroscopic model be obtained by
coarse-graining a microscopic process? No, not if the microscopic process is a finite
Markov chain with rational transition probabilities, as in counting statistics. Every
lumped transition probability is then a finite sum of rationals, hence rational.
Consider the phenomenological two-state system
\begin{equation}
\hat{\sf P}=\begin{pmatrix} 0.5 & 0.5 \\ \frac{1}{\pi} & 1-\frac{1}{\pi}\end{pmatrix}.
\end{equation}
It is a perfectly valid stochastic matrix, but it cannot arise from lumping any such
chain, because its transition probability contains the transcendental number $\pi$. (This
does not exclude microscopic models with irrational probabilities or continuous
microscopic dynamics, for example deterministic maps with a Markov partition.)

Another point is worth noting. Detailed balance,
$\hat\pi_i\hat p_{ij}=\hat\pi_j\hat p_{ji}$, is the signature of reversibility of the
stationary process. Macroscopic detailed balance does not guarantee a reversible
microscopic world. To illustrate, we construct a 4-state microscopic chain with one-way
fluxes that coarse-grains into a two-state chain obeying detailed balance. (A stationary two-state chain always obeys detailed balance, since stationarity forces
the single net current $\hat\pi_A\hat p_{AB}-\hat\pi_B\hat p_{BA}$ to vanish. The point
of the example is therefore that the microscopic currents
$J_{ij}=\tilde\pi^*_i\tilde p_{ij}-\tilde\pi^*_j\tilde p_{ji}$ do not vanish, even though
their aggregate does.)

Let $\Omega_{\rm micro}=\{1,2,3,4\}$ with
\begin{equation}
\tilde{\sf P}=\begin{pmatrix}
0.1 & 0.4 & 0.5 & 0.0 \\
0.0 & 0.5 & 0.0 & 0.5 \\
0.5 & 0.0 & 0.2 & 0.3 \\
0.0 & 0.5 & 0.4 & 0.1
\end{pmatrix}.\label{eq:4stateex}
\end{equation}
The chain is irreducible, with stationary distribution
$\tilde{\boldsymbol\pi}^*=(20,63,36,47)/166>0$. It has directed fluxes: $\tilde p_{12}=0.4$
but $\tilde p_{21}=0$, so $\tilde\pi^*_1\tilde p_{12}\neq\tilde\pi^*_2\tilde p_{21}$, and
detailed balance fails.

Let's take the partition $A=\{1,2\}$, $B=\{3,4\}$. The block transition probabilities are the
same for every microstate in a block:
\begin{align}
P(1\to B)&=\tilde p_{13}+\tilde p_{14}=0.5,\,\, P(2\to B)=\tilde p_{23}+\tilde p_{24}=0.5,\\
P(3\to A)&=\tilde p_{31}+\tilde p_{32}=0.5,\,\, P(4\to A)=\tilde p_{41}+\tilde p_{42}=0.5,
\end{align}
and the probabilities of staying in a block follow from the row sums. Strong lumpability
holds, with
\begin{equation}
\hat{\sf P}=\begin{pmatrix}0.5&0.5\\0.5&0.5\end{pmatrix},\,\, 
\hat\pi_A=\hat\pi_B=\tfrac12,\,\, 
\hat\pi_A\hat p_{AB}=\hat\pi_B\hat p_{BA}=\tfrac14 .
\end{equation}
An observer restricted to macroscopic observables would conclude that the system obeys
detailed balance, unaware that the microstates carry directed probability currents whose
net effect between the two blocks vanishes upon coarse-graining.

\subsection{Macroscopic Irreversibility}
\label{sec:relaxation_irreversibility}
Let's consider Eq.~\eqref{eq:macro-evolution} again. Having established the macroscopic
transition matrix $\hat{\sf P}$, we now derive the exact relaxation law for the expected
occupancy of Urn A. The conditional expectation of $m_{t+1}$ given $m_t$,
$\mathbb{E}[m_{t+1}\mid m_t]=\sum_{m'}m'\,P(m_{t+1}=m'\mid m_t)$, is the mean of
$m_{t+1}$ for a given value of $m_t$; it is a function of $m_t$, and hence a random
variable. Using the transition probabilities $p_{m,m-1}=m/N$ and
$p_{m,m+1}=(N-m)/N$,
\begin{align}
    \mathbb{E}[m_{t+1}\mid m_t]
    &=
    (m_t-1)\,p_{m_t,m_t-1} + (m_t+1)\,p_{m_t,m_t+1} \nonumber\\
    &=
    (m_t-1)\frac{m_t}{N} + (m_t+1)\frac{N-m_t}{N} \nonumber\\
    &=
    \left(1-\frac{2}{N}\right)m_t+1.
\label{eq:conditional-mean-simplified}
\end{align}
Averaging Eq.~\eqref{eq:conditional-mean-simplified} over the distribution of $m_t$, using
the \emph{law of total expectation}, $\mathbb{E}[Y]=\mathbb{E}\big[\mathbb{E}[Y\mid X]\big]$
with $Y=m_{t+1}$ and $X=m_t$, yields the recurrence relation
$\langle m_{t+1}\rangle=\left(1-\frac{2}{N}\right)\langle m_t\rangle+1$. Introducing the
\emph{deviation of the mean occupancy} from its equilibrium value,
$\mu_t\equiv\langle m_t\rangle-\frac{N}{2}$, the constant terms cancel, leaving
$\mu_{t+1}=\left(1-\frac{2}{N}\right)\mu_t$. Iteration from $t=0$ gives the exact
discrete-time relaxation law:
\begin{equation}
    \mu_t=\left(1-\frac{2}{N}\right)^t\mu_0 .
\label{eq:dmo}
\end{equation}
For $N>2$, the factor $1-2/N$ lies strictly between zero and one, so the mean occupancy
relaxes monotonically toward $N/2$. (For $N=2$ the factor vanishes and the mean reaches
$N/2$ in one step.)

This framework transitions smoothly into a continuous-time formulation by assigning independent exponential clocks to the balls. Adopting a normalisation where the total event rate is one (each ball flips urns at rate $1/N$), the continuous-time transition rates are $R_{m\to m-1}=\frac{m}{N}$ and $R_{m\to m+1}=\frac{N-m}{N}$. Let $p_m(t)\equiv P(m(t)=m)$. The continuous-time forward master equation becomes:
\begin{align}
    \frac{\partial p_m(t)}{\partial t}
    &=
    \frac{m+1}{N}\,p_{m+1}(t) + \frac{N-m+1}{N}\,p_{m-1}(t) - p_m(t),\label{eq:oume}
\end{align}
with boundary conditions $p_{-1}(t)=p_{N+1}(t)=0$. Unlike the discrete chain, this process is aperiodic because the exponentially distributed waiting times remove the parity oscillation, allowing pointwise convergence to the binomial stationary distribution $\pi^*_m$. 

The mean $\langle m(t)\rangle = \sum_{m=0}^{N} m\,p_m(t)$ should follow the net drift due to transition rates. Indeed, the expected instantaneous change of the occupancy is
\begin{align}
    \frac{d}{dt}\langle m(t)\rangle
    &=
    \left\langle (+1)R_{m\to m+1} + (-1)R_{m\to m-1}\right\rangle\nonumber\\
    &=
    \left\langle \frac{N-m}{N}-\frac{m}{N}\right\rangle=
    1-\frac{2}{N}\langle m(t)\rangle .
\end{align}
The equation is exact and requires no
moment closure. In terms of the deviation $\mu(t) = \langle m(t)\rangle - \frac{N}{2}$, this yields the differential equation $\frac{d\mu}{dt} = -\frac{2}{N}\mu(t)$, whose solution is the continuous-time exponential decay:
\begin{equation}
    {
    \mu(t) = \mu(0)e^{-2t/N}
    }.
\label{eq:continuous_relaxation}
\end{equation}
This matches the large $N$ limit of the discrete formulation [Eq.~\eqref{eq:dmo}], as $\left(1-\frac{2}{N}\right)^t \simeq \exp(-2t/N)$.


Equations governing this monotonic relaxation---whether the discrete $\mu_t = (1-2/N)^t\mu_0$ or the continuous $\mu(t) = \mu(0)e^{-2t/N}$---profoundly illustrate the emergence of \emph{macroscopic irreversibility} from reversible microscopic laws. So we see that the macroscopic dynamics relaxes to the highest entropy state---the macrostate with maximum number of microstates.

As established previously, the microscopic transition matrix $\tilde{\sf P}$ obeys
detailed balance, Eq.~\eqref{eq:micro-detailed-balance}, because it is symmetric and the
stationary distribution is uniform. The stationary microscopic process is therefore
reversible: if one were to watch a film of a single particle trajectory at the micro-level,
playing it forward or backward in time would yield statistically indistinguishable
sequences of steps. However, the relaxation equation~\eqref{eq:continuous_relaxation}
introduces a definitive ``arrow of time''. The macroscopic observable $\mu(t)$ decays
deterministically and monotonically to zero. A film of the {mean} macrostate $\mu(t)$
evolving from a nonequilibrium state (e.g., all balls in Urn A) to equilibrium looks
drastically different when played in reverse; a macroscopic system does not
spontaneously un-relax to a highly ordered state.

This apparent Loschmidt's paradox is resolved by understanding the role of coarse-graining. The exact relaxation equation demonstrates that irreversibility is not a property of the fundamental mechanics, but an emergent statistical feature born from information loss. Because the equilibrium macrostate $m=N/2$ possesses an overwhelmingly larger statistical weight (${}^N C_{N/2}$) than the peripheral states, any stochastic step taken in the micro-world overwhelmingly favours macroscopic flux towards $m=N/2$. The relaxation equation thus mathematically encodes the Second Law of Thermodynamics: macroscopic irreversibility is simply the deterministic manifestation of a reversible system wandering into the vast, dominant combinatorial space of equilibrium.

Before we discuss the resolution of the paradox in more detail, it is worth reflecting
that a careful account of the arrow of time rests on separating three notions of
reversibility. \emph{Microscopic reversibility} means that every individual labelled
transition has an equally probable reverse. \emph{Macroscopic Markov reversibility}
means that the coarse-grained chain satisfies detailed balance with respect to its
binomial stationary distribution, so that equilibrium flows between macrostates exactly
cancel. \emph{Macroscopic irreversibility} means that a low-entropy macrostate typically
relaxes toward equilibrium, while the reverse, though not forbidden, is overwhelmingly
atypical.

The upshot is that macroscopic detailed balance does not abolish the statistical arrow
of time. It only tells us that the net macroscopic currents vanish when the system is
already described by the correct stationary ensemble. The irreversibility we actually
observe belongs to nonequilibrium, low-entropy preparations and is expressed through
typicality and multiplicity, the sheer imbalance between the number of ways to approach
equilibrium and to leave it. In the Ehrenfest model it does not come from any asymmetry
in the transition rules. This is exactly the distinction between properties of the
dynamics and properties of the ensemble: the rule is symmetric, but the states we
prepare and the states we typically observe are not.

Microscopic reversibility, together with strong lumpability, yields macroscopic Markov
reversibility: the partition must respect the dynamics, so that microstates in the same
macrostate share the same block-transition statistics. The converse fails. A
macroscopically reversible Markov chain can arise from many microscopic dynamics,
including irreversible ones, as in the four-state example of Eq.~\eqref{eq:4stateex},
because detailed balance constrains only the aggregate macrostate-to-macrostate
probabilities, not the individual trajectories composing them.

This is why the meaning of detailed balance must not be overread. It is a property of
the transition rule {together with} a specific equilibrium measure, asserting that
in equilibrium each flux is matched by its reverse. It does not require the actual,
time-dependent distribution of the system to equal that measure. The same rule governs
the dynamics whether the initial distribution is the equilibrium one or a constrained,
low-entropy one. Detailed balance describes the {rule}, not the instantaneous
condition of the system, and that distinction is precisely what resolves Loschmidt's
paradox: the reversibility of the rule and the irreversibility of typical evolution
answer two different questions (the balance of fluxes in equilibrium, and the fate of an
atypical initial distribution), with no contradiction between them.

Much of the force of Loschmidt's paradox arises from an implicit
identification of two different probabilistic objects: the
conditional probability of a trajectory given its initial
microstate, and the unconditional probability of observing
that trajectory under a specified ensemble preparation. Keeping
these two objects distinct is, as we shall see, the whole of the
formal resolution, with the physically decisive input entering only
afterward through the multiplicities of the Ehrenfest macrostates.

Consider a microscopic trajectory of the Ehrenfest urn model of
length \(T\),
\begin{equation}
    \gamma=(\omega_0,\omega_1,\ldots,\omega_T),
    \quad \omega_t\in\{A,B\}^N,
\end{equation}
where each \(\omega_t\) is a labelled assignment of the \(N\) balls
to the two urns, and define its reversed trajectory by
\begin{equation}
    \gamma^\dagger
    =
    (\omega_T,\omega_{T-1},\ldots,\omega_0).
\end{equation}
The conditional probability of the forward trajectory, given its
initial microstate, is
\begin{equation}
    P(\gamma\mid\omega_0)
    =
    \prod_{t=0}^{T-1}
    \tilde{p}_{\omega_t\omega_{t+1}}.
\end{equation}
This expression contains only transition probabilities of the
microscopic chain \(\tilde{\sf P}\); it does not specify how the
initial microstate \(\omega_0\) was selected. Because each Ehrenfest
step moves a single ball, the only nonzero factors correspond to
neighbouring vertices of the hypercube,
\(d_H(\omega_t,\omega_{t+1})=1\), each carrying weight \(1/N\).

The microscopic Ehrenfest chain is symmetric,
\begin{equation}
    \tilde{p}_{\omega_t\omega_{t+1}}
    =
    \tilde{p}_{\omega_{t+1}\omega_t}
\end{equation}
for every allowed transition, since transferring a given ball from
one urn to the other and the reverse transfer occur with the same
rate \(1/N\). Hence
\begin{align}
    P(\gamma\mid\omega_0)
    =
    \prod_{t=0}^{T-1}
    \tilde{p}_{\omega_t\omega_{t+1}}
=
    \prod_{t=0}^{T-1}
    \tilde{p}_{\omega_{t+1}\omega_t}
=
    P(\gamma^\dagger\mid\omega_T).
\end{align}
This is microscopic reversibility at the level of complete
trajectories. Every entropy-increasing microscopic history of the
urns has an entropy-decreasing reversed history of exactly the same
conditional probability, provided each trajectory is conditioned on
its own initial microstate. Were this the entire probabilistic
description, Loschmidt's objection would prevail.

Macroscopic observations, however, concern unconditional path
probabilities, which depend both on the transition dynamics and on
the preparation of the initial ensemble. Suppose the forward
experiment begins with a distribution \(\rho_0\) on the microstate
space, while the reversed experiment begins with a distribution
\(\rho_T\). The corresponding unconditional path probabilities are
\begin{align}
    P_{\rho_0}(\gamma)
    &=
    \rho_0(\omega_0)\,
    P(\gamma\mid\omega_0),
    \\
    P_{\rho_T}(\gamma^\dagger)
    &=
    \rho_T(\omega_T)\,
    P(\gamma^\dagger\mid\omega_T).
\end{align}
Taking their ratio and applying the conditional path equivalence,
all dynamical factors cancel and one obtains
\begin{equation}
    {
    \begin{split}
    \frac{P_{\rho_0}(\gamma)}
         {P_{\rho_T}(\gamma^\dagger)}
    =
    \frac{\rho_0(\omega_0)}
         {\rho_T(\omega_T)}
    \end{split}
    }.
\label{eq:path-ratio}
\end{equation}
The physical meaning of Eq.~\eqref{eq:path-ratio} deserves emphasis.
The asymmetry between the forward and the reversed observations is
determined entirely by the distributions used to prepare their
initial microstates. Microscopic reversibility does not imply that
forward and reversed trajectories are equally probable under
arbitrary ensembles; it implies only that their conditional
probabilities are equal when each trajectory is conditioned on the
appropriate initial endpoint. The arrow of time, insofar as it
appears in the relative frequency of observed paths, is therefore
carried by the preparation distributions \(\rho_0\) and \(\rho_T\),
not by the microscopic transition matrix \(\tilde{\sf P}\). This
resolves the formal part of Loschmidt's objection: conditional
reversibility, a property of the dynamics, must not be confused with
unconditional ensemble symmetry, which depends on both the dynamics
and the preparation. A further question remains, however: why are
physically realised preparations systematically asymmetric? 

To
answer it, suppose the
forward ensemble is prepared uniformly within a low-entropy
macrostate \(A\) of occupancy \(m_A\), so that
\begin{equation}
    \rho_0(\omega_0)
    =
    \frac{1}{\Omega(A)}
    =
    \frac{1}{{}^{N}C_{m_A}}
    \qquad(\omega_0\in A),
\end{equation}
while the reverse experiment begins from the full microscopic
equilibrium measure,
\begin{equation}
    \rho_T(\omega_T)=2^{-N}.
\end{equation}
Equation~\eqref{eq:path-ratio} then gives
\begin{align}
    \frac{P_{\rho_0}(\gamma)}
         {P_{\rho_T}(\gamma^\dagger)}=
    \frac{\rho_0(\omega_0)}
         {\rho_T(\omega_T)}
=
    \frac{2^N}{\Omega(A)}
=
    \frac{2^N}{{}^{N}C_{m_A}}.\label{eq:zces}
\end{align}
For an occupancy \(m_A\) macroscopically separated from \(N/2\),
this ratio is exponentially large in \(N\). 

It quantifies the
asymmetry between observing ordinary relaxation from a
deliberately prepared low-entropy state of the urns and observing
the corresponding anti-relaxation history after sampling an
initial microstate from equilibrium. The arrow of time is
therefore not a dynamical asymmetry of the transition
matrix; it is a statistical and combinatorial asymmetry,
transmitted to observable path probabilities through the initial
ensemble. However, we must not forget that this asymmetry of preparation is not itself derived from
reversible statistical mechanics. In
laboratory settings such low-entropy states are produced through
external preparation procedures, and in cosmology the claim that
the universe began in an exceptionally low-entropy condition is
often called the \emph{past hypothesis}, whose ultimate explanation
lies outside the present stochastic model. The temporal direction of entropy
increase is inherited from the temporal asymmetry of the boundary
condition, not from any failure of microscopic reversibility.

The path-probability ratio of Eq.~\eqref{eq:path-ratio} is the building block of a broader class of results, the \emph{fluctuation theorems}. Their central quantity is the \emph{dissipation} of a trajectory,
\begin{equation}
  \Sigma_T(\gamma)=\ln\frac{P_{\rho_0}(\gamma)}{P_{\rho_T}(\gamma^\dagger)},
\end{equation}
which measures how much more likely $\gamma$ is than its reverse $\gamma^\dagger$, and is physically the entropy produced along $\gamma$, in units of $k_B$. Evaluated on the random trajectory of an experiment, $\Sigma_T$ becomes a random variable. For the Ehrenfest urn, Eq.~\eqref{eq:zces} gives $\Sigma_T=(S_{\rm eq}-S_A)/k_B$. Let $P_{\rm f}(\Sigma_T=a)$ be the probability that the forward experiment, prepared with $\rho_0$, yields dissipation $a$, and $P_{\rm r}$ the corresponding probability for the reversed experiment, prepared with $\rho_T$. By definition, every trajectory with $\Sigma_T(\gamma)=a$ satisfies $P_{\rho_0}(\gamma)=e^{a}P_{\rho_T}(\gamma^\dagger)$, and its reverse $\gamma^\dagger$ has dissipation $-a$ in the reversed experiment; summing over all such trajectories gives $P_{\rm f}(\Sigma_T=a)=e^{a}P_{\rm r}(\Sigma_T=-a)$. When the two experiments are statistically identical, $P_{\rm f}=P_{\rm r}\equiv P$, and this becomes the \emph{Evans--Searles fluctuation theorem}:
\begin{equation}
  \frac{P(\Sigma_T=a)}{P(\Sigma_T=-a)}=e^{a}.
\end{equation}
Entropy-decreasing trajectories thus occur, as Loschmidt insisted, but are exponentially suppressed, and the second law holds only on average over trajectories, $\avg{\Sigma_T}\ge0$. The \emph{Crooks fluctuation theorem} and the \emph{Jarzynski equality} extend this reasoning to systems driven by external forces between equilibrium states.

Finally we note that the analogue of Zermelo's paradox is also present in Ehrenfest Urn model rather transparently.
The microscopic process is a finite irreducible Markov chain.
Consequently, every microscopic state is positive recurrent and is
visited infinitely often with probability one. (This is the
stochastic analogue of Poincar\'e recurrence.) Hence, entropy cannot increase forever---as the arrow of time seems to suggest---bringing the paradox to the fore. The resolution of the paradox lies in the realization that the existence of recurrence, however, says nothing by itself about
the timescale on which recurrence occurs. Formally, \emph{Kac's lemma} resolves this apparent contradiction by quantifying the
mean recurrence time. For an irreducible, positive recurrent chain, the mean time to
return to a state $\omega$, starting from $\omega$, is the inverse of its stationary
probability (periodicity plays no role):
\begin{equation}
    \mathbb{E}\big[\tau_\omega\mid X_0=\omega\big]
    =\frac{1}{\tilde{\pi}^*_\omega}=2^N .
\end{equation}
For macroscopic $N$ this time is astronomically large, far exceeding the age of the
universe. The same holds for a low-entropy macrostate: the mean time to return to
$A_m$, starting from the stationary distribution restricted to it, is
$1/\hat\pi^*_m=2^N/{}^{N}C_{m}$. This is exponentially large in $N$ whenever
$m/N\neq\tfrac12$, and it is the same combinatorial factor as the ratio of path
probabilities above.

\subsection{Fokker--Planck \& It\^o Equations}
\label{sec:continuum}

For large $N$ the discrete jump process is well approximated by a continuous diffusion.
Introduce the rescaled variable $x=m/N\in[0,1]$, with lattice spacing $\Delta x=1/N$, and
the probability density $p(x,t)=N\,p(m,t)$, so that $p(m,t)=p(x,t)\,\Delta x$ is the
probability mass in an interval of length $\Delta x$ and $\int_0^1p\,dx=1$. Measuring
time in units of $1/N$ (the mean time between steps, each ball flipping at unit rate),
Taylor-expanding the right-hand side of Eq.~\eqref{eq:oume} in powers of $1/N$ and
retaining terms up to second order yields the forward Fokker--Planck equation
\begin{equation}
\frac{\partial p(x,t)}{\partial t}
=-\frac{\partial}{\partial x}\big[a(x)\,p\big]+\frac12\frac{\partial^2}{\partial x^2}\big[b^2\,p\big],
\end{equation}
where $a(x)=1-2x$ and $b^2=1/N$.

This Fokker--Planck equation governs the evolution of the probability density $p(x, t)$. An equivalent description at the level of individual sample paths is provided by an It\^{o} stochastic differential equation (SDE) whose ensemble of realisations reproduces $p(x, t)$.  Because the diffusion coefficient is constant, the It\^{o} and Stratonovich interpretations coincide and no calculus ambiguity arises. The corresponding SDE is the Ornstein--Uhlenbeck (OU) equation:
\begin{equation}
dx(t) = [1 - 2x(t)] \, dt + \frac{1}{\sqrt{N}} \, dW(t).
\end{equation}
The drift restores the trajectory linearly toward the equilibrium value $x^* = 1/2$ with rate 2, while the noise amplitude $1/\sqrt{N}$ encodes the discreteness of the underlying urn dynamics and vanishes in the thermodynamic limit $N \to \infty$.

The standard form of the Ornstein--Uhlenbeck (OU) process is typically expressed as
\begin{equation}
dx(t) = \theta[\mu - x(t)]dt + \sigma dW(t),\label{eq:ouisde}
\end{equation}
where $\theta$ is the rate of mean reversion (strictly greater than zero). It dictates how
fast the process pulls back to the long-term mean, $\mu$ is the long-term mean or
equilibrium level to which the process tends to revert over time, and $\sigma$ is the
volatility or diffusion coefficient (strictly greater than zero) determining the magnitude
of the random shocks to the process. (Henceforth, we shall take $\mu=0$ for convenience.)
This process is widely used in physics, mathematics, and quantitative finance because it
captures the behaviour of a variable that fluctuates randomly but eventually tends to
return to a central, stable value (mean-reverting behaviour). It is also a very important process because, in the context of continuous-time
stochastic processes, \emph{Doob's theorem} singles it out: a stationary Gaussian
Markov process with continuous covariance is, up to an affine change of the state
variable and a rescaling of time, the stationary OU process. 

Let's solve Eq.~\eqref{eq:ouisde}. It is Markov by construction, as it is an It\^o SDE. To
solve it, we use an integrating factor, $e^{\theta t}$. We apply It\^{o}'s Lemma to the
function $f(t,x(t))=x(t)e^{\theta t}$; since $f$ is linear in $x$, the second-derivative
term vanishes and
\begin{equation}
d[x(t) e^{\theta t}] = \theta x(t) e^{\theta t} dt + e^{\theta t} dx(t).
\end{equation}
Substituting the SDE for $dx(t)$ (with $\mu=0$),
\begin{eqnarray}
&&d[x(t) e^{\theta t}] = \theta x(t) e^{\theta t} dt + e^{\theta t} (-\theta x(t) dt + \sigma dW(t))\nonumber\\
\implies\,&&d[x(t) e^{\theta t}] = \sigma e^{\theta t} dW(t).
\end{eqnarray}
Integrating both sides from time $0$ to $t$ and rearranging, we obtain the solution
\begin{equation}
x(t) = x_0 e^{-\theta t} + \sigma \int_0^t e^{-\theta(t-t')} dW(t').\label{eq:x(t)i}
\end{equation}
Since the It\^o integral of a deterministic function against the Wiener process is
Gaussian, the OU process is Gaussian for a deterministic initial condition, or for a
Gaussian initial condition independent of $W$. Now, to check stationarity, we calculate the
mean, the variance, and the covariance.

To find the mean (the first moment), we take the expected value of both sides of
Eq.~\eqref{eq:x(t)i}. Since the It\^{o} integral of a deterministic function with respect
to a Wiener process has zero expected value, we get
\begin{equation}
\mathbb{E}[x(t)] = \mathbb{E}[x_0] e^{-\theta t}=x_0 e^{-\theta t},
\end{equation}
assuming a deterministic initial condition. Note that the expectation taken above is
actually a double expectation, one over the noise and the other over the initial
conditions.

For the variance (the second central moment), we use It\^{o}'s isometry
[Eq.~(\ref{eq:ii})], so that the variance of the It\^{o} integral is
\begin{eqnarray}
\text{Var}[x(t)]& =& \mathbb{E}\left[\left(\sigma \int_0^t e^{-\theta(t-t')} dW(t')\right)^2\right]\nonumber\\
&=&  \sigma^2 \int_0^t e^{-2\theta(t-t')} dt'= \frac{\sigma^2}{2\theta} (1 - e^{-2\theta t}).\qquad\qquad
\end{eqnarray}

Finally, the covariance. For a deterministic $x_0$ the term $x_0e^{-\theta t}$ is
non-random, so it drops out of the covariance and only the two stochastic integrals remain:
\begin{eqnarray}
&&\text{Cov}[x(t'), x(t)] \nonumber\\
&&= \mathbb{E}\left[\left(\sigma \int_0^{t'} e^{-\theta(t'-u)} dW(u)\right)\left(\sigma \int_0^t e^{-\theta(t-v)} dW(v)\right)\right].\nonumber\\
\end{eqnarray}
Assume, without loss of generality, that $t>t'$. Because the increments of the Wiener
process are independent, only the overlapping interval $(0,t')$ contributes. By the
two-integrand form of It\^{o}'s isometry,
$\mathbb{E}\big[\int f\,dW\int g\,dW\big]=\int fg\,du$ over the common interval,
\begin{eqnarray}
\text{Cov}[x(t'), x(t)] &=& \sigma^2 \int_0^{t'} e^{-\theta(t'-u)} e^{-\theta(t-u)} du\nonumber\\
&= &\frac{\sigma^2}{2\theta} \left(e^{-\theta(t-t')} - e^{-\theta(t+t')}\right).\label{eq:sdecoveq}
\end{eqnarray}

As $t', t \to \infty$ with $t-t'$ held finite, the term $e^{-\theta(t+t')}$ vanishes and the
covariance simplifies to
\begin{equation}
\text{Cov}[x(t'), x(t)] = \frac{\sigma^2}{2\theta} e^{-\theta|t-t'|}.
\end{equation}
In the same limit the mean goes to zero and the variance tends to $\sigma^2/2\theta$, both
independent of time. Since a Gaussian process is fixed by its first two moments, a
time-independent mean and a covariance depending only on $t-t'$ together imply
stationarity: the OU process becomes stationary at long times. It is exactly stationary
if started from the Gaussian distribution $\mathcal{N}(0,\sigma^2/2\theta)$, independent
of $W$. 

Above conclusions can also be seen by solving the corresponding forward Fokker--Planck equation,
\begin{equation}
\frac{\partial p(x, t)}{\partial t} = \theta \frac{\partial}{\partial x} \big(x p(x, t)\big) + \frac{\sigma^2}{2} \frac{\partial^2 p(x, t)}{\partial x^2}.\label{eq:OUfpE}
\end{equation}
The deterministic initial condition used above means that
\begin{equation}
p(x, 0) = \delta(x - x_0).
\end{equation}

To solve this equation, we apply the spatial Fourier transform,
\begin{equation}
\hat{p}(k, t) = \int_{-\infty}^{\infty} e^{-ikx} p(x, t) dx,
\end{equation}
which is the characteristic function $\mathbb{E}[e^{ikx}]$ evaluated at $-k$. (With the
opposite sign convention, $k\to-k$ throughout; nothing else changes.) Using the standard
properties of Fourier transforms,
\begin{align}
\mathcal{F}\left[\frac{\partial^2 p}{\partial x^2}\right] &= (ik)^2 \hat{p} = -k^2 \hat{p}, \\
\mathcal{F}\left[\frac{\partial}{\partial x}(xp)\right] &= ik \cdot \mathcal{F}[xp] = ik \left(i \frac{\partial \hat{p}}{\partial k}\right) = -k \frac{\partial \hat{p}}{\partial k}.
\end{align}
Substituting these into Eq.~\eqref{eq:OUfpE} yields a first-order linear partial
differential equation in Fourier space:
\begin{equation}
\frac{\partial \hat{p}}{\partial t} + \theta k \frac{\partial \hat{p}}{\partial k} = -\frac{\sigma^2}{2} k^2 \hat{p}.\label{eq:OUfpEf}
\end{equation}

We introduce a moving coordinate system by defining a characteristic curve $k(t)$ such that
\begin{equation}
\frac{dk}{dt} = \theta k \implies k(t) = \xi e^{\theta t} \quad \text{or} \quad \xi = k e^{-\theta t},
\end{equation}
where $\xi$ is a constant of integration. Along this characteristic curve, the total
derivative of $\hat{p}$ with respect to time is
\begin{equation}
\frac{d\hat{p}}{dt} = \frac{\partial \hat{p}}{\partial t} + \frac{\partial \hat{p}}{\partial k} \frac{dk}{dt} = \frac{\partial \hat{p}}{\partial t} + \theta k \frac{\partial \hat{p}}{\partial k}.
\end{equation}
This matches the left side of Eq.~\eqref{eq:OUfpEf}, allowing us to rewrite it as an
ordinary differential equation (ODE):
\begin{equation}
\frac{d\hat{p}}{dt} = -\frac{\sigma^2}{2} k^2 \hat{p}.
\end{equation}
Substituting $k = \xi e^{\theta t}$,
\begin{equation}
\frac{1}{\hat{p}} d\hat{p} = -\frac{\sigma^2}{2} \xi^2 e^{2\theta t} dt.
\end{equation}
Integrating both sides with respect to $t$,
\begin{equation}
\ln \hat{p} = -\frac{\sigma^2}{4\theta} \xi^2 e^{2\theta t} + C(\xi).
\end{equation}
Restoring $\xi = k e^{-\theta t}$,
\begin{equation}
\ln \hat{p}(k, t) = -\frac{\sigma^2}{4\theta} k^2 + C(k e^{-\theta t}).\label{eq:OUlog}
\end{equation}

The initial condition is $p(x, 0) = \delta(x - x_0)$, whose Fourier transform is
\begin{equation}
\hat{p}(k, 0) = e^{-ikx_0}.
\end{equation}
Evaluating our general solution at $t = 0$,
\begin{equation}
\ln \hat{p}(k, 0) = -\frac{\sigma^2}{4\theta} k^2 + C(k) = -ikx_0.
\end{equation}
Solving for the unknown function $C(k)$,
\begin{equation}
C(k) = -ikx_0 + \frac{\sigma^2}{4\theta} k^2.
\end{equation}

Substituting $C(k e^{-\theta t})$ back into Eq.~\eqref{eq:OUlog}, grouping by $k$, and
taking the exponential gives the final characteristic function---
\begin{equation}
\hat{p}(k, t) = \exp\left(-ik\mu(t) - \frac{1}{2} k^2 \sigma_x^2(t)\right),
\end{equation}
where $\mu(t) = x_0 e^{-\theta t}$ and $\sigma_x^2(t) = \frac{\sigma^2}{2\theta} (1 - e^{-2\theta t})$.

The inverse Fourier transform yields the exact transition probability density,
\begin{equation}
p(x,t\mid x_0,0) = \frac{1}{\sqrt{2\pi \sigma_x^2(t)}} \exp\left(-\frac{(x - \mu(t))^2}{2\sigma_x^2(t)}\right),\label{eq:tptpou}
\end{equation}
where we have made the initial condition explicit. The time-dependent mean
$\mu(t)=x_0e^{-\theta t}$ and variance
$\sigma_x^2(t)=\frac{\sigma^2}{2\theta}(1-e^{-2\theta t})$ could also be found without
solving the equation fully, by multiplying the forward Fokker--Planck equation by $x^n$
($n=1,2$), integrating over $x\in(-\infty,\infty)$, and solving the resulting ODEs.

The covariance can likewise be obtained from the transition probability. For $t'<t$ it
relies on the Markov property and on the conditional mean read off from
Eq.~\eqref{eq:tptpou}, $\mathbb{E}[x(t)\mid x(t')]=x(t')\,e^{-\theta(t-t')}$. Conditioning
on $x(t')$ (the Markov property allows conditioning on $x(t')$ alone, instead of the whole
history), using the law of total expectation,
$\mathbb{E}[Y]=\mathbb{E}\big[\mathbb{E}[Y\mid X]\big]$, and taking out the known factor
$x(t')$,
\begin{align}
\mathbb{E}[x(t')x(t)]
&=\mathbb{E}\Big[\mathbb{E}\big[x(t')\,x(t)\mid x(t')\big]\Big]\nonumber\\
&=\mathbb{E}\Big[x(t')\,\mathbb{E}\big[x(t)\mid x(t')\big]\Big]\nonumber\\
&=\langle x^2(t')\rangle\,e^{-\theta(t-t')}.
\end{align}
Subtracting the product of the means,
$\langle x(t')\rangle\langle x(t)\rangle=\langle x(t')\rangle^2e^{-\theta(t-t')}$,
\begin{align}
\mathrm{Cov}[x(t'),x(t)]
&=\mathrm{Var}[x(t')]\,e^{-\theta(t-t')}\nonumber\\
&=\frac{\sigma^2}{2\theta}\left(e^{-\theta(t-t')}-e^{-\theta(t+t')}\right),
\end{align}
which is Eq.~\eqref{eq:sdecoveq}, obtained earlier from the SDE.

These results fit Doob's theorem. In the stationary regime ($t,t'\to\infty$ with $t-t'$
fixed) the mean is constant (zero) and the covariance has the exponential form
\begin{equation}
\mathrm{Cov}[x(t'),x(t)]=C(0)\,e^{-\theta|t-t'|},\qquad C(0)=\frac{\sigma^2}{2\theta},
\end{equation}
with $\theta>0$ (the trivial constant case $\theta=0$ is excluded). The theorem states
that this exponential covariance is forced for any stationary Gaussian Markov process
with continuous covariance: the Markov property gives $C(s+t)\,C(0)=C(s)\,C(t)$, whose
only continuous solutions are $C(t)=C(0)\,e^{-\theta|t|}$.

\begin{center}
	$\symking\symqueen\symrook\symknight\symbishop\sympawn\epsdice{1}\,\epsdice{2}\,\epsdice{3}\,\epsdice{4}\,\epsdice{5}\,\epsdice{6}\,\spadesuit\heartsuit\clubsuit\diamondsuit$
\end{center}
\newpage
\newpage
\clearpage
\onecolumngrid
\section{Detailed Balance: Equilibrium's Signature}
\epigraph{\textsf{To keep your balance, you must keep moving.}}{Albert Einstein}
\twocolumngrid

\indent {\color{deepsea}{\LARGE \textgoth{C}}}onceptual origins of the principle of detailed balance trace back to the late 19th century and the foundational work of James Clerk Maxwell (1867) and Ludwig Boltzmann (1872) on the kinetic theory of gases. In developing his celebrated $H$-theorem, Boltzmann explicitly used the concept of \emph{microscopic reversibility}, assuming that at thermal equilibrium, the rate of any collision process between molecules is exactly balanced by the rate of its reverse process. This assumption was essential to explain how an ideal gas maintains its stable, most probable state (the Maxwell--Boltzmann distribution).

The principle was first explicitly formulated for complex chemical networks in 1901 by the Austrian chemist Rudolf Wegscheider. Wegscheider realized that overall macroscopic equilibrium requires each individual, cyclic step in a multi-step chemical reaction mechanism to be independently balanced, laying down mathematical constraints on rate constants that prevented unphysical, perpetual chemical loops. Shortly after, between 1916 and 1917, Albert Einstein famously relied on this exact balancing concept to derive his $A$ and $B$ coefficients for the quantum theory of radiation. The principle was further clarified and generalized for chemical systems by Gilbert N. Lewis in 1925. Lewis introduced what he called the \emph{principle of entire equilibrium}, which stated that in a system at chemical equilibrium, every individual elementary reaction is balanced by its own reverse reaction. 

A few years later, in 1931, Lars Onsager provided a more general and mathematically rigorous formulation with his celebrated reciprocal relations. Onsager showed that for systems near equilibrium, the matrix of coefficients relating thermodynamic forces (like temperature gradients) and fluxes (like heat flow) must be symmetric. This symmetry is a direct consequence of the time-reversal invariance of the underlying microscopic dynamics---the very heart of detailed balance. Onsager's work, which earned him the Nobel Prize, connected detailed balance to the broader field of irreversible thermodynamics and solidified its role as a fundamental pillar of statistical physics, defining the unique nature of the equilibrium state.

The specific term \emph{principle of detailed balance} was officially coined and given its standard, modern mathematical formulation by physicist Richard C. Tolman in his influential 1938 textbook, ``{The Principles of Statistical Mechanics}''. Tolman's work rigorously tied the principle to the time-reversal invariance of quantum and classical physical laws, popularizing the exact terminology that is used universally today.

\subsection{Concept \& Definition}
In certain Fokker--Planck descriptions, particularly for overdamped dynamics in configuration space, an equilibrium stationary solution may have an identically vanishing probability current. This is one manifestation of the more general physical principle of \emph{detailed balance}. In an isolated classical system whose Hamiltonian is invariant under time reversal, the
equilibrium measure is likewise invariant under the reversal operation.
Moreover, the probability of any dynamical transition is equal to the
probability of the corresponding reversed transition when both are evaluated
in the same stationary state. In equilibrium statistical mechanics, detailed balance is equivalent to the equilibrium condition in the dynamical sense: the probability of each transition equals that of its time-reversed counterpart, so there is no net irreversible probability current or entropy production. By contrast, a nonequilibrium steady state is stationary but violates detailed balance. It is maintained by external driving and is characterized by persistent irreversible probability currents, unequal forward and backward transition rates, and positive entropy production. Thus, stationarity alone does not imply equilibrium: equilibrium requires detailed balance, whereas a nonequilibrium steady state is a stationary state with broken detailed balance.

To formulate mathematical condition of detailed balance, consider a gas described by particle positions
$\boldsymbol q$ and momenta $\boldsymbol p$. Suppose that the system evolves
from $(\boldsymbol q,\boldsymbol p)$ at time $t$ to
$(\boldsymbol q',\boldsymbol p')$ at time $t+\tau$:
\begin{equation}
    (\boldsymbol q,\boldsymbol p,t)
    \longrightarrow
    (\boldsymbol q',\boldsymbol p',t+\tau).
\end{equation}
The reverse process is obtained by reversing the physical trajectory. Since
momentum changes sign under time reversal, the corresponding reversed
transition is
\begin{equation}
    (\boldsymbol q',-\boldsymbol p',t)
    \longrightarrow
    (\boldsymbol q,-\boldsymbol p,t+\tau).
\end{equation}
Thus, reversing a transition involves more than simply exchanging the
initial and final states: all time-odd variables must also be transformed.

For a stationary system, detailed balance states that the joint probability
of the forward transition equals that of its time-reversed counterpart:
\begin{equation}
    p_s(\boldsymbol q',\boldsymbol p',\tau;
    \boldsymbol q,\boldsymbol p,0)
    =
    p_s(\boldsymbol q,-\boldsymbol p,\tau;
    \boldsymbol q',-\boldsymbol p',0).
    \label{eq:db_phase_space_joint}
\end{equation}
If the dynamics form a homogeneous Markov process, the joint probability can
be factorized into a conditional transition probability and a stationary
one-time density. Eq.~\eqref{eq:db_phase_space_joint} then becomes
\begin{align}
    p(\boldsymbol q',\boldsymbol p',\tau \mid \boldsymbol q,\boldsymbol p,0)\, p_s(\boldsymbol q,\boldsymbol p) =& p(\boldsymbol q,-\boldsymbol p,\tau \mid \boldsymbol q',-\boldsymbol p',0)\nonumber\\
    &\times p_s(\boldsymbol q',-\boldsymbol p').
    \label{eq:db_phase_space_conditional}
\end{align}

Let's digress a bit to showcase an important caveat: Strictly speaking, an isolated Hamiltonian system evolves deterministically in
the full phase space. Its transition kernel is therefore singular and may be
written as
\begin{equation}
    P_\tau(\boldsymbol z\mid\boldsymbol z')
    =
    \delta\!\left[
    \boldsymbol z-\Phi_\tau(\boldsymbol z')
    \right],
    \label{eq:hamiltonian_transition_kernel}
\end{equation}
where  $\boldsymbol z\equiv(\boldsymbol q,\boldsymbol p)$ and $\Phi_\tau$ denotes the Hamiltonian flow. The evolution is nevertheless
Markovian in the formal sense, since
\begin{equation}
    \Phi_{\tau_1+\tau_2}
    =
    \Phi_{\tau_1}\circ\Phi_{\tau_2}.
    \label{eq:hamiltonian_semigroup}
\end{equation}
It is important, however, to distinguish this deterministic Markov property
from a stochastic Markov process with a smooth transition density. Let ${\hat{{T}}}\boldsymbol z
    =
    (\boldsymbol q,-\boldsymbol p)$.
If the Hamiltonian is invariant under time reversal,
\begin{equation}
    H({\hat{{T}}}\boldsymbol z)
    =
    H(\boldsymbol z),
    \label{eq:time_reversal_invariant_hamiltonian}
\end{equation}
then the Hamiltonian flow satisfies
\begin{equation}
    {\hat{{T}}}\Phi_\tau
    =
    \Phi_{-\tau}{\hat{{T}}}.
    \label{eq:reversible_hamiltonian_flow}
\end{equation}
This identity expresses the reversibility of the deterministic trajectories:
the time reverse of a trajectory generated for a duration $\tau$ is a
trajectory generated backward in time for the same duration. Consequently, the detailed-balance relation for an isolated Hamiltonian system
should be understood as a statement about the stationary phase-space measure
and the reversibility of the deterministic flow. Any stationary density of the
form
\begin{equation}
    p_s(\boldsymbol z)
    =
    Z^{-1}F\!\left(H(\boldsymbol z)\right)
    \label{eq:energy_based_stationary_density}
\end{equation}
is invariant under time reversal, because $p_s({\hat{{T}}}\boldsymbol z)
    =
    p_s(\boldsymbol z)$. The canonical density,
\begin{equation}
    p_{\mathrm{eq}}(\boldsymbol z)
    =
    Z^{-1}\exp\!\left[-\beta H(\boldsymbol z)\right],
    \label{eq:canonical_density}
\end{equation}
is a particular example, provided that the partition function $Z$ is finite.
For an isolated system, one may instead use the microcanonical measure
\begin{equation}
    p_E(\boldsymbol z)
    =
    \Omega(E)^{-1}
    \delta\!\left(E-H(\boldsymbol z)\right),
    \label{eq:microcanonical_density}
\end{equation}
which is also invariant under both Hamiltonian evolution and time reversal.
Thus, Hamiltonian reversibility provides the deterministic analogue of the
stochastic detailed-balance condition. The preceding Hamiltonian argument applies to the complete phase-space state
$\boldsymbol z=(\boldsymbol q,\boldsymbol p)$. A projection onto only a subset
of variables, such as $\boldsymbol q$ or $\boldsymbol p$, need not itself be a
Markov process.  Accordingly, detailed balance of the full Hamiltonian dynamics does not by
itself imply detailed balance for an arbitrary reduced description. 

Coming back to the characterization of the detailed balance, we envisage that the similar construction as embodied in Eq.~\eqref{eq:db_phase_space_conditional} can be adopted for a general collection of physical variables
\[
\boldsymbol x=(x_1,x_2,\ldots,x_n).
\]
Each component has a definite time-reversal parity, represented by
\begin{equation}
    x_i\longmapsto \varepsilon_i x_i,\,\,
    \varepsilon_i=
    \begin{cases}
        +1, & \text{if }x_i\text{ is even under time reversal},\\
        -1, & \text{if }x_i\text{ is odd under time reversal}.
    \end{cases}
    \label{eq:variable_parities}
\end{equation}
We collect these transformations in the notation
\begin{equation}
    {\hat{{T}}}\boldsymbol x
    \equiv
    \boldsymbol{\varepsilon x}
    =
    (\varepsilon_1x_1,\varepsilon_2x_2,\ldots,
    \varepsilon_nx_n).
    \label{eq:time_reversed_state}
\end{equation}
For the phase-space example, $\boldsymbol q$ is even and
$\boldsymbol p$ is odd.

In the absence of any time-reversal-odd external parameter, the general
detailed-balance condition is
\begin{equation}
    p_s(\boldsymbol x,t+\tau;\boldsymbol x',t)
    =
    p_s({\hat{{T}}}\boldsymbol x',t+\tau;
    {\hat{{T}}}\boldsymbol x,t).
    \label{eq:general_db_joint}
\end{equation}
For a Markov process, this is equivalently written as
\begin{equation}
    p(\boldsymbol x,\tau\mid\boldsymbol x',0)\, p_s(\boldsymbol x') = p({\hat{{T}}}\boldsymbol x',\tau \mid {\hat{{T}}}\boldsymbol x,0)\, p_s({\hat{{T}}}\boldsymbol x).
    \label{eq:general_db_conditional}
\end{equation}
The stationary distribution must also be invariant under time reversal:
\begin{equation}
    p_s(\boldsymbol x)
    =
    p_s({\hat{{T}}}\boldsymbol x).
    \label{eq:stationary_time_reversal_symmetry}
\end{equation}
This condition follows by considering the equal-time limit of
Eq.~\eqref{eq:general_db_conditional}. In that limit, the two-time probability
reduces to
\begin{equation}
    \delta(\boldsymbol x-\boldsymbol x')p_s(\boldsymbol x')
    =
    \delta({\hat{{T}}}\boldsymbol x-
    {\hat{{T}}}\boldsymbol x')p_s({\hat{{T}}}\boldsymbol x).
    \label{eq:equal_time_db}
\end{equation}
Because time reversal is an invertible transformation with unit absolute
Jacobian, the delta distributions on both sides represent the same
coincidence condition. Therefore, one arrives at Eq.~\eqref{eq:stationary_time_reversal_symmetry}; finally, Eq.~\eqref{eq:general_db_conditional} can be cast as follows:
\begin{equation}
    p(\boldsymbol x,\tau\mid\boldsymbol x',0)\, p_s(\boldsymbol x') = p({\hat{{T}}}\boldsymbol x',\tau \mid {\hat{{T}}}\boldsymbol x,0)\, p_s(\boldsymbol x)
    \label{eq:general_db_conditionalfinal}
\end{equation}
---the detailed balance condition.

The preceding formulation is the strict form of detailed balance, in which
the forward and reversed transitions are compared within the same stationary
process. A more general situation arises when the dynamics depend on one or
more fixed external parameters that are odd under time reversal. Let
$\boldsymbol\lambda$ denote such a collection of parameters, satisfying: $
    \boldsymbol\lambda
    \longmapsto
    -\boldsymbol\lambda$. A magnetic field is a standard example. An imposed angular velocity may
provide another example, depending on the physical description of the
system.

When these parameters are present, time reversal maps the process at
$\boldsymbol\lambda$ onto a process at $-\boldsymbol\lambda$. The appropriate
two-time relation is consequently
\begin{equation}
    p^{\boldsymbol\lambda}
    (\boldsymbol x,t+\tau;\boldsymbol x',t)
    =
    p^{-\boldsymbol\lambda}
    ({\hat{{T}}}\boldsymbol x',t+\tau;
    {\hat{{T}}}\boldsymbol x,t).
    \label{eq:parameter_reversal_joint}
\end{equation}
In conditional-probability notation, this becomes
\begin{equation}
    p^{\boldsymbol\lambda} (\boldsymbol x,\tau\mid\boldsymbol x',0)\, p_s^{\boldsymbol\lambda}(\boldsymbol x') = p^{-\boldsymbol\lambda} ({\hat{{T}}}\boldsymbol x',\tau \mid {\hat{{T}}}\boldsymbol x,0)\, p_s^{-\boldsymbol\lambda} ({\hat{{T}}}\boldsymbol x).
    \label{eq:parameter_reversal_conditional}
\end{equation}

Equation~\eqref{eq:parameter_reversal_conditional} compares two different
processes: the forward process with parameter $\boldsymbol\lambda$ and the
reversed process with parameter $-\boldsymbol\lambda$. It therefore does not,
in general, establish detailed balance for a single stationary process at a
fixed value of $\boldsymbol\lambda$. For this reason, the broader relation is
more accurately called \emph{time-reversal invariance} or
\emph{generalized detailed balance}. The term \emph{detailed balance} is
reserved for the stricter case in which each transition is balanced by its
time reverse within the same stationary state and with the external
parameters held fixed under time reversal.

\subsection{Irreversible Current}

In this section, \(z\) denotes a generic one-dimensional stochastic
variable. The symbols \(x\) and \(v\) are reserved for position and
velocity, respectively. This distinction is important because position is
even under time reversal, whereas velocity is odd.

For a one-dimensional stochastic variable \(z\), the Fokker--Planck
equation is
\begin{equation}
\frac{\partial p(z,t)}{\partial t}
=
-\frac{\partial}{\partial z}\bigl[A(z)p(z,t)\bigr]
+\frac{1}{2}\frac{\partial^2}{\partial z^2}
\bigl[B(z)p(z,t)\bigr],
\label{eq:FPE_general}
\end{equation}
where \(A(z)\) and \(B(z)\) are the drift and diffusion coefficients,
respectively. They are purposefully taken to be time-independent so that the equation allows for stationary distributions. The probability current is
\begin{equation}
J(z,t)
=
A(z)p(z,t)
-\frac{1}{2}\frac{\partial}{\partial z}
\bigl[B(z)p(z,t)\bigr],
\label{eq:current_general}
\end{equation}
so that
\begin{equation}
\frac{\partial p(z,t)}{\partial t}
=
-\frac{\partial J(z,t)}{\partial z}.
\end{equation}

At stationarity,
\begin{equation}
\frac{\partial p_s(z)}{\partial t}=0,
\end{equation}
and therefore
\begin{equation}
\frac{\partial J_s(z)}{\partial z}=0.
\end{equation}
Thus, in one dimension, \(J_s(z)\) is a constant. Under natural boundary
conditions, this constant vanishes:
\begin{equation}
J_s(z)=0.
\label{eq:one_dimensional_zero_current}
\end{equation}

However, \(J_s(z)=0\) is only a stationarity condition. It is not, in
general, a sufficient condition for detailed balance. The distinction
becomes important when the stochastic variables have different
time-reversal parities.

Let $\varepsilon=\pm1$ denote the time-reversal parity of $z$:
$\varepsilon=+1$ for an even variable, such as position, and
$\varepsilon=-1$ for an odd variable, such as velocity.
The drift is decomposed as
\begin{align}
A^{\mathrm{rev}}(z)
&=
\frac{1}{2}
\left[A(z)-\varepsilon A(\varepsilon z)\right],
\label{eq:Arev_general}\\
A^{\mathrm{irr}}(z)
&=
\frac{1}{2}
\left[A(z)+\varepsilon A(\varepsilon z)\right],
\label{eq:Airr_general}
\end{align}
so that $A(z) = A^{\mathrm{rev}}(z)+A^{\mathrm{irr}}(z)$.
The nomenclature is rooted in the transformation properties under time
reversal. Consider the deterministic equation part of the corresponding SDE:
\begin{equation}
    d{z} = A(z)dt.
    \label{eq:det_ode}
\end{equation}
Under $t \mapsto -t$, the left-hand side transforms as
$\dot{z} \mapsto -\varepsilon \dot{z}$,
while the right-hand side transforms as $A(z)\mapsto A(\varepsilon z)$.
Invariance of Eq.~\eqref{eq:det_ode} therefore requires
\begin{equation}
    A(z) = -\varepsilon A(\varepsilon z),
    \label{eq:TR_invariance}
\end{equation}
which is precisely the property satisfied by $A^{\mathrm{rev}}$,
as is immediate from its definition~\eqref{eq:Arev_general}.
The irreversible part satisfies the opposite relation,
\begin{equation}
    A^{\mathrm{irr}}(z) = +\varepsilon A^{\mathrm{irr}}(\varepsilon z),
    \label{eq:Airr_symmetry}
\end{equation}
and therefore breaks time-reversal invariance.

The probability current is correspondingly decomposed as
\begin{align}
J^{\mathrm{rev}}(z,t)
&=
A^{\mathrm{rev}}(z)\,p(z,t),
\label{eq:Jrev_general}\\
J^{\mathrm{irr}}(z,t)
&=
A^{\mathrm{irr}}(z)\,p(z,t)
-\frac{1}{2}\frac{\partial}{\partial z}
\bigl[B(z)\,p(z,t)\bigr].
\label{eq:Jirr_general}
\end{align}
The assignment of diffusion term explicitly with $J^{\rm irr}$ reflects the usual time-reversal convention: diffusion represents stochastic fluctuations associated with environmental noise and is therefore grouped with irreversible dynamics. We note that at stationarity (when probability density is $t$-independent), the diffusive contribution $\frac{1}{2}\frac{\partial}{\partial z}[B(z)\,p(z)]=D(z)$ (say) for the reversed trajectories can be written as $D(\varepsilon z)=\frac{1}{2}\frac{\partial}{\partial \varepsilon z}[B(\varepsilon z)\,p(\varepsilon z)]$. $B$ measures the strength of the fluctuations and is
associated with the noise covariance rather than with the direction of
the stochastic increment; since a variance is unchanged when the sign
of a fluctuating increment is reversed, the diffusion coefficient is
time-reversal even: $B(\varepsilon z)=B(z)$. Furthermore in detailed balance: $p_s(\varepsilon z)=p_s(z)$. Therefore, in the specialized situation of detailed balance, at least, clearly: $D(z)=\varepsilon D(\varepsilon z)$---just the way $A^{\mathrm{irr}}(z)$ transforms.

It is obvious that, by construction,
\begin{equation}
J(z,t)
=
J^{\mathrm{rev}}(z,t)+J^{\mathrm{irr}}(z,t).
\end{equation}
Within the reversible/irreversible Fokker--Planck current decomposition, the local condition associated with detailed balance is
\begin{equation}
J_s^{\mathrm{irr}}(z)=0
\label{eq:DB_current}
\end{equation}
pointwise in the stationary state. In contrast, \(J_s(z)=0\) may result
from cancellation between nonzero reversible and irreversible currents. Let us proceed to witness this through an example.

Consider the Ornstein--Uhlenbeck process for the generic stochastic
variable \(z\):
\begin{equation}
dz(t)
=
\theta\bigl[\mu-z(t)\bigr]dt
+\sigma\,dW(t),
\qquad
z(0)=z_0,
\label{eq:OU}
\end{equation}
where \(\theta>0\), \(\mu\) is constant, and \(W(t)\) is a Wiener process.

The associated Fokker--Planck equation is
\begin{equation}
\frac{\partial p(z,t)}{\partial t}
=
-\frac{\partial}{\partial z}
\bigl[\theta(\mu-z)p(z,t)\bigr]
+\frac{1}{2}\frac{\partial^2}{\partial z^2}
\bigl[\sigma^2p(z,t)\bigr].
\label{eq:FPE_OU}
\end{equation}

The probability current is
\begin{equation}
J(z,t)
=
\theta(\mu-z)p(z,t)
-\frac{\sigma^2}{2}
\frac{\partial p(z,t)}{\partial z}.
\label{eq:current_OU}
\end{equation}

Under natural boundary conditions, for the initial condition---$p(z,0)=\delta(z-z_0)$, the transition probability is
\begin{equation}
p(z,t|z_0,0)
=
\frac{\exp\left[
-\frac{
\theta\left[z-\mu-(z_0-\mu)e^{-\theta t}\right]^2
}{
\sigma^2\bigl(1-e^{-2\theta t}\bigr)
}
\right]}{\sqrt{
{\frac{1}{\theta}[\pi\sigma^2\bigl(1-e^{-2\theta t}\bigr)}
]}}.
\label{eq:transition}
\end{equation}
The corresponding stationary distribution, achieved in the limit $t\to\infty$, is
\begin{equation}
p_s(z)
=
\sqrt{\frac{\theta}{\pi\sigma^2}}
\exp\left[
-\frac{\theta(z-\mu)^2}{\sigma^2}
\right].
\label{eq:ps_OU}
\end{equation}
This can be obtained by putting Eq.~\eqref{eq:current_OU} to constant (which is zero owing to natural boundary conditions) and solving for $p_s(z)$ [which replaces $p(z,t)$ under imposed condition: $\partial_tp(z,t)=0$].

Since
\begin{equation}
\frac{dp_s(z)}{dz}
=
-\frac{2\theta(z-\mu)}{\sigma^2}p_s(z),
\end{equation}
the stationary total current is
\begin{equation}
J_s(z)
=
\theta(\mu-z)p_s(z)
-\frac{\sigma^2}{2}\frac{dp_s(z)}{dz}
=0.
\label{eq:Js_OU}
\end{equation}
The vanishing of \(J_s(z)\) follows from stationarity and the
one-dimensional boundary conditions. It does not, by itself, establish
detailed balance. In what follows, let us explicitly witness this fact by contrasting two cases: $z$ as an odd and an even variable.

The microscopic detailed-balance condition is expressed through the
stationary two-time transition probabilities:
\begin{equation}
p(z,\tau|z',0)p_s(z')
=
p(\varepsilon z',\tau|\varepsilon z,0)
p_s(\varepsilon z).
\label{eq:DB}
\end{equation}

When the $z$-variable is interpreted as a position \(x\), the process is
\begin{equation}
dx(t)
=
\theta\bigl[\mu-x(t)\bigr]dt
+\sigma\,dW(t),
\qquad
x(0)=x_0.
\label{eq:OU_position}
\end{equation}
Position is even under time reversal, so \(\varepsilon=+1\). The
detailed-balance condition becomes
\begin{equation}
p(x,\tau|x',0)p_s(x')
=
p(x',\tau|x,0)p_s(x).
\label{eq:DBeven}
\end{equation}

We define
\begin{equation}
\xi=x-\mu,
\quad
\xi'=x'-\mu,
\quad
\alpha=e^{-\theta\tau},
\quad
s^2=\frac{\sigma^2}{2\theta}(1-\alpha^2).
\end{equation}
The exponent on the left-hand side of Eq.~\eqref{eq:DBeven} is
\begin{align}
-\frac{(\xi-\alpha\xi')^2}{2s^2}
-\frac{\theta{\xi'}^2}{\sigma^2}
&=
-\frac{\xi^2-2\alpha\xi\xi'+{\xi'}^2}{2s^2},
\end{align}
where
\begin{equation}
\frac{\alpha^2}{2s^2}
+\frac{\theta}{\sigma^2}
=
\frac{1}{2s^2}.
\end{equation}
This expression is symmetric under
\(\xi\leftrightarrow\xi'\), so detailed balance holds for the position
process.

In this case where we are dealing with an  even variable ($\varepsilon=+1$),
\begin{align}
A^{\mathrm{rev}}(x)&=\frac{1}{2}\bigl[A(x)-A(x)\bigr]=0,
\qquad\\
A^{\mathrm{irr}}(x)&=\frac{1}{2}\bigl[A(x)+A(x)\bigr]=\theta(\mu-x).
\end{align}
Consequently,
\begin{equation}
J_s^{\mathrm{rev}}(x)=0,
\qquad
J_s^{\mathrm{irr}}(x)=J_s(x)=0.
\end{equation}
Thus, detailed balance holds in this particular position interpretation.
Nevertheless, the equality \(J_s(x)=0\) alone should not be regarded as a
general definition of detailed balance. In this one-dimensional example,
it agrees with detailed balance because the irreversible current also
vanishes.

Next interpret the $z$-variable as a velocity \(v\):
\begin{equation}
dv
=\theta\bigl(\mu-v\bigr)dt
+\sigma\,dW.
\label{eq:OU_velocity}
\end{equation}
Velocity is odd under time reversal, so \(\varepsilon=-1\). Detailed
balance, Eq.~\eqref{eq:DB}, therefore requires
\begin{equation}
p(v,\tau|v',0)p_s(v')
=
p(-v',\tau|-v,0)p_s(-v).
\label{eq:DBodd}
\end{equation}

Let's define
\begin{equation}
\eta=v-\mu,
\qquad
\eta'=v'-\mu.
\end{equation}
For the left-hand side of the detailed-balance relation~\eqref{eq:DBodd},
the exponent is
\begin{equation}
E_L
=
-\frac{(\eta-\alpha\eta')^2}{2s^2}
-\frac{\theta{\eta'}^2}{\sigma^2}.
\end{equation}
On the right-hand side, however, the time-reversed arguments are
$-v'$ and $-v$. Their deviations from the stationary mean $\mu$ are
\begin{equation}
(-v')-\mu=-(v'+\mu),
\qquad
(-v)-\mu=-(v+\mu),
\end{equation}
which are not equal to $-\eta'$ and $-\eta$ unless $\mu=0$.
Consequently, the right-hand-side exponent of Eq.~\eqref{eq:DBodd} is
\begin{equation}
E_R
=
-\frac{\left[-v'-\mu-\alpha(-v-\mu)\right]^2}{2s^2}
-\frac{\theta(v+\mu)^2}{\sigma^2}.
\end{equation}
A direct simplification gives
\begin{equation}
E_L-E_R
=
\frac{2\mu(1-\alpha)(v+v')}{s^2}.
\end{equation}
Thus, detailed balance holds for all $v$ and $v'$ only when $\mu=0$.

For the velocity variable,
\begin{align}
A^{\mathrm{rev}}(v)
&=
\frac{1}{2}\bigl[A(v)+A(-v)\bigr]
=
\theta\mu,
\label{eq:Arev_velocity}\\
A^{\mathrm{irr}}(v)
&=
\frac{1}{2}\bigl[A(v)-A(-v)\bigr]
=
-\theta v,
\label{eq:Airr_velocity}
\end{align}
as $\varepsilon=-1$. The stationary reversible current is
\begin{equation}
J_s^{\mathrm{rev}}(v)
=
\theta\mu\,p_s(v),
\label{eq:Jrev_velocity}
\end{equation}
whereas the irreversible current is
\begin{align}
J_s^{\mathrm{irr}}(v)
=
-\theta v\,p_s(v)
-\frac{\sigma^2}{2}\frac{dp_s(v)}{dv}=
-\theta\mu\,p_s(v).
\label{eq:Jirr_velocity}
\end{align}
Therefore,
\begin{equation}
J_s(v)
=
J_s^{\mathrm{rev}}(v)+J_s^{\mathrm{irr}}(v)
=0,
\end{equation}
even though
\begin{equation}
J_s^{\mathrm{rev}}(v)
=
-J_s^{\mathrm{irr}}(v)
\neq0
\qquad
(\mu\neq0).
\end{equation}
This example demonstrates explicitly that $J_s(v)=0$ does not imply detailed balance. The total current vanishes because the
reversible current generated by the constant force is canceled by the
irreversible current associated with friction and diffusion. The state is
therefore a nonequilibrium steady state when \(\mu\neq0\).

For the one-dimensional variable \(z\), stationarity implies $\frac{dJ_s(z)}{dz}=0.$ Under natural boundary conditions, this forces $J_s(z)=0.$ Consequently, one-dimensional systems cannot generally distinguish
between a genuinely equilibrium state and a nonequilibrium steady state
whose reversible and irreversible currents cancel. However, for the position interpretation of the Ornstein--Uhlenbeck process, detailed balance holds
and both the total and irreversible stationary currents vanish; for the
velocity interpretation with \(\mu\neq0\), the total stationary current
still vanishes, but the reversible and irreversible currents are separately
nonzero and cancel each other. Therefore, although we mentioned that the detailed balance condition should be identified with $J_s^{\mathrm{irr}}=0$, it is not really explicit from 1D cases discussed above: one could have equally well erroneously concluded that the condition is that $J_s^{\mathrm{rev}}=0$. Only in higher dimensions can circulating divergence-free currents make
the distinction between total, reversible, and irreversible currents
transparent.

In higher dimensions, let \(\boldsymbol{z}\) denote the state vector. The
Fokker--Planck equation becomes
\begin{equation}
\frac{\partial p(\boldsymbol{z},t)}{\partial t}
=
-\boldsymbol{\nabla}_{\boldsymbol{z}}
\cdot\boldsymbol{J}(\boldsymbol{z},t).
\end{equation}
At stationarity,
\begin{equation}
\boldsymbol{\nabla}_{\boldsymbol{z}}
\cdot\boldsymbol{J}_s(\boldsymbol{z})=0.
\label{eq:divergence_free}
\end{equation}
Unlike in one dimension, a divergence-free multidimensional current need
not vanish:
\begin{equation}
\boldsymbol{\nabla}_{\boldsymbol{z}}
\cdot\boldsymbol{J}_s(\boldsymbol{z})=0
\quad\not\Rightarrow\quad
\boldsymbol{J}_s(\boldsymbol{z})=\boldsymbol{0}.
\end{equation}
The current may instead circulate in state space.

For example, an equilibrium system may possess a nonzero reversible
circulating current while its irreversible current vanishes:
\begin{equation}
\boldsymbol{J}_s^{\mathrm{irr}}(\boldsymbol{z})
=\boldsymbol{0},
\qquad
\boldsymbol{J}_s^{\mathrm{rev}}(\boldsymbol{z})\neq\boldsymbol{0},
\qquad
\boldsymbol{\nabla}_{\boldsymbol{z}}
\cdot\boldsymbol{J}_s^{\mathrm{rev}}(\boldsymbol{z})=0.
\end{equation}
Thus, the following conditions are distinct in multidimensional systems:
\begin{equation}
\boldsymbol{J}_s(\boldsymbol{z})=\boldsymbol{0},
\qquad
\boldsymbol{J}_s^{\mathrm{irr}}(\boldsymbol{z})=\boldsymbol{0},
\qquad
\boldsymbol{\nabla}_{\boldsymbol{z}}
\cdot\boldsymbol{J}_s(\boldsymbol{z})=0.
\end{equation}
The condition associated with detailed balance is the absence of
irreversible stationary flow:
\begin{equation}
{
\boldsymbol{J}_s^{\mathrm{irr}}(\boldsymbol{z})
=
\boldsymbol{0}.
}
\label{eq:DB_multidimensional}
\end{equation}
Higher-dimensional models are therefore needed to distinguish genuine
equilibrium, reversible circulation, irreversible probability flow, and
the weaker condition of stationarity.

\subsection{Multivariate Current}

The \(n\)-dimensional multivariate It\^o stochastic differential equation
is
\begin{equation}
    d\boldsymbol{x}(t)
    =
    \boldsymbol{a}\bigl(\boldsymbol{x}(t),t\bigr)\,dt
    +
    {\sf b}\bigl(\boldsymbol{x}(t),t\bigr)\,d\boldsymbol{W}(t),
    \label{eq:vector_ito_sde}
\end{equation}
where the real-valued vectors---
\begin{equation}
    \boldsymbol{x}
    =
    \begin{pmatrix}
        x_1\\
        \vdots\\
        x_n
    \end{pmatrix},\,\,
    \boldsymbol{a}
    =
    \begin{pmatrix}
        a_1\\
        \vdots\\
        a_n
    \end{pmatrix},\,\,\&\,\,
        d\boldsymbol{W}
    =
    \begin{pmatrix}
        dW_1\\
        \vdots\\
        dW_m
    \end{pmatrix}
\end{equation}
---are respectively, the state vector, the drift vector, and the vector of \(m\) independent Wiener increments satisfying
\begin{equation}
    \left\langle d\boldsymbol{W}(t) \right\rangle
    =
    \boldsymbol{0},
    \qquad
    \left\langle
        d\boldsymbol{W}(t)
        d\boldsymbol{W}^{\mathsf T}(t)
    \right\rangle
    =
    {\sf I}_m\,dt.
\end{equation}
Here, ${\sf I}_m$ is the $m$-dimensional identity matrix.

The matrix
\({\sf b}(\boldsymbol{x},t)\in\mathbb{R}^{n\times m}\) is the
noise-amplitude matrix. In component form, Eq.~\eqref{eq:vector_ito_sde}
is
\begin{equation}
    dx_i(t)
    =
    a_i\bigl(\boldsymbol{x}(t),t\bigr)\,dt
    +
    \sum_{\alpha=1}^{m}
    b_{i\alpha}\bigl(\boldsymbol{x}(t),t\bigr)\,dW_\alpha(t),
    \label{eq:component_ito_sde}
\end{equation}
where \(i=1,\ldots,n\).

The diffusion matrix is defined by
\begin{equation}
    {\sf B}(\boldsymbol{x},t)
    =
    {\sf b}(\boldsymbol{x},t)
    {\sf b}^{\mathsf T}(\boldsymbol{x},t),
\end{equation}
or, in components,
\begin{equation}
    B_{ij}(\boldsymbol{x},t)
    =
    \sum_{\alpha=1}^{m}
    b_{i\alpha}(\boldsymbol{x},t)
    b_{j\alpha}(\boldsymbol{x},t).
    \label{eq:diffusion_matrix_components}
\end{equation}
Consequently,
\begin{equation}
    \left\langle
        \bigl({\sf b}\,d\boldsymbol{W}\bigr)
        \bigl({\sf b}\,d\boldsymbol{W}\bigr)^{\mathsf T}
    \right\rangle
    =
    {\sf B}(\boldsymbol{x},t)\,dt.
\end{equation}

Because \({\sf B}={\sf b}{\sf b}^{\mathsf T}\), the diffusion matrix is
symmetric:
\begin{equation}
    B_{ij}(\boldsymbol{x},t)
    =
    B_{ji}(\boldsymbol{x},t).
    \label{eq:diffusion_symmetry}
\end{equation}
It is also positive semidefinite, since for every
\(\boldsymbol{v}\in\mathbb{R}^{n}\),
\begin{equation}
    \boldsymbol{v}^{\mathsf T}{\sf B}\boldsymbol{v}
    =
    \left\|
        {\sf b}^{\mathsf T}\boldsymbol{v}
    \right\|^2
    \geq 0.
\end{equation}

Let \(p(\boldsymbol{x},t)\) denote the probability density of
\(\boldsymbol{x}(t)\). The corresponding forward Fokker--Planck equation
is
\begin{align}
    \frac{\partial p(\boldsymbol{x},t)}{\partial t}
    ={}&
    -\sum_{i=1}^{n}
    \frac{\partial}{\partial x_i}
    \left[
        A_i(\boldsymbol{x},t)p(\boldsymbol{x},t)
    \right]
    \nonumber\\
    &+
    \frac{1}{2}
    \sum_{i,j=1}^{n}
    \frac{\partial^2}{\partial x_i\,\partial x_j}
    \left[
        B_{ij}(\boldsymbol{x},t)p(\boldsymbol{x},t)
    \right].
    \label{eq:fokker_planck_equation}
\end{align}
As is commonly found in the literature, we now denote the drift vector by
\(\boldsymbol{A}\), rather than \(\boldsymbol{a}\). This change is purely
notational. More generally, if a matrix \(B_{ij}\) appears in the second-order Fokker--Planck operator without being explicitly constructed as
\({\sf b}{\sf b}^{\mathsf T}\), only its symmetric part contributes,
provided the relevant derivatives are sufficiently regular. Hence, without
loss of generality, one may take \(B_{ij}=B_{ji}\).

In passing, to summarily mention the structural correspondence and uniqueness properties between this multivariate It\^o SDE and its corresponding forward Fokker--Planck equation, we note that while a well-behaved multivariate It\^o SDE uniquely determines a single forward Fokker--Planck equation via its drift vector $\boldsymbol{A}(\boldsymbol{x},t) = \boldsymbol{a}(\boldsymbol{x},t)$ and diffusion matrix $B_{ij}(\boldsymbol{x},t) = (\mathsf{b}\mathsf{b}^{\mathsf{T}})_{ij}$, the reverse mapping is fundamentally non-unique. Specifically, although the deterministic drift vector of the underlying SDE is uniquely recovered from the first-order advection coefficients such that $\boldsymbol{a}(\boldsymbol{x},t) = \boldsymbol{A}(\boldsymbol{x},t)$, recovering the noise-amplitude matrix $\mathsf{b}(\boldsymbol{x},t)$ from a given diffusion matrix $B_{ij}(\boldsymbol{x},t)$ admits an infinite family of solutions. Any matrix $\mathsf{b}(\boldsymbol{x},t) \in \mathbb{R}^{n \times m}$ satisfying the algebraic relation $\mathsf{b}\mathsf{b}^{\mathsf{T}} = \mathsf{B}$ is entirely valid, allowing distinct microscopic SDEs with varying noise configurations and dimensions $m$ to map to the exact same macroscopic Fokker--Planck equation.

The non-uniqueness of $\mathsf{b}$ given $\mathsf{B}$ can be made explicit as follows. If $\mathsf{b}(\boldsymbol{x},t)\in\mathbb{R}^{n\times m}$ is any particular solution of $\mathsf{b}\mathsf{b}^T = \mathsf{B}$, then for any orthogonal matrix $\mathsf{Q}\in O(m)$ satisfying $\mathsf{Q}\mathsf{Q}^T = \mathsf{I}_m$, the transformed matrix $\mathsf{b}' = \mathsf{b}\mathsf{Q}$ is equally valid, since
\begin{equation}
    \mathsf{b}'\mathsf{b}'^T
    =
    (\mathsf{b}\mathsf{Q})(\mathsf{b}\mathsf{Q})^T
    =
    \mathsf{b}\,\mathsf{Q}\mathsf{Q}^T\mathsf{b}^T
    =
    \mathsf{b}\mathsf{b}^T
    =
    \mathsf{B}.
\end{equation}
Since $O(m)$ is a continuous \emph{Lie group} of dimension $m(m-1)/2$, this generates an infinite family of noise-amplitude matrices all consistent with the same diffusion matrix $\mathsf{B}$ and hence the same Fokker--Planck equation. Physically, this reflects the fact that the Fokker--Planck equation is insensitive to rotations among the noise channels $dW_\alpha$, as only their covariance structure $\langle d\boldsymbol{W}\,d\boldsymbol{W}^T\rangle = \mathsf{I}_m\,dt$ enters the macroscopic description.

Next, we observe that the Fokker--Planck equation~ \eqref{eq:fokker_planck_equation} can be equivalently written as the
continuity equation:
\begin{equation}
    \frac{\partial p(\boldsymbol{x},t)}{\partial t}
    =
    -\sum_{i=1}^{n}
    \frac{\partial J_i(\boldsymbol{x},t)}{\partial x_i},
    \label{eq:continuity_equation}
\end{equation}
where the probability-current components are
\begin{equation}
    J_i(\boldsymbol{x},t)
    =
    A_i(\boldsymbol{x},t)p(\boldsymbol{x},t)
    -
    \frac{1}{2}
    \sum_{j=1}^{n}
    \frac{\partial}{\partial x_j}
    \left[
        B_{ij}(\boldsymbol{x},t)
        p(\boldsymbol{x},t)
    \right].
    \label{eq:current_components}
\end{equation}

Recalling notations introduced in Eq.~\eqref{eq:time_reversed_state}, the drift is decomposed into reversible and irreversible parts:
\begin{equation}
    A_i(\boldsymbol{x})
    =
    A_i^{\mathrm{rev}}(\boldsymbol{x})
    +
    A_i^{\mathrm{irr}}(\boldsymbol{x}),
    \label{eq:drift_decomposition}
\end{equation}
where the two components are defined by
\begin{align}
    A_i^{\mathrm{rev}}(\boldsymbol{x})
    &=
    \frac{1}{2}
    \left[
        A_i(\boldsymbol{x})
        -
        \varepsilon_i
        A_i(\hat{T}\boldsymbol{x})
    \right],
    \label{eq:reversible_drift}
    \\
    A_i^{\mathrm{irr}}(\boldsymbol{x})
    &=
    \frac{1}{2}
    \left[
        A_i(\boldsymbol{x})
        +
        \varepsilon_i
        A_i(\hat{T}\boldsymbol{x})
    \right].
    \label{eq:irreversible_drift}
\end{align}
Just like in the 1D case discuss earlier, the reversible drift preserves time-invariance but the irreversible drift is breaks it. Note we have switched to $t$-independent drift and diffusion coefficients so that we can discuss stationary solutions of the Fokker--Planck equation.

The probability current is correspondingly decomposed as
\begin{align}
    J_i^{\mathrm{rev}}(\boldsymbol{x},t)
    &=
    A_i^{\mathrm{rev}}(\boldsymbol{x})
    p(\boldsymbol{x},t),
    \label{eq:reversible_current}
    \\
    J_i^{\mathrm{irr}}(\boldsymbol{x},t)
    &=
    A_i^{\mathrm{irr}}(\boldsymbol{x})
    p(\boldsymbol{x},t)
    -
    \frac{1}{2}
    \sum_{j=1}^{n}
    \frac{\partial}{\partial x_j}
    \left[
        B_{ij}(\boldsymbol{x})
        p(\boldsymbol{x},t)
    \right].
    \label{eq:irreversible_current}
\end{align}
Thus,
\begin{equation}
    J_i(\boldsymbol{x},t)
    =
    J_i^{\mathrm{rev}}(\boldsymbol{x},t)
    +
    J_i^{\mathrm{irr}}(\boldsymbol{x},t).
    \label{eq:current_decomposition}
\end{equation}

Assume that a stationary distribution \(p_s(\boldsymbol{x})\) exists and
is invariant under time reversal:
\begin{equation}
    p_s(\hat{T}\boldsymbol{x})
    =
    p_s(\boldsymbol{x}).
    \label{eq:stationary_parity}
\end{equation}
The stationary continuity equation is
\begin{equation}
    \sum_{i=1}^{n}
    \frac{\partial J_{s,i}(\boldsymbol{x})}{\partial x_i}
    =
    0.
    \label{eq:stationary_continuity}
\end{equation}

Since detailed balance means {equilibrium with no probability flow that
breaks time-reversal symmetry}; therefore, the irreversible part of the
stationary probability current must vanish at every point in state space:
\begin{equation}
    J_{s,i}^{\mathrm{irr}}(\boldsymbol{x})=0,
    \qquad i=1,\ldots,n.
    \label{eq:detailed_balance_current}
\end{equation}
This detailed balance condition equivalently can be cast as a set of three conditions in terms of drift vector and diffusion matrix. They are presented below.

\textit{Condition 1: Vanishing Irreversible Stationary Current:} Using Eq.~\eqref{eq:irreversible_current}, the condition
\(J_{s,i}^{\mathrm{irr}}=0\) gives
\begin{equation}
    A_i^{\mathrm{irr}}(\boldsymbol{x})p_s(\boldsymbol{x})
    -
    \frac{1}{2}
    \sum_{j=1}^{n}
    \frac{\partial}{\partial x_j}
    \left[
        B_{ij}(\boldsymbol{x})p_s(\boldsymbol{x})
    \right]
    =
    0.
\end{equation}
Therefore, the first detailed-balance condition is
\begin{equation}
    A_i^{\mathrm{irr}}(\boldsymbol{x})p_s(\boldsymbol{x})
    =
    \frac{1}{2}
    \sum_{j=1}^{n}
    \frac{\partial}{\partial x_j}
    \left[
        B_{ij}(\boldsymbol{x})p_s(\boldsymbol{x})
    \right]
    \label{eq:condition_one}
\end{equation}
for \(i=1,\ldots,n\).

\emph{Condition 2: Reversible Stationary-Current Incompressibility:} The stationary continuity equation is
\begin{equation}
    \sum_{i=1}^{n}
    \frac{\partial J_{s,i}}{\partial x_i}
    =
    0.
\end{equation}
Using
\[
J_{s,i}
=
J_{s,i}^{\mathrm{rev}}
+
J_{s,i}^{\mathrm{irr}}
\]
and imposing detailed balance,
\(J_{s,i}^{\mathrm{irr}}=0\), yields
\begin{equation}
    \sum_{i=1}^{n}
    \frac{\partial J_{s,i}^{\mathrm{rev}}(\boldsymbol{x})}
         {\partial x_i}
    =
    0.
\end{equation}
Since
\[
J_{s,i}^{\mathrm{rev}}(\boldsymbol{x})
=
A_i^{\mathrm{rev}}(\boldsymbol{x})p_s(\boldsymbol{x}),
\]
the second condition is
\begin{equation}
    \sum_{i=1}^{n}
    \frac{\partial}{\partial x_i}
    \left[
        A_i^{\mathrm{rev}}(\boldsymbol{x})
        p_s(\boldsymbol{x})
    \right]
    =
    0
    \label{eq:condition_two}
\end{equation}
This states that the reversible stationary current must be
divergence-free.

Detailed balance does not generally require the total
stationary current to vanish. Indeed,
\(
J_{s,i}
=
J_{s,i}^{\mathrm{rev}}
\)
may be nonzero even though
\(
J_{s,i}^{\mathrm{irr}}=0.
\)
The stronger condition
\(
J_{s,i}=0
\)
would require the reversible current to vanish as well.

\emph{Condition 3: Diffusion Tensor Parity:} For a system to be perfectly time-reversible, its probability current must behave like a purely kinematic flux, reversing its direction under time reversal such that \(J_i(\boldsymbol{\varepsilon x}) = -\varepsilon_i J_i(\boldsymbol{x})\). The parity-relation of $A_i^{\mathrm{rev}}$ from Eq.~\eqref{eq:reversible_drift} combines with the even parity of $p_s$ from Eq.~\eqref{eq:stationary_parity} to yield the precise kinematic parity that a time-reversible stationary current, \(J_{s,i}^{\mathrm{rev}}\), must satisfy.

However, the stationary irreversible current \(J_{s,i}^{\mathrm{irr}}\) is constructed from the dissipative drift and the diffusive flux, both of which are associated with entropy production. Because of this, \(J_{s,i}^{\mathrm{irr}}\) fails to reverse direction properly and is instead expected to transform with an time-reversal parity:
\begin{equation}
J_{s,i}^{\mathrm{irr}}(\boldsymbol{\varepsilon x}) = \varepsilon_i J_{s,i}^{\mathrm{irr}}(\boldsymbol{x}).\label{eq:beechme}
\end{equation}
It is precisely this anomalous parity that identifies \(J_{s,i}^{\mathrm{irr}}\) as the source of macroscopic time-symmetry breaking, explaining why it must vanish identically under detailed balance. We note that for the drift part of $J_{s,i}^{\mathrm{irr}}$, the parity-relation of $A_i^{\mathrm{irr}}$ from Eq.~\eqref{eq:irreversible_drift} combines with the even parity of $p_s$ from Eq.~\eqref{eq:stationary_parity} to yield parity relation akin to Eq.~\eqref{eq:beechme}. We need to impose the same parity-relation on the diffusion part of the current. 

Evaluating \(J_{s,i}^{\mathrm{irr}}(\boldsymbol{\varepsilon x})\) using \(p_s(\boldsymbol{\varepsilon x}) = p_s(\boldsymbol{x})\) and \(\partial / \partial(\varepsilon_j x_j) = \varepsilon_j \partial / \partial x_j\):
\begin{equation}
J_{s,i}^{\mathrm{irr}}(\boldsymbol{\varepsilon x}) = \varepsilon_i A_i^{\mathrm{irr}}(\boldsymbol{x}) p_s(\boldsymbol{x}) - \frac{1}{2} \sum_j \varepsilon_j \frac{\partial}{\partial x_j} \left[ B_{ij}(\boldsymbol{\varepsilon x}) p_s(\boldsymbol{x}) \right].
\end{equation}
Equating this to \(\varepsilon_i J_{s,i}^{\mathrm{irr}}(\boldsymbol{x})\):
\begin{align}
&\varepsilon_i A_i^{\mathrm{irr}}(\boldsymbol{x}) p_s(\boldsymbol{x}) - \frac{1}{2} \sum_j \varepsilon_j \frac{\partial}{\partial x_j} \left[ B_{ij}(\boldsymbol{\varepsilon x}) p_s(\boldsymbol{x}) \right] \nonumber \\
&= \varepsilon_i A_i^{\mathrm{irr}}(\boldsymbol{x}) p_s(\boldsymbol{x}) - \frac{1}{2} \sum_j \varepsilon_i \frac{\partial}{\partial x_j} \left[ B_{ij}(\boldsymbol{x}) p_s(\boldsymbol{x}) \right].
\end{align}
Canceling the drift terms and multiplying by \(\varepsilon_j\) (noting \(\varepsilon_j^2 = 1\)) yields:
\begin{equation}
\sum_j \frac{\partial}{\partial x_j} \left[ B_{ij}(\boldsymbol{\varepsilon x}) p_s(\boldsymbol{x}) \right] = \sum_j \varepsilon_i \varepsilon_j \frac{\partial}{\partial x_j} \left[ B_{ij}(\boldsymbol{x}) p_s(\boldsymbol{x}) \right].
\end{equation}
For this relation to hold identically for arbitrary \(p_s(\boldsymbol{x})\), the third condition follows, alongside the standard spatial matrix symmetry:
\begin{equation}
B_{ij}(\boldsymbol{\varepsilon x}) = \varepsilon_i \varepsilon_j B_{ij}(\boldsymbol{x}) \quad \text{and} \quad B_{ij}(\boldsymbol{x}) = B_{ji}(\boldsymbol{x}).
\label{eq:cond3}
\end{equation}

\subsection{Kramers' Dynamics \& Brownian Gyrator}
To illustrate the general framework of probability currents and detailed balance in concrete physical settings, we examine two paradigmatic stochastic systems. The first is Kramers' underdamped dynamics, which describes a particle moving through a viscous medium under both a conservative potential and thermal noise; this system reaches thermal equilibrium and serves as the baseline for understanding reversible versus irreversible currents. The second is the Brownian gyrator, an overdamped two-dimensional particle coupled to two thermal reservoirs at generally different temperatures; this system can sustain a genuine nonequilibrium steady state (NESS). Together, these two examples sharply contrast the roles of reversible and irreversible probability currents. The main result is that detailed balance does not require the total stationary current to vanish: in Kramers' dynamics, a nonzero but entropy-neutral reversible current (Hamiltonian circulation) persists at equilibrium, while the irreversible current vanishes; in the Brownian gyrator, there is no reversible current at all, and it is the nonvanishing of the irreversible current---driven by the temperature difference between the two reservoirs---that signals the breakdown of detailed balance and the onset of nonequilibrium circulation.

\emph{Kramers' Model of Brownian Motion}: As the first example, consider a particle of mass $m$ moving in a potential
$U(\boldsymbol{q})$, where $\boldsymbol{q}\in\mathbb{R}^{d}$ is the
position and $\boldsymbol{v}\in\mathbb{R}^{d}$ is the velocity.  Kramers'
underdamped It\^o stochastic differential equation is
\begin{align}
    d\boldsymbol{q}
    &=\boldsymbol{v}\,dt,
    \label{eq:q_sde}
    \\
    d\boldsymbol{v}
    &=
    \left[
        -\frac{\gamma}{m}\boldsymbol{v}
        -\frac{1}{m}{\boldsymbol{\nabla}} U(\boldsymbol{q})
    \right]dt
    +
    \frac{\sqrt{2\gamma k_{\rm B}T}}{m}
    \,d\boldsymbol{W}(t),
    \label{eq:v_sde}
\end{align}
where $\gamma>0$ is the friction coefficient, $T$ is the temperature,
$k_{\rm B}$ is Boltzmann's constant, and
$d\boldsymbol{W}(t)$ is a $d$-dimensional Wiener increment satisfying
\begin{equation}
    \left\langle dW_\alpha(t)\right\rangle=0,
    \qquad
    \left\langle dW_\alpha(t)dW_\delta(t)\right\rangle
    =\delta_{\alpha\delta}\,dt.
\end{equation}
Eq.~\eqref{eq:q_sde} states that the velocity is the instantaneous
rate of change of position.  Eq.~\eqref{eq:v_sde} contains the
conservative acceleration $-{\boldsymbol{\nabla}} U/m$, the viscous friction
$-\gamma\boldsymbol{v}/m$, and a random force generated by collisions
with a thermal environment.  The noise amplitude is related to the
friction coefficient through the fluctuation--dissipation relation.  In
particular, the velocity diffusion coefficient is
\begin{equation}
    D_v=\frac{\gamma k_{\rm B}T}{m^2}.
\end{equation}
This relation ensures relaxation toward thermal equilibrium at
temperature $T$.

Introducing the phase-space vector
\begin{equation}
    \boldsymbol{x}
    =
    \begin{pmatrix}
        \boldsymbol{q}\\
        \boldsymbol{v}
    \end{pmatrix},
\end{equation}
the stochastic equation can be written as
\begin{equation}
    d\boldsymbol{x}
    =
    \boldsymbol{A}(\boldsymbol{x})\,dt
    +
    {\sf b}\,d\boldsymbol{W},
\end{equation}
where
\begin{equation}
    \boldsymbol{A}(\boldsymbol{x})
    =
    \begin{pmatrix}
        \boldsymbol{v}\\[2mm]
        -\dfrac{\gamma}{m}\boldsymbol{v}
        -\dfrac{1}{m}{\boldsymbol{\nabla}} U(\boldsymbol{q})
    \end{pmatrix},
    \qquad
    {\sf b}
    =
    \begin{pmatrix}
        {\sf 0}_{d\times d}\\[1mm]
        \dfrac{\sqrt{2\gamma k_{\rm B}T}}{m}{\sf I}_d
    \end{pmatrix}.
\end{equation}
Consequently, the diffusion matrix is
\begin{equation}
    {\sf B}={\sf b}{\sf b}^{\mathsf T}
    =
    \begin{pmatrix}
        {\sf 0}_{d\times d} & {\sf 0}_{d\times d}\\
        {\sf 0}_{d\times d}
        &
        \dfrac{2\gamma k_{\rm B}T}{m^2}{\sf I}_d
    \end{pmatrix}.
    \label{eq:diffusion_matrix}
\end{equation}
The diffusion matrix is degenerate because the noise acts directly on
the velocity but not directly on the position.

Let $p(\boldsymbol{q},\boldsymbol{v},t)$ denote the probability density
in phase space.  The forward Fokker--Planck equation corresponding to
Eqs.~\eqref{eq:q_sde} and~\eqref{eq:v_sde} is
\begin{align}
    \frac{\partial p}{\partial t}
    ={}&
    -{\boldsymbol{\nabla}}_{\boldsymbol{q}}\cdot
    \left(\boldsymbol{v}p\right)
    \nonumber\\
    &+
    {\boldsymbol{\nabla}}_{\boldsymbol{v}}\cdot
    \left[
        \left(
            \frac{\gamma}{m}\boldsymbol{v}
            +
            \frac{1}{m}{\boldsymbol{\nabla}} U
        \right)p
    \right]
    +
    \frac{\gamma k_{\rm B}T}{m^2}
    {\boldsymbol{\nabla}}_{\boldsymbol{v}}^2p.
    \label{eq:kramers_fpe}
\end{align}
The first term represents deterministic transport in position space.
The second term contains conservative acceleration and viscous
friction in velocity space, while the last term represents thermal
diffusion of the velocity.

For a general multivariate It\^o process, the probability current in
phase space has components
\begin{equation}
    J_i(\boldsymbol{x},t)
    =
    A_i(\boldsymbol{x},t)p(\boldsymbol{x},t)
    -
    \frac{1}{2}
    \sum_j
    \frac{\partial}{\partial x_j}
    \left[
        B_{ij}(\boldsymbol{x},t)p(\boldsymbol{x},t)
    \right].
    \label{eq:general_current}
\end{equation}
Under time reversal, positions are even and velocities are odd, therefore the reversible and irreversible parts of the
drift are
\begin{align}
    \boldsymbol{A}^{\rm rev}
    &=
    \begin{pmatrix}
        \boldsymbol{v}\\[1mm]
        -\dfrac{1}{m}{\boldsymbol{\nabla}} U(\boldsymbol{q})
    \end{pmatrix},
    \label{eq:Arev}
    \\
    \boldsymbol{A}^{\rm irr}
    &=
    \begin{pmatrix}
        \boldsymbol{0}\\[1mm]
        -\dfrac{\gamma}{m}\boldsymbol{v}
    \end{pmatrix}.
    \label{eq:Airr}
\end{align}
The corresponding currents are
\begin{align}
    \boldsymbol{J}^{\rm rev}
    &=
    \begin{pmatrix}
        \boldsymbol{v}p\\[1mm]
        -\dfrac{1}{m}{\boldsymbol{\nabla}} U\,p
    \end{pmatrix},
    \label{eq:Jrev_general}
    \\
    \boldsymbol{J}^{\rm irr}
    &=
    \begin{pmatrix}
        \boldsymbol{0}\\[1mm]
        -\dfrac{\gamma}{m}\boldsymbol{v}p
        -\dfrac{\gamma k_{\rm B}T}{m^2}
        {\boldsymbol{\nabla}}_{\boldsymbol{v}}p
    \end{pmatrix}.
    \label{eq:Jirr_general}
\end{align}

Assume that $U(\boldsymbol{q})$ is sufficiently confining for the
stationary density to be normalizable.  The (Maxwell--Boltzmann)
stationary distribution is
\begin{equation}
    p_s(\boldsymbol{q},\boldsymbol{v})
    =
    \frac{1}{Z}
    \exp\left[
        -\frac{
        \dfrac{m}{2}\lvert\boldsymbol{v}\rvert^2
        +U(\boldsymbol{q})}
        {k_{\rm B}T}
    \right],
    \label{eq:boltzmann}
\end{equation}
where $Z$ is the normalizing factor.
This result follows directly by substituting the candidate density into
the stationary Fokker--Planck equation.  Its derivatives are
\begin{equation}
    {\boldsymbol{\nabla}}_{\boldsymbol{q}}p_s
    =
    -\frac{{\boldsymbol{\nabla}} U}{k_{\rm B}T}p_s,
    \qquad
    {\boldsymbol{\nabla}}_{\boldsymbol{v}}p_s
    =
    -\frac{m\boldsymbol{v}}{k_{\rm B}T}p_s.
    \label{eq:stationary_derivatives}
\end{equation}
The stationary irreversible current is therefore
\begin{align}
    \boldsymbol{J}^{\rm irr}_s
    =
    \begin{pmatrix}
        \boldsymbol{0}\\[1mm]
        -\dfrac{\gamma}{m}\boldsymbol{v}p_s
        -\dfrac{\gamma k_{\rm B}T}{m^2}
        {\boldsymbol{\nabla}}_{\boldsymbol{v}}p_s
    \end{pmatrix}=
    \boldsymbol{0}.
    \label{eq:stationary_irreversible_current}
\end{align}
Thus, frictional transport and thermal diffusion exactly cancel at
equilibrium.

The stationary reversible current is
\begin{equation}
    \boldsymbol{J}^{\rm rev}_s
    =
    \begin{pmatrix}
        \boldsymbol{v}p_s\\[1mm]
        -\dfrac{1}{m}{\boldsymbol{\nabla}} U(\boldsymbol{q})p_s
    \end{pmatrix}.
    \label{eq:stationary_reversible_current}
\end{equation}
It is generally nonzero.  Its phase-space divergence vanishes because
the conservative dynamics preserves the Maxwell--Boltzmann density:
\begin{equation}
    {\boldsymbol{\nabla}}_{\boldsymbol{q}}\cdot
    \left(\boldsymbol{v}p_s\right)
    +
    {\boldsymbol{\nabla}}_{\boldsymbol{v}}\cdot
    \left(
        -\frac{1}{m}{\boldsymbol{\nabla}} U\,p_s
    \right)
    =0.
\end{equation}
Physically, $\boldsymbol{J}^{\rm rev}_s$ represents deterministic
Hamiltonian motion through phase space along surfaces of constant
energy.  It is a circulating, entropy-neutral current: it does not
indicate nonequilibrium dissipation and is compatible with equilibrium
time reversal, since the velocity changes sign under
$\boldsymbol{v}\mapsto-\boldsymbol{v}$.

By contrast, $\boldsymbol{J}^{\rm irr}$ contains the frictional and
thermal contributions.  The friction term
$-\gamma\boldsymbol{v}p/m$ removes kinetic energy from the particle,
whereas the diffusive term
$-\gamma k_{\rm B}T{\boldsymbol{\nabla}}_{\boldsymbol{v}}p/m^2$ represents the
injection of random thermal fluctuations.  Away from equilibrium these
two effects need not cancel, and a nonzero irreversible current signals
dissipation, probability transport, and entropy production.  At
equilibrium, however, Eq.~\eqref{eq:stationary_irreversible_current}
shows that
\begin{equation}
    \boldsymbol{J}^{\rm irr}_s=\boldsymbol{0}.
\end{equation}
This is the detailed-balance condition for the irreversible dynamics.

Therefore, detailed balance for Kramers' dynamics does not generally
require the total stationary phase-space current to vanish.  Instead,
the irreversible stationary current must vanish, while the reversible
stationary current may remain nonzero and divergence-free.  The total
stationary current is consequently
\begin{equation}
    \boldsymbol{J}_s
    =
    \boldsymbol{J}^{\rm rev}_s
    +
    \boldsymbol{J}^{\rm irr}_s
    =
    \boldsymbol{J}^{\rm rev}_s,
\end{equation}
and it describes equilibrium Hamiltonian circulation rather than
irreversible entropy-producing transport.

\emph{Brownian gyrator}: Next consider a different example that, unlike the Kramer's dynamics, can sustain an NESS---a two-dimensional Brownian particle with position vector
\[
    \boldsymbol{q}
    =
    \begin{pmatrix}
        x\\
        y
    \end{pmatrix},
\]
moving in a harmonic potential
\begin{equation}
    U(x,y)
    =
    \frac{1}{2}
    \left(
        k_x x^2
        +
        k_y y^2
        +
        2u xy
    \right),
    \label{eq:gyrator_potential}
\end{equation}
where $k_x>0$, $k_y>0$, and $k_xk_y-u^2>0$ ensure that the potential is
confining. The coupling parameter $u$ rotates the principal axes of the
potential relative to the Cartesian axes.

The two coordinates are coupled to thermal reservoirs with generally
different temperatures $T_x$ and $T_y$. The overdamped Langevin
equations for the Brownian gyrator are
\begin{align}
    dx
    &=
    -\frac{k_xx+uy}{\gamma_x}\,dt
    +
    \sqrt{\frac{2k_{\rm B}T_x}{\gamma_x}}\,dW_x(t),
    \label{eq:gyrator_x_explicit}
    \\
    dy
    &=
    -\frac{ux+k_yy}{\gamma_y}\,dt
    +
    \sqrt{\frac{2k_{\rm B}T_y}{\gamma_y}}\,dW_y(t).
    \label{eq:gyrator_y_explicit}
\end{align}

The stochastic equation can be written as
\begin{equation}
    d\boldsymbol{q}
    =
    \boldsymbol{A}(\boldsymbol{q})\,dt
    +
    {\sf b}\,d\boldsymbol{W},
    \label{eq:gyrator_vector_sde}
\end{equation}
where
\begin{equation}
    \boldsymbol{A}(\boldsymbol{q})
    =
    -
    \begin{pmatrix}
        \dfrac{k_xx+uy}{\gamma_x}\\[3mm]
        \dfrac{ux+k_yy}{\gamma_y}
    \end{pmatrix},
    \qquad
    {\sf b}
    =
    \begin{pmatrix}
        \sqrt{\dfrac{2k_{\rm B}T_x}{\gamma_x}} & 0\\[3mm]
        0 & \sqrt{\dfrac{2k_{\rm B}T_y}{\gamma_y}}
    \end{pmatrix}.
\end{equation}

The diffusion matrix is therefore
\begin{equation}
    {\sf B}
    =
    {\sf b}{\sf b}^{\mathsf T}
    =
    \begin{pmatrix}
        \dfrac{2k_{\rm B}T_x}{\gamma_x} & 0\\[3mm]
        0 & \dfrac{2k_{\rm B}T_y}{\gamma_y}
    \end{pmatrix}.
    \label{eq:gyrator_diffusion_matrix}
\end{equation}
Unlike the underdamped Kramers process, the Brownian gyrator is
overdamped, and the diffusion matrix is nondegenerate provided that
$T_x,T_y>0$.

Let $p(x,y,t)$ denote the probability density in the $(x,y)$ plane.
The forward Fokker--Planck equation corresponding to
Eqs.~\eqref{eq:gyrator_x_explicit} and
\eqref{eq:gyrator_y_explicit} is
\begin{align}
    \frac{\partial p}{\partial t}
    ={}&
    \frac{\partial}{\partial x}
    \left[
        \frac{k_xx+uy}{\gamma_x}p
    \right]
    +
    \frac{\partial}{\partial y}
    \left[
        \frac{ux+k_yy}{\gamma_y}p
    \right]
    \nonumber\\
    &+
    \frac{k_{\rm B}T_x}{\gamma_x}
    \frac{\partial^2p}{\partial x^2}
    +
    \frac{k_{\rm B}T_y}{\gamma_y}
    \frac{\partial^2p}{\partial y^2}.
    \label{eq:gyrator_fpe}
\end{align}


For the overdamped Brownian gyrator, both $x$ and $y$ are even under
time reversal. Thus, the drift has no reversible component:
\begin{equation}
    \boldsymbol{A}^{\rm rev}
    =
    \boldsymbol{0},
    \label{eq:gyrator_Arev}
\end{equation}
while the full drift is irreversible:
\begin{equation}
    \boldsymbol{A}^{\rm irr}
    =
    \boldsymbol{A}
    =
    -
    \begin{pmatrix}
        \dfrac{k_xx+uy}{\gamma_x}\\[3mm]
        \dfrac{ux+k_yy}{\gamma_y}
    \end{pmatrix}.
    \label{eq:gyrator_Airr}
\end{equation}
Accordingly,
\begin{equation}
    \boldsymbol{J}^{\rm rev}
    =
    \boldsymbol{0},
    \label{eq:gyrator_Jrev}
\end{equation}
and
\begin{equation}
    \boldsymbol{J}^{\rm irr}
    =
    \begin{pmatrix}
        -\dfrac{k_xx+uy}{\gamma_x}p
        -
        \dfrac{k_{\rm B}T_x}{\gamma_x}
        \dfrac{\partial p}{\partial x}
        \\[4mm]
        -\dfrac{ux+k_yy}{\gamma_y}p
        -
        \dfrac{k_{\rm B}T_y}{\gamma_y}
        \dfrac{\partial p}{\partial y}
    \end{pmatrix}.
    \label{eq:gyrator_Jirr}
\end{equation}
Therefore, in the overdamped Brownian gyrator, the total current is
identical to the irreversible current:
\begin{equation}
    \boldsymbol{J}
    =
    \boldsymbol{J}^{\rm irr}.
    \label{eq:gyrator_total_current}
\end{equation}

It is useful to introduce the drift matrix ${\sf K}$ and the
matrix ${\sf D}$ according to
\begin{equation}
    d\boldsymbol{q}
    =
    -{\sf K}\boldsymbol{q}\,dt
    +
    \sqrt{2{\sf D}}\,d\boldsymbol{W},
    \label{eq:gyrator_ou_form}
\end{equation}
where
\begin{equation}
    {\sf K}
    =
    \begin{pmatrix}
        \dfrac{k_x}{\gamma_x} & \dfrac{u}{\gamma_x}\\[3mm]
        \dfrac{u}{\gamma_y} & \dfrac{k_y}{\gamma_y}
    \end{pmatrix},
    \qquad
    {\sf D}
    =
    \begin{pmatrix}
        \dfrac{k_{\rm B}T_x}{\gamma_x} & 0\\[3mm]
        0 & \dfrac{k_{\rm B}T_y}{\gamma_y}
    \end{pmatrix}.
    \label{eq:gyrator_K_D}
\end{equation}
Thus, the drift is
\begin{equation}
    \boldsymbol{A}(\boldsymbol{q})
    =
    -{\sf K}\boldsymbol{q},
\end{equation}
and the diffusion matrix appearing in the Fokker--Planck equation is
\begin{equation}
    {\sf B}=2{\sf D}.
\end{equation}

Because the drift is linear and the noise is Gaussian, the stationary
state is a Gaussian distribution. The origin is the stationary mean,
since the deterministic force vanishes at $\boldsymbol{q}=\boldsymbol{0}$.
Therefore,
\begin{equation}
    p_s(\boldsymbol{q})
    =
    \frac{1}{2\pi\sqrt{\det{\sf C}}}
    \exp\left[
        -\frac{1}{2}
        \boldsymbol{q}^{\mathsf T}
        {\sf C}^{-1}
        \boldsymbol{q}
    \right],
    \label{eq:gyrator_stationary_gaussian}
\end{equation}
where
\begin{equation}
    {\sf C}
    =
    \left\langle
        \boldsymbol{q}\boldsymbol{q}^{\mathsf T}
    \right\rangle_s
    =
    \begin{pmatrix}
        C_{xx} & C_{xy}\\
        C_{xy} & C_{yy}
    \end{pmatrix}
    \label{eq:gyrator_covariance}
\end{equation}
is the stationary covariance matrix.

In other words, because the Brownian gyrator is a stable linear stochastic system with
additive Gaussian noise, it is a two-dimensional
Ornstein--Uhlenbeck process. Doob's theorem therefore guarantees that
its transition probabilities and stationary distribution are Gaussian.
The covariance matrix of the stationary Gaussian is determined by the
stationary second-moment equation, which is the continuous Lyapunov
equation as explicitly shown below.

To determine ${\sf C}$, consider the covariance matrix at an arbitrary
time,
\begin{equation}
    {\sf C}(t)
    =
    \left\langle
        \boldsymbol{q}(t)\boldsymbol{q}(t)^{\mathsf T}
    \right\rangle.
\end{equation}
Using the It\^o product rule,
\begin{equation}
    d\left(
        \boldsymbol{q}\boldsymbol{q}^{\mathsf T}
    \right)
    =
    (d\boldsymbol{q})\boldsymbol{q}^{\mathsf T}
    +
    \boldsymbol{q}(d\boldsymbol{q})^{\mathsf T}
    +
    d\boldsymbol{q}\,d\boldsymbol{q}^{\mathsf T},
\end{equation}
together with
\begin{equation}
    d\boldsymbol{q}\,d\boldsymbol{q}^{\mathsf T}
    =
    2{\sf D}\,dt,
\end{equation}
gives
\begin{equation}
    \frac{d{\sf C}}{dt}
    =
    -{\sf K}{\sf C}
    -
    {\sf C}{\sf K}^{\mathsf T}
    +
    2{\sf D}.
    \label{eq:gyrator_covariance_evolution}
\end{equation}
At stationarity, the covariance is time independent, so
$d{\sf C}/dt=0$. Consequently, ${\sf C}$ satisfies the continuous
Lyapunov equation
\begin{equation}
    {\sf K}{\sf C}
    +
    {\sf C}{\sf K}^{\mathsf T}
    =
    2{\sf D}.
    \label{eq:gyrator_lyapunov_clear}
\end{equation}

Looking component-wise, the Lyapunov equation gives
the three scalar equations solving which gives
\begin{equation}
    C_{xy}
    =
    \frac{
        u k_{\rm B}
        \left(
            \dfrac{T_x}{\gamma_y k_x}
            +
            \dfrac{T_y}{\gamma_x k_y}
        \right)
    }{
        \dfrac{k_x}{\gamma_x}
        +
        \dfrac{k_y}{\gamma_y}
        +
        \dfrac{u^2}{\gamma_y k_x}
        +
        \dfrac{u^2}{\gamma_x k_y}
    },
    \label{eq:gyrator_Cxy_explicit}
\end{equation}
and thus,
\begin{equation}
    {\sf C}
    =
    \begin{pmatrix}
        \dfrac{k_{\rm B}T_x-uC_{xy}}{k_x}
        &
        C_{xy}
        \\[4mm]
        C_{xy}
        &
        \dfrac{k_{\rm B}T_y-uC_{xy}}{k_y}
    \end{pmatrix}.
    \label{eq:gyrator_covariance_explicit}
\end{equation}
The covariance matrix is positive definite under the stability
condition $k_xk_y-u^2>0$. Therefore, Eq.~\eqref{eq:gyrator_stationary_gaussian}
defines a normalizable stationary density.

For equal reservoir temperatures,
\begin{equation}
    T_x=T_y=T,
\end{equation}
the stationary distribution reduces to the Boltzmann distribution
\begin{equation}
    p_s(x,y)
    =
    \frac{1}{Z}
    \exp\left[
        -\frac{U(x,y)}{k_{\rm B}T}
    \right],
    \label{eq:gyrator_equilibrium_boltzmann}
\end{equation}
with $Z$ being the normalization factor. Its derivatives are
\begin{equation}
    \frac{\partial p_s}{\partial x}
    =
    -\frac{1}{k_{\rm B}T}
    \frac{\partial U}{\partial x}p_s,
    \qquad
    \frac{\partial p_s}{\partial y}
    =
    -\frac{1}{k_{\rm B}T}
    \frac{\partial U}{\partial y}p_s.
    \label{eq:gyrator_equilibrium_derivatives}
\end{equation}
Substitution into Eq.~\eqref{eq:gyrator_Jirr} gives
\begin{equation}
    \boldsymbol{J}^{\rm irr}_s
    =
    \boldsymbol{0}.
    \label{eq:gyrator_equilibrium_current}
\end{equation}
Thus, when both reservoirs have the same temperature, the Brownian
gyrator satisfies detailed balance and there is no stationary
probability circulation.

When $T_x\neq T_y$, the stationary density is generally not of the
Boltzmann form associated with a single temperature. The stationary
current is then nonzero:
\begin{equation}
    \boldsymbol{J}_s
    =
    \boldsymbol{J}^{\rm irr}_s
    \neq
    \boldsymbol{0}.
    \label{eq:gyrator_nonequilibrium_current}
\end{equation}
This current is divergence-free in the stationary state,
\begin{equation}
    {\boldsymbol{\nabla}}_{\boldsymbol{q}}\cdot\boldsymbol{J}_s=0,
    \label{eq:gyrator_stationary_divergence}
\end{equation}
but it does not vanish. The probability therefore circulates around
the origin in the $(x,y)$ plane. The direction of this circulation is
determined by the signs of $u$ and $T_x-T_y$.

For completeness, the stationary current can be written in matrix form
as
\begin{equation}
    \boldsymbol{J}_s(\boldsymbol{q})
    =
    \left(
        {\sf D}{\sf C}^{-1}
        -
        {\sf K}
    \right)
    \boldsymbol{q}\,p_s(\boldsymbol{q}).
    \label{eq:gyrator_matrix_current}
\end{equation}
The current vanishes if and only if
\begin{equation}
    {\sf K}{\sf C}={\sf D},
    \label{eq:gyrator_detailed_balance_condition}
\end{equation}
which is the detailed-balance condition for this overdamped process.
For the coupled potential considered here, this condition is satisfied
when $T_x=T_y$, or trivially when $u=0$. If $u\neq0$ and
$T_x\neq T_y$, the two thermal reservoirs drive a persistent
nonequilibrium circulation. We emphasize that the condition
$u=0$ gives zero stationary current even when $T_x\neq T_y$, but this
case has no gyrator rotation because the coordinates are uncoupled. It is
a nonequilibrium system with heat conduction between independent
degrees of freedom, rather than a circulating Brownian gyrator.

The physical origin of the Brownian-gyrator effect is the combination
of a tilted coupling potential and unequal thermal fluctuations. Noise
in the $x$ direction is generated at temperature $T_x$, whereas noise
in the $y$ direction is generated at temperature $T_y$. The coupling
term $uxy$ transfers fluctuations between the two coordinates. Because
the potential axes are not aligned with the temperature axes, this
transfer produces a rotational probability current.

Therefore, in contrast to underdamped Kramers dynamics, where a
nonzero reversible Hamiltonian current may exist at equilibrium, the
standard overdamped Brownian gyrator has no reversible phase-space
current. At equal temperatures its total stationary current vanishes.
At unequal temperatures, however, the total stationary current is a
nonzero irreversible circulating current:
\begin{equation}
    \boldsymbol{J}_s
    =
    \boldsymbol{J}^{\rm irr}_s
    \neq
    \boldsymbol{0}.
\end{equation}
This circulation is the signature of broken detailed balance and
nonequilibrium entropy production.

\begin{center}
	$\symking\symqueen\symrook\symknight\symbishop\sympawn\epsdice{1}\,\epsdice{2}\,\epsdice{3}\,\epsdice{4}\,\epsdice{5}\,\epsdice{6}\,\spadesuit\heartsuit\clubsuit\diamondsuit$
\end{center}
\newpage
\newpage
\clearpage
\onecolumngrid
\section{Linear Responses: Encashing Balance}
\epigraph{\textsf{That which goes far returns.}}{Lao Tzu (Laozi)}
\twocolumngrid

\indent {\color{deepsea}{\LARGE \textgoth{A}}}lthough a statement about equilibrium, the detailed balance condition carries within it the seeds of something more powerful. When a system is weakly perturbed away from equilibrium, its response cannot be independent of its equilibrium fluctuations---the same microscopic dynamics that enforces detailed balance also governs how the system reacts to a small external field. This connection is made precise by \emph{linear response theory}, which expresses the response of an observable to a perturbation entirely in terms of equilibrium correlation functions, revealing that dissipation and fluctuation are not separate phenomena but two faces of the same underlying dynamics.

Linear response theory and \emph{linear irreversible thermodynamics} are deeply intertwined, describing the same near-equilibrium transport phenomena from complementary viewpoints: the microscopic and the macroscopic. While irreversible thermodynamics characterizes transport through phenomenological \emph{Onsager coefficients} that link thermodynamic fluxes linearly to their driving forces, linear response theory provides the statistical foundation for these coefficients via the \emph{fluctuation-dissipation theorem}. This connection---mathematically realized through the \emph{Green--Kubo relations}---establishes that a system's dissipative response to an external perturbation is governed by the same dynamics as the decay of its spontaneous thermal fluctuations at equilibrium. Linear response theory thus serves as the microscopic underpinning of linear irreversible thermodynamics, validating its macroscopic transport laws and grounding the \emph{Onsager reciprocal relations} in the time-reversal symmetry of the underlying molecular dynamics.

\subsection{Thermodynamics \& Response Functions}
Thermodynamics is mainly the study of energy, work done, and heat exchanges in macroscopic systems where the microscopic details of the systems are ignored while adopting a coarse-grained pragmatic description for them. A thermodynamic system can be \emph{isolated}, \emph{closed}, or \emph{open}, respectively, meaning that the walls bounding the system are \emph{adiabatic} (allow transfer of neither heat or matter), \emph{diathermal} (allow transfer of only heat), or neither (i.e., allow transfer of both heat and matter). The state of such a system is characterized by thermodynamic variables/coordinates: \emph{extensive} variables/coordinates (e.g., energy, volume, polarization, magnetization, and number of particles that are proportional to the size of the system) and \emph{intensive} variables/coordinates (e.g., pressure, electric field, magnetic field and chemical potential that are independent of the size of the system). Let us consider only \emph{homogeneous} system such that value of any of the thermodynamic variables is unchanged across the system. The \emph{first law of thermodynamics}, which essentially is statement of the conservation of energy, states: $dE=\textit{\dj}Q+\textit{\dj}W$, where $dE$ is the internal energy, $\textit{\dj}Q$ is the heat supplied to the system, and $\textit{\dj}W$ is the (generalized) work done on the system.The symbol `$\textit{\dj}$\,' is used to denote inexact differential: the total heat transferred and the total work done depend on the details/path of the process. 

There is a special thermodynamical state, called \emph{equilibrium state}, such that its thermodynamic coordinates do not appear to change over time scales of observation period that is much greater than microscopic time scales but naturally not infinitely long. When in \emph{thermodynamic equilibrium}, a system is simultaneously in \emph{thermal equilibrium} (uniform constant temperature), \emph{mechanical equilibrium} (balanced mechanical forces), \emph{chemical equilibrium} (constant chemical composition), and \emph{phase equilibrium} (no phase changes like melting and evaporation). It may appear a bit of a circular argument that the concept of thermal equilibrium (defined using the concept of temperature) in the light of the \emph{zeroth law of thermodynamics}, expounds the existence of temperature $T$ as a state variable. The zeroth law states that the thermal equilibrium among thermodynamic systems is a transitive property; thus, two systems in thermal equilibrium have identical temperatures. In theoretical thermodynamics, the temperature is expressed in \emph{thermodynamic scale} (in the unit of Kelvin) by choosing the triple point of water-ice-steam system to be $273.16$ Kelvin. The thermodynamic temperature must always be non-negative; otherwise it can be shown to lead to the violation of the \emph{second law of thermodynamics} which we state below.

Suppose a \emph{quasistatic} process on a system is performed. It means that the process is so slow that each step of the process the system is given enough time to settle down practically into an equilibrium state and consequently, the specific values of the thermodynamic variables can define it unambiguously. In such a process, the infinitesimal work done, $\textit{\dj}W$, due to change in the value of an extensive variable $\mc{D}_i$ under the influence of an intensive variable $\mc{F}_i$ (not constructed by dividing an extensive variable by system's volume or mass) can be written as $\textit{\dj}W=\mc{F}_id\mc{D}_i$; thus, $\mc{D}_i$ may be seen as a \emph{generalized displacement} and $\mc{F}_i$ as its \emph{conjugate generalized force}. However, what about the conjugate variable of temperature $T$, an intensive variable? We consider a further refinement of quasistatic process: \emph{reversible process}. A reversible process is a quasistatic process that when run backward in time from the final state leads to the unchanged initial state. (While a reversible process is quasistatic, the converse need not be true.) It was established by Clausius that for such a process $\textit{\dj}Q/T$ is an exact differential, $dS$, where $S$ is an extensive variable and called it entropy; we specifically call it \emph{Clausius entropy} in order to differentiate it from the other definitions of entropy. Therefore, the first law can be written as: $dE=TdS+\sum_i\mc{F}_id\mc{D}_i$. For any arbitrary process (not necessarily reversible) leading to heat $\textit{\dj Q}$ supplied to the system, $dS\ge \textit{\dj Q}/T$ where $T$ is the temperature at which the heat is exchanged; since $S$ is a state function (defined only for the equilibrium states), $dS$ is evaluated along any reversible path connecting the same initial and final equilibrium states. Also note that equality holds only for reversible processes. 

The futile endeavours of inventing perfect engine (which transforms heat completely into work) and perfect refrigerator  (which transfers heat from cold body to hot body without any energy consumption) led to an empirical law---\emph{the second law of thermodynamics}---that is often expressed using the Clausius entropy: Entropy of an isolated thermodynamic system can not decrease in any (spontaneous) process, i.e., $dS\ge0$. The entropy of a system with its isolated subsystems at respective distinct equilibria, goes to an equilibrium state of higher entropy, following irreversible exchanges of heat inside it once all the adiabatic walls are removed between the subsystems. The intermediate non-equilibrium states are captured by the time-varying values of $\mc{D}_i$'s. Thus, stated another way, in the final equilibrium state, the entropy of an isolated system at the fixed total energy is maximized by the set of $\mc{D}_i$'s that take their equilibrium values in the final state. This is the \emph{maximum entropy principle}.

In passing, we observe that the Clausius entropy is defined only up to an additive constant. In this context, we mention that there is one more law---the \emph{third law of thermodynamics}---that states that the entropy of any system is a universal constant (conveniently taken as zero) at $T=0$. It can be shown that the third law implies unattainability (in a finite number of steps) of the absolute zero temperature, and also implies vanishing of thermal expansivities and heat capacities of the system at $T=0$. However, since at very low temperature the classical mechanics is not strictly valid, the third law explicitly requires quantum mechanics for its justification.

The Clausius entropy for a system in equilibrium may be envisaged to be determinable from $E$ and $\mc{D}_i$'s, i.e., $S=S(E,\{\mc{D}_i\})$---called \emph{(entropic) fundamental relation}---whose specific form would depend on the system in hand. All the intensive variables can, thus, be obtained as first order partial derivatives of $S$ with respect to $\mc{D}_i$'s. The relations relating the intensive with the extensive variables are called the \emph{equation of states}. Furthermore, inverting the preceding equation to write the \emph{(energetic)  fundamental relation}, $E=E(S,\{\mc{D}_i\})$, we can mathematically write the assumed extensivity of energy as $E(\lambda S,\{\lambda\mc{D}_i\})=\lambda E(S,\{\mc{D}_i\})$ ($\lambda>0$) which is valid in case there is short-range interactions between constituent particles of the system because then the energy is additive and extensive (in general, additivity and extensivity need not imply each other). Along with the first law, it implies the \emph{Euler relation}, $E=TS+\sum_i\mc{F}_i\mc{D}_i$, that in turn leads to the \emph{Gibbs--Duhem relation}: $SdT+\sum_i\mc{D}_id\mc{F}_i$=0. It should be kept in mind that the fundamental relations are not derived from thermodynamic considerations, but are rather found phenomenologically using insights from experiments performed on the thermodynamic system under question. A basic assumption of thermodynamics is that only a few coordinates are enough to specify thermodynamic equilibrium.

At this juncture, it is of use to comment that the maximum entropy principle is equivalent to the \emph{minimum energy principle}: In the final equilibrium state, the energy of a closed system at the fixed total entropy is minimized by the set of $\mc{D}_i$'s that take their equilibrium values in the final state. In practice, the tuneable thermodynamic variables may not be (some of) the extensive variables and so $E$ or $S$ may not be the right quantities for finding the equilibrium state variables. Therefore, with the Euler relation in the back of ones mind, one uses the Legendre transformation to define \emph{enthalpy} $\mc{H}\equiv E-\sum_i\mc{F}_i\mc{D}_i$, \emph{Helmholtz free energy} ${F}\equiv E-TS$, \emph{Gibbs free energy} $G\equiv E-TS-\sum_i\mc{F}_i\mc{D}_i$, and \emph{grand potential} $\mc{G}\equiv E-TS-\sum_i\mu_iN_i$. Note here we are not considering the pairs $(\mu_i,N_i)$---the chemical potential of a species and the corresponding number of particles---in the set of the pairs $(\mc{F}_i,\mc{D}_i)$; in other words, we are separating chemical work from mechanical work. ${\mc H}$, $F$, $G$ and ${\mc G}$ are respectively minimized during adiabatic transformation with mechanical work at constant generalized forces, isothermal transformation in absence of mechanical work, isothermal transformation with mechanical work at constant generalized forces, and isothermal transformation with chemical work at constant chemical potentials. In view of the afore-discussed minimization schemes, $E$, $\mc{H}$, $F$, $G$, and $\mc{G}$ are appositely called \emph{thermodynamic potentials} in analogy with the role of the potential energy in the Newtonian dynamics.

Once the fundamental relation is known in terms of the Clausius
entropy---or, equivalently, in terms of the appropriate thermodynamic
potential---all macroscopic equilibrium properties of the system can, in
principle, be determined. Among the most important of these properties
are response coefficients, commonly called \emph{thermodynamic} or
\emph{equilibrium susceptibilities}. They quantify the change in an
equilibrium observable produced by an infinitesimal variation of its
conjugate external control parameter. Since these changes can be
measured experimentally, susceptibilities provide a direct link between
thermodynamic theory and observation.

Let \(h\) be an externally controlled generalized field and let \(X\) be
the microscopic observable conjugate to it. Their coupling is represented
by the Hamiltonian
\begin{equation}
H(z;h)=H_0(z)-hX(z),
\label{eq:hamiltonian-coupling}
\end{equation}
where \(z\) denotes a complete microscopic state of the system. For a
classical system, \(z\) consists of all particle positions and momenta
and can therefore be viewed as a point in phase space.

Although \(X\) is defined at the microscopic level, its equilibrium
average is a macroscopic thermodynamic quantity. This provides the
standard connection between equilibrium statistical mechanics and
thermodynamics. In the usual
thermodynamic setting, \(X\) represents the microscopic counterpart of an
extensive thermodynamic variable, while \(h\) is its conjugate intensive
field.  To make this connection explicit, consider the
canonical partition function
\begin{equation}
Z(T,h)
=
\int dz\,\exp[-\beta H(z;h)],
\qquad
\beta=\frac{1}{k_{\mathrm B}T}.
\label{eq:partition-function}
\end{equation}
For a system with discrete microscopic states, the phase-space integral
in Eq.~\eqref{eq:partition-function} is replaced by a sum over those
states.

The partition function is appropriate because it incorporates the
statistical contributions of all microscopic states and therefore
contains the equilibrium thermodynamic information of the system. The
equilibrium average of \(X\) is
\begin{equation}
\langle X\rangle_h
=
\frac{1}{Z(T,h)}
\int dz\,X(z)\exp[-\beta H(z;h)] .
\label{eq:equilibrium-average}
\end{equation}

This average is the value obtained by repeatedly measuring \(X\) after
the system has equilibrated at temperature \(T\) and applied field \(h\).
Thus, \(\langle X\rangle_h\) is the macroscopic equilibrium value
associated with the microscopic observable \(X\). For a sufficiently
large system, the relative fluctuations around this value are generally
small, so the measured macroscopic quantity is well represented by
\(\langle X\rangle_h\).

The Helmholtz free energy is defined in terms of the partition function
by
\begin{equation}
F(T,h)=-k_{\mathrm B}T\ln Z(T,h).
\label{eq:helmholtz-free-energy}
\end{equation}
Since
\(
\frac{\partial H}{\partial h}=-X,
\),
differentiation of Eq.~\eqref{eq:helmholtz-free-energy} with respect to
\(h\) gives
\begin{equation}
\langle X\rangle_h
=
-\left(\frac{\partial F}{\partial h}\right)_T .
\label{eq:thermodynamic-average}
\end{equation}

Equation~\eqref{eq:thermodynamic-average} explains why
\(\langle X\rangle_h\) is a thermodynamic quantity: it is obtained as a
derivative of the macroscopic Helmholtz free energy, even though \(X\)
itself is defined microscopically.

The corresponding static susceptibility, which in equilibrium is also
called the equilibrium or thermodynamic susceptibility, is
\begin{equation}
\chi_{\mathrm{eq}}
=
\left(\frac{\partial\langle X\rangle_h}{\partial h}\right)_T
=
-\left(\frac{\partial^2F}{\partial h^2}\right)_T .
\label{eq:equilibrium-susceptibility}
\end{equation}

Thus, \(\chi_{\mathrm{eq}}\) measures the response of the equilibrium
value of \(X\) to its conjugate external field. Equivalently, it measures
the curvature of the equilibrium free energy with respect to that field.
A large susceptibility means that a small change in \(h\) produces a
comparatively large change in \(\langle X\rangle_h\). For example, the magnetic susceptibility is $\left(\frac{\partial M}{\partial B}\right)_T$, 
where \(M\) is the equilibrium magnetization and \(B\) is the applied
magnetic field. Similarly, if \(\mathcal P\) denotes the electric
polarization and \(\mathcal E\) the electric field, then $\left(\frac{\partial\mathcal P}{\partial\mathcal E}\right)_T$, Another equilibrium susceptibility is the isothermal compressibility is $-\frac{1}{V}
\left(\frac{\partial V}{\partial P}\right)_T$ where the minus sign accounts for the usual decrease of volume as the pressure increases.

The partition function also makes the connection between susceptibility
and equilibrium fluctuations explicit. Substituting the microscopic
Hamiltonian from Eq.~\eqref{eq:hamiltonian-coupling} into
Eq.~\eqref{eq:equilibrium-average} and differentiating with respect to
\(h\) gives
\begin{equation}
\left(\frac{\partial\langle X\rangle_h}{\partial h}\right)_T
=
\beta\left[
\langle X^2\rangle_h-\langle X\rangle_h^2
\right].
\label{eq:fluctuation-response}
\end{equation}
At zero field limit, $h\to0$, Eq.~\eqref{eq:fluctuation-response} becomes
\begin{equation}
\chi_{\mathrm{eq}}
=
\beta\left[
\langle X^2\rangle_0-\langle X\rangle_0^2
\right]
=
\beta\left\langle(\delta X)^2\right\rangle_0 ,
\label{eq:equilibrium-fluctuation-susceptibility}
\end{equation}
where $\delta X=X-\langle X\rangle_0$.

Therefore, in the classical canonical ensemble, the zero-field
equilibrium susceptibility is proportional to the spontaneous
equilibrium fluctuations of the conjugate observable. Eq.~\eqref{eq:equilibrium-fluctuation-susceptibility} is the
fluctuation--susceptibility relation. In systems with spontaneous symmetry breaking or other singular
behaviour, the zero-field limit must be specified carefully. In
particular, the limits \(h\to0^+\), \(h\to0^-\), and the thermodynamic
limit may not commute. In such cases, the susceptibility can depend on
the order in which these limits are taken.

Note that the fluctuation--susceptibility relation also determines the
coefficient of the \emph{linear equilibrium response}. For a sufficiently
small, time-independent field \(h\) applied about zero field,
\begin{equation}
\langle X\rangle_h-\langle X\rangle_0
=
\chi_{\mathrm{eq}}(T)\,h+\mathcal{O}(h^2),
\label{eq:zero-field-linear-response}
\end{equation}
where $\chi_{\mathrm{eq}}=
\beta\left\langle(\delta X)^2\right\rangle_0$,
for the classical canonical ensemble. Hence, the spontaneous equilibrium
fluctuations of \(X\) determine the leading-order response to a weak
external field. 

The static equilibrium response considered in Eq.~\eqref{eq:zero-field-linear-response} is the time-independent special case of the more general framework of \emph{linear response theory}. In the equilibrium setting, a small change in the external field produces a proportional change in the equilibrium average of its conjugate observable, with the proportionality coefficient given by the equilibrium susceptibility. Linear response theory extends this idea to time-dependent perturbations and describes how the corresponding observables respond in time. The equilibrium relations derived above therefore provide the static foundation for the dynamical response functions.  

To make this connection more explicit, let the external field be
\(h(t)\), and let \(\langle X\rangle_{\rm eq}\) denote the equilibrium
average in the absence of the perturbation. For a sufficiently weak
time-dependent field, the change in the expectation value, to first
order in \(h\), may be expressed as
\begin{equation}
\delta\langle X(t)\rangle
\equiv
\langle X(t)\rangle-\langle X\rangle_{\rm eq}
=
\int_{-\infty}^{t}\!dt'\,\chi(t-t')\,h(t'),
\label{eq:linear-response}
\end{equation}
where the angular brackets without the subscript `eq' denote the
nonequilibrium average in the presence of the time-dependent
perturbation, and \(\chi(\tau)\) is the  \emph{causal dynamic response
function}. Causality requires \(\chi(\tau)=0\) for \(\tau<0\), i.e.,
\(\chi(\tau)=\Theta(\tau)\,\chi(\tau)\), where \(\Theta(\tau)\) is the
Heaviside step function,
\begin{equation}
\Theta(\tau)=
\begin{cases}
0, & \tau<0,\\
1, & \tau>0.
\end{cases}
\end{equation}
Its value at \(\tau=0\) is convention-dependent and does not affect
the response for times away from the switching instant. Causality
allows the upper limit in Eq.~(\ref{eq:linear-response}) to be
extended to infinity:
\begin{equation}
\delta\langle X(t)\rangle
=
\int_{-\infty}^{\infty}\!dt'\,\chi(t-t')\,h(t').
\label{eq:convolution}
\end{equation}
This convolution form also assumes that the unperturbed equilibrium
state is stationary, so that the response depends only on the time
difference \(t-t'\).

The static susceptibility is a special limit of the above definition.
Suppose that a constant field is switched on at \(t=0\),
\begin{equation}
h(t)=h\,\Theta(t).
\end{equation}
Substituting into Eq.~(\ref{eq:convolution}) and writing
\(\chi(t-t')=\Theta(t-t')\,\chi(t-t')\), we obtain, for \(t>0\),
\begin{align}
\delta\langle X(t)\rangle
&=h\int_{-\infty}^{\infty}\!dt'\,
  \Theta(t-t')\,\Theta(t')\,\chi(t-t')
  \nonumber\\
&=h\int_{0}^{t}\!dt'\,\chi(t-t')
  \nonumber\\
&=h\int_{0}^{t}\!d\tau\,\chi(\tau),
\end{align}
where in the second line we used
\(\Theta(t')\,\Theta(t-t')=1\) only for \(0<t'<t\) (and zero
otherwise), and in the third line we substituted \(\tau=t-t'\).

If the system relaxes to equilibrium and the integral converges, the
long-time value of the response is
\begin{equation}
\delta\langle X\rangle_{\infty}
\equiv
\lim_{t\to\infty}\delta\langle X(t)\rangle
=
h\int_0^\infty\!d\tau\,\chi(\tau),
\end{equation}
with
\(\delta\langle X\rangle_{\infty}=\langle X\rangle_h-\langle X\rangle_{\rm eq}\),
where \(\langle X\rangle_h\) denotes the equilibrium average in the
presence of the static field \(h\). Consequently, the static
susceptibility is
\begin{equation}
\chi_{\mathrm{stat}}
=
\left.\frac{\partial\langle X\rangle_h}{\partial h}\right|_{h=0}
=
\int_0^\infty\!d\tau\,\chi(\tau).
\end{equation}
Under the usual assumptions of equilibrium, linear response, and
relaxation to the perturbed equilibrium state, this static
susceptibility equals the thermodynamic susceptibility,
\(\chi_{\mathrm{stat}}=\chi_{\mathrm{eq}}\)
[see Eq.~(\ref{eq:equilibrium-susceptibility})].
\subsection{Affinity \& Flux}
The response of systems that are out of equilibrium---but sufficiently close to equilibrium---to applied thermodynamic forces can be characterized and quantified by \emph{kinetic coefficients}. These coefficients can, in turn, be related directly to appropriate response functions. As in equilibrium statistical mechanics, where thermodynamic response functions can be derived from microscopic principles, near-equilibrium nonequilibrium phenomena are described by two complementary frameworks: \emph{irreversible thermodynamics} and its microscopic foundation in the statistical mechanics of fluctuations. In this section, we shall be concerned with irreversible thermodynamics.

Irreversible thermodynamics extends the thermodynamics to consider the rates of physical processes. To be able to do so, in addition to the basic principles of equilibrium thermodynamics, it adopts the principle of time reversal symmetry of physical laws. Such a formulation, by construction, deals with thermodynamic variables that change with space and time in the system of interest. In this context, one assumes that the system can be subdivided into many subsystems each being at \emph{local equilibrium} so that locally thermodynamic variables can be defined. Suppose the subsystem is of size ${l}$, then a local equilibrium of a thermodynamic variable $X$ is assumed to be maintained if $\frac{l|\boldsymbol{\nabla}X|}{X}<\frac{\Delta X}{X}\ll 1$, where ${l\boldsymbol{\nabla} X}$ is the external gradient induced change over the subsystem and $\Delta X$ is the equilibrium fluctuation. In addition, the thermodynamic variables are assumed to be slow variables: their characteristic relaxation times are much longer than the relaxation times of the microscopic degrees of freedom.

To bring forward the concepts of irreversible thermodynamics straightforwardly, we consider an isolated system [to be denoted by `tot' (total system)] composed of two weakly coupled subsystems (to be denoted by primes and no primes). Let $X_i$ be the \emph{conserved} extensive thermodynamic variable---any change in the variable in one subsystem is compensated by change in the variable in the other subsystem. Obviously,
\begin{equation}
X_i+X'_i=X_i^{\rm tot}={\rm constant}.
\end{equation}
Due to additivity of entropy, on invoking the entropic fundamental relation, we can write:
\begin{equation}
S(X_i)+S'(X'_i)=S^{\rm tot}.
\end{equation}
The equilibrium values of $X_i$ and $X'_i$ are determined by maximizing $S^{\rm tot}$, i.e., through the vanishing of
\begin{eqnarray}
\mathcal{A}_i\equiv\left.\frac{\partial S^{\rm tot}}{\partial X_i}\right|_{X_i^{\rm tot}}&=&\left.\frac{\partial [S(X_i)+S'(X'_i)]}{\partial X_i}\right|_{X_i^{\rm tot}}\label{eq:aftot}\\
&=&\frac{\partial S}{\partial X_i}+\frac{\partial S'}{\partial X'_i}\frac{\partial X'_i}{\partial X_i}\nonumber\\
&=&A_i-A'_i,
\end{eqnarray}
where we have denoted \emph{entropy-representation intensive variable} in the two subsystems as $A_i\equiv\frac{\partial S}{\partial X_i}$ and $A'_i\equiv\frac{\partial S'}{\partial X'_i}$. Note that, very strictly speaking, the {entropy-representation intensive variable} is intensive in nature but may be called conjugate---in the traditional sense of equilibrium thermodynamics---to the extensive variable under consideration only within a multiplicative factor of $1/T$. $\mathcal{A}_i$ is called \emph{affinity}---a generalized entropy-producing force that drives the flux of $X_i$  and hence the irreversible nonequilibrium process. In general, affinities can depend on time.

The response of the system to this force is characterized by the rate of change of $X_i$:
\begin{equation}
J_i=\frac{dX_i}{dt};
\end{equation}
which is termed as \emph{flux} conjugate to $X_i$. So in general, if there are many $X_i$s so that $S^{\rm tot}=S^{\rm tot}({X}_1^{\rm tot},{X}_2^{\rm tot},\cdots,{X}_i^{\rm tot},\cdots)=S^{\rm tot}(\boldsymbol{X}^{\rm tot})$---more compactly, we have the rate of entropy production due to irreversible process as:
\begin{equation}
\frac{dS^{\rm tot}}{dt}=\sum_{i}\left.\frac{\partial S^{\rm tot}}{\partial X_i}\right|_{\boldsymbol{X}^{\rm tot}}\frac{dX_i}{dt}=\sum_i\mathcal{A}_iJ_i.\label{eq:stot2}
\end{equation}
So the product of flux and its conjugate affinity determines the rate of entropy production. At equilibrium, the affinities and the fluxes need to vanish so that there is no entropy production. In general, fluxes can depend on time.

As a simple illustration, consider two weakly coupled bodies that can exchange
energy but are otherwise isolated from their surroundings. Let $E$ and $E'$ be
their internal energies, so that
\begin{equation}
E+E'=E^{\rm tot}=\mathrm{constant}.
\end{equation}
The entropy-representation intensive variables conjugate to the energies are
\begin{equation}
A=\left(\frac{\partial S}{\partial E}\right)_{\! \{X_{j\neq E}\}}
=\frac{1}{T},
\qquad
A'=\left(\frac{\partial S'}{\partial E'}\right)_{\! \{X'_{j\neq E}\}}
=\frac{1}{T'}.
\end{equation}
Consequently, the thermal affinity is
\begin{equation}
\mathcal{A}_E=A-A'
=\frac{1}{T}-\frac{1}{T'}.
\end{equation}
Naturally, we have the relevant flux,
\begin{equation}
J_E=\frac{dE}{dt},
\end{equation}
that denotes the energy flux into the first subsystem; energy
conservation requires: ${dE'}/{dt}=-J_E$.
The total entropy-production rate is therefore
\begin{equation}
\frac{dS^{\rm tot}}{dt}
=
\mathcal{A}_EJ_E
=
\left(\frac{1}{T}-\frac{1}{T'}\right)J_E.
\end{equation}

During the relaxation process, $T$ and $T'$ generally depend on time, and
therefore both $\mathcal{A}_E(t)$ and $J_E(t)$ are time-dependent. If
$T>T'$, heat flows from the hotter first subsystem to the colder second
subsystem, so that $J_E<0$ under the convention
$J_E=dE/dt$. Thus,
\begin{equation}
\frac{dS^{\rm tot}}{dt}
=
\left(\frac{1}{T}-\frac{1}{T'}\right)J_E
\geq 0.
\end{equation}
At equilibrium, the temperatures become equal:
\(T=T',
\)
and consequently,
\begin{equation}
\mathcal{A}_E=J_E=\frac{dS^{\rm tot}}{dt}=0.
\end{equation}

Now let us transit from two subsystems to a three-dimensional continuous system which may be treated a formed of infinitely many subsystems at local equilibrium. We introduce small-case notations $s$ and $x_i$ to denote the per-unit-volume counterparts of local entropy, $S$, and local extensive variable, $X_i$. Therefore, we have
\begin{align}
&dS(\boldsymbol{r},t)=\sum_iA_i(\boldsymbol{r},t)dX_i(\boldsymbol{r},t)\nonumber\\
\implies&ds(\boldsymbol{r},t)=\sum_iA_i(\boldsymbol{r},t)dx_i(\boldsymbol{r},t).\label{eq:ds=sum}
\end{align}
Obviously, $A_i=\frac{\partial s}{\partial x_i}$ and none of the $x_i$s would correspond to volume. Note here we explicitly showcase the spatial and temporal dependences of variable through $\boldsymbol{r}$ and $t$, respectively.

Now we should $\boldsymbol{j}_i$ for per unit volume $\boldsymbol{J}_i$ flux; the symbols are boldfaced because now they must be three dimensional spatial vectors. We introduce a definition of entropy-flux density $\boldsymbol{j}_s$ as:
\begin{align}
\boldsymbol{j}_s(\boldsymbol{r},t)\equiv\sum_iA_i(\boldsymbol{r},t)\boldsymbol{j}_i(\boldsymbol{r},t).\label{eq:itj}
\end{align}
Using this we can write
\begin{align}
\frac{ds(\boldsymbol{r},t)}{dt}=\frac{\partial s(\boldsymbol{r},t)}{\partial t}+\boldsymbol{\nabla}\cdot \boldsymbol{j}_s(\boldsymbol{r},t);\label{eq:its}
\end{align}
in words, the rate of production of entropy at a local spatial region is rate of increase of entropy in this region plus rate at which the amount of entropy is entering the region.

Now from Eq.~(\ref{eq:ds=sum}), we have
\begin{align}
\frac{\partial s(\boldsymbol{r},t)}{\partial t}=\sum_iA_i(\boldsymbol{r},t)\frac{\partial x_i(\boldsymbol{r},t)}{\partial t},\label{eq:it1}
\end{align}
and by definition, we have
\begin{align}
\frac{\partial x_i(\boldsymbol{r},t)}{\partial t}+\boldsymbol{\nabla}\cdot \boldsymbol{j}_i(\boldsymbol{r},t)=0\label{eq:it2}
\end{align}
for any conversed variable. Finally, substituting Eq.~\eqref{eq:itj}, Eq.~\eqref{eq:it1} and Eq.~\eqref{eq:it2} in Eq.~\eqref{eq:its}, we arrive at:
\begin{align}
\frac{ds(\boldsymbol{r},t)}{dt}=\sum_i\boldsymbol{\nabla}A_i(\boldsymbol{r},t)\cdot \boldsymbol{j}_i(\boldsymbol{r},t).
\end{align}
In other words, the affinity in continuous system must be recognized as the gradient of the corresponding entropy-representation intensive variable: $\boldsymbol{\mathcal{A}}_i\equiv\boldsymbol{\nabla}A_i(\boldsymbol{r},t)$---the boldfaced affinity denotes that it has three spatial components. So, if we let $\alpha$ denote both nature of the extensive variable and the coordinate of a point in space, then notationally:
\begin{align}
\frac{ds(\boldsymbol{r},t)}{dt}=\sum_\alpha\mathcal{A}_\alpha(\boldsymbol{r},t){j}_\alpha(\boldsymbol{r},t).
\end{align}
It may be compared with Eq.~\eqref{eq:stot2}.

As an illustration of the continuous-system formulation, consider a material
whose local energy density is denoted by $e(\boldsymbol{r},t)$. 
The entropy-representation intensive variable conjugate to $e$ is
\begin{equation}
A_e(\boldsymbol{r},t)
=
\left(\frac{\partial s}{\partial e}\right)_{\!\{x_{j\neq e}\}}
=
\frac{1}{T(\boldsymbol{r},t)}.
\end{equation}
The corresponding flux $\boldsymbol{j}_e$ is the local energy-current
density, which we denote by $\boldsymbol{q}$, so that the energy-continuity equation is
\begin{equation}
\frac{\partial e(\boldsymbol{r},t)}{\partial t}
+
\boldsymbol{\nabla}\cdot\boldsymbol{q}(\boldsymbol{r},t)
=
0.
\end{equation}

According to the continuous-system definition, the thermal affinity is the
spatial gradient of the entropy-representation intensive variable:
\begin{equation}
\boldsymbol{\mathcal{A}}_e(\boldsymbol{r},t)
=
\boldsymbol{\nabla}A_e(\boldsymbol{r},t)
=
\boldsymbol{\nabla}\left(\frac{1}{T(\boldsymbol{r},t)}\right).
\end{equation}
Consequently, the local entropy-production rate associated with heat
conduction is
\begin{equation}
\frac{ds(\boldsymbol{r},t)}{dt}
=
\boldsymbol{\mathcal{A}}_e(\boldsymbol{r},t)
\cdot\boldsymbol{q}(\boldsymbol{r},t)
=
\boldsymbol{\nabla}\left(\frac{1}{T(\boldsymbol{r},t)}\right)
\cdot\boldsymbol{q}(\boldsymbol{r},t).\label{eq:aiillus}
\end{equation}

For ordinary heat conduction, the energy-current density is related to the
temperature field by \emph{Fourier's law},
\begin{equation}
\boldsymbol{q}(\boldsymbol{r},t)
=
-\kappa(\boldsymbol{r},t)
\boldsymbol{\nabla}T(\boldsymbol{r},t),
\end{equation}
where $\kappa$ is the \emph{thermal conductivity}. The local entropy-production rate, Eq.~\eqref{eq:aiillus}, becomes
\begin{align}
\frac{ds(\boldsymbol{r},t)}{dt}
&=
\boldsymbol{\nabla}\left(\frac{1}{T}\right)
\cdot\left[-\kappa\boldsymbol{\nabla}T\right] \\
&=
\frac{\kappa(\boldsymbol{r},t)}
{T^2(\boldsymbol{r},t)}
\left|\boldsymbol{\nabla}T(\boldsymbol{r},t)\right|^2
\geq 0,
\end{align}
provided that $\kappa\geq 0$.

Thus, heat flows from regions of higher temperature toward regions of lower
temperature, while the resulting temperature gradient produces nonnegative
entropy locally. During a transient process, both
$\boldsymbol{\mathcal{A}}_e(\boldsymbol{r},t)$ and
$\boldsymbol{q}(\boldsymbol{r},t)$ may depend on position and time. In a
steady state, they become time-independent but may still vary spatially. At
uniform equilibrium temperature,
\begin{equation}
\boldsymbol{\nabla}T=0,
\qquad
\boldsymbol{\mathcal{A}}_e=\boldsymbol{q}
=
\frac{ds}{dt}=0.
\end{equation}
\subsection{Linear Irreversible Thermodynamics}
In the last example, Fourier's law is somewhat special, but it is
representative of a wide class of irreversible processes. It has two
important characteristics:

\begin{itemize}
    \item The flux at a given instant depends only on the affinity at that
    same instant. The system is therefore \emph{purely resistive}.
    
    \item The flux is linearly proportional to the affinity. The system is
    therefore \emph{linear} in its response to generalized thermodynamic
    forces.
\end{itemize}

Linear irreversible thermodynamics focuses on systems possessing these
properties, at least sufficiently close to equilibrium. We now formulate
this description more generally.

As before, let Greek indices denote both the type of extensive variable and
the spatial component associated with it. In the most general local
description, a flux may depend on several affinities, as well as on the
local thermodynamic state:
\begin{equation}
J_\alpha =
J_\alpha\!\left(
\mathcal{A}_1,\mathcal{A}_2,\ldots,\mathcal{A}_\beta,\ldots;
A_1,A_2,\ldots,A_\beta,\ldots
\right).
\label{eq:general_flux_dependence}
\end{equation}
Thus, a flux may respond not only to its conjugate affinity but also to
other affinities. 

\begin{table*}[t]
\centering
\caption{Direct and cross effects in linear irreversible thermodynamics.
The diagonal entries represent direct effects, in which a flux is driven by
its corresponding affinity. The off-diagonal entries represent cross
effects, in which one flux is driven by a different affinity.}
\label{tab:dce}

\renewcommand{\arraystretch}{1.25}
\setlength{\tabcolsep}{4pt}
\newcommand*{\tcell}[2]{\parbox[t]{#1}{\raggedright\hyphenpenalty=10000\exhyphenpenalty=10000\strut #2\strut}}

\resizebox{\textwidth}{!}{%
\begin{tabular}{l|l|l|l|l|l|}
\tcell{2.65cm}{} &
\tcell{2.75cm}{\textbf{Temperature gradient}} &
\tcell{2.75cm}{\textbf{Electric potential gradient}} &
\tcell{2.75cm}{\textbf{Chemical potential gradient}} &
\tcell{2.75cm}{\textbf{Pressure gradient}} &
\tcell{4.2cm}{\textbf{Magnetic field gradient}}
\\
\hline

\tcell{2.65cm}{\textbf{Heat flux}} &
\tcell{2.75cm}{Fourier heat conduction} &
\tcell{2.75cm}{Peltier effect} &
\tcell{2.75cm}{Dufour effect} &
\tcell{2.75cm}{Mechanocaloric effect} &
\tcell{4.2cm}{Thermomagnetic effect}
\\
\hline

\tcell{2.65cm}{\textbf{Electric current flux}} &
\tcell{2.75cm}{Seebeck effect} &
\tcell{2.75cm}{Ohmic electrical conduction} &
\tcell{2.75cm}{Electrodiffusion} &
\tcell{2.75cm}{Electro-osmosis} &
\tcell{4.2cm}{Magnetoelectric effect}
\\
\hline

\tcell{2.65cm}{\textbf{Mass diffusion flux}} &
\tcell{2.75cm}{Soret effect} &
\tcell{2.75cm}{Electromigration} &
\tcell{2.75cm}{Fickian mass diffusion} &
\tcell{2.75cm}{Pressure diffusion} &
\tcell{4.2cm}{Magnetodiffusion}
\\
\hline

\tcell{2.65cm}{\textbf{Volume or fluid flux}} &
\tcell{2.75cm}{Thermo-osmosis} &
\tcell{2.75cm}{Streaming potential effect} &
\tcell{2.75cm}{Osmotic flow} &
\tcell{2.75cm}{Viscous or pressure-driven flow} &
\tcell{4.2cm}{Magnetohydrodynamic flow}
\\
\hline

\tcell{2.65cm}{\textbf{Spin or magnetic flux}} &
\tcell{2.75cm}{Thermomagnetic spin transport} &
\tcell{2.75cm}{Spin-dependent electrical transport} &
\tcell{2.75cm}{Spin diffusion} &
\tcell{2.75cm}{Spin pumping by fluid flow} &
\tcell{4.2cm}{Magnetic relaxation}
\\
\hline
\end{tabular}%
}
\end{table*}

Assuming that all fluxes vanish when all affinities vanish, the fluxes can
be expanded for sufficiently small affinities:
\begin{equation}
J_\alpha =
\sum_{\beta}
L_{\alpha\beta}
\bigl(A_1,A_2,\ldots,A_\beta,\ldots\bigr)
\mathcal{A}_\beta
+
\mathcal{O}(\mathcal{A}^2),
\label{eq:jLa}
\end{equation}
where
\begin{equation}
L_{\alpha\beta}
\equiv
\left.
\frac{\partial J_\alpha}
{\partial \mathcal{A}_\beta}
\right|_{\mathcal{A}_\gamma=0\;\forall\gamma}.
\label{eq:kinetic_coefficients}
\end{equation}
The quantities \(L_{\alpha\beta}\) are called \emph{Onsager/phenomenological/transport/kinetic coefficients}. They are evaluated
in the reference equilibrium state, or in a local-equilibrium state when
the system is spatially nonuniform.

Thus, within the linear-response regime, with terms of second and higher order in the
affinities neglected, we have:
\begin{equation}
J_\alpha
=
\sum_\beta L_{\alpha\beta}\mathcal{A}_\beta.
\label{eq:linear_constitutive_relation}
\end{equation}
The diagonal coefficients describe \emph{direct effects}, whereas the
off-diagonal coefficients describe \emph{cross-effects}; see Table~\ref{tab:dce} for real physical examples.

Linearity alone,
however, does not impose any symmetry on the matrix \(L_{\alpha\beta}\). The possible couplings between fluxes and affinities are first restricted
by the \emph{Curie principle}. In an isotropic medium, quantities belonging
to different tensorial classes cannot couple linearly. Thus, a scalar
affinity can produce only a scalar flux, a vector affinity only a vector
flux, and a second-rank tensor affinity only a second-rank tensor flux of
the appropriate symmetry. These restrictions depend on the symmetry of the medium. The preceding
statement applies most directly to an isotropic medium in the absence of
external fields or other influences that reduce its symmetry. Anisotropy,
magnetic fields, rotation, and similar effects can permit additional
couplings that would be forbidden in an isotropic medium.

The \emph{linear constitutive equations}, i.e., Eq.~\eqref{eq:linear_constitutive_relation}, are phenomenological relations describing
the macroscopic response of a system close to equilibrium. Further
restrictions on the kinetic coefficients follow from microscopic
reversibility.

Let \(\epsilon_\alpha=\pm1\) denote the time-reversal parity of the extensive
state variable \(X_\alpha\) associated with the process labelled by
\(\alpha\). In the absence of time-reversal-odd external fields (such as a
magnetic field or an overall rotation), microscopic reversibility leads to
\begin{equation}
L_{\alpha\beta}
=
\epsilon_\alpha\epsilon_\beta L_{\beta\alpha}.
\label{eq:onsager_general}
\end{equation}
For example, energy \(E\), particle number \(N\), and temperature \(T\) are
time-reversal even, whereas momentum \(\mathbf p\), velocity \(\mathbf v\),
and angular momentum \(\mathbf L\) are time-reversal odd. Because the flux
\(J_\alpha=\dot X_\alpha\) is a time derivative, and time reversal changes the
sign of \(d/dt\), the flux has parity \(-\epsilon_\alpha\). The affinity, being
the derivative of the time-reversal-even entropy with respect to \(X_\alpha\),
has the same parity \(\epsilon_\alpha\) as the state variable.

Both cases occur in ordinary fluids. Heat conduction is associated with the
even variable energy: the heat flux \(\mathbf J_q\) is time-reversal odd,
while its affinity \(\nabla(1/T)\) is time-reversal even. The same holds for
diffusion, associated with the even variable particle number, with an odd
particle flux and an even affinity \(-\nabla(\mu/T)\). Viscous flow, by
contrast, is associated with the odd variable momentum density
\(\rho\mathbf v\), where \(\mathbf v(\mathbf r,t)\) is the local fluid velocity,
i.e., the mass-weighted average of the molecular velocities near \(\mathbf r\).
Its flux, the viscous stress tensor \(\Pi_{ij}\), is time-reversal even: its
kinetic contribution is built from products
\(m\,(u_i-v_i)(u_j-v_j)\), where \(\mathbf u\) is the velocity of an individual
molecule and \(\mathbf u-\mathbf v\) its velocity relative to the local flow,
and such products are unchanged when all molecular velocities are reversed.
Its affinity, the velocity gradient \(-T^{-1}\nabla\mathbf v\), is
time-reversal odd.
Consequently, couplings among even-variable processes, such as
thermo-diffusion, have symmetric coefficients, \(L_{\alpha\beta}=L_{\beta\alpha}\),
whereas a coupling between an even-variable and an odd-variable process would
be antisymmetric, \(L_{\alpha\beta}=-L_{\beta\alpha}\).

For processes with the same time-reversal parity, equation
\eqref{eq:onsager_general} reduces to the familiar  \emph{Onsager reciprocity relation}
\begin{equation}
L_{\alpha\beta}=L_{\beta\alpha}.
\label{eq:onsager_relation}
\end{equation}

If the system is subject to a time-reversal-odd external parameter
\(\mathcal{B}\), such as a magnetic field \(\mathbf B\) or an angular
velocity \(\boldsymbol{\Omega}\), the appropriate relation is the
\emph{Onsager--Casimir relation}:
\begin{equation}
L_{\alpha\beta}(\mathcal{B})
=
\epsilon_\alpha\epsilon_\beta
L_{\beta\alpha}(-\mathcal{B}).
\label{eq:onsager_casimir_relation}
\end{equation}
Under time reversal,
\begin{equation}
\mathbf B\longrightarrow-\mathbf B,
\qquad
\boldsymbol{\Omega}\longrightarrow-\boldsymbol{\Omega}.
\end{equation}
If several time-reversal-odd external parameters are present, all of them
must be reversed simultaneously.

For variables with equal time-reversal parity,
\begin{equation}
L_{\alpha\beta}(\mathcal{B})
=
L_{\beta\alpha}(-\mathcal{B}),
\end{equation}
whereas for variables with opposite parity,
\begin{equation}
L_{\alpha\beta}(\mathcal{B})
=
- L_{\beta\alpha}(-\mathcal{B}).
\end{equation}
At \(\mathcal{B}=0\), Eq.~\eqref{eq:onsager_casimir_relation} becomes
Eq.~\eqref{eq:onsager_general}. The familiar symmetric form is
therefore a special case and does not follow from linearity alone.

Using the linear constitutive relation \eqref{eq:linear_constitutive_relation},
the  \emph{local entropy-production rate} is
\begin{equation}
\sigma
=
\sum_\alpha J_\alpha\mathcal{A}_\alpha
=
\sum_{\alpha,\beta}
\mathcal{A}_\alpha L_{\alpha\beta}\mathcal{A}_\beta.
\label{eq:entropy_production_linear}
\end{equation}
Decompose the phenomenological coefficient matrix into symmetric and
antisymmetric parts:
\begin{equation}
L_{\alpha\beta}
=
L^{(s)}_{\alpha\beta}
+
L^{(a)}_{\alpha\beta},
\label{eq:L_decomposition}
\end{equation}
where
\begin{align}
L^{(s)}_{\alpha\beta}\equiv
\frac{1}{2}
\left(
L_{\alpha\beta}+L_{\beta\alpha}
\right)\,\,\&\,\,
L^{(a)}_{\alpha\beta}
\equiv
\frac{1}{2}
\left(
L_{\alpha\beta}-L_{\beta\alpha}
\right).
\end{align}
Because \(\mathcal{A}_\alpha\mathcal{A}_\beta\) is symmetric under
interchange of \(\alpha\) and \(\beta\), the antisymmetric part does not
contribute:
\begin{equation}
\sigma
=
\sum_{\alpha,\beta}
\mathcal{A}_\alpha
L^{(s)}_{\alpha\beta}
\mathcal{A}_\beta.
\label{eq:symmetric_entropy_production}
\end{equation}
The second law therefore requires the symmetric, dissipative part of the
phenomenological coefficient matrix to be positive semidefinite:
\begin{equation}
\sum_{\alpha,\beta}
\mathcal{A}_\alpha
L^{(s)}_{\alpha\beta}
\mathcal{A}_\beta
\geq 0
\quad
\text{for every admissible set of affinities}.
\label{eq:positive_semidefinite}
\end{equation}

As an example, two coupled processes, written as
\begin{equation}
\begin{pmatrix}
J_1\\
J_2
\end{pmatrix}
=
\begin{pmatrix}
L_{11} & L_{12}\\
L_{21} & L_{22}
\end{pmatrix}
\begin{pmatrix}
\mathcal{A}_1\\
\mathcal{A}_2
\end{pmatrix},
\label{eq:two_processes}
\end{equation}
may be considered. Since the antisymmetric part of \(L\) does not contribute to the quadratic form,
Eq.~\eqref{eq:positive_semidefinite} reads, for two processes,
\begin{equation}
\sigma
=
L_{11}\mathcal A_1^2
+2L^{(s)}_{12}\mathcal A_1\mathcal A_2
+L_{22}\mathcal A_2^2
\geq 0,
\end{equation}
with $L^{(s)}_{12}=\tfrac12\left(L_{12}+L_{21}\right).$
Setting \(\mathcal A_2=0\) or \(\mathcal A_1=0\) gives \(L_{11}\geq0\) and
\(L_{22}\geq0\). For \(L_{11}>0\), completing the square,
\begin{equation}
\sigma
=
L_{11}\left(\mathcal A_1+\frac{L^{(s)}_{12}}{L_{11}}\mathcal A_2\right)^2
+\frac{L_{11}L_{22}-\big(L^{(s)}_{12}\big)^2}{L_{11}}\,\mathcal A_2^2,
\end{equation}
and choosing \(\mathcal A_1\) so that the first term vanishes shows that the
coefficient of \(\mathcal A_2^2\) must be non-negative. (For \(L_{11}=0\),
\(\sigma\) is linear in \(\mathcal A_1\) and stays non-negative only if
\(L^{(s)}_{12}=0\), which the same inequality enforces.) Thus
\begin{equation}
L_{11}\geq 0,
\qquad
L_{22}\geq 0,
\qquad
L_{11}L_{22}
-
\left(L^{(s)}_{12}\right)^2
\geq 0.
\label{eq:two_process_positivity}
\end{equation}
When ordinary Onsager reciprocity applies,
\(L_{12}=L_{21}\), this reduces to
\begin{equation}
L_{11}L_{22}-L_{12}^{\,2}\geq 0.
\label{eq:onsager_two_process_positivity}
\end{equation}
The results above describe the linear-response regime and the restrictions
on the phenomenological coefficients.  

The celebrated  \emph{Prigogine's minimum-entropy-production theorem} is a more
restricted consequence of these relations. It applies when the system is
sufficiently close to equilibrium, remains in a stationary regime, and is subject to fixed external constraints, and the phenomenological coefficients obey Onsager reciprocity. At thermodynamic equilibrium, all thermodynamic affinities vanish; the linear phenomenological relations therefore imply that all
corresponding fluxes vanish. Consequently, the entropy-production rate is zero at equilibrium. This expresses the absence of irreversible processes at equilibrium. A stationary nonequilibrium state (NESS), by contrast, may have nonzero affinities
and fluxes, and therefore nonzero entropy-production rate, because these processes are
maintained by fixed external constraints.
Let
\(\mathcal{A}_\alpha^{(0)}\), \(J_\alpha^{(0)}\), and
\(\sigma^{(0)}\) denote the affinities, fluxes, and entropy-production rate
of the stationary nonequilibrium state. Consider small admissible
variations about this state,
\begin{equation}
\mathcal{A}_\alpha
=
\mathcal{A}_\alpha^{(0)}
+
\delta\mathcal{A}_\alpha,
\qquad
J_\alpha
=
J_\alpha^{(0)}
+
\delta J_\alpha.
\end{equation}
In the linear regime the coefficients \(L_{\alpha\beta}\) are independent of
the affinities, so the same linear law holds in both the stationary and the
varied state:
\begin{equation}
J^{(0)}_\alpha=\sum_\beta L_{\alpha\beta}\mathcal{A}^{(0)}_\beta,
\qquad
J_\alpha=\sum_\beta L_{\alpha\beta}\mathcal{A}_\beta .
\end{equation}
Subtracting the first relation from the second gives:
\begin{equation}
\delta J_\alpha
=
\sum_\beta L_{\alpha\beta}\,
\delta\mathcal{A}_\beta .\label{eq:deljalpha}
\end{equation}
The variations of the fluxes are therefore related to the variations of the
affinities by the same coefficient matrix that governs the fluxes themselves.

The first variation of Eq.~\eqref{eq:entropy_production_linear}, i.e., the
part of \(\sigma-\sigma^{(0)}\) linear in the variations, is
\begin{align}
\delta\sigma
&=
\sum_\alpha
\left(
J_\alpha^{(0)}\delta\mathcal{A}_\alpha
+
\mathcal{A}_\alpha^{(0)}\delta J_\alpha
\right)
\nonumber\\
&=
\sum_\alpha J_\alpha^{(0)}\delta\mathcal{A}_\alpha
+
\sum_\alpha \mathcal{A}_\alpha^{(0)}
\sum_\beta L_{\alpha\beta}\,\delta\mathcal{A}_\beta
\nonumber\\
&=
\sum_\alpha J_\alpha^{(0)}\delta\mathcal{A}_\alpha
+
\sum_\beta
\Big(\sum_\alpha L_{\alpha\beta}\,\mathcal{A}_\alpha^{(0)}\Big)
\delta\mathcal{A}_\beta
\nonumber\\
&=
\sum_\alpha J_\alpha^{(0)}\delta\mathcal{A}_\alpha
+
\sum_\beta
\Big(\sum_\alpha L_{\beta\alpha}\,\mathcal{A}_\alpha^{(0)}\Big)
\delta\mathcal{A}_\beta
\nonumber\\
&=
\sum_\alpha J_\alpha^{(0)}\delta\mathcal{A}_\alpha
+
\sum_\beta J_\beta^{(0)}\,\delta\mathcal{A}_\beta
\nonumber\\
&=
2\sum_\alpha J_\alpha^{(0)}\,\delta\mathcal{A}_\alpha .
\label{eq:first_variation_sigma}
\end{align}
In the second line we have used Eq.~\eqref{eq:deljalpha}; in the
third, interchanged the order of the two sums; in the fourth, used Onsager
reciprocity, \(L_{\alpha\beta}=L_{\beta\alpha}\); and in the fifth, recognised
the bracket as the stationary flux,
\(J_\beta^{(0)}=\sum_\alpha L_{\beta\alpha}\mathcal{A}_\alpha^{(0)}\), so that
the two sums coincide.

Fixed external constraints divide the affinities into two groups. The first
group, \(\alpha=1,\dots,k\), consists of constrained affinities imposed from
outside, for example by reservoirs held at fixed temperatures or chemical
potentials. Since every admissible variation must respect the same
constraints, these affinities cannot change, \(\delta\mathcal{A}_\alpha=0\),
although their fluxes \(J^{(0)}_\alpha\) are in general nonzero and are
sustained by the reservoirs. The second group, \(\alpha=k+1,\dots,n\),
consists of free affinities that are not imposed from outside but are set by
the system itself, for example a concentration difference across a cell
whose walls are impermeable to matter. Such affinities are not fixed by the
boundary conditions; they are functions of the internal state, here the
concentrations on either side of the cell. Because their conjugate quantities
are conserved and neither supplied nor removed from outside, a nonzero flux
redistributes them within the system, which alters the affinity, which in
turn alters the flux. This feedback continues until the flux vanishes, since
only then does the state stop changing. Hence \(J^{(0)}_\alpha=0\) for these
processes, and the stationary value of the affinity is determined by this
condition rather than imposed from outside. Thermo-diffusion in a closed cell
between two heat reservoirs contains both kinds: the thermal affinity is
constrained, being fixed by the reservoir temperatures, while the
concentration affinity is free, building up until diffusion balances the
thermally driven flow and the particle flux vanishes.

In each term of Eq.~\eqref{eq:first_variation_sigma} one factor therefore
vanishes: \(\delta\mathcal{A}_\alpha\) for the constrained processes and
\(J^{(0)}_\alpha\) for the free ones. Consequently,
\begin{equation}
\delta\sigma\big|_{\rm stationary}=0.
\label{eq:stationary_first_variation}
\end{equation}
The stationary state is therefore an extremum of the entropy production with
respect to all variations compatible with the external constraints. This
first-order condition alone does not tell us whether the extremum is a
minimum, a maximum, or a saddle point; to decide this, one must examine the
second variation \(\delta^2\sigma\).

The quadratic variation of $\sigma$ is determined by the symmetric part of the kinetic
coefficient matrix:
\begin{align}
\delta^2\sigma
=
2\sum_\alpha \delta J_\alpha\,\delta\mathcal{A}_\alpha
=
2\sum_{\alpha,\beta}
\delta\mathcal{A}_\alpha
L^{(s)}_{\alpha\beta}
\delta\mathcal{A}_\beta \geq0.
\label{eq:second_variation_sigma}
\end{align}
Here, we have used that, on substituting
\(\mathcal{A}_\alpha=\mathcal{A}_\alpha^{(0)}+\delta\mathcal{A}_\alpha\) and
\(J_\alpha=J_\alpha^{(0)}+\delta J_\alpha\) into
\(\sigma=\sum_\alpha J_\alpha\mathcal{A}_\alpha\), the part quadratic in the
variations is \(\sum_\alpha\delta J_\alpha\,\delta\mathcal{A}_\alpha\); and
that with the convention \(\sigma=\sigma^{(0)}+\delta\sigma+\tfrac12\delta^2\sigma\),
the second variation is twice this quantity. By the positive-semidefiniteness condition [see
Eq.~\eqref{eq:positive_semidefinite}], $\delta^2\sigma\geq0.$ 

Since \(\sigma\) is exactly quadratic in the affinities in the linear regime,
the expansion terminates at second order, and the stationary state is the
minimum of the entropy-production rate over all admissible states,
\begin{equation}
\sigma^{(0)}
=
\min_{\rm admissible}\sigma ,
\label{eq:prigogine_minimum}
\end{equation}
which is unique when the coefficient matrix restricted to the free processes
is positive definite.

This result is {Prigogine's minimum-entropy-production theorem}. It is not a
statement that entropy production decreases during an arbitrary
nonequilibrium evolution, but a local variational property of the stationary
state: among all states compatible with fixed external constraints, such as
imposed temperatures or chemical potentials, the stationary
state is the one of minimum entropy production, the comparison being made
over variations that preserve those constraints. Its validity rests on a
local-equilibrium (or otherwise near-equilibrium) description, linear
phenomenological relations with constant coefficients, and Onsager
reciprocity, \(L_{\alpha\beta}=L_{\beta\alpha}\). At equilibrium, where both
affinities and fluxes vanish, it reduces to the trivial case
\(\sigma_{\rm eq}=0\). Far from equilibrium, where nonlinear response,
instabilities, or additional dynamical effects may become important, it does
not generally apply.

\subsection{Deriving Onsager Reciprocity Relation}
The linchpin in the derivation of the Onsager reciprocity relations is \emph{Onsager's regression hypothesis}. The hypothesis began as an intuitive and largely unproven leap of faith. Onsager proposed that the way a spontaneous fluctuation arising in an equilibrium system gradually decays should, on average, follow the same laws as the relaxation of a small disturbance deliberately introduced from outside. Although these two situations appear different---one produced by random microscopic motion and the other by a controlled macroscopic intervention---Onsager argued that, once created, a fluctuation should regress toward equilibrium independently of its origin. This daring identification provided a conceptual bridge between reversible microscopic dynamics and irreversible macroscopic behaviour. The hypothesis was later placed on a firmer theoretical foundation through linear-response theory, although the term ``hypothesis'' remained attached to it.

We now provide a rather straightforward mathematical realization of the hypothesis, we consider a simple set up: Let us assume a thermodynamic system with time-dependent Hamiltonian:
\begin{equation}
H=H_0-\sum _ih_i(t)X_i,\label{eq:HOd2GK}
\end{equation}
where $X_i$s are thermodynamic observables and $h_i$s corresponding conjugate external perturbations. We assume that each $h_i(t)=h_i$ (a small constant) when $t\in (\infty,0)$; at $t=0$, all perturbations are switched off and the system relaxes to equilibrium as $t\to\infty$ under $H_0$. So essentially, we had an unperturbed system under $H_0$ which was perturbed by $h_i$s to be in new equilibrium under $H_0-\sum_ih_iX_i$ up to $t=0^-$, then on suddenly turning off the external forces, the system is thrown off equilibrium and follows a nonequilibrium relaxation towards a new equilibrium. 

During this relaxation, an $X_i(t)$ is explicitly time-dependent. If we are interested in calculating its (time-dependent) expected value, then we need to take into consideration the fact that ultimately it---or more precisely, its argument $z$ (microscopic coordinates)---is undergoing deterministic Hamiltonian evolution; hence, its various realizations are deterministic time-series of various realizations of the initial conditions it started with. (As earlier, \(z\) denotes a complete microscopic state of the system: in classical system, \(z\) consists of all particle positions and momenta placing $z$ as a point in the phase space.) The probability distribution of the initial conditions, i.e., equivalently how the microscopic states are distributed at $t=0$, is simply determined by the canonical distribution function. Mathematically, thus,
\begin{eqnarray}
\langle X_i(t)\rangle=\frac{\int dz X_i(t)e^{-\beta(H_0(z)-\sum _jh_jX_j(0))}}{\int dz\,e^{-\beta(H_0(z)-\sum _jh_jX_j(0))}}\,\,{\rm for}\,\, t>0,\qquad
\end{eqnarray}
where denominator is just the normalization factor.

Now, assuming $h$ to be small, we proceed further keeping only the terms up to linear order in $h$ and denoting equilibrium average as $\langle \cdot\rangle_{0}$:

\begin{eqnarray}
&&\langle X_i(t)\rangle=\frac{\int dz X_i(t)e^{-\beta H_0(z)}(1+\beta\sum _jh_jX_j(0))}{\int dz e^{-\beta H_0(z)}(1+\beta\sum _jh_jX_j(0))}\qquad\\
\implies&&\langle X_i(t)\rangle\approx\frac{\left\langle X_i(t)\left[1+\beta\sum _jh_jX_j(0)\right]\right\rangle_{0}}{\left\langle \left[1+\beta\sum _jh_jX_j(0)\right]\right\rangle_{0}}\quad\\
\implies&&\langle X_i(t)\rangle\approx{\left\langle X_i(t)\left[1+\beta\sum _jh_jX_j(0)\right]\right\rangle_{0}}\quad\nonumber\\
&&\phantom{\langle X_i(t)\rangle=}\times{ \left[1-\beta\sum _jh_j\left\langle X_j(0)\right\rangle_{0}\right]}\quad\\
\implies&&\langle X_i(t)\rangle-\langle X_i(t)\rangle_{0}=\beta \sum _jh_j\left[\langle X_j(0)X_i(t)\rangle_0\right.-\nonumber\\
&&\phantom{\langle X_i(t)\rangle-\langle X_i(t)\rangle_{0}=}\left.\langle X_j(0)\rangle_0\langle X_i(t)\rangle_0\right]+\mathcal{O}(h^2)\quad\\
\implies&&\langle X_i(t)\rangle=\beta \sum _jh_j\langle X_j(0)X_i(t)\rangle_0+\mathcal{O}(h^2).\label{eq:ORH}
\end{eqnarray}
where in the final equation, without any prejudice of generality, we have taken $\langle X_i\rangle_0=0$ for all $i$. 

The final equation carries profound physical content. It tells us that the average  deviation of a macroscopic observable from its unperturbed equilibrium evolution is 
linear in the perturbation field $h_j$, with the proportionality constant given by the {equilibrium time correlation function}. But the equation is even richer than it first appears---hidden inside it is the 
\emph{Onsager regression hypothesis}. As mentioned earlier, Onsager proposed, on purely physical 
grounds, that the relaxation of a small spontaneous fluctuation from equilibrium follows 
{exactly the same dynamics} as the relaxation of a macroscopic deviation induced 
by an external perturbation---the system has no memory of how the deviation was 
created. Eq.~\eqref{eq:ORH} makes this intuition precise and quantitative: the nonequilibrium 
response $\langle X_i(t)\rangle - 0$ is governed entirely by the 
equilibrium time correlation function $\langle X_j(0)X_i(t)\rangle_0$, a 
quantity that characterizes nothing but the spontaneous fluctuations of the unperturbed 
system. The decay of the response and the decay of equilibrium fluctuations are, in this 
sense, one and the same---which is precisely what Onsager asserted. The derivation 
above thus provides the microscopic statistical-mechanical {justification} for 
why the regression hypothesis works: it is a direct consequence of expanding the 
perturbed Boltzmann distribution to first order in $h_j$.

Moreover, this is precisely the \emph{fluctuation-dissipation theorem} in its most transparent 
form: the response of a system to a weak external perturbation---the ``dissipation'' 
side---is entirely determined by the spontaneous equilibrium fluctuations encoded in 
the correlator---the ``fluctuation'' side. The Onsager hypothesis captures the 
qualitative essence of this structure; the fluctuation-dissipation theorem gives it its 
full quantitative and microscopic content. 

The theorem shows up in a remarkably wide range of physical systems. Take a
pollen grain suspended in water. The water molecules knock it about at
random, so that it jitters and slowly wanders away by diffusion; these
random kicks are the fluctuation. When the grain is pushed, the same
molecules resist its motion and turn its kinetic energy into heat; this
viscous drag is the dissipation. Because one process produces both, they
are not independent: how far the grain wanders is fixed by how strongly it
is held back, through the \emph{Einstein relation}---\(D=k_BT/\gamma\). The same
story holds in an electrical circuit. The electrons in a resistor, or in
the resistive channel of a Josephson junction, scatter randomly at any
finite temperature. Even when no current is driven, this makes the voltage
across it fluctuate; this \emph{Johnson--Nyquist noise} is the fluctuation. When a
current does flow, the same scattering converts electrical energy into
heat, and the resistance that measures this loss is the dissipation. The
size of the noise is set by the resistance itself, just as the random kicks
on the pollen grain are set by its friction. Thermal radiation provides a
third example. A body at finite temperature spontaneously emits radiation,
because the charges within it move randomly; this thermal emission is the
fluctuation. The same body absorbs radiation falling on it and turns that
energy into heat; this absorption is the dissipation. In equilibrium with
its surroundings, the body must emit exactly as much as it absorbs, or it
would heat up or cool down. A surface that absorbs strongly therefore also
glows strongly, which is \emph{Kirchhoff's law of thermal radiation}, and the blackbody spectrum is the
result of this balance. In each case the lesson is the same: a system that
can lose energy into its surroundings must also receive random kicks from
them. Fluctuation and dissipation are not separate phenomena but two
inseparable aspects of the same coupling to a thermal environment. It is worth noting, however, that 
the fluctuation-dissipation relation goes strictly beyond the regression hypothesis to connect to response functions. 

The next result needed for the Onsager relations is the equilibrium
ensemble average of the product of an extensive observable \(X_i\) and an
affinity \(\mathcal{A}_j\). Take the system, together with any reservoir it
exchanges energy with, to be isolated, so that its total energy, volume and
particle number are fixed and all accessible microstates are equally
probable. The \(X_i\) are then internal variables, such as the share of
energy in one part of the system, which fluctuate subject to these
constraints, and \(S\) is the total entropy. Let \(\boldsymbol{X}^{\rm eq}\)
denote the equilibrium values of the \(X_i\), with entropy
\(S(\boldsymbol{X}^{\rm eq})\), and \(S(\boldsymbol{X})\) the entropy when
the system is displaced from equilibrium. Since the entropy is maximal at
equilibrium, the first-order term of its Taylor expansion vanishes, and to
second order
\begin{equation}
    S(\boldsymbol{X})
    =
    S(\boldsymbol{X}^{\rm eq})
    -
    \frac{1}{2}\sum_{i,j}\sigma_{ij}
    \bigl(X_i-X_i^{\rm eq}\bigr)\bigl(X_j-X_j^{\rm eq}\bigr),
    \label{eq:tayex}
\end{equation}
where we have defined: $\sigma_{ij}\equiv-\frac{\partial^2 S}{\partial X_i\partial X_j}|_{\boldsymbol{X}=\boldsymbol{X}^{\rm eq}}$; note that $\sigma_{ij}=\sigma_{ji}$. Subsequently, from Eq.~\eqref{eq:tayex}, affinity $\mathcal{A}_i$ corresponding to observable $X_i$, can be found as [see Eq.~\eqref{eq:aftot}]:
\begin{equation}
    \mathcal{A}_i
    =
    \frac{\partial S}{\partial X_i}
    =
    -\sum_{j}\sigma_{ij}\bigl(X_j-X_j^{\rm eq}\bigr),
    \label{eq:Atotfluc}
\end{equation}
which vanishes at equilibrium and is nonzero whenever the system is
displaced from it.

Now, in line with the profound hypothesis of Boltzmann, the thermodynamic entropy can be equivalently seen as derived from microscopic considerations so that the equilibrium macrostate's entropy $S(\boldsymbol{X}^{\rm eq})=k_B\ln\Omega^{\rm eq}$ where $\Omega$ is the number of distinct microstates corresponding to a given macrostate. Perturbations lead to perturbations in entropy of the system, of course, through perturbations in  the number of distinct microstates: In general, hence, we write $S(\boldsymbol{X})=k_B\ln\Omega$---valid even when the system is not in equilibrium. Therefore, since $\Delta S=S(\boldsymbol{X})-S(\boldsymbol{X}^{\rm eq})=k_B\ln\Omega-k_B\ln\Omega^{\rm eq}$, the probability of being in perturbed state---due to stochastic detailed balanced fluctuations at equilibrium---being proportional to corresponding $\Omega$ (all microstates being equally probable) effectively is proportional to $\exp[\{S(\boldsymbol{X})-S(\boldsymbol{X}^{\rm eq})\}/k_B]$.

Consequently, the equilibrium ensemble average of $X_k$,
using Eq.~\eqref{eq:tayex}, is:
\begin{align}
\langle X_k\rangle_{\rm eq}=X_k^{\rm eq}=\frac{\int_{-\infty}^{\infty}X_k\,
e^{\Delta S(\boldsymbol{X})/k_B}
\prod_idX_i}
{\int_{-\infty}^{\infty}
e^{\Delta S(\boldsymbol{X})/k_B}
\prod_idX_i},
\label{eq:fluint}
\end{align}
where we defined $\Delta S(\boldsymbol{X})\equiv
S(\boldsymbol{X})-S(\boldsymbol{X}^{\rm eq})$ for compactness, and
the limits have been extended to $\pm\infty$ because large
fluctuations contribute negligibly to the integral. We now
differentiate both sides of Eq.~\eqref{eq:fluint} with respect to
$X_l^{\rm eq}$. The left-hand side gives immediately:
\begin{align}
\frac{\partial X_k^{\rm eq}}{\partial X_l^{\rm eq}}=\delta_{kl}.
\label{eq:lhs}
\end{align}
For the right-hand side, $X_l^{\rm eq}$ enters solely through
$\Delta S(\boldsymbol{X})$. Differentiating Eq.~\eqref{eq:tayex}
with respect to $X_l^{\rm eq}$ and exploiting $\sigma_{mn}=\sigma_{nm}$:
\begin{align}
\frac{\partial\,\Delta S}{\partial X_l^{\rm eq}}
&= \frac{1}{2}\sum_{m,n}\sigma_{mn}
\bigl[\delta_{ml}(X_n-X_n^{\rm eq})+(X_m-X_m^{\rm eq})\delta_{nl}\bigr]\nonumber\\
&=\sum_{m}\sigma_{ml}(X_m-X_m^{\rm eq})\nonumber\\
&=-\mathcal{A}_l(\boldsymbol{X}),
\label{eq:dSdXeq}
\end{align}
where the last step uses Eq.~\eqref{eq:Atotfluc} with
$\sigma_{ml}=\sigma_{lm}$. This is a useful relation of affinity which will come in handy in the calculation below.

Denoting the normalization integral
$\mathcal{Z}\equiv\int_{-\infty}^{\infty}
e^{\Delta S/k_B}\prod_i dX_i$
and applying the quotient rule to Eq.~\eqref{eq:fluint},
the derivative of the right-hand side becomes:
\begin{align}
&\frac{1}{\mathcal{Z}}
\frac{\partial}{\partial X_l^{\rm eq}}
\int_{-\infty}^{\infty} X_k\,
e^{\Delta S/k_B}\prod_i dX_i
-\frac{X_k^{\rm eq}}{\mathcal{Z}}
\frac{\partial \mathcal{Z}}{\partial X_l^{\rm eq}}\nonumber\\
&=-\frac{1}{k_B\mathcal{Z}}
\int_{-\infty}^{\infty}X_k\,\mathcal{A}_l\,
e^{\Delta S/k_B}\prod_i dX_i\nonumber\\
&\quad
+\frac{X_k^{\rm eq}}{k_B\mathcal{Z}}
\int_{-\infty}^{\infty}\mathcal{A}_l\,
e^{\Delta S/k_B}\prod_i dX_i\nonumber\\
&=-\frac{\langle X_k\mathcal{A}_l\rangle_{\rm eq}}{k_B}
+\frac{X_k^{\rm eq}}{k_B}
\langle\mathcal{A}_l\rangle_{\rm eq}.
\label{eq:rhs}
\end{align}
The second term vanishes identically since affinities average to
zero at equilibrium, $\langle\mathcal{A}_l\rangle_{\rm eq}=0$
(no thermodynamic forces persist at equilibrium).
Equating Eqs.~\eqref{eq:lhs} and \eqref{eq:rhs} yields the central result:
\begin{align}
{\langle X_k\,\mathcal{A}_l\rangle_{\rm eq}=-k_B\,\delta_{kl}.}
\label{eq:XAresult}
\end{align}
This compact result states that the fluctuation of an extensive
variable $X_k$ is correlated with the thermodynamic force
$\mathcal{A}_l$ only for its own conjugate pair ($k=l$), with
a universal strength set purely by $k_B$, independently of
any microscopic detail of the system.

Equipped with Eqs.~\eqref{eq:ORH} and \eqref{eq:XAresult}, we can now derive Onsager relation. We first point out that each $X_i$ is supposed to have definite parity under time-reversal $\hat {T}$: $\hat {T}X_i=\varepsilon_iX_i$ ($\varepsilon_i=\pm1$); the Hamiltonian of the system is supposed to have time translation and reversal symmetry.  Since $\hat{T}H = H$, the equilibrium measure is invariant under time-reversal. Combined with the trajectory-reversal property: $X_i(\Gamma(t)) = [\hat{T}X_i](\hat{T}\Gamma(-t))$ ($\Gamma$ being the trajectory), the equilibrium average is unchanged when all observables are time-reversed and $t\to -t$. Using these:
\begin{align}
\langle X_i(t)X_j(0)\rangle_{\rm eq}&=\langle [\hat {T}X_i(-t)][\hat {T}X_j(0)]\rangle_{\rm eq}\nonumber\\
\implies\,\langle X_i(t)X_j(0)\rangle_{\rm eq}&=\langle [\varepsilon_iX_i(-t)][\varepsilon_jX_j(0)]\rangle_{\rm eq}\nonumber\\
\implies\,\langle X_i(t)X_j(0)\rangle_{\rm eq}&=\varepsilon_i\varepsilon_j\langle X_i(0)X_j(t)\rangle_{\rm eq},
\label{eq:hatTXp}
\end{align}
owing to stationarity of the equilibrium state. 

Now in the linear nonequilibrium regime, the mean value of an isolated fluctuating observable $X_i$ away from its equilibrium mean value (that may be assumed to be zero without any loss of generality) must relax exponentially because that is what its equilibrium correlation functions do. So, $\frac{d\langle X_i\rangle}{dt}\propto-\langle X_i\rangle$. Thus, in the presence of many fluctuating $X_i$s, the relaxation ansatz is:
\begin{align}
J_i=\frac{d\langle X_i\rangle}{dt}=-\sum_jM_{ij}\langle X_j\rangle,\label{eq:Mij}
\end{align}
where $M_{ij}$ is \emph{phenomenological relaxation coefficient} that depends on the nature of the $X_i$s and conditions that lead system out of equilibrium in the first place. However, rather than trying to give more detailed reason behind the ansatz, for our goal we justify it \emph{a posteriori} through the fact that this leads to linear relationship between flux and affinity----a relationship whose proportionality constants' symmetry-property is what we are after: $J_i=\dot{X}_i(t)=\sum_kL_{ik}\mathcal{A}_k(t)$. To see this explicitly, we multiply this relation by $X_j(0)$ and take equilibrium average:
\begin{align}
&\frac{d}{dt}\langle X_i(t)X_j(0)\rangle_{\rm eq}=\sum_kL_{ik}\langle\mathcal{A}_k(t)X_j(0)\rangle_{\rm eq}\nonumber\\
\implies\,&-\sum_kM_{ik}\langle X_k(t)X_j(0)\rangle_{\rm eq}=\sum_kL_{ik}\langle\mathcal{A}_k(t)X_j(0)\rangle_{\rm eq}\nonumber\\
\implies\,&\sum_kM_{ik}\langle X_k(0)X_j(0)\rangle_{\rm eq}=-\sum_kL_{ik}\langle\mathcal{A}_k(0)X_j(0)\rangle_{\rm eq}\nonumber\\
\implies\,&\sum_kM_{ik}\langle X_k(0)X_j(0)\rangle_{\rm eq}=k_BL_{ij},\label{eq:LxM}
\end{align}
using Eq.~\eqref{eq:XAresult}; the second implication above has been obtained by substituting
Eq.~\eqref{eq:ORH} into Eq.~\eqref{eq:Mij}:
\begin{align}
&\beta\sum_j h_j\frac{d}{dt}\langle X_i(t)X_j(0)\rangle_{\rm eq}\nonumber\\
=&-\beta\sum_j h_j\sum_k M_{ik}\langle X_k(t)X_j(0)\rangle_{\rm eq}.
\end{align}
Since the fields $h_j$ are independent and arbitrary (and $M_{ik}$,
being a linear-response coefficient, does not depend on them), the
coefficients of each $h_j$ on the two sides must be equal, giving
\begin{align}
\frac{d}{dt}\langle X_i(t)X_j(0)\rangle_{\rm eq}=-\sum_kM_{ik}\langle X_k(t)X_j(0)\rangle_{\rm eq}.\label{eq:Mijt}
\end{align}
Thus, we find $L_{ij}$ by construction from $M_{ij}$.

As the final step, differentiating Eq.~\eqref{eq:hatTXp} with respect to time and using Eq.~\eqref{eq:Mijt}, we arrive at:
\begin{align}
&\sum_kM_{ik}\langle X_k(t)X_j(0)\rangle_{\rm eq}=\varepsilon_i\varepsilon_j\sum_kM_{jk}\langle X_k(t)X_i(0)\rangle_{\rm eq}\nonumber\\
&\implies\,L_{ij}=\varepsilon_i\varepsilon_jL_{ji},
\end{align}
which is obtained by putting $t=0$ and recalling Eq.~\eqref{eq:LxM}. This is the Onsager reciprocity relation.

This derivation can be easily adapted to derive the Onsager--Casimir relation as well. To this end, note that the assumption $\hat{T}H=H$ breaks down when the system is subject to an external magnetic field $\mathbf{B}$ or placed in a rotating frame with angular velocity $\boldsymbol{\Omega}$. Both $\mathbf{B}$ and $\boldsymbol{\Omega}$ are axial vectors that are odd under time-reversal:
\begin{align}
\hat{T}\mathbf{B}=-\mathbf{B},\qquad\hat{T}\boldsymbol{\Omega}=-\boldsymbol{\Omega},
\end{align}
since $\mathbf{B}$ is sourced by currents and $\boldsymbol{\Omega}$ is proportional to angular momentum, both of which reverse under $\hat{T}$. Consequently, $\hat{T}H(\mathbf{B},\boldsymbol{\Omega})=H(-\mathbf{B},-\boldsymbol{\Omega})$: time-reversing the dynamics is equivalent to reversing the external field. The equilibrium measure at field $\mathbf{B}$ is therefore mapped by $\hat{T}$ onto the equilibrium measure at $-\mathbf{B}$, and Eq.~\eqref{eq:hatTXp} is accordingly modified to:
\begin{align}
\langle X_i(t)X_j(0)\rangle_{\rm eq}|_{\mathbf{B},\boldsymbol{\Omega}}=\varepsilon_i\varepsilon_j\langle X_j(t)X_i(0)\rangle_{\rm eq}|_{-\mathbf{B},-\boldsymbol{\Omega}}.
\label{eq:hatTXpB}
\end{align}
Carrying this modified identity through the remainder of the derivation in place of Eq.~\eqref{eq:hatTXp} --- differentiating with respect to $t$, applying the relaxation equation on both sides with $M_{ij}$ now evaluated at ($\mathbf{B}$, $\boldsymbol{\Omega}$), and setting $t=0$---one arrives at: 
\begin{align}
{L_{ij}(\mathbf{B},\boldsymbol{\Omega})=\varepsilon_i\varepsilon_jL_{ji}(-\mathbf{B},-\boldsymbol{\Omega})}
\end{align}
---the Onsager--Casimir relation.

\subsection{Green--Kubo Relations}
We now turn to a systematic derivation of transport coefficients directly
from equilibrium fluctuations. The physical setting is the same as before:
the system is governed by the Hamiltonian
[cf.\ Eq.~\eqref{eq:HOd2GK}]:
\begin{equation}
H(t)=H_0-\sum_j h_j(t)X_j,
\label{eq:Hamiltonian}
\end{equation}
with the external fields $h_j(t)$ taken to be weak and
time-dependent rather than static; but now we assume that the fields are
switched on adiabatically from the remote past, $h_j(t)\to0$ as
$t\to-\infty$, and that before this the system is in thermal
equilibrium with respect to the unperturbed Hamiltonian $H_0$ at
inverse temperature $\beta=1/k_{\mathrm B}T$. The response to the
fields is then measured relative to this initial equilibrium state.

The phase-space distribution $\rho(z,t)$ satisfies the Liouville equation
\begin{equation}
\frac{\partial \rho}{\partial t}
=
\{H,\rho\},
\label{eq:Liouville}
\end{equation}
where $z=(q,p)$ denotes a point in phase space---$q$ and $p$ being all the generalised coordinates and momenta, respectively---and $\{A,B\}$ is the \emph{Poisson bracket}, defined for any two phase-space functions $A(z)$ and $B(z)$ as
\begin{equation}
\{A,B\}
\equiv
\sum_k
\left(
\frac{\partial A}{\partial q_k}\frac{\partial B}{\partial p_k}
-
\frac{\partial A}{\partial p_k}\frac{\partial B}{\partial q_k}
\right).
\label{eq:Poisson_bracket_def}
\end{equation}
Hamilton's equations of motion, $\frac{dq_k}{dt}=\partial H/\partial p_k$ and $\frac{dp_k}{dt}=-\partial H/\partial q_k$, imply that the total time derivative of any phase-space function $A(z(t))$ along a Hamiltonian trajectory is $\frac{dA}{dt}=\{A,H\}$. 

We write the nonequilibrium distribution as
\begin{equation}
\rho(z,t)
=
\rho_0(z)+\delta\rho(z,t),
\label{eq:distribution_decomposition}
\end{equation}
where $\delta\rho$ is assumed to be first order in the external fields $h_j(t)$. The equilibrium distribution associated with the unperturbed Hamiltonian is
\begin{equation}
\rho_0(z)=\frac{e^{-\beta H_0(z)}}{Z_0},
\label{eq:rhonotup}
\end{equation}
which is the state of the system before the fields are switched on, so that $\delta\rho(z,-\infty)=0$.
Since \(\rho_0\) depends on \(z\) only through \(H_0\), which is conserved under the unperturbed motion, \(\{H_0,\rho_0\}=0\). Substituting Eq.~\eqref{eq:distribution_decomposition} into Eq.~\eqref{eq:Liouville} and retaining terms to first order in the perturbing fields gives
\begin{equation}
\frac{\partial\,\delta\rho}{\partial t}
=
\{H_0,\delta\rho\}
-\sum_j h_j(t)\{X_j,\rho_0\}.
\label{eq:linearized_Liouville_noop}
\end{equation}
We note that in the light of Eq.~\eqref{eq:rhonotup},
\begin{equation}
\{X_j,\rho_0\}
=
-\beta\rho_0\{X_j,H_0\}
=
-\beta\rho_0\dot{X}_j,
\label{eq:rho0_bracket}
\end{equation}
where $\dot{X}_j\equiv\{X_j,H_0\}$ is the time derivative of $X_j$ along the unperturbed dynamics. Substituting into Eq.~\eqref{eq:linearized_Liouville_noop},
\begin{equation}
\frac{\partial\,\delta\rho}{\partial t}
=
\{H_0,\delta\rho\}
+\beta\rho_0\sum_j h_j(t)\dot{X}_j.
\label{eq:linearized_explicit}
\end{equation}

We solve this equation by seeking $\delta\rho$ in the form
\begin{equation}
\delta\rho(z,t)
=
-\beta\rho_0(z)\,\phi(z,t),
\label{eq:ansatz}
\end{equation}
where $\phi(z,t)$ is to be determined. Substituting into Eq.~\eqref{eq:linearized_explicit} and using $\{H_0,-\beta\rho_0\phi\}=-\beta\rho_0\{H_0,\phi\}$ (since $\rho_0$ is a function of $H_0$ alone and hence commutes with it under the Poisson bracket), we arrive at:
\begin{eqnarray}
\frac{\partial\phi}{\partial t}+\{\phi,H_0\}
=
-\sum_j h_j(t)\dot{X}_j.
\label{eq:phi_equation}
\end{eqnarray}
Introducing the unperturbed linear Liouville operator
$\mathcal{L}_0A\equiv\{A,H_0\}$, we define the time-evolved phase-space
function
\begin{equation}
A(s)\equiv e^{s\mathcal{L}_0}A ,
\label{eq:time_evolved_def}
\end{equation}
i.e., the solution of $\partial_s A(s)=\{A(s),H_0\}$ with $A(0)=A$.
Eq.~\eqref{eq:phi_equation} then reads
$\partial_t\phi+\mathcal{L}_0\phi=-\sum_j h_j(t)\dot X_j$, whose
solution satisfying $\phi(-\infty)=0$ is
\begin{equation}
\phi(t)
=
-\sum_j\int_{-\infty}^{t}dt'\,h_j(t')\,
e^{(t'-t)\mathcal{L}_0}\dot{X}_j .
\label{eq:phi_solution}
\end{equation}
Hence, by Eq.~\eqref{eq:ansatz},
\begin{align}
\delta\langle O(t)\rangle
&=\int dz\,O\delta\rho=
-\beta\!\int dz\,\rho_0\,O\,\phi\nonumber\\
&=
\beta\sum_j\int_{-\infty}^{t}dt'\,h_j(t')
\int dz\,\rho_0\,O\,
\bigl(e^{(t'-t)\mathcal{L}_0}\dot{X}_j\bigr)\nonumber\\
&=
\beta\sum_j\int_{-\infty}^{t}dt'\,h_j(t')
\langle O(0)\,\dot X_j(t'-t)\rangle_0.
\label{eq:obs_change_noop}
\end{align}
Since both the ensemble $\rho_0$ and the propagator $e^{s\mathcal{L}_0}$
are generated by $H_0$, the integral in Eq.~\eqref{eq:obs_change_noop}
is an equilibrium correlation, and by stationarity
\begin{align}
\langle O(0)\,\dot X_j(t'-t)\rangle_0
=\langle O(t-t')\,\dot X_j(0)\rangle_0.
\end{align}
With $\dot X_j=J_j$ (the flux conjugate to $X_j$),
\begin{equation}
\delta\langle O(t)\rangle
=
\beta\sum_j\int_{-\infty}^{t}\!dt'\,
\langle O(t-t')\,J_j(0)\rangle_0\,h_j(t').
\label{eq:obs_response_noop}
\end{equation}
The causal response function [recall Eq.~\eqref{eq:linear-response}] is therefore
\begin{equation}
\chi_{Oj}(\tau)
=
\beta\,\Theta(\tau)\,
\langle O(\tau)\,J_j(0)\rangle_0,
\label{eq:response_function_noop}
\end{equation}
in terms of which
\begin{equation}
\delta\langle O(t)\rangle
=
\sum_j\int_0^\infty d\tau\,
\chi_{Oj}(\tau)\,h_j(t-\tau).
\label{eq:response_convolution_noop}
\end{equation}
Of particular interest is the response of the fluxes themselves,
$O=J_i$. In equilibrium the fluxes have zero mean,
$\langle J_i\rangle_0=\frac{d}{dt}\langle X_i\rangle_0=0$,
so $J_i$ is itself a fluctuation. Setting $O=J_i$ in
Eq.~\eqref{eq:response_function_noop} then gives the
\emph{fluctuation--dissipation relation}
\begin{equation}
\chi_{ij}(\tau)
=
\beta\,\Theta(\tau)\,
\langle J_i(\tau)\,J_j(0)\rangle_0,
\label{eq:FDR_noop}
\end{equation}
i.e., the dissipative response of the fluxes is fixed by their spontaneous
equilibrium fluctuations. 

Consider now a field switched on in the remote past and held constant
thereafter, $h_j(t)=h_j$. Since the system starts in equilibrium with
respect to $H_0$ [cf.\ the condition $\delta\rho(-\infty)=0$], the
initial transients die out and the fluxes settle to steady values.
Eq.~\eqref{eq:response_convolution_noop} then gives
\begin{equation}
\delta\langle J_i\rangle
=
\sum_j\left[\int_0^\infty d\tau\,\chi_{ij}(\tau)\right]h_j .
\label{eq:steady_flux_h}
\end{equation}
To cast this in the language of irreversible thermodynamics, we
express the fields through the corresponding affinities. The field
$h_j$ does work $h_j\,dX_j$ on the system [cf.\ the coupling in
Eq.~\eqref{eq:Hamiltonian}]. In the steady state this work is
dissipated as heat into the surroundings at temperature $T$, so the
total entropy is produced at the rate
\begin{equation}
\frac{dS^{\rm tot}}{dt}
=
\frac{1}{T}\sum_j h_j\,\langle J_j\rangle
\equiv
\sum_j\mathcal{A}_j\,\langle J_j\rangle .
\label{eq:entropy_production_h}
\end{equation}
The affinity conjugate to $X_j$ [cf.\ Eq.~\eqref{eq:aftot}] is
therefore
\begin{equation}
\mathcal{A}_j=\frac{\partial S^{\rm tot}}{\partial X_j}
=\frac{h_j}{T}.
\label{eq:affinity_field}
\end{equation}
Substituting $h_j=T\mathcal{A}_j$ in Eq.~\eqref{eq:steady_flux_h}
yields the linear constitutive
relation~\eqref{eq:linear_constitutive_relation},
$\delta\langle J_i\rangle=\sum_jL_{ij}\mathcal{A}_j$, with
$L_{ij}=T\int_0^\infty d\tau\,\chi_{ij}(\tau)$. Using
Eq.~\eqref{eq:FDR_noop}, this becomes
\begin{equation}
L_{ij}
=
\frac{1}{k_{\mathrm B}}
\int_0^\infty d\tau\,
\langle J_i(\tau)\,J_j(0)\rangle_0 ,
\label{eq:Green_Kubo_noop}
\end{equation}
the \emph{Green--Kubo relation}. The Onsager coefficients are thus
determined entirely by the time integral of the equilibrium flux
correlations.

The long-time limit in Eq.~\eqref{eq:Green_Kubo_noop} requires some
care. In a finite system the coordinates $X_i$ are bounded, so a
constant field cannot sustain a steady flux. Instead, the system
eventually relaxes to the new equilibrium state
$\rho\propto e^{-\beta(H_0-\sum_j h_jX_j)}$. In that state the fluxes
vanish, by the same argument that gave $\langle J_i\rangle_0=0$. A
finite conductor, for instance, merely polarizes: charge accumulates
at its surfaces until the current stops. Correspondingly, the integral
in Eq.~\eqref{eq:Green_Kubo_noop} taken over {all} times is zero
for any finite system.

Genuine transport arises from a separation of time scales. Let
$\tau_c$ be the microscopic relaxation time of the flux correlations,
and $\tau_B$ the time over which the system's boundaries make
themselves felt. For $\tau_c\ll\tau\ll\tau_B$, the running integral
$\int_0^\tau ds\,\langle J_i(s)J_j(0)\rangle_0$ reaches a plateau. The
transport coefficient is this plateau value. Since $\tau_B$ grows
without bound with system size, the plateau extends to arbitrarily
long times in the thermodynamic limit.
Eq.~\eqref{eq:Green_Kubo_noop} is therefore to be understood as
\begin{align}
L_{ij}
&=
\lim_{\tau\to\infty}\,
\lim_{\substack{N,V\to\infty\\ N/V\ \mathrm{fixed}}}\left[
\frac{1}{k_{\mathrm B}}
\int_0^{\tau}ds\,
\langle J_i(s)\,J_j(0)\rangle_0 \right],
\label{eq:Green_Kubo_limits}
\end{align}
with the thermodynamic limit taken {before} the long-time limit.
Taken in the opposite order, the limits give $L_{ij}=0$.

Let us, as illustration, write down the Green--Kubo formula for the
diffusion coefficient. Before doing so, it is worth making explicit
the two levels of description that the formula connects. The variables
$X_j$ in Eq.~\eqref{eq:Hamiltonian}, and hence the fluxes
$J_j=\dot X_j=\{X_j,H_0\}$, are phase-space functions: microscopic
dynamical variables that fluctuate along each trajectory. Only their
ensemble averages, $\langle X_j\rangle$ and $\langle J_j\rangle$, are
thermodynamic quantities. The Onsager coefficient $L_{ij}$ belongs to
the macroscopic level, since it relates averaged fluxes to affinities.
Yet Eq.~\eqref{eq:Green_Kubo_noop} expresses it entirely through
correlations of the microscopic fluxes in the unperturbed equilibrium
state. This is the essential content of the Green--Kubo relation: a
macroscopic coefficient is fixed by microscopic fluctuations.

A related point concerns the choice of $X$. The derivation never
required $X$ to be extensive; any phase-space function will do. For
self-diffusion, the natural choice is the position $x_\alpha$ of a
single tagged particle, a microscopic variable with no thermodynamic
counterpart. Alternatively, one may consider $N_t$ tagged particles
and take the genuinely extensive variable
$X_\alpha=\sum_{n}x^{(n)}_\alpha$ [cf.\ Eq.~\eqref{eq:stot2}], with
flux $J_\alpha=\sum_{n}v^{(n)}_\alpha$. Its autocorrelation then
contains, besides the self terms
$\langle v^{(n)}_\alpha(\tau)\,v^{(n)}_\alpha(0)\rangle_0$, cross terms
between different particles. For dilute, mutually non-interacting
tagged particles the cross terms vanish, the collective coefficient is
simply $N_t$ times the single-particle one, and the two views agree.
For interacting particles the cross terms survive, and the extensive
variable yields the \emph{collective diffusion coefficient}, which in
general differs from the  \emph{self-diffusion coefficient} considered here.

We therefore take $X=x_\alpha$ for a single tagged particle, whose
conjugate flux is the $\alpha$-component of its velocity,
\begin{equation}
J_\alpha \equiv \dot{x}_\alpha = v_\alpha .
\label{eq:flux_diffusion}
\end{equation}
Since $\langle v_\alpha\rangle_0=0$ in equilibrium,
Eq.~\eqref{eq:Green_Kubo_noop} gives
\begin{eqnarray}
L_{\alpha\alpha}
=
\frac{1}{k_{\mathrm B}}
\int_0^\infty d\tau\,C_\alpha(\tau),\,\,\,\, C_\alpha(\tau)\equiv\langle v_\alpha(\tau)\,v_\alpha(0)\rangle_0;\,\qquad
\label{eq:Laa_vv}
\end{eqnarray}
$C_\alpha(\tau)$
is the velocity autocorrelation function. It is the diagonal component
$C_{\alpha\alpha}$ of the tensor
$C_{\alpha\beta}(\tau)=\langle v_\alpha(\tau)\,v_\beta(0)\rangle_0$;
likewise, the diffusion coefficient $D_\alpha$ introduced below stands
for $D_{\alpha\alpha}$, and no summation over repeated indices is
implied. 

To give $L_{\alpha\alpha}$ a physical meaning, apply a weak constant force $F_\alpha$ to the tagged
particle. The perturbed Hamiltonian $H=H_0-F_\alpha x_\alpha$ has the
form of Eq.~\eqref{eq:Hamiltonian} with $X=x_\alpha$ and $h=F_\alpha$,
so by Eq.~\eqref{eq:affinity_field} the corresponding affinity is
$\mathcal{A}_\alpha=F_\alpha/T$. The steady response is a drift
velocity $\langle v_\alpha\rangle=L_{\alpha\alpha}\mathcal{A}_\alpha
=L_{\alpha\alpha}F_\alpha/T$. The \emph{mobility}, i.e., the drift
velocity per unit applied force, is therefore
\begin{equation}
\mu_{\rm mob}=\frac{\langle v_\alpha\rangle}{F_\alpha}
=
\frac{L_{\alpha\alpha}}{T}
=
\frac{1}{k_{\mathrm B}T}\int_0^\infty d\tau\,C_\alpha(\tau).
\label{eq:mobility}
\end{equation}
Mobility and diffusion are linked by the \emph{Einstein relation},
$D_\alpha=\mu_{\rm mob}\,k_{\mathrm B}T$. Taking it for granted for the
moment, Eq.~\eqref{eq:mobility} immediately yields the Green--Kubo
formula for the diffusion coefficient,
\begin{equation}
D_\alpha
=
\int_0^\infty d\tau\,
\langle v_\alpha(\tau)\,v_\alpha(0)\rangle_0 .
\label{eq:GK_diffusion}
\end{equation}
Einstein originally obtained the relation by requiring that, in a
static external potential, the drift current produced by the force
balance the diffusive current, so that the Boltzmann distribution is
stationary. That argument, however, presupposes that Fick's law holds
unchanged in the presence of the force. We now show instead that
Eq.~\eqref{eq:GK_diffusion} follows directly from the equilibrium
dynamics, with no applied force at all. Combined with
Eq.~\eqref{eq:mobility}, this derives the Einstein relation.

The diffusion coefficient is defined without reference
to any applied force. The fluid remains in equilibrium throughout; we
simply follow the displacement
$\Delta x_\alpha(t)\equiv x_\alpha(t)-x_\alpha(0)$ of the tagged
particle. Owing to translational invariance, we may place the origin
at the particle's initial position, $x_\alpha(0)=0$, so that its
position and its displacement coincide,
$x_\alpha(t)=\Delta x_\alpha(t)$. Because the initial microstate is
drawn from the equilibrium ensemble $\rho_0$, this displacement is a
random variable. Let $c(x_\alpha,t)$ denote its probability density,
so that $c\,dx_\alpha$ is the probability that the particle lies in
$[x_\alpha,x_\alpha+dx_\alpha]$ at time $t$. Operationally, $c$ is the
normalized histogram of displacements obtained by repeating the
observation many times from equilibrium initial conditions. Since the
particle starts at the origin, $c(x_\alpha,0)=\delta(x_\alpha)$.

For a single particle, $c$ is thus a probability density rather than a
concentration. It can nevertheless be read as one. Consider a dilute
cloud of identical, mutually independent tagged particles released at
the origin. Their normalized concentration at time $t$ coincides with
$c$, since each particle independently samples $c$. This equivalence is
what allows macroscopic transport laws, originally formulated for
concentrations, to be applied to $c$.

Being a distribution over the equilibrium ensemble, $c$ has moments
that are equilibrium averages. In particular, $\int dx_\alpha\,c=1$,
and the mean-square displacement is
$\langle\Delta x_\alpha^2(t)\rangle_0=\int dx_\alpha\,x_\alpha^2\,c$.
Since there is no preferred direction,
$\langle\Delta x_\alpha\rangle_0=0$, so the mean-square displacement is
the variance of $c$ and measures its spreading. On coarse-grained time
scales, $c$ obeys \emph{Fick's law},
$\partial_t c=D_\alpha\,\partial_\alpha^2 c$, where
$\partial_\alpha\equiv\partial/\partial x_\alpha$. This is an instance
of Onsager's regression hypothesis: spontaneous equilibrium
fluctuations decay according to the macroscopic transport law.
Integrating by parts twice, and using the normalization of $c$, we get
\begin{equation}
\frac{d}{dt}\langle\Delta x_\alpha^2\rangle_0
=\int dx_\alpha\,x_\alpha^2\,\partial_tc=
D_\alpha\int dx_\alpha\,x_\alpha^2\,\partial_\alpha^2 c
=
2D_\alpha .
\label{eq:fick_moment}
\end{equation}
Since $\langle\Delta x_\alpha^2(0)\rangle_0=0$, the mean-square
displacement grows linearly:
\begin{equation}
\langle\Delta x_\alpha^2(t)\rangle_0\simeq 2D_\alpha t .
\label{eq:msd_fick}
\end{equation}
The approximate sign reflects that Fick's law holds only on time
scales long compared with the velocity-correlation time. This
long-time growth defines $D_\alpha$.

To connect Eq.~\eqref{eq:msd_fick} with the microscopic dynamics, we
express the displacement through the velocity within the same
equilibrium average. Writing
$\Delta x_\alpha(t)=\int_0^t dt_1\,v_\alpha(t_1)$, squaring, and
averaging, the integrand becomes
$\langle v_\alpha(t_1)v_\alpha(t_2)\rangle_0$. By stationarity this
depends only on $t_1-t_2$ and equals $C_\alpha(t_1-t_2)$, with
$C_\alpha(s)=\langle v_\alpha(s)\,v_\alpha(0)\rangle_0$. Moreover,
$C_\alpha$ is even: shifting both times by $s$ and using that
classical observables commute,
$C_\alpha(-s)=\langle v_\alpha(-s)\,v_\alpha(0)\rangle_0
=\langle v_\alpha(0)\,v_\alpha(s)\rangle_0=C_\alpha(s)$,
so that $C_\alpha(t_1-t_2)=C_\alpha(|t_1-t_2|)$.
The two triangles $t_1>t_2$ and $t_1<t_2$ of the integration square
therefore contribute equally. Changing variables to $s=t_1-t_2\ge 0$
on the first triangle and then exchanging the order of integration,
we find
\begin{align}
\langle\Delta x_\alpha^2(t)\rangle_0
&=
\int_0^t\!dt_1\!\int_0^t\!dt_2\,C_\alpha(|t_1-t_2|)
\nonumber\\
&=
2\int_0^t\!dt_1\!\int_0^{t_1}\!ds\,C_\alpha(s)
\nonumber\\
&=
2\int_0^t\!dt_1\!\int_0^{t}\!ds\,\Theta(t_1-s)\,C_\alpha(s)
\nonumber\\
&=
2\int_0^t\!ds\,C_\alpha(s)\int_s^t\!dt_1
\nonumber\\
&=
2t\int_0^t ds\,C_\alpha(s)
-2\int_0^t ds\,s\,C_\alpha(s).
\label{eq:msd}
\end{align}
We now divide by $2t$ and let $t\to\infty$. The first term tends to
$\int_0^\infty C_\alpha\,ds$. The second term, divided by $t$,
vanishes provided this integral converges, i.e., provided $C_\alpha$
decays fast enough at long times (see the caveats below). Comparing
with Eq.~\eqref{eq:msd_fick}, we arrive at
\begin{equation}
D_\alpha
=
\lim_{t\to\infty}\frac{\langle\Delta x_\alpha^2(t)\rangle_0}{2t}
=
\int_0^\infty d\tau\,
\langle v_\alpha(\tau)\,v_\alpha(0)\rangle_0 ,
\nonumber
\end{equation}
which is Eq.~\eqref{eq:GK_diffusion}, now obtained from the
spontaneous wandering of the particle alone. Comparing with
Eq.~\eqref{eq:mobility} then establishes the Einstein relation,
\begin{equation}
D_\alpha=\mu_{\rm mob}\,k_{\mathrm B}T .
\label{eq:Einstein}
\end{equation}
Both sides of this result are equilibrium quantities, and derived this
way the relation holds for the full Hamiltonian dynamics,
independently of any model of the surrounding medium. Mobility
measures the response to a systematic push, and diffusion measures
spontaneous wandering. Both are governed by how long the particle
remembers its velocity, with $k_{\mathrm B}T$ converting between them.

For an isotropic fluid the three diagonal components coincide, and
\begin{equation}
D
=
\frac{1}{3}
\int_0^\infty d\tau\,
\langle \mathbf{v}(\tau)\cdot\mathbf{v}(0)\rangle_0 .
\label{eq:GK_diffusion_isotropic}
\end{equation}
The total velocity autocorrelation
$C(\tau)\equiv\langle\mathbf{v}(\tau)\cdot\mathbf{v}(0)\rangle_0$
measures how long the particle retains memory of its initial direction
of motion. It starts from $C(0)=\langle v^2\rangle_0=3k_{\mathrm B}T/m$
by equipartition, and collisions destroy most of this memory within a
time of order $\tau_c$. Since the area under $C(\tau)$ is $3D$, long
velocity memory means fast diffusion, and rapid decorrelation means
slow diffusion. The connection to the fluctuation--dissipation theorem
is thus manifest: $D$ is fixed entirely by an equilibrium fluctuation
of the unperturbed system, with no reference to any external
perturbation.

Two caveats concern the long-time behaviour of $C(\tau)$. First, as
discussed after Eq.~\eqref{eq:Green_Kubo_limits}, the integral must be
evaluated in the thermodynamic limit. In a finite box the particle's
displacement is bounded, and the area under $C(\tau)$ taken over all
times vanishes. Second, even in an infinite fluid, the residual memory
beyond $\tau_c$ decays only algebraically, as $\tau^{-d/2}$ in $d$
dimensions. This is the \emph{Alder--Wainwright long-time tail}, caused
by momentum returned to the particle by the flow it creates. The
integral therefore converges in three dimensions but diverges
logarithmically in two, where the mean-square displacement grows
faster than linearly and no finite self-diffusion coefficient exists.
In practice this divergence is cut off by the system size, by the
observation time, or by coupling to a surrounding medium, yielding an
effective, scale-dependent $D$.
\begin{center}
	$\symking\symqueen\symrook\symknight\symbishop\sympawn\epsdice{1}\,\epsdice{2}\,\epsdice{3}\,\epsdice{4}\,\epsdice{5}\,\epsdice{6}\,\spadesuit\heartsuit\clubsuit\diamondsuit$
\end{center}
\newpage
\newpage
\clearpage
\onecolumngrid
\section{Backward Equations: Future's Expectations}
\epigraph{\textsf{Not all those who wander are lost.}}{J. R. R. Tolkien}
\twocolumngrid

\indent {\color{deepsea}{\LARGE \textgoth{I}}}n the last few pages, we dealt with systems
at or close to equilibrium. Detailed balance characterised the stationary
state, and linear response described small departures from it, in which
fluxes are proportional to the forces that drive them. Beyond this regime
lies a wide range of questions, from nonequilibrium steady states sustained
by external driving to the relaxation of systems far from equilibrium. 

One important class among them concerns rare and strongly nonlinear events: a
particle escaping from a deep potential well, a chemical reaction, a switch
between metastable states, or the extinction of a population. Such events are driven by large fluctuations rather than small ones, and
their timing cannot be read off from any stationary distribution, whether
that is the equilibrium one or the trivial state in which all probability
has been absorbed. For these problems the relevant question is not how the
probability distribution evolves, but when, and where, a trajectory first
reaches a given boundary. 

The natural tool for this class of problems is the
\emph{backward Fokker--Planck equation}, more commonly called the
\emph{backward Kolmogorov equation}. Whereas the forward equation evolves
the distribution from a fixed initial condition, the backward equation
describes how the future behaviour of a stochastic system depends on its
initial state. This makes it the direct route to first-passage quantities
such as absorption probabilities, mean first-passage times and escape
rates, with applications ranging from activated escape and chemical
reactions to neuronal firing, transport through channels and population
extinction. 

\subsection{Continuous-Time Backward Equations}

The forward Fokker--Planck equation answers the question:
Given the starting point $x_0$, how does the probability distribution
of the process evolve at later times? It describes the motion and spreading of an ensemble of particles. If
many particles are released from the same initial position, then
$p(x,t\mid x_0,t_0)$ gives their spatial density at time $t$. An equivalent complementary question is: How does the probability of reaching a specified final state depend on where and when the process was started? This is conveniently answered by what is known as the \emph{backward master equation}---and its approximation, \emph{backward Fokker--Planck Equation}---which is what we will discuss now.

Let us first derive the intended equation: The derivation follows closely the derivation we did for the forward Fokker--Planck equation. To this end, we recall the Chapman--Kolmogorov equation in the following form:
\begin{align}
p(x,t\mid x_0,t_0)
&=
\int dx'\,
p(x,t\mid x',t')
p(x',t'\mid x_0,t_0).
\label{eq:CK}
\end{align}
We choose $t'=t_0+\Delta t_0$ and, thus, for a sufficiently short time interval, the transition probability
can be written as
\begin{align}
p(x',t_0+\Delta t_0\mid x_0,t_0)
&=
\delta(x'-x_0)
\left[
1-a^{(0)}(x_0,t_0)\Delta t_0
\right]
\nonumber\\
+&R_{t_0}(x'\mid x_0)\Delta t_0
+\mathcal{O}(\Delta t_0^2);\quad
\label{eq:short-time-transition}
\end{align}
henceforth we drop  the terms of order $\Delta t_0^2$ and higher. Probability conservation requires
\begin{equation}
a^{(0)}(x_0,t_0)
=
\int dx'\,R_{t_0}(x'\mid x_0).
\label{eq:a0}
\end{equation}
Substituting Eq.~\eqref{eq:short-time-transition} into
Eq.~\eqref{eq:CK} gives
\begin{align}
p(x,t\mid x_0,t_0)
&=
p(x,t\mid x_0,t_0+\Delta t_0)
\left[
1-a^{(0)}(x_0,t_0)\Delta t_0
\right]
\nonumber\\
&
+\Delta t_0\int dx'\,
p(x,t\mid x',t_0+\Delta t_0)
R_{t_0}(x'\mid x_0).
\end{align}
Therefore,
\begin{align}
&p(x,t\mid x_0,t_0)
-
p(x,t\mid x_0,t_0+\Delta t_0)
\nonumber\\
&=
-\Delta t_0\,a^{(0)}(x_0,t_0)
p(x,t\mid x_0,t_0+\Delta t_0)
\nonumber\\
&\quad
+\Delta t_0\int dx'\,
p(x,t\mid x',t_0+\Delta t_0)
R_{t_0}(x'\mid x_0).\label{eq:daalde}
\end{align}
Using Eq.~\eqref{eq:a0}, the first term on the R.H.S. can
be written as
\begin{align}
&-\Delta t_0\,a^{(0)}(x_0,t_0)
p(x,t\mid x_0,t_0+\Delta t_0)
\nonumber\\
&=
-\Delta t_0
p(x,t\mid x_0,t_0+\Delta t_0)
\int dx'\,R_{t_0}(x'\mid x_0).
\end{align}
Hence, putting this in Eq.~(\ref{eq:daalde}) and dividing the equation by $\Delta t_0$ on which the limit
$\Delta t_0\rightarrow 0$ is imposed, yields:
\begin{align}
-\frac{\partial}{\partial t_0}
p(x,t\mid x_0,t_0)
&=
\int dx'\,R_{t_0}(x'\mid x_0)
p(x,t\mid x',t_0)
\nonumber\\
&\quad
-
\int dx'\,R_{t_0}(x'\mid x_0)
p(x,t\mid x_0,t_0).
\label{eq:BME}
\end{align}
Eq.~\eqref{eq:BME} is the \emph{backward master equation}. The backward equation does not describe a particle literally moving backward in time. Instead, it describes how a prediction about the future changes when the starting point or starting time is changed.

Next to get to the backward version of the Fokker--Planck equation, we introduce
\begin{equation}
R_{t_0}(x'\mid x_0)
=
R_{t_0}(x_0,\Delta),
\qquad
\Delta\equiv x'-x_0.
\end{equation}
Since
\(
x'=x_0+\Delta\) and \(dx'=d\Delta\), the backward master equation becomes
\begin{align}
-\frac{\partial}{\partial t_0}
p(x,t\mid x_0,t_0)
&=
\int d\Delta\,
R_{t_0}(x_0,\Delta)
p(x,t\mid x_0+\Delta,t_0)
\nonumber\\
&\quad
-
\int d\Delta\,
R_{t_0}(x_0,\Delta)
p(x,t\mid x_0,t_0).
\label{eq:BME-Delta}
\end{align}
We expand the transition probability in powers of $\Delta$:
\begin{align}
p(x,t\mid x_0+\Delta,t_0)
&=
\sum_{k=0}^{\infty}
\frac{\Delta^k}{k!}
\frac{\partial^k}{\partial x_0^k}
p(x,t\mid x_0,t_0).
\label{eq:Taylor-backward}
\end{align}
The $k=0$ term cancels the second integral in
Eq.~\eqref{eq:BME-Delta}. Consequently, the Kramers--Moyal expansion comes out as
\begin{align}
-\frac{\partial p}{\partial t_0}
&=
\sum_{k=1}^{\infty}
\frac{1}{k!}
\left[
\int d\Delta\,
\Delta^k R_{t_0}(x_0,\Delta)
\right]
\frac{\partial^k}{\partial x_0^k}
p(x,t\mid x_0,t_0).
\label{eq:KM-backward}
\end{align}
Finally, by defining the first two jump moments by
\begin{align}
a^{(1)}(x_0,t_0)
&\equiv
\int d\Delta\,
\Delta R_{t_0}(x_0,\Delta),
\\
a^{(2)}(x_0,t_0)
&\equiv
\int d\Delta\,
\Delta^2 R_{t_0}(x_0,\Delta),
\label{eq:jump-moments}
\end{align}
 and truncating the Kramers--Moyal expansion after the second term gives
\begin{align}
-\frac{\partial}{\partial t_0}
p(x,t\mid x_0,t_0)
&=
a^{(1)}(x_0,t_0)
\frac{\partial}{\partial x_0}
p(x,t\mid x_0,t_0)
\nonumber\\
&\quad
+\frac{1}{2}a^{(2)}(x_0,t_0)
\frac{\partial^2}{\partial x_0^2}
p(x,t\mid x_0,t_0)
\label{eq:BFPE}
\end{align}
as the well-known \emph{backward Fokker--Planck Equation}.

In the special case when the process is \emph{time-homogeneous}---that is,
when its transition density is invariant under uniform time translations---%
$p$ depends on $(t,t_0)$ only through the elapsed time $\tau \equiv t - t_0$:
\begin{equation}
p(x,t\mid x_0,t_0) = \tilde{p}(x,\tau\mid x_0),
\label{eq:reduction}
\end{equation}
and the jump moments lose their explicit time dependence, i.e.,
$a^{(n)}(x_0,t_0) \to a^{(n)}(x_0)$.
Since $\partial\tau/\partial t_0 = -1$ at fixed $t$, the chain rule
converts Eq.~\eqref{eq:BFPE} into a forward evolution equation for
$\tilde p(x,\tau\mid x_0)$ in the elapsed time $\tau$. Subsequently,
since $\partial_\tau = \partial_t$ at fixed $t_0$, this equation
can be recast in the original two-time notation as
\begin{align}
\frac{\partial p(x,t\mid x_0,t_0)}{\partial t}
=& \; a^{(1)}(x_0)\,\frac{\partial p(x,t\mid x_0,t_0)}{\partial x_0}\nonumber\\
&+ \tfrac{1}{2}\,a^{(2)}(x_0)\,\frac{\partial^2 p(x,t\mid x_0,t_0)}{\partial x_0^2}.\label{eq:thbfpe}
\end{align}
This is the backward Fokker--Planck equation for a time-homogeneous
Markov process.

To physically understand the contrast between the forward and the backward equations, consider the example of Brownian motion with constant drift:
\begin{equation}
dX(t)
=
v\,dt+\sqrt{2D}\,dW(t),
\label{eq:drift-diffusion-SDE}
\end{equation}
with $X(t_0)=x_0$. Here $v$ is the drift velocity and $D$ is the diffusion coefficient. Recalling the correspondence with the It\^o SDE:
the forward Fokker--Planck equation becomes
\begin{equation}
\frac{\partial p}{\partial t}
=
-v\frac{\partial p}{\partial x}
+
D\frac{\partial^2p}{\partial x^2},
\label{eq:forward-example}
\end{equation}
with the initial condition:
\begin{equation} \lim_{t \to t_0^+} p(x,t \mid x_0,t_0) = \delta(x - x_0).\label{eq:initial-condition} \end{equation}
This equation describes a probability packet whose mean position is
\begin{equation}
\langle X_t\rangle
=
x_0+v(t-t_0),
\label{eq:mean-example}
\end{equation}
and whose variance is
\begin{equation}
\operatorname{Var}(X_t)
=
2D(t-t_0).
\label{eq:variance-example}
\end{equation}
The backward Fokker--Planck equation is:
\begin{equation}
-\frac{\partial p}{\partial t_0}
=
v\frac{\partial p}{\partial x_0}
+
D\frac{\partial^2p}{\partial x_0^2}.
\label{eq:backward-example}
\end{equation}
Both the forward and backward equations produce the same transition density:
\begin{align}
p(x,t\mid x_0,t_0)
&=
\frac{1}{\sqrt{4\pi D(t-t_0)}}
\nonumber\\
&\quad\times
\exp\left[
-\frac{
\left[x-x_0-v(t-t_0)\right]^2
}{
4D(t-t_0)
}
\right].
\label{eq:Gaussian-transition-density}
\end{align}
While the forward equation describes how this Gaussian packet evolves as a
function of its final position $x$ and final time $t$, the backward
equation describes how the same Gaussian changes when the initial
position $x_0$ or initial time $t_0$ is changed.

We note that because the Dirac delta function is symmetric, $\delta(x - x_0) = \delta(x_0 - x)$, this single mathematical boundary condition, Eq.~\eqref{eq:initial-condition}, serves both equations, but it is interpreted differently depending on perspective:
\begin{itemize}
    \item {For the forward equation:} It acts as a true {initial condition} in final coordinates. We fix the source point $x_0$ at time $t_0$, giving us a sharp spike at $x = x_0$. As $t$ increases ($t > t_0$), this spike immediately broadens into the drifting Gaussian packet, Eq.~\eqref{eq:Gaussian-transition-density}.
    \item {For the backward equation:} It acts as a \emph{terminal condition} in initial coordinates. We fix the destination point $x$ at time $t$. We then run the equation {backwards} in time ($t_0 < t$). As $t_0$ rolls backward away from $t$, we see which initial positions $x_0$ could have successfully propagated to that final target.
\end{itemize}

It is, therefore, natural that the backward equation is particularly useful for calculating expected future observables. Let
\begin{equation}
u(x_0,t_0)
=
\mathbb{E}_{x_0,t_0}
\left[f(X(t))\right]
=
\int dx\,
f(x)\,p(x,t\mid x_0,t_0),
\label{eq:expected-observable}
\end{equation}
where $f$ is a function of the state at the later time $t$, and $\mathbb{E}_{x_0,t_0}$ denotes the average over all realizations of the noise with the starting point held fixed at $X(t_0)=x_0$. Operationally, one launches many walkers from the same point $x_0$ at time $t_0$ and averages $f$ over their positions at time $t$; the average is over the random endpoint $X(t)$, not over $x_0$. Since $f$ does not depend on $(x_0,t_0)$, applying the backward equation \eqref{eq:BFPE} under the integral gives
\begin{align}
-\frac{\partial u(x_0,t_0)}{\partial t_0}
=&
a^{(1)}(x_0,t_0)
\frac{\partial u(x_0,t_0)}{\partial x_0}\nonumber\\
&+\frac{1}{2}a^{(2)}(x_0,t_0)
\frac{\partial^2u(x_0,t_0)}{\partial x_0^2},\qquad\quad
\label{eq:backward-observable-equation}
\end{align}
to be solved backward from the {terminal} condition
\begin{equation}
u(x_0,t)=f(x_0),
\label{eq:terminal-condition}
\end{equation}
which holds because a walker starting at $t_0=t$ has no time to move. For a point detector, $f(x)=\delta(x-x_1)$, we recover $u=p(x_1,t\mid x_0,t_0)$ and its terminal condition $p(x_1,t\mid x_0,t)=\delta(x_1-x_0)$ discussed above. 

Thus, the two equations are not independent physical laws, but they serve different purposes: the forward equation tracks the evolution of probability densities, whereas the backward equation tracks the evolution of future expectations and event probabilities.
\subsection{Forward--Backward Duality}
In fact, we can formalize the aforementioned duality by mathematically showing that the two versions of the Fokker--Planck equations are adjoint descriptions of the same stochastic dynamics. Let us now proceed to elaborate on this. 

In compact notations: The forward Fokker--Planck equation is
\begin{equation}
\frac{\partial}{\partial t}
p(x,t\mid x_0,t_0)
=
{\mathcal{L}}(x,t)p(x,t\mid x_0,t_0),
\label{eq:FFPEL}
\end{equation}
whereas the backward Fokker--Planck equation is
\begin{equation}
-\frac{\partial}{\partial t_0}
p(x,t\mid x_0,t_0)
=
{\mathcal{L}}^\dagger(x_0,t_0)
p(x,t\mid x_0,t_0),
\label{eq:BFPEL}
\end{equation}
where, in the shortened notation---
\begin{align}
A\equiv a^{(1)}(x,t),\quad B\equiv a^{(2)}(x,t),
\end{align}
the forward (linear) operator is
\begin{align}
{\mathcal{L}}(x,t)
=
-\frac{\partial}{\partial x}
A
+
\frac{1}{2}
\frac{\partial^2}{\partial x^2}B,
\label{eq:L-forward}
\end{align}
and the backward (linear) operator is
\begin{equation}
{\mathcal{L}}^\dagger(x,t)
=
A\,\frac{\partial }{\partial x}
+
\frac{1}{2}B\,
\frac{\partial^2 }{\partial x^2}.
\label{eq:L-backward}
\end{equation}
Note, we anticipate that ${\mathcal{L}}^\dagger$ is the formal adjoint of ${\mathcal{L}}$---hence, the notation. We show this fact explicitly as follows.

Let $f$ and $g$ be smooth functions on the real line with compact support, i.e., there exists a finite $R_0>0$ such that $f(x)=g(x)=0$ $\forall |x|>R_0$; equivalently, both functions vanish outside the interval $[-R_0,R_0]$. We define
\begin{equation}
\langle f,g\rangle
\equiv
\int_{-\infty}^{\infty}dx\,f(x)g(x).
\end{equation}
Then
\begin{align}
\langle f,{\mathcal{L}}g\rangle
=
\int dx\,f(x)
\left\{
-\frac{\partial}{\partial x}
\left[A\,g(x)\right]
+
\frac{1}{2}
\frac{\partial^2}{\partial x^2}
\left[B\,g(x)\right]
\right\}.
\end{align}
Integrating the first term by parts gives
\begin{align}
-\int dx\,f\frac{\partial}{\partial x}(Ag)
=
\int dx\,A\frac{\partial f}{\partial x}g,
\end{align}
because the boundary term vanishes. For the diffusion term, similarly integrating by parts twice gives
\begin{align}
\frac{1}{2}\int dx\,f
\frac{\partial^2}{\partial x^2}(Bg)
=
\frac{1}{2}\int dx\,
\frac{\partial^2 f}{\partial x^2}Bg.
\end{align}
Therefore,
\begin{eqnarray}
\langle f,{\mathcal{L}}g\rangle
&=
\int dx\,
\left[
A\,\frac{\partial f}{\partial x}
+
\frac{1}{2}B\,
\frac{\partial^2f}{\partial x^2}
\right]g(x)=
\langle {\mathcal{L}}^\dagger f,g\rangle.\qquad
\label{eq:adjoint-property}
\end{eqnarray}
Therefore, ${\mathcal{L}}^\dagger$ is the formal \emph{adjoint} of ${\mathcal{L}}$. This proof can be extended in situations with weaker conditions on $f$ and $g$.

A useful deduction from this adjointness property is that the solutions of the backward and the forward equations are formally exactly same, as we have already witnessed in Eq.~\eqref{eq:Gaussian-transition-density} as an example. Let's see it now in general terms.

The forward Fokker--Planck equation, Eq.~\eqref{eq:FFPEL}, can be written in integral form as
\begin{align}
p(x,t\mid x_0,t_0)
&=
\delta(x-x_0)
\nonumber\\
&\quad
+\int_{t_0}^{t}dt_1\,
{\mathcal{L}}(x,t_1)p(x,t_1\mid x_0,t_0),
\label{eq:forward-integral}
\end{align}
where the initial condition has been fixed as
\begin{equation}
p(x,t_0\mid x_0,t_0)=\delta(x-x_0).\label{eq:ffpi}
\end{equation}
Iterating Eq.~\eqref{eq:forward-integral}---i.e., by replacing $p(x,t_1\mid x_0,t_0)$ with an expression similar to Eq.~\eqref{eq:forward-integral}---gives the \emph{Dyson series}:
\begin{align}
p(x,t\mid x_0,t_0)
&=
\biggl[
1
+\int_{t_0}^{t}dt_1\,{\mathcal{L}}(x,t_1)
\nonumber\\
&\quad
+\int_{t_0}^{t}dt_1
\int_{t_0}^{t_1}dt_2\,
{\mathcal{L}}(x,t_1){\mathcal{L}}(x,t_2)
+\cdots
\biggr]
\nonumber\\
&\quad\times
\delta(x-x_0).
\label{eq:dyson-series}
\end{align}

To express the nested integrals in exponential form, we introduce the
arrowed time-ordering operator, $\overleftarrow{\mathcal{T}}$, that orders operators with later times to
the left, i.e., 
\begin{align}
\overleftarrow{\mathcal{T}}\,
\bigl[{\mathcal{L}}(x,t_1){\mathcal{L}}(x,t_2)\bigr]
&=
\begin{cases}
{\mathcal{L}}(x,t_1){\mathcal{L}}(x,t_2), & t_1>t_2,\\[4pt]
{\mathcal{L}}(x,t_2){\mathcal{L}}(x,t_1), & t_2>t_1.
\end{cases}
\label{eq:time-ordering-two}
\end{align}
For the second-order term in Eq.~\eqref{eq:dyson-series}, one has
\begin{align}
&\int_{t_0}^{t}dt_1
\int_{t_0}^{t_1}dt_2\,
{\mathcal{L}}(x,t_1){\mathcal{L}}(x,t_2)
\nonumber\\
&\qquad=
\frac{1}{2!}
\int_{t_0}^{t}dt_1
\int_{t_0}^{t}dt_2\,
\overleftarrow{\mathcal{T}}\,
\bigl[
{\mathcal{L}}(x,t_1){\mathcal{L}}(x,t_2)
\bigr]\nonumber\\
&\qquad=
\overleftarrow{\mathcal{T}}
\biggl[
\frac{1}{2!}
\left(
\int_{t_0}^{t}dt'\,{\mathcal{L}}(x,t')
\right)^2\,\biggr].
\label{eq:second-order-ordered}
\end{align}
The factor $1/2!$ appears because the square integration domain
$[t_0,t]\times[t_0,t]$ is divided into the two regions
$t_1>t_2$ and $t_2>t_1$. The time-ordering operator gives the same
ordered product in both regions.

Extending this understanding to the more general order nested integrals present in the above Dyson series, Eq.~\eqref{eq:dyson-series}, yields the complete iterated solution cast as the arrowed time-ordered exponential:
\begin{align}
p(x,t\mid x_0,t_0)
&=
\overleftarrow{\mathcal{T}}
\exp\left[
\int_{t_0}^{t}dt'\,{\mathcal{L}}(x,t')
\right]
\delta(x-x_0)
\nonumber\\
&\equiv
\hat{O}(x)\delta(x-x_0).
\label{eq:O-forward}
\end{align}

For the backward Fokker--Planck equation, Eq.~\eqref{eq:BFPEL}, along with the 
with terminal condition same as Eq.~\eqref{eq:ffpi}, we similarly arrive at:
\begin{align}
p(x,t\mid x_0,t_0)
&=
\delta(x-x_0)
\nonumber\\
&\quad
+
\int_{t}^{t_0}dt_1\,
[-{\mathcal{L}}^\dagger(x_0,t_1)]
p(x,t\mid x_0,t_1).
\label{eq:backward-integral}
\end{align}
Now we do the manipulations similar to those undertaken in the case of forward Fokker--Planck equation: Iterating this equation, and absorbing the negative sign in flipping the integral limits, yields the Dyson series in the exponential form:
\begin{align}
p(x,t\mid x_0,t_0)
&=
\overrightarrow{\mathcal{T}}
\exp\left[
\int_{t_0}^{t}dt'\,
{\mathcal{L}}^\dagger(x_0,t')
\right]
\delta(x-x_0)
\nonumber\\
&\equiv
\hat{O}^\dagger(x_0)\delta(x-x_0),
\label{eq:O-backward}
\end{align}
where $\overrightarrow{\mathcal{T}}$ orders earlier times to the left. It is obvious that $\hat{O}^\dagger$ is adjoint of $\hat{O}$---and hence the notation---because of the property that $(UV)^\dagger=V^\dagger U^\dagger$, i.e., adjoint reverses the order of multiplication.

Finally, to show that Eqs.~\eqref{eq:O-forward} and \eqref{eq:O-backward} give the same transition probability, let
$f(x)$ be an arbitrary smooth test function. Then
\begin{align}
\int dx\,f(x)
\hat{O}(x)\delta(x-x_0)
&=
\int dx\,
\left[
\hat{O}^\dagger(x)f(x)
\right]\delta(x-x_0)
\nonumber\\
&=
\hat{O}^\dagger(x_0)f(x_0)\nonumber\\
&=
\int dx\,f(x)
\hat{O}^\dagger(x_0)\delta(x-x_0),\nonumber\
\label{eq:operator-adjoint}
\end{align}
using the defining property of the Dirac-delta function. Consequently,
\begin{eqnarray}
\int dx\,f(x)
\left[
\hat{O}(x)\delta(x-x_0)
-
\hat{O}^\dagger(x_0)\delta(x-x_0)
\right]=0,\qquad\,\,
\end{eqnarray}
for every test function $f$. Thus, it follows, in the sense of distributions, that
\begin{equation}
{
\hat{O}(x)\delta(x-x_0)
=
\hat{O}^\dagger(x_0)\delta(x-x_0),
}
\end{equation}
proving that the forward and backward equations produce the same
transition probability.
\subsection{Discrete-Time Backward Equation}

Here, we shall confine ourselves to the simple case of finite state space, $\mathcal X$. Consider a time-homogeneous discrete-time Markov chain
$\{X_n\}_{n\geq 0}$. Let its one-step transition probabilities be $
p_{ij}
=
P(X_{n+1}=j\mid X_n=i)$ and the state-probability row vector at time $n$ be $\boldsymbol{\pi}(n)
=
\bigl(\pi_i(n)\bigr)_{i\in\mathcal X}=\bigl( P(X_n=i)\bigr)_{i\in\mathcal X}$. The discrete-time master equation is
\begin{equation}
\boldsymbol{\pi}(n+1)
=
\boldsymbol{\pi}(n)\mathsf P,
\label{eq:forward_master_equation_duality}
\end{equation}
which, for unambiguity for the matter of this section, will be called the \emph{forward discrete-time master equation}.

In this section, an \emph{observable} is any state-dependent quantity represented by a function,
\(
f:\mathcal{X}\to\mathbb{R}.
\)
For each state \(i\in\mathcal{X}\), the value \(f(i)\) specifies the quantity of interest when the Markov chain occupies state \(i\). Since probability distributions are represented by row vectors, the observable is represented by the corresponding column vector,
\(
\boldsymbol{f}
=
\bigl(f(i)\bigr)_{i\in\mathcal{X}}^{\mathsf T},
\)
so that, if \(\boldsymbol{\pi}\) is a row probability vector, the expectation of \(f\) is
\begin{equation}
\mathbb{E}_{\boldsymbol{\pi}}[f(X)]
=
\sum_{i\in\mathcal{X}}\pi_i f(i).
\end{equation}
Thus, the term ``observable'' refers to the state-dependent quantity being measured, whereas its column-vector form is simply the representation compatible with the row-vector convention for probability distributions.

To follow how expected values of \(f\) evolve, we introduce the time-indexed column vector \(\boldsymbol{u}(n)\), with \(\boldsymbol{u}(0)=\boldsymbol{f}\), evolved by
\begin{equation}
\boldsymbol{u}(n+1)
=
\mathsf P\boldsymbol{u}(n),
\label{eq:backward_master_equation_duality}
\end{equation}
or, in component form,
\begin{equation}
u_i(n+1)
=
\sum_{j\in\mathcal X}
p_{ij}u_j(n).
\label{eq:backward_component_duality}
\end{equation}
Its solution, \(\boldsymbol{u}(n)=\mathsf P^n\boldsymbol{f}\), gives
\begin{equation}
u_i(n)
=
\sum_{j\in\mathcal{X}}P(X_n=j\mid X_0=i)\,f(j)
=
\mathbb{E}[f(X_n)\mid X_0=i],\label{eq:citethis1}
\end{equation}
i.e., the expected value of \(f\) after \(n\) steps, given that the chain starts in state \(i\).

The transformation is termed ``backward'' because of the algebraic duality---to be elaborated below---with the forward master equation: there, \(\mathsf P\) multiplies probability row vectors from the right, whereas here it multiplies the observable column vector from the left. The name also has a direct probabilistic meaning. The forward equation extends a trajectory by one step at its `end': the distribution at time $n$ is pushed one step into the future, $\pi_j(n+1)=\sum_i\pi_i(n)p_{ij}$. The backward equation instead extends it by one step at its `beginning':
\begin{align}
&u_i(n+1)
=
\mathbb E[f(X_{n+1})\mid X_0=i]
\nonumber\\
&=
\sum_{j\in\mathcal X}P(X_1=j\mid X_0=i)\,
\mathbb E[f(X_{n+1})\mid X_1=j,\,X_0=i]
\nonumber\\
&=
\sum_{j\in\mathcal X}p_{ij}\,
\mathbb E[f(X_{n+1})\mid X_1=j]
\nonumber\\
&=
\sum_{j\in\mathcal X}p_{ij}\,u_j(n),
\end{align}
where the second line follows from the law of total expectation, conditioning on the first step; the third from the Markov property; and the last from time-homogeneity, $\mathbb E[f(X_{n+1})\mid X_1=j]=\mathbb E[f(X_n)\mid X_0=j]$. Throughout, $f$ is evaluated at the same time $n+1$; from the second line onward, however, the first step has already been taken, so the chain starts at time $1$ in state $j$ with only $n$ steps remaining. Thus, the argument of $\boldsymbol{u}$ is not a clock time but the number of steps remaining until $f$ is evaluated. Calling the time at which $f$ is evaluated the \emph{observation time}, the added step moves the start one step into the past while the observation time is held fixed. To make this explicit, fix the observation time $N$ and define $v_i(m)=\mathbb E[f(X_N)\mid X_m=i]$ for $m\le N$, so that $u_i(n)=v_i(N-n)$. Then $\boldsymbol v(m)=\mathsf P\,\boldsymbol v(m+1)$, which is solved for $m=N-1,N-2,\dots,0$, i.e., literally backward in time, starting from the terminal condition $\boldsymbol v(N)=\boldsymbol f$; the latter holds because $v_i(N)=\mathbb E[f(X_N)\mid X_N=i]=f(i)$.

This terminal-condition viewpoint is the natural one for boundary-value and absorption problems, such as finite-time absorption or fixation probabilities, where $\boldsymbol f$ encodes the target event at the observation time.

Now, we shed light on the duality of the forward and the backward equations. The natural pairing between a probability row vector and an observable column vector is defined by
\begin{equation}
\left\langle
\boldsymbol{\pi},\boldsymbol{u}
\right\rangle
\equiv
\boldsymbol{\pi}\boldsymbol{u}
=
\sum_{i\in\mathcal X}\pi_i u_i.
\label{eq:natural_row_column_pairing}
\end{equation}
When $\boldsymbol{\pi}$ is a probability distribution, this pairing is the expectation of the observable $\boldsymbol{u}$ with respect to $\boldsymbol{\pi}$:
\begin{equation}
\left\langle
\boldsymbol{\pi},\boldsymbol{u}
\right\rangle
=
\mathbb E_{\boldsymbol{\pi}}[u(X)].
\end{equation}

The forward and the backward actions of $\mathsf P$ are dual with respect to the pairing in Eq.~\eqref{eq:natural_row_column_pairing}. Indeed, for every row vector $\boldsymbol{\pi}$ and every column vector $\boldsymbol{u}$,
\begin{align}
\left\langle
\boldsymbol{\pi}\mathsf P,\boldsymbol{u}
\right\rangle
=
(\boldsymbol{\pi}\mathsf P)\boldsymbol{u}
=
\boldsymbol{\pi}(\mathsf P\boldsymbol{u})
=
\left\langle
\boldsymbol{\pi},\mathsf P\boldsymbol{u}
\right\rangle.
\label{eq:forward_backward_operator_duality}
\end{align}

Using the forward equation \eqref{eq:forward_master_equation_duality} and the backward equation \eqref{eq:backward_master_equation_duality}, we obtain
\begin{align}
\left\langle
\boldsymbol{\pi}(n+1),\boldsymbol{u}(0)
\right\rangle
=
\left\langle
\boldsymbol{\pi}(n)\mathsf P,\boldsymbol{u}(0)
\right\rangle
&=
\left\langle
\boldsymbol{\pi}(n),\mathsf P\boldsymbol{u}(0)
\right\rangle
\nonumber\\
&=
\left\langle
\boldsymbol{\pi}(n),\boldsymbol{u}(1)
\right\rangle.
\label{eq:one_step_forward_backward_duality}
\end{align}

Eq.~\eqref{eq:one_step_forward_backward_duality} states that the expected value of an observable after one forward transition can be calculated in two equivalent ways. First, one may evolve the probability distribution from $\boldsymbol{\pi}(n)$ to $\boldsymbol{\pi}(n+1)$ and then evaluate the original observable $\boldsymbol{u}(0)$. Second, one may keep $\boldsymbol{\pi}(n)$ fixed, evolve the observable backward from $\boldsymbol{u}(0)$ to $\boldsymbol{u}(1)=\mathsf P\boldsymbol{u}(0)$, and evaluate it with respect to $\boldsymbol{\pi}(n)$. Iterating, $\langle\boldsymbol{\pi}(n),\boldsymbol{u}(0)\rangle=\langle\boldsymbol{\pi}(0),\boldsymbol{u}(n)\rangle$, i.e., $\mathbb E[f(X_n)]$ may be obtained either by evolving the distribution forward for $n$ steps or by evolving the observable backward for $n$ steps. Thus, forward evolution of the probability distribution and backward evolution of the observable produce the same expected value. They are not two different dynamical laws; rather, they are two dual representations of the same Markov transition. This is the discrete-time counterpart of the adjoint relation between the forward and the backward generators discussed earlier for continuous state spaces.

As an illustration of the backward master equation, we delve into the case of
absorption probability. Let $\mathcal B\subseteq\mathcal X$ be a target absorbing set.
Define the \emph{first hitting time} of $\mathcal B$ by
\begin{equation}
\tau_{\mathcal B}=\inf\{n\geq 0:X_n\in\mathcal B\},
\end{equation}
with the convention $\inf\emptyset=\infty$, so that $\tau_{\mathcal B}=\infty$ if the
chain never reaches $\mathcal B$ and $\tau_{\mathcal B}=0$ if it starts there. This
differs from the \emph{first passage time} of a state $j$,
$\inf\{n\ge1:X_n=j\}$, which excludes the starting instant; the two coincide when the
chain starts outside $j$. The \emph{return} (recurrence) \emph{time} is the special case
of the first passage time for a chain started in $X_0=i$ and returning to $i$.

The \emph{finite-time absorption probability}, conditional on the initial state
\(X_0=i\), is defined as the probability that the Markov chain has entered
the absorbing target set \(\mathcal B\) by time \(n\):
\begin{equation}
u_i(n)
=
P(\tau_{\mathcal B}\leq n\mid X_0=i).
\label{eq:finite_absorption_probability}
\end{equation}
Note that we are slightly abusing the notation as we are now using $u$ not as a general variable but as a very specific one.

To derive the backward equation, condition on the state occupied after one step:
\begin{align}
u_i(n+1)
&=
\sum_{j\in\mathcal X}
P(\tau_{\mathcal B}\leq n+1,X_1=j\mid X_0=i)
\nonumber\\
&=
\sum_{j\in\mathcal X}
P(\tau_{\mathcal B}\leq n+1
\mid X_1=j,X_0=i)\nonumber\\
&\quad\qquad\times
P(X_1=j\mid X_0=i),
\end{align}
where the second line follows from the chain rule of probability. For $i\notin\mathcal B$, we have $\tau_{\mathcal B}\geq1$, so the event $\{\tau_{\mathcal B}\leq n+1\}$ means that, starting from $X_1=j$, the chain enters $\mathcal B$ within the remaining $n$ steps. By the Markov property, the evolution after the first step is independent of the initial state $i$, and, by time-homogeneity, it depends only on the number of remaining steps. Hence, $P(\tau_{\mathcal B}\leq n+1\mid X_1=j,X_0=i)=u_j(n)$, and with $P(X_1=j\mid X_0=i)=p_{ij}$ we obtain
\begin{equation}
u_i(n+1)
=
\sum_{j\in\mathcal X}p_{ij}u_j(n),
\qquad i\notin\mathcal B,
\label{eq:backward_master_component}
\end{equation}
which is the time-dependent backward master equation. It is to be solved with $u_i(n)=1$ for $i\in\mathcal B$ and the initial condition $u_i(0)=\mathbf 1_{\mathcal B}(i)$, where
\[
\mathbf{1}_{\mathcal{B}}(i) = \begin{cases}
1, & i\in\mathcal B,\\
0, & i\notin\mathcal B,
\end{cases}
\]
is the indicator function of $\mathcal B$.

This example is a special case of the general framework with the observable $f=\mathbf{1}_{\mathcal B}$, so that $\boldsymbol{u}(0)=\boldsymbol{f}$ is precisely the above initial condition. Indeed, since $\mathcal B$ is absorbing, $\{X_n\in\mathcal B\}=\{\tau_{\mathcal B}\leq n\}$, and therefore
\begin{align}
\mathbb{E}\!\left[\mathbf{1}_{\mathcal B}(X_n)\mid X_0=i\right]
&=
P\!\left(X_n\in\mathcal B\mid X_0=i\right)
\nonumber\\
&=
P(\tau_{\mathcal B}\leq n\mid X_0=i)
\nonumber\\
&=
u_i(n),
\end{align}
which is in line with Eq.~\eqref{eq:citethis1}.

Finally, the \emph{ultimate absorption probability} is defined, whenever the limit exists, by
\begin{equation}
\phi_i
=
\lim_{n\to\infty}u_i(n).
\label{eq:ultimate_absorption_probability}
\end{equation}
Taking the limit $n\to\infty$ in Eq.~\eqref{eq:backward_master_component} gives
\begin{equation}
{
\phi_i
=
\sum_{j\in\mathcal X}p_{ij}\phi_j
}
\label{eq:backward_master_stationary}
\end{equation}
or, in vector notation,
\begin{equation}
{
\boldsymbol{\phi}
=
\mathsf P\boldsymbol{\phi}.
}
\label{eq:backward_vector_stationary}
\end{equation}
The vector $\boldsymbol{\phi}$ describes an ultimate state-dependent quantity, such as the probability of eventual absorption in a specified absorbing set. It should not be confused with a stationary state-probability distribution $\boldsymbol{\pi}^*$; $\boldsymbol{\phi}$ is a column vector describing an invariant observable.

In this context, another important quantity of interest is the \emph{mean first hitting time} from state $i$,
\begin{equation}
T_i
=
\mathbb E_i[\tau_{\mathcal B}]
\equiv
\mathbb E[\tau_{\mathcal B}\mid X_0=i],
\label{eq:mean_first_hitting_time}
\end{equation}
where $\mathbb E_i$ denotes the expectation conditional on the chain starting in state $i$. If $i\in\mathcal B$, then $\tau_{\mathcal B}=0$, so that
\begin{equation}
T_i=0,
\qquad i\in\mathcal B.
\label{eq:mfht_boundary}
\end{equation}
For $i\notin\mathcal B$, the first step takes one unit of time and moves the chain to some state $j$ with probability $p_{ij}$. By the Markov property and time-homogeneity, the expected additional time needed to reach $\mathcal B$ from $j$ is then $T_j$. Averaging over all possible $j$ gives
\begin{equation}
T_i
=
1+\sum_{j\in\mathcal X}p_{ij}T_j
=
1+\sum_{j\notin\mathcal B}p_{ij}T_j,
\qquad i\notin\mathcal B,
\label{eq:mfht}
\end{equation}
where the second equality uses Eq.~\eqref{eq:mfht_boundary}. Eq.~\eqref{eq:mfht} has the same structure as the backward master equation~\eqref{eq:backward_master_component}, except for the additional constant $1$, which counts the first step out of state $i$; all subsequent steps are accounted for in $T_j$. Labelling the non-target states as $1,\ldots,m$, let $\boldsymbol T=(T_1,\ldots,T_m)^{\mathsf T}$, let $\mathsf Q=(p_{ij})_{i,j\notin\mathcal B}$ be the restriction of $\mathsf P$ to the non-target states, and let $\boldsymbol 1$ be the column vector of $m$ ones. Then Eq.~\eqref{eq:mfht} reads $\boldsymbol T=\boldsymbol 1+\mathsf Q\boldsymbol T$, whose solution is
\begin{equation}
\boldsymbol{T}
=
(\mathsf{I}-\mathsf{Q})^{-1}\boldsymbol{1},
\label{eq:mfht_matrix}
\end{equation}
provided $\mathsf I-\mathsf Q$ is invertible. This holds when $\mathcal B$ is reached with probability one from every non-target state; otherwise, $T_i=\infty$ for those states from which $\mathcal B$ may never be reached.

\subsection{Moran Process: An Example}
As an application of the above formalism, we consider a central question
in population genetics: what is the probability that an allele introduced
into a finite population eventually reaches \emph{fixation}? Let $A$ and $B$
denote two alternative alleles at the same locus, and refer to individuals
carrying them as types $A$ and $B$, respectively. In a haploid population,
the number of type-$A$ individuals is therefore equal to the number of
copies of allele $A$. An allele is said to be \emph{fixed} when it is
carried by every individual in the population, so that its frequency is
one and the alternative allele has been eliminated; conversely, it is said
to be \emph{lost} when its frequency falls to zero. The outcome depends on
the \emph{fitness} of each type, which quantifies its reproductive
success: the expected number of offspring an individual of that type
contributes to the next generation, or, in continuous-time and
event-based models, its rate of reproduction. Fitter types are copied
more often, so differences in fitness bias the population towards one
allele, whereas in the absence of such differences, allele frequencies
change only through random sampling.

In the \emph{Moran process}, we consider a finite, haploid, well-mixed,
asexual population of fixed size $N$. Reproduction occurs through
discrete birth--death events with overlapping generations. At each
event, one individual is selected to reproduce with probability
proportional to its fitness, while another individual is selected
uniformly for removal. The offspring is an exact clonal copy of its
parent and immediately replaces the removed individual. We assume no
mutation, migration, recombination, or variation in population size.
Selection is introduced by assigning the two types constant fitnesses
$w_A$ and $w_B$, and the relative fitness of allele $A$ is defined as
\(
r=\frac{w_A}{w_B}.
\)
Since, as shown below, only this ratio enters the dynamics of the
allele's copy number, $r$ alone specifies the selective regime: $r=1$
corresponds to neutrality, $r>1$ to positive selection for allele $A$,
and $0<r<1$ to negative selection.

In this model, fixation of allele $A$ occurs when all $N$ individuals
carry allele $A$. If the process begins with $i$ copies of allele $A$, we
denote the probability of its fixation by $\phi_i$. To compute it, let
\[
X_n\in\{0,1,\ldots,N\}
\]
denote the number of individuals carrying allele $A$ after the $n$th
birth--death event. The states $0$ and $N$ are absorbing:
\[
X_n=0
\quad\Longleftrightarrow\quad
\text{allele $A$ is lost},
\]
and
\[
X_n=N
\quad\Longleftrightarrow\quad
\text{allele $A$ is fixed}.
\]

More generally, fixation probability is a special case of an ultimate
absorption probability. Writing $\tau_k\equiv\tau_{\{k\}}$ for the first
hitting time of a single state $k$, and taking the absorbing set to be
\(
\mathcal{B}=\{N\}
\)
in Eq.~\eqref{eq:ultimate_absorption_probability}, we obtain
\(
\phi_i=P(\tau_N<\infty\mid X_0=i).
\)
Since state $0$ is also absorbing, a trajectory that reaches $0$ can
never reach $N$; hence $\tau_N<\infty$ holds precisely when $N$ is
reached before $0$, and
\(
\phi_i=P(\tau_N<\tau_0\mid X_0=i),
\)
the probability that allele $A$ reaches fixation before being lost.

If $X_n=i$, the population contains $i$ individuals of type $A$ and
$N-i$ of type $B$. A birth--death event changes the copy number of
allele $A$ only if the reproducing and the removed individuals are of
different types, and the two selections are made independently. The
copy number increases by one when a type-$A$ individual reproduces and a
type-$B$ individual is removed, so that
\begin{equation}
	b_i
	=
	{P}(X_{n+1}=i+1\mid X_n=i)
	=
	\frac{i w_A}{i w_A+(N-i)w_B}
	\frac{N-i}{N}.
	\label{eq:moran_birth_probability}
\end{equation}
The first factor is the probability that the reproducing individual is
of type $A$: each individual is chosen with probability proportional to
its fitness, so the $i$ type-$A$ individuals together carry weight
$i w_A$ out of the total fitness $i w_A+(N-i)w_B$ of the population.
The second factor is the probability that the removed individual is of
type $B$, since removal is uniform over all $N$ individuals.

Conversely, the copy number decreases by one when a type-$B$ individual
reproduces and a type-$A$ individual is removed:
\begin{equation}
	d_i
	=
	{P}(X_{n+1}=i-1\mid X_n=i)
	=
	\frac{(N-i)w_B}{i w_A+(N-i)w_B}
	\frac{i}{N},
	\label{eq:moran_death_probability}
\end{equation}
where the first factor is now the fitness-weighted probability of
choosing a type-$B$ parent and the second the probability of removing
a type-$A$ individual.

Because removal is uniform over the whole population, it may also pick
an individual of the same type as the parent, possibly the parent
itself; the offspring then replaces an individual of its own type and
the copy number is unchanged. These same-type events, a type-$A$ parent
with a type-$A$ removal or a type-$B$ parent with a type-$B$ removal,
exhaust the remaining possibilities, so the probability that the copy
number stays at $i$ is
\begin{equation}
	1-b_i-d_i.
	\label{eq:moran_stay_probability}
\end{equation}

The common denominator, the total fitness, cancels in the ratio
\begin{equation}
	\frac{d_i}{b_i}
	=
	\frac{(N-i)w_B\, i}{i w_A\,(N-i)}
	=
	\frac{w_B}{w_A}
	=
	\frac{1}{r},
	\label{eq:moran_ratio}
\end{equation}
which is independent of $i$. Thus selection biases each step towards
the fitter type by the same factor at every copy number, while the
probability that a step occurs at all, $b_i+d_i$, carries the factor
$i(N-i)$ and therefore vanishes at both absorbing states. In the
neutral case $r=1$ the two probabilities coincide,
$b_i=d_i=i(N-i)/N^2$, and the copy number performs an unbiased random
walk whose steps are most frequent at intermediate $i$.

Let's find $\phi_i$---the probability that allele $A$ eventually fixes given that its
initial copy number is $i$. The boundary conditions are
\begin{equation}
\phi_0=0,
\qquad
\phi_N=1.
\label{eq:moran_fixation_boundary_conditions}
\end{equation}
The fixation probabilities obey the stationary backward master equation
\eqref{eq:backward_master_stationary}, $\boldsymbol\phi=\mathsf P\boldsymbol\phi$,
with the absorbing set $\mathcal B=\{N\}$. The Moran chain is a
birth--death chain: from state $i$ it can move only to $i\pm1$ or stay,
so its transition matrix is tridiagonal, with
$p_{i,i+1}=b_i$, $p_{i,i-1}=d_i$ and $p_{ii}=1-b_i-d_i$ for
$1\leq i\leq N-1$, and $p_{00}=p_{NN}=1$. The rows $i=0$ and $i=N$ of
Eq.~\eqref{eq:backward_vector_stationary} reduce to identities; the
boundary values \eqref{eq:moran_fixation_boundary_conditions} follow
instead from the initial condition $\boldsymbol u(0)=\mathbf 1_{\{N\}}$,
since an absorbing chain started at $0$ never reaches $N$, while one
started at $N$ is already there. For $1\leq i\leq N-1$ the sum over $j$
retains only three terms, which is the first-step analysis conditioned
on the outcome of the first birth--death event:
\begin{align}
\phi_i
&=
b_i\phi_{i+1}
+
d_i\phi_{i-1}
+
(1-b_i-d_i)\phi_i,
\end{align}
{which can be rearranged as}
\begin{align}
\Delta_{i+1}
=
\frac{d_i}{b_i}\Delta_i,
\label{eq:moran_ratio_difference_equation}
\end{align}
where we have defined the successive differences:
\(
\Delta_i\equiv\phi_i-\phi_{i-1},
\,
1\leq i\leq N.
\)
Consequently,
\begin{align}
\Delta_{i}
=
\Delta_1\prod_{j=1}^{i-1}\frac{d_j}{b_j}.
\end{align}
Because of this and because $\phi_0=0$, summing the successive differences gives
\begin{equation}
\phi_i
=
\sum_{k=1}^{i}\Delta_k
=
\Delta_1
\sum_{k=1}^{i}
\prod_{j=1}^{k-1}\frac{d_j}{b_j}
=
\Delta_1
\sum_{k=0}^{i-1}
\prod_{j=1}^{k}\frac{d_j}{b_j},
\end{equation}
where the last form follows on shifting $k\to k+1$, and the empty
product (for $k=1$ in the middle expression, $k=0$ in the last) is
defined to be $1$. Applying the
boundary condition $\phi_N=1$ yields
\begin{equation}
1
=
\Delta_1
\sum_{k=0}^{N-1}
\prod_{j=1}^{k}\frac{d_j}{b_j}.
\end{equation}
Therefore, the general fixation probability is
\begin{equation}
\phi_i
=
\frac{
\displaystyle
\sum_{k=0}^{i-1}
\prod_{j=1}^{k}\frac{d_j}{b_j}
}{
\displaystyle
\sum_{k=0}^{N-1}
\prod_{j=1}^{k}\frac{d_j}{b_j}
}.
\label{eq:moran_general_fixation_probability}
\end{equation}
For the Moran process, the ratio $d_j/b_j$ is constant and equal to
$1/r$; see Eq.~\eqref{eq:moran_ratio}. Hence,
\begin{equation}
\phi_i
=
\frac{
\displaystyle
\sum_{k=0}^{i-1}r^{-k}
}{
\displaystyle
\sum_{k=0}^{N-1}r^{-k}
}\implies
{
\phi_i=
\begin{cases}
\displaystyle
\frac{1-r^{-i}}{1-r^{-N}},
& r\neq 1,\\[2ex]
\displaystyle
\frac{i}{N},
& r=1.
\end{cases}
}
\label{eq:moran_fixation_probability_cases}
\end{equation}
The case $r=1$ corresponds to neutrality and can also be obtained by taking the
limit of the expression for $r\neq 1$ as $r\to 1$.

Thus, under neutrality, the fixation probability of an allele is equal
to its initial frequency. In particular, a single copy of a neutral
allele in a population of size $N$ has fixation probability: \(
\phi_1=\frac{1}{N}.
\)
The neutral case illustrates a central principle of \emph{Kimura's neutral
theory of molecular evolution}. When $r=1$, allele $A$ has no selective
advantage over allele $B$, yet its frequency can still change through
random genetic drift. In particular, a neutral allele initially present
in $i$ copies fixes with probability $i/N$, and a single newly arisen
neutral copy fixes with probability $1/N$. Thus, the absence of
selection does not imply the absence of evolutionary change. Rather,
random sampling in finite populations can cause alleles to be lost or
to become fixed, so that stochasticity alone can produce evolution at
the population level.

In the neutral evolution case, the fact that $\phi_i=i/N$ can be equivalently noted when the problem is appropriately and approximately cast in terms of the Fokker--Planck language. We let $N$ become very large, such that the states---now represented by $\frac{0}{N},\,\frac{1}{N}\,,\cdots\,,\frac{N}{N}$---appears to belong to continuous state space $[0,1]$ with $x\equiv i/N$. In this language, 
\begin{align}
\phi_i=p(x=1,t=\infty\mid x_0={i}/{N}, t_0=0).
\end{align}
 The fixation probability satisfies the time-independent backward Fokker--Planck equation that can be written in the form:
\begin{align}
 \left[A(x_0)\frac{\partial}{\partial x_0}+\frac{B(x_0)}{2}\frac{\partial^2}{\partial x_0^2}\right]p(1,\infty\mid x_0, t_0)=0.
 \end{align} The obvious absorbing boundary conditions are: $$p(1,\infty\mid x_0=0, t_0)=0\quad\&\quad p(1,\infty\mid x_0=1, t_0)=1.$$ Solving the second-order time-independent backward Fokker--Planck equation, thus, yields:
\begin{align}
p(1,\infty\mid x_0, t_0)=\frac{\int_0^{x_0}\exp\left[-\int_0^z\frac{2A(y)}{B(y)}dy\right]dz}{\int_0^1\exp\left[-\int_0^z\frac{2A(y)}{B(y)}dy\right]dz}.
\end{align} 
In the evolution process, the drift term, $A(x_0)$, corresponds to the directional processes like selection, and the diffusion term, $B(x_0)$, corresponds to the non-directional processes like neutral drift. So, when $r=1$, $A(x_0)=0$ and straightforwardly, $p(1,\infty\mid x_0, t_0)=x_0$, or in other notations, $\phi_i=i/N$---the exact known result.

We conclude by commenting on the mean first hitting time. The absorbing states are $0$ and $N$, so we define the absorption time
\begin{align}
\tau_{\{0,N\}}=\inf\{n\geq 0:X_n\in\{0,N\}\},
\end{align}
and its mean
\begin{align}
T_i=\mathbb E_i[\tau_{\{0,N\}}].
\end{align}
Since both $0$ and $N$ are absorbing, i.e.,
\begin{align}
{
T_0=0,\qquad T_N=0.\label{eq:mpbc}
}
\end{align}
For $1\leq i\leq N-1$, the first-hitting-time equation \eqref{eq:mfht} becomes
\begin{align}
&T_i
=
1+b_iT_{i+1}
+d_iT_{i-1}
+(1-b_i-d_i)T_i\nonumber\\
\implies &b_i(T_{i+1}-T_i)
+
d_i(T_{i-1}-T_i)
=
-1.\label{eq:mfhtmp}
\end{align}

For a neutral Moran process,
\(
b_i=d_i=\frac{i(N-i)}{N^2},
\)
and Eq.~\eqref{eq:mfhtmp} becomes
\begin{equation}
T_{i+1}-2T_i+T_{i-1}
=
-\frac{N^2}{i(N-i)},
\qquad
1\leq i\leq N-1,
\label{eq:neutral_mfht}
\end{equation}
a \emph{discrete Poisson equation} subject to the \emph{Dirichlet conditions}
\eqref{eq:mpbc}. It can be solved by Green's function method; we only give the final answer: For
$1\leq i\leq N-1$,
\begin{align}
{
T_i
=
N\left[
(N-i)\bigl(H_{N-1}-H_{N-i-1}\bigr)
+
i\bigl(H_{N-1}-H_{i}\bigr)
\right],
}
\end{align}
where
\(
H_m=\sum_{k=1}^{m}\frac1k
\)
is the $m$th harmonic number. In particular,
\begin{align}
{
T_1=N H_{N-1},
}
\end{align}
so that, for large $N$, owing to properties of harmonic number,  $T_1\simeq N\ln N$ birth--death events.

Here $T_i$ is the mean time to absorption at either $0$ or $N$, not the
mean time conditional on fixation at $N$. If the target is only state
$N$, the relevant hitting time is
\(
\tau_{\{N\}}=\inf\{n\geq 0:X_n=N\}.
\)
Because $0$ is also absorbing, a process starting from $1\leq i\leq N-1$
can reach $0$ before reaching $N$, and once it reaches $0$ it can never
reach $N$. Therefore,
\(
P(\tau_{\{N\}}=\infty\mid X_0=i)=1-\phi_i>0,
\)
and hence the unconditional mean hitting time is
\(
\mathbb{E}[\tau_{\{N\}}\mid X_0=i]=\infty
\)
for $1\leq i\leq N-1$.

\subsection{Absorption in Fokker--Planck Dynamics}
Let us now find the absorption probability and the mean first hitting time in the
time-homogeneous Fokker--Planck dynamics. We consider the system, a particle, to be
confined in a one dimensional interval $[{x_a},{x_b}]$. As can be guessed, specifying
boundary conditions for the backward Fokker--Planck equation, which is the primary tool
in our present endeavour, is a technical necessity. So first let's do that. The boundary
conditions derived below hold for general drift $A(x,t)$ and diffusion $B(x,t)$; the
assumption that these are time-independent will be used only afterwards, for the
absorption probability and the mean first hitting time.

With $p=p(x,t\mid x_0,t_0)$, let the forward Fokker--Planck equation---
\begin{equation}
\partial_t p=\mathcal{L} p=-\partial_x J,
\qquad
J=A(x,t)p-\frac12\partial_x\!\left[B(x,t)p\right],
\label{eq:fp}
\end{equation}
be defined on the domain \(D=[{x_a},{x_b}]\). The corresponding backward operator, acting on the initial coordinate
\(x_0\), is
\begin{equation}
\mathcal{L}^\dagger
=
A(x_0,t_0)\partial_{x_0}
+\frac12 B(x_0,t_0)\partial_{x_0}^2.
\label{eq:backward_operator}
\end{equation}

We call $p(x,t)$ a \emph{forward function} if it evolves under the forward equation and
$f(x,t)$ a \emph{backward function} if it evolves under the backward equation; for such a pair, Eq.~\eqref{eq:adjoint-property} implies that
\begin{eqnarray}
0&=&\langle  f,\mathcal{L}p \rangle-\langle  \mathcal{L}^{\dagger}f,p \rangle\nonumber\\
&=&\int_{x_a}^{x_b}
\left[f\,\mathcal{L}p
-p\,\mathcal{L}^{\dagger}f\right]dx\nonumber\\
&=&
\left[
-fJ-\frac12 Bp\,\partial_x f
\right]_{x_a}^{x_b},
\label{eq:boundary_identity}
\end{eqnarray}
where integration by parts has been used. Thus, adjointness requires the total boundary term to vanish. For standard local boundary conditions, this is enforced by making the boundary term vanish separately at each endpoint---\(x=x_a\) and \(x=x_b\). The reason is that the endpoint values are normally independent. One can choose functions $p$ and $f$ whose behaviour near $x_a$ is changed without changing their behaviour near $x_b$, and vice versa. If the adjoint identity must hold for every admissible pair $(p,f)$, cancellation between unrelated endpoints cannot be relied upon. Local conditions are therefore the natural way to guarantee adjointness. 

From Eq.~\eqref{eq:boundary_identity} it follows that if the forward (absorbing) condition is $p=0$, then---since the current \(J\) need not vanish---the adjoint condition is $f=0$. Similarly, if the forward (reflecting) condition is zero probability flux ($J=0$), the corresponding adjoint condition is $\partial_x f=0$ (assuming $B\ne0$ at the boundary). The conditions derived for $f$ apply directly to $p(x,t\mid x_0,t_0)$, because $p$, viewed as a function of its initial arguments $(x_0,t_0)$ at fixed $(x,t)$, satisfies the backward equation and is thus itself an admissible choice of $f$. Concluding, the backward boundary conditions are:
\begin{eqnarray}
p(x,t\mid x_a,t_0)=0
\end{eqnarray}
for absorbing boundary at $x_a$, and
\begin{eqnarray}
\left.\partial_{x_0}p(x,t\mid x_0,t_0)\right|_{x_0=x_a}=0
\end{eqnarray}
for reflecting boundary at $x_a$.

From now on we take the drift and diffusion to be time-independent, $A=A(x)$ and
$B=B(x)$. Now, for the calculation of absorption probability, we consider both $x_a$
and $x_b$ to be absorbing; then the probability of absorption is, of course, $1$.
However, the probability that the particle under drift-diffusion will exit from $x_b$
(or equivalently will get absorbed at $x_b$) can be found by an argument similar to
the one for the fixation probability in the Moran process. So, starting at
$x\in[{x_a},{x_b}]$, the ultimate absorption probability at $x_b$ is simply
\begin{align}
\phi_x^{(b)}=\frac{\int_{x_a}^{x}\exp\left[-\int_{x_a}^z\frac{2A(y)}{B(y)}dy\right]dz}{\int_{x_a}^{x_b}\exp\left[-\int_{x_a}^z\frac{2A(y)}{B(y)}dy\right]dz},
\end{align}
and the similar probability to exit at $x_a$ is $\phi_x^{(a)}=1-\phi_x^{(b)}$. If instead $x_a$ is reflecting and $x_b$ absorbing, the condition
$\partial_x\phi=0$ at $x_a$ forces $\phi_x^{(b)}=1$: absorption at $x_b$ is certain,
and the only nontrivial question is how long it takes.

Next, we calculate the average time for the particle to first leave the domain
$[{x_a},{x_b}]$, i.e., the mean first hitting time of the absorbing boundary set
$\mathcal{B}$: $\{x_a,x_b\}$ if both ends are absorbing, $\{x_a\}$ if $x_a$ is absorbing
and $x_b$ is reflecting, and $\{x_b\}$ if $x_b$ is absorbing and $x_a$ is reflecting.
This is also the mean first-passage time (or mean exit time) out of the domain, for
two reasons: First, the paths of the diffusion are continuous, so the particle cannot
jump over an absorbing end: to leave the domain it must reach the boundary, and the
first time it reaches $\mathcal{B}$ is the first time it passes out. Second, for a
starting point $x\in(x_a,x_b)$ the particle does not begin in $\mathcal{B}$, so the
zero-step case that separates hitting from first passage in the discrete-time
definitions cannot occur. We use the name first hitting time below.

Let $\tau_\mathcal{B}=\inf\{t\ge0:X(t)\in\mathcal{B}\}$ be the first time the particle,
started at $x$ at time $t_0=0$, reaches $\mathcal{B}$. Let $p$ be the transition density
of the process with absorption at the absorbing ends and reflection at the reflecting
ends, so that probability is lost only through absorption. For $x\in(x_a,x_b)$, the
probability that the particle has not yet reached $\mathcal{B}$ by time $t$ is then
\begin{align}
G(x,t)\equiv\int_{x_a}^{x_b}p(x',t\mid x,0)\,dx'=P(\tau_\mathcal{B}>t).\label{eq:Gxt}
\end{align}
The last equality follows from recalling the notation that $\tau_\mathcal{B}$ is the random variable when the domain is left starting at $x$. Since the process is homogeneous, Eq.~\eqref{eq:thbfpe} is applicable to our case; integrating that over $dx'$, we obtain:
\begin{align}
\partial_tG(x,t)=A(x)\partial_xG(x,t)+\frac{B(x)}{2}\partial_x^2G(x,t).\label{eq:Gpde}
\end{align}
Since $p(x',0\mid x,0)=\delta(x'-x)$, we have the initial condition---$G(x,0)=1$ for $x\in[{x_a},{x_b}]$ [recall Eq.~\eqref{eq:Gxt}]. Also, directly from the definitions, the boundary conditions are obvious: $G(x_i,t)=0$ if $x_i$ is absorbing and $\partial_xG(x,t)|_{x=x_i}=0$ if $x_i$ is reflecting; $i$ is either $a$ or $b$.

Now, the cdf of $\tau_\mathcal{B}$, 
\begin{align}
F_{\tau_\mathcal{B}}(t)=P(\tau_\mathcal{B}\le t)=1-G(x,t),
\end{align}
owing to Eq.~\eqref{eq:Gxt}. Therefore, the mean $\tau_\mathcal{B}$---which we denote as $T(x)$---is given by
\begin{align}
T(x)=\int_{0}^{\infty}tdF_{\tau_\mathcal{B}}(t)&=-\int_{0}^{\infty}t \partial_tG(x,t)dt\nonumber\\
&=\int_{0}^{\infty}G(x,t)dt,\label{eq:TxG}
\end{align}
where the last equality follows by integration by parts. The boundary term
$-tG(x,t)\big|_0^\infty$ vanishes at $t=0$ because of the factor $t$, and at $t\to\infty$
because $G(x,t)$ decays exponentially (finite interval, $B>0$, at least one absorbing
end), so $tG(x,t)\to0$. In particular, $G(x,\infty)=0$: the noise eventually carries the particle to an absorbing
end, so it cannot remain in the domain forever.

Next, integrating Eq.~\eqref{eq:Gpde} over time from $0$ to $\infty$, we arrive at:
\begin{align}
   &-1=G(x,\infty)-G(x,0)=A(x)\partial_xT(x)+\frac{B(x)}{2}\partial_x^2T(x)\qquad\nonumber\\
   &\implies\partial_x^2T(x)+\frac{2A(x)}{B(x)}\partial_xT(x)=-\frac{2}{B(x)}.
\label{eq:Txpde}
\end{align}
Again to solve this differential equation, we require boundary conditions in terms of $T(x)$. From Eq.~\eqref{eq:TxG}, we gather that at absorbing boundary, $T(x)$ vanishes due to the same condition that $G(x,t)$ has to obey; similarly, at reflecting boundary $\partial_xT(x)$ must vanish.  Solving we get, when the left end of $[{x_a},{x_b}]$ is reflecting and the right end is absorbing:
\begin{equation}
T(x)
=
2\int_x^{x_b} \frac{dy}{\psi(y)}
\int_{x_a}^y \frac{\psi(z)}{B(z)}\,dz;\label{eq:lrrA}
\end{equation}
when the left end of $[{x_a},{x_b}]$ is absorbing and the right end is reflecting:
\begin{equation}
T(x)
=
2\int_{x_a}^x \frac{dy}{\psi(y)}
\int_y^{x_b} \frac{\psi(z)}{B(z)}\,dz;
\end{equation}
and when both ends of $[{x_a},{x_b}]$ are absorbing:
\begin{align}
T(x)
=
\frac{2}{ \int_{x_a}^{x_b} \frac{dw}{\psi(w)}}
\left[
\left(\int_{x_a}^x \frac{dw}{\psi(w)}\right)
\left(\int_x^{x_b} \frac{dy}{\psi(y)}
      \int_{x_a}^{y} \frac{\psi(z)}{B(z)}\,dz\right)\right.\nonumber\\
-
\left.\left(\int_x^{x_b} \frac{dw}{\psi(w)}\right)
\left(\int_{x_a}^x \frac{dy}{\psi(y)}
      \int_{x_a}^{y} \frac{\psi(z)}{B(z)}\,dz\right)
\right].\phantom{x}
\end{align}
Here $\psi(z)\equiv \exp\left[\int_{x_a}^z\frac{2A(y)}{B(y)}dy\right]$ which is the integrating factor of the backward operator:
\(\mathcal{L}^\dagger f=\tfrac12B\,\psi^{-1}\partial_x\bigl(\psi\,\partial_xf\bigr)\).

The reflecting--absorbing result~\eqref{eq:lrrA} has an important
physical application: thermally activated escape from a metastable state.
This is a fundamental mechanism in statistical physics, chemical kinetics,
condensed matter physics and nucleation theory. At low temperature the
escape time is exponentially large compared with microscopic relaxation
times, as described by the \emph{Arrhenius law}. We now derive this law from
Eq.~\eqref{eq:lrrA}, for an overdamped Brownian particle in a
one-dimensional metastable potential; the same method is applied to
nucleation later, where the reaction coordinate is the size of a nucleus of
the new phase.

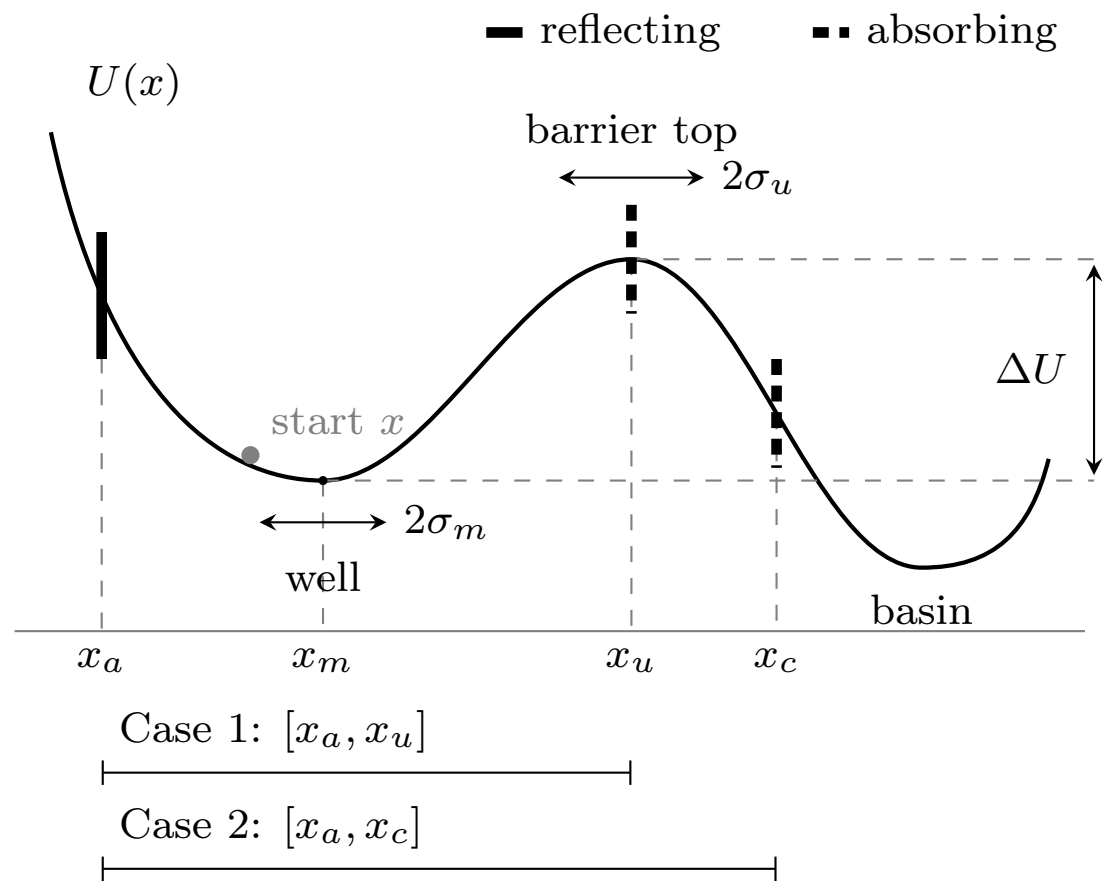
\begin{figure}[h]
\centering
\begin{tikzpicture}[x=1.2cm,y=1.2cm,>=stealth,font=\footnotesize]

  \draw[gray] (0.6,0) -- (6.5,0);

  \draw[thick]
    (0.8,2.75) .. controls (1.1,1.35) and (1.7,0.83) .. (2.3,0.83)
               .. controls (2.9,0.83)  and (3.4,2.05) .. (4.0,2.05)
               .. controls (4.6,2.05)  and (5.0,0.35) .. (5.6,0.35)
               .. controls (6.0,0.35)  and (6.2,0.55) .. (6.3,0.95);
  \node[anchor=south west] at (0.9,2.8) {$U(x)$};

  \foreach \x/\y in {1.08/1.86, 2.3/0.83, 4.0/2.05, 4.8/1.20}
    \draw[dashed,gray] (\x,\y) -- (\x,0);
  \draw[dashed,gray] (2.3,0.83) -- (6.55,0.83);
  \draw[dashed,gray] (4.0,2.05) -- (6.55,2.05);

  \draw[<->] (6.55,0.86) -- (6.55,2.02);
  \node[left] at (6.5,1.45) {$\Delta U$};

  \draw[line width=2pt] (1.08,2.2) -- (1.08,1.5);                               
  \draw[line width=2pt,dash pattern=on 3pt off 2pt] (4.0,2.35) -- (4.0,1.75);   
  \draw[line width=2pt,dash pattern=on 3pt off 2pt] (4.8,1.5)  -- (4.8,0.9);    

  \draw[<->] (3.6,2.5) -- (4.4,2.5);
  \node[right] at (4.4,2.5) {$2\sigma_u$};
  \draw[<->] (1.95,0.6) -- (2.65,0.6);
  \node[right] at (2.65,0.6) {$2\sigma_m$};

  \fill[gray] (1.9,0.97) circle (0.05) node[above right=1pt] {start $x$};
  \fill (2.3,0.83) circle (0.025);
  \node[above] at (4.0,2.55) {barrier top};
  \node at (2.3,0.3)         {well};
  \node at (5.6,0.12)        {basin};

  \node[below] at (1.08,0) {$x_a$};
  \node[below] at (2.3,0)  {$x_m$};
  \node[below] at (4.0,0)  {$x_u$};
  \node[below] at (4.8,0)  {$x_c$};

  \draw[line width=2pt] (3.2,3.3) -- (3.4,3.3);
  \node[right] at (3.4,3.3) {reflecting};
  \draw[line width=2pt,dash pattern=on 3pt off 2pt] (5.0,3.3) -- (5.2,3.3);
  \node[right] at (5.2,3.3) {absorbing};

  \node[anchor=west] at (1.08,-0.55) {Case 1: $[x_a,x_u]$};
  \draw[|-|] (1.08,-0.78) -- (4.0,-0.78);
  \node[anchor=west] at (1.08,-1.08) {Case 2: $[x_a,x_c]$};
  \draw[|-|] (1.08,-1.31) -- (4.8,-1.31);

\end{tikzpicture}
\caption{\small Geometry of the Kramers problem (not to scale). The particle starts in
the metastable well of $U(x)$ (minimum $x_m$); the barrier top is at $x_u$ and
$\Delta U=U(x_u)-U(x_m)$. Solid bar: reflecting boundary $x_a$, placed far up the left
wall, $x_m-x_a\gg\sigma_m$. Dashed bars: absorbing boundary, either $x_u$ (Case~1) or
$x_c$ (Case~2). The arrows of length $2\sigma_m$ and $2\sigma_u$ show the Gaussian
widths $\pm\sigma$ about the well minimum and the barrier top, with
$\sigma_m=[\beta U''(x_m)]^{-1/2}$ and $\sigma_u=[\beta|U''(x_u)|]^{-1/2}$; for
$\beta\Delta U\gg1$ both are much smaller than shown. In Case~1 the absorbing bar at
$x_u$ removes the right half of the Gaussian, so only half contributes. In Case~2, $x_c$ lies beyond the top,
$x_c-x_u\gg\sigma_u$, with $U<U(x_u)$ on $(x_u,x_c]$; the Gaussian is then two-sided.}
\label{fig:kramers-landscape}
\end{figure}

Consider the overdamped Langevin equation
\begin{equation}
    \gamma\,dX
    =
    -U'(X)\,dt
    +
    \sqrt{2\gamma\kB\Theta}\,dW,
    \label{eq:overdamped-langevin}
\end{equation}
where \(U\) is the potential, \(\gamma\) the friction coefficient,
\(\Theta\) the temperature and \(W\) a standard Wiener process; we write
\(\beta=1/\kB\Theta\). [We denote the temperature by \(\Theta\) rather than
the usual \(T\), which is reserved here for the mean first hitting time
\(T(x)\).] In the notation \(dX=A\,dt+\sqrt{B}\,dW\) used above, the drift
and diffusion coefficients are \(A=-U'/\gamma\) and
\(B=2\kB\Theta/\gamma\). Let \(U\) have a metastable minimum at \(x_m\) and
a neighbouring barrier top at \(x_u\), so that \(U'(x_m)=U'(x_u)=0\),
\(U''(x_m)>0\) and \(U''(x_u)<0\), with barrier height
\(\Delta U=U(x_u)-U(x_m)>0\). Refer Fig.~\ref{fig:kramers-landscape}.

The \emph{Kramers problem} is to find the rate at which a particle, initially in the
metastable \emph{well} around \(x_m\), escapes over the \emph{barrier} at \(x_u\) into
the neighbouring \emph{basin}. Such a rate is well defined only when escape is a rare
event, i.e.\ in the high-barrier (low-temperature) regime \(\beta\Delta U\gg1\), since
temperature and barrier enter only through this ratio. Linearizing about the minimum, \(\gamma\,\dot\delta=-U''(x_m)\,\delta\) with
\(\delta=x-x_m\), gives \(\delta(t)=\delta(0)\,e^{-t/\tau_{\mathrm{rel}}}\), where
\(\tau_{\mathrm{rel}}=\gamma/U''(x_m)\) is the relaxation time within the well.
Successive intervals of this length act as independent attempts, each drawing afresh
from the local Boltzmann distribution and reaching the barrier with probability of order
\(e^{-\beta\Delta U}\). Escape is therefore a memoryless process, whose rate
\(k_{\mathrm{esc}}\) is the success probability per attempt divided by the attempt
duration. The mean escape time \(1/k_{\mathrm{esc}}\) is then of order
\(\tau_{\mathrm{rel}}e^{\beta\Delta U}\gg\tau_{\mathrm{rel}}\), during which the particle
samples the local Boltzmann distribution many times and forgets its starting point.
Hence \(k_{\mathrm{esc}}\) is independent of the initial position in the well. For
\(\beta\Delta U\lesssim1\) this separation of time scales fails: the particle crosses
before equilibrating, the escape time depends on where it started, and no single rate
exists.

To turn this estimate into a result, we place a reflecting boundary at
\(x_a<x_m\) and an absorbing boundary at \(x_u\), which is exactly the setting
of Eq.~\eqref{eq:lrrA} with \(x_b=x_u\). Here \(2A/B=-\beta U'\) and
\(2/B=\beta\gamma\), so \(\psi(z)=e^{-\beta[U(z)-U(x_a)]}\). The constant factors
\(e^{\pm\beta U(x_a)}\) cancel between \(1/\psi(y)\) and \(\psi(z)\), and
Eq.~\eqref{eq:lrrA} becomes
\begin{equation}
    T(x)
    =
    \beta\gamma
    \int_x^{x_u}dy\,e^{\beta U(y)}
    \int_{x_a}^{y}dz\,e^{-\beta U(z)} .
    \label{eq:exact-mfpt}
\end{equation}
This result is exact. For \(\beta\Delta U\gg1\) both integrands are sharply peaked and
the integrals follow from Laplace's method, with corrections of relative order
\(1/\beta\Delta U\). Near the minimum and the barrier top,
\(e^{-\beta U(z)}\simeq e^{-\beta U(x_m)}e^{-(z-x_m)^2/2\sigma_m^2}\) and
\(e^{\beta U(y)}\simeq e^{\beta U(x_u)}e^{-(y-x_u)^2/2\sigma_u^2}\), Gaussians whose
thermal widths \(\sigma_m=[\beta U''(x_m)]^{-1/2}\) and
\(\sigma_u=[\beta|U''(x_u)|]^{-1/2}\) are the distances at which \(U\) differs from its
extremum by \(\kB\Theta/2\), since
\(\tfrac12U''(x_m)\sigma_m^2=\tfrac12|U''(x_u)|\sigma_u^2=\kB\Theta/2\). 
We assume \(x_m-x_a\gg\sigma_m\) and \(x_u-x\gg\sigma_u\), so that the lower limits of the two integrals lie far from the peaks and
exclude no significant part of either Gaussian. The result is then
independent of the artificial boundary \(x_a\) and of the starting point \(x\).
 The inner integral is then
the full Gaussian, for every \(y\) beyond the well and in particular near \(x_u\), where
the outer integrand is largest:
\begin{equation}
    \int_{x_a}^{y}e^{-\beta U(z)}\,dz
    \simeq
    e^{-\beta U(x_m)}
    \sqrt{\frac{2\pi}{\beta U''(x_m)}} .
    \label{eq:inner-laplace}
\end{equation}
Since \(x_u\) is the upper limit of the outer integral, only half of its Gaussian is
covered:
\begin{equation}
    \int_x^{x_u}e^{\beta U(y)}\,dy
    \simeq
    e^{\beta U(x_u)}
    \sqrt{\frac{\pi}{2\beta|U''(x_u)|}} .
    \label{eq:outer-laplace}
\end{equation}
Inserting both into Eq.~\eqref{eq:exact-mfpt}, the powers of \(\beta\)
cancel against the prefactor, and
\begin{equation}
    T(x)
    \simeq
    \frac{\pi\gamma}{\sqrt{U''(x_m)|U''(x_u)|}}\,
    e^{\Delta U/\kB\Theta},
    \label{eq:arrhenius-barrier-boundary}
\end{equation}
independent of the starting point \(x\) in the well, as anticipated.

Eq.~\eqref{eq:arrhenius-barrier-boundary} is the mean time to reach the
barrier top.  If escape is instead defined as crossing the barrier and entering the neighbouring
basin, the absorbing boundary is placed at some $x_c>x_u$, which we take to lie
several barrier widths beyond the top, $x_c-x_u\gg(\beta|U''(x_u)|)^{-1/2}$, with
$U<U(x_u)$ on $(x_u,x_c]$. Then $x_u$ becomes an interior point of the outer integral
in Eq.~\eqref{eq:exact-mfpt}, with upper limit $x_c$, and its Gaussian contribution is
two-sided. The mean escape time $ T_{\mathrm{esc}}$ is then twice as long, i.e., $2T(x)$; the corresponding escape rate is
\begin{equation}
    k_{\mathrm{esc}}
    =
    \frac{1}{T_{\mathrm{esc}}}
    \simeq
    \frac{\sqrt{U''(x_m)|U''(x_u)|}}{2\pi\gamma}\,
    e^{-\Delta U/\kB\Theta},
    \label{eq:kramers-rate}
\end{equation}
the overdamped \emph{Kramers formula}. Its exponential factor is the \emph{Arrhenius law}:
it is the probability of a thermal fluctuation large enough to carry the
particle over the barrier \(\Delta U\). The prefactor sets the attempt
frequency. The curvature \(U''(x_m)\) measures the stiffness of the
metastable well, \(|U''(x_u)|\) the sharpness of the barrier top, and the
friction \(\gamma\) the dissipative time scale. We note that the
mean escape time may be cast as
\(T_{\mathrm{esc}}\simeq2\pi\sqrt{U''(x_m)/|U''(x_u)|}\,
\tau_{\mathrm{rel}}e^{\beta\Delta U}\sim\tau_{\mathrm{rel}}e^{\beta\Delta U}\). This confirms the estimate
\(\tau_{\mathrm{rel}}e^{\beta\Delta U}\) made at the outset.

The mean escape time is thus a prefactor times the Arrhenius exponential: the barrier
height alone sets the exponential, while the friction and the curvatures of the well
and the barrier top set the prefactor. This division survives beyond the overdamped
limit, where inertial effects modify only the prefactor.

\begin{center}
	$\symking\symqueen\symrook\symknight\symbishop\sympawn\epsdice{1}\,\epsdice{2}\,\epsdice{3}\,\epsdice{4}\,\epsdice{5}\,\epsdice{6}\,\spadesuit\heartsuit\clubsuit\diamondsuit$
\end{center}
\newpage
\newpage
\clearpage
\onecolumngrid
\section{Far From Equilibrium: Case Studies}
\epigraph{\textsf{I am content with a universe that is diverse.}}{Freeman Dyson}
\twocolumngrid

\indent {\color{deepsea}{\LARGE \textgoth{C}}}ollective behaviour---the emergence of system-wide patterns from the interactions of many particles---is one of the central concerns of statistical mechanics. We often describe such patterns in terms of phases. But phases are not confined to systems in equilibrium: systems driven far from equilibrium can also display striking forms of collective organization, even when they do not satisfy detailed balance. Identifying and understanding these phases usually requires studying each system on its own terms. And even when a system approaches an equilibrium phase, the process by which it gets there is itself dynamical and typically out of equilibrium.

To explore these ideas, we will build on a few canonical examples of nonequilibrium statistical mechanics. We will begin with \emph{nucleation} and \emph{kinetic Ising models}, then turn to the \emph{totally asymmetric simple exclusion process (TASEP)}. Together, these examples offer a way to study how collective behaviour arises, how it changes over time, and how to analyze it beyond the equilibrium framework. Of course, as with the other topics in these lecture notes, we will only be scratching the surface; this cannot be a comprehensive account of everything discussed here.

\subsection{Nucleation}

Nucleation is the process by which a new, thermodynamically preferred phase
first appears inside a metastable parent phase. The new phase does not form
everywhere at once. Instead, thermal fluctuations continually create small,
localised regions of it, called \emph{nuclei}, most of which dissolve again. Only
a nucleus that by chance exceeds a critical size grows irreversibly and
converts the system. When the nuclei form in the bulk of a pure phase,
rather than on walls or impurities, the process is called \emph{homogeneous
nucleation}, and it is the case considered here. Nucleation initiates the
condensation of a supersaturated vapour into droplets, the freezing of
supercooled water, the crystallisation of melts and solutions, precipitation
in alloys, and the reversal of the magnetisation of a magnet whose field has
been reversed.

What makes nucleation subtle is that it is a rare event. A metastable phase
is locally stable: every small fluctuation towards the new phase costs free
energy, so the system can remain in the wrong phase for times exponentially
longer than any microscopic time scale. Small droplets of pure water, for instance, can be cooled to about
\(-38\,^{\circ}\)C at atmospheric pressure before ice nucleates
homogeneously. The central quantity is therefore the \emph{nucleation rate}, and these notes ask
two questions about it: why a metastable phase can survive for so long,
and which features of its decay rate are fixed by thermodynamics and which
by kinetics. 

A ferromagnet whose field is suddenly reversed illustrates the problem in
its simplest form. It does not reverse at once, but lingers in its old
magnetisation until a thermal fluctuation creates a large enough seed of the
new one. The field quench of the two-dimensional Ising model is thus the
natural starting point, since there the energy of every seed can be counted
exactly.

Consider a square lattice of \(N\) sites with unit spacing and periodic
boundary conditions. Each site \(i\) carries a spin \(s_i\in\{-1,+1\}\),
representing a magnetic moment pointing up (\(s_i=+1\)) or down
(\(s_i=-1\)) along the axis on which the external magnetic field is
applied. The energy of a configuration \(\mathbf{s}=(s_1,\dots,s_N)\) is
\begin{equation}
    H(\mathbf{s})
    =
    -J\sum_{\langle ij\rangle}s_is_j
    -
    h\sum_{i} s_i ,
    \label{eq:ising}
\end{equation}
where the first sum runs over bonds, that is, pairs \(\langle ij\rangle\) of
nearest-neighbour sites, each counted once. Every site has four nearest
neighbours, and each bond is shared by its two end sites, so there are
\(4N/2=2N\) bonds. The exchange coupling \(J>0\) is ferromagnetic: a bond
contributes \(-J\) when its spins are parallel and \(+J\) when they are
antiparallel, so breaking a bond costs \(2J\). The second term is the
energy of the moments in a uniform field \(h\) along the same axis,
measured in units of energy; \(h>0\) points up and favours up spins, and
\(h<0\) points down and favours down spins.

The model has a phase transition, located exactly by Onsager in his
celebrated solution of the two-dimensional Ising model, at the critical
temperature \(\kB T_c=2J/\ln(1+\sqrt2)\simeq2.269\,J\). Above \(T_c\)
thermal agitation wins: the spins point up and down at random, and the
magnetisation vanishes in zero field. Below \(T_c\) the coupling wins:
most spins align with their neighbours, and the system acquires a
spontaneous magnetisation per site \(m_{\mathrm{s}}(T)\), the average
spin \(\tfrac1N\sum_is_i\) that survives without any field. It approaches
1 as \(T\to0\), when all spins are aligned, and falls to 0 as
\(T\to T_c\). At \(h=0\) the magnetisation can point either way, giving
two ordered phases, one with \(+m_{\mathrm{s}}\) (mostly up) and one with
\(-m_{\mathrm{s}}\) (mostly down), of exactly equal free energy, since
reversing every spin leaves the coupling energy unchanged. A field breaks
this tie: \(h>0\) lowers the energy of the up phase, which is therefore
the one selected.

The field quench proceeds as follows. The system is prepared in the up
phase at \(T<T_c\) in a field \(h>0\), and the field is then suddenly
reversed to \(h=-|h|\). At the instant of reversal the configuration is
unchanged but its energy is not:
\begin{itemize}
    \item all-up state before the reversal: \(H=-2JN-|h|N\);
    \item all-up state after the reversal: \(H=E_{\uparrow}\equiv-2JN+|h|N\), so the quench
          raises its energy by \(2|h|N\);
    \item all-down state after the reversal: \(H=-2JN-|h|N\), so the down
          phase now lies \(2|h|N\) below the up phase, both measured in the
          reversed field.
\end{itemize}
The up phase is therefore no longer the stable one. The question is how it
decays.

The Hamiltonian fixes the energies but not the dynamics. Physically, spins
are flipped by exchanging energy with the surroundings (the heat bath,
e.g., lattice vibrations), and each flip is a local, random event. The
configuration therefore changes gradually, through local rearrangements,
rather than all at once: the system cannot jump directly from the up phase
to the down phase, but must pass through intermediate configurations
containing small seeds of the down phase.

Which changes actually occur is decided by the energy change they cause.
For any step from one configuration to another we write
\begin{equation}
    \delta E = H_{\mathrm{after}} - H_{\mathrm{before}} .
    \label{eq:step-energy}
\end{equation}
A step with \(\delta E<0\) is \emph{downhill}, and one with \(\delta E>0\)
is \emph{uphill}. In equilibrium with the heat bath, every step and its
reverse must balance (detailed balance), which fixes the ratio of the rate
of a step to that of its reverse at \(e^{-\delta E/\kB T}\), but not the
rates themselves. The natural choice consistent with this is that a
downhill step happens readily, at the bare attempt rate, since the energy
it releases is simply absorbed by the bath, whereas an uphill step must
borrow \(\delta E\) from the bath and is suppressed by the Boltzmann
factor \(e^{-\delta E/\kB T}\), which is small when \(\delta E\gg\kB T\).
A change made of downhill steps is therefore \emph{spontaneous}, while one
that requires climbing happens only as a rare thermal fluctuation. The
decay of the up phase thus hinges on one question: can the system reach
the down phase by downhill steps, or must it first climb?

After the reversal, Eq.~\eqref{eq:ising} reads
\begin{equation}
    H
    =
    -J\sum_{\langle ij\rangle}s_is_j
    +
    |h|\sum_i s_i .
    \label{eq:ising-2d-reversed}
\end{equation}
Consider a configuration with \(n\) down spins and \(B\) broken bonds,
that is, nearest-neighbour pairs with \(s_is_j=-1\). Every parallel bond
contributes \(+1\) to the bond sum and every broken bond \(-1\), so
\(\sum_{\langle ij\rangle}s_is_j=(2N-B)-B=2N-2B\); likewise
\(\sum_is_i=(N-n)-n=N-2n\). Hence
\begin{equation}
    H
    =
    E_{\uparrow}+2J\,B-2|h|\,n,
    \qquad
    E_{\uparrow}=-2JN+|h|N,
    \label{eq:ising-2d-count}
\end{equation}
where \(E_\uparrow\) is the energy of the all-up state after the reversal.
Measuring the energy of any configuration from \(E_\uparrow\),
\begin{equation}
    \Delta E \equiv H-E_\uparrow = 2J\,B-2|h|\,n .
    \label{eq:ising-2d-DeltaE}
\end{equation}
Each broken bond thus raises the energy by \(2J\), and each down spin
lowers it by \(2|h|\) through the reversed field; the all-down state
(\(n=N\), \(B=0\)) has \(\Delta E=-2|h|N\), the lowest of all. 

The energy
\(\Delta E\) must not be confused with \(\delta E\): \(\Delta E\) belongs to
a configuration and says where the system is, measured from the all-up
state, whereas \(\delta E\) belongs to a step and decides whether that step
is spontaneous. The two are related by
\(\delta E=\Delta E_{\mathrm{after}}-\Delta E_{\mathrm{before}}\), since the
reference \(E_\uparrow\) cancels, so a configuration with \(\Delta E>0\)
costs energy relative to the all-up state, but whether the system moves
towards it or away from it is decided by the sign of \(\delta E\) for the
next step.

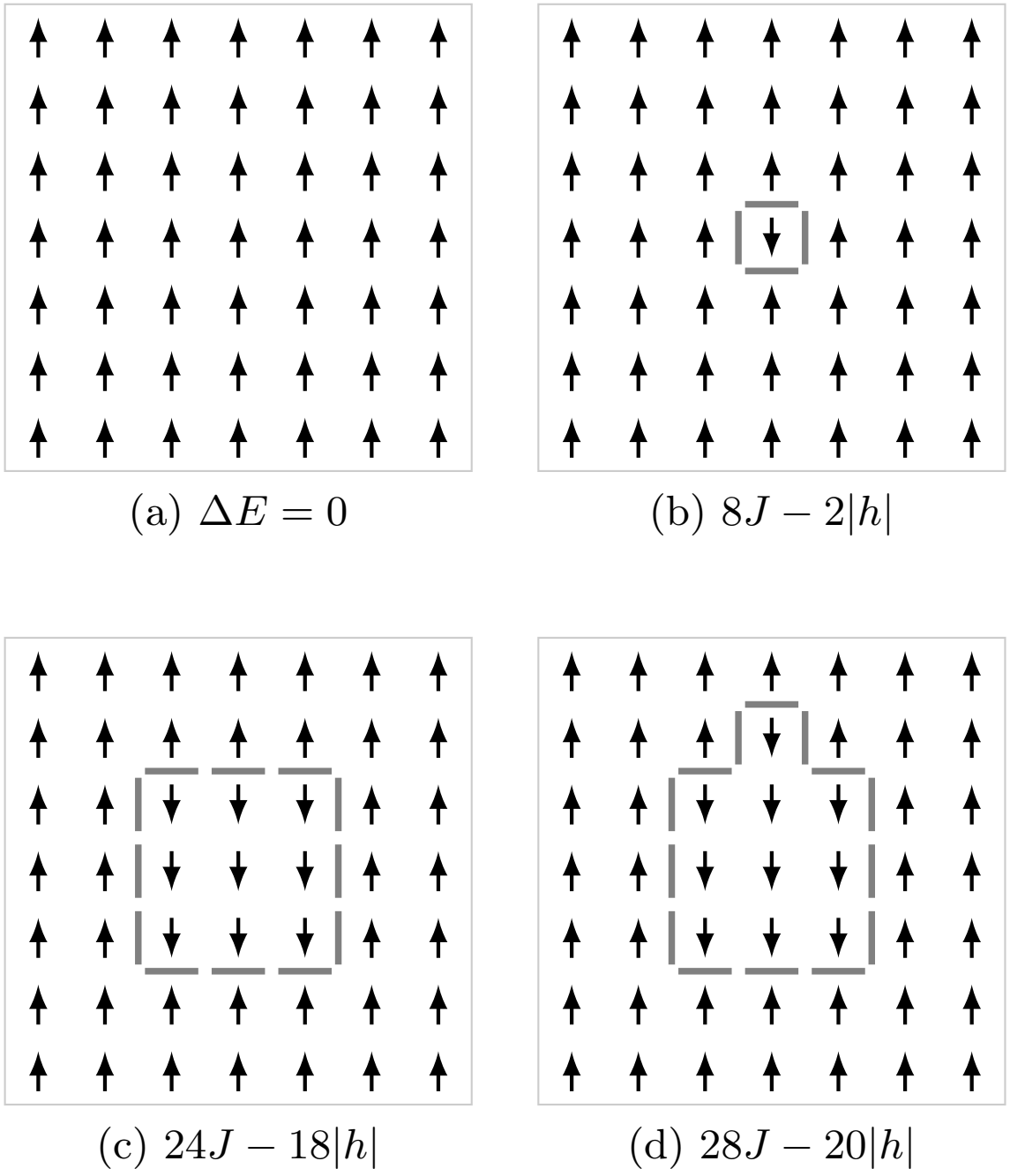
\begin{figure}
\centering
\begin{tikzpicture}[x=0.5cm,y=0.5cm,
  upsp/.style={-latex,thick,black},
  dnsp/.style={-latex,thick,black},
  bb/.style={gray,line width=1.4pt},
  every node/.style={font=\small}]
\draw[gray!40] (-0.5,-0.5) rectangle (6.5,6.5);
\foreach \x/\y in {0/6,1/6,2/6,3/6,4/6,5/6,6/6,0/5,1/5,2/5,3/5,4/5,5/5,6/5,0/4,1/4,2/4,3/4,4/4,5/4,6/4,0/3,1/3,2/3,3/3,4/3,5/3,6/3,0/2,1/2,2/2,3/2,4/2,5/2,6/2,0/1,1/1,2/1,3/1,4/1,5/1,6/1,0/0,1/0,2/0,3/0,4/0,5/0,6/0} \draw[upsp] (\x,\y-0.3)--(\x,\y+0.3);
\node[below] at (3,-0.6) {(a) \(\Delta E=0\)};
\draw[gray!40] (7.5,-0.5) rectangle (14.5,6.5);
\foreach \x/\y in {8/6,9/6,10/6,11/6,12/6,13/6,14/6,8/5,9/5,10/5,11/5,12/5,13/5,14/5,8/4,9/4,10/4,11/4,12/4,13/4,14/4,8/3,9/3,10/3,12/3,13/3,14/3,8/2,9/2,10/2,11/2,12/2,13/2,14/2,8/1,9/1,10/1,11/1,12/1,13/1,14/1,8/0,9/0,10/0,11/0,12/0,13/0,14/0} \draw[upsp] (\x,\y-0.3)--(\x,\y+0.3);
\foreach \x/\y in {11/3} \draw[dnsp] (\x,\y+0.3)--(\x,\y-0.3);
\draw[bb] (10.6,3.5)--(11.4,3.5);
\draw[bb] (10.5,2.6)--(10.5,3.4);
\draw[bb] (11.5,2.6)--(11.5,3.4);
\draw[bb] (10.6,2.5)--(11.4,2.5);
\node[below] at (11,-0.6) {(b) \(8J-2|h|\)};
\begin{scope}[shift={(-16,-9.5)}]
\draw[gray!40] (15.5,-0.5) rectangle (22.5,6.5);
\foreach \x/\y in {16/6,17/6,18/6,19/6,20/6,21/6,22/6,16/5,17/5,18/5,19/5,20/5,21/5,22/5,16/4,17/4,21/4,22/4,16/3,17/3,21/3,22/3,16/2,17/2,21/2,22/2,16/1,17/1,18/1,19/1,20/1,21/1,22/1,16/0,17/0,18/0,19/0,20/0,21/0,22/0} \draw[upsp] (\x,\y-0.3)--(\x,\y+0.3);
\foreach \x/\y in {18/4,19/4,20/4,18/3,19/3,20/3,18/2,19/2,20/2} \draw[dnsp] (\x,\y+0.3)--(\x,\y-0.3);
\draw[bb] (17.6,4.5)--(18.4,4.5);
\draw[bb] (18.6,4.5)--(19.4,4.5);
\draw[bb] (19.6,4.5)--(20.4,4.5);
\draw[bb] (17.5,3.6)--(17.5,4.4);
\draw[bb] (20.5,3.6)--(20.5,4.4);
\draw[bb] (17.5,2.6)--(17.5,3.4);
\draw[bb] (20.5,2.6)--(20.5,3.4);
\draw[bb] (17.5,1.6)--(17.5,2.4);
\draw[bb] (17.6,1.5)--(18.4,1.5);
\draw[bb] (18.6,1.5)--(19.4,1.5);
\draw[bb] (20.5,1.6)--(20.5,2.4);
\draw[bb] (19.6,1.5)--(20.4,1.5);
\node[below] at (19,-0.6) {(c) \(24J-18|h|\)};
\draw[gray!40] (23.5,-0.5) rectangle (30.5,6.5);
\foreach \x/\y in {24/6,25/6,26/6,27/6,28/6,29/6,30/6,24/5,25/5,26/5,28/5,29/5,30/5,24/4,25/4,29/4,30/4,24/3,25/3,29/3,30/3,24/2,25/2,29/2,30/2,24/1,25/1,26/1,27/1,28/1,29/1,30/1,24/0,25/0,26/0,27/0,28/0,29/0,30/0} \draw[upsp] (\x,\y-0.3)--(\x,\y+0.3);
\foreach \x/\y in {27/5,26/4,27/4,28/4,26/3,27/3,28/3,26/2,27/2,28/2} \draw[dnsp] (\x,\y+0.3)--(\x,\y-0.3);
\draw[bb] (26.6,5.5)--(27.4,5.5);
\draw[bb] (26.5,4.6)--(26.5,5.4);
\draw[bb] (25.6,4.5)--(26.4,4.5);
\draw[bb] (27.5,4.6)--(27.5,5.4);
\draw[bb] (27.6,4.5)--(28.4,4.5);
\draw[bb] (25.5,3.6)--(25.5,4.4);
\draw[bb] (28.5,3.6)--(28.5,4.4);
\draw[bb] (25.5,2.6)--(25.5,3.4);
\draw[bb] (28.5,2.6)--(28.5,3.4);
\draw[bb] (25.5,1.6)--(25.5,2.4);
\draw[bb] (25.6,1.5)--(26.4,1.5);
\draw[bb] (26.6,1.5)--(27.4,1.5);
\draw[bb] (28.5,1.6)--(28.5,2.4);
\draw[bb] (27.6,1.5)--(28.4,1.5);
\node[below] at (27,-0.6) {(d) \(28J-20|h|\)};
\end{scope}
\end{tikzpicture}
\caption{\small A patch of the square-lattice Ising model after the field is
reversed to \(h=-|h|\). Up spins (\(s=+1\)) and down spins (\(s=-1\)) are
shown as arrows, and broken bonds (\(\uparrow\downarrow\)) as grey
segments. (a)~The all-up state, now metastable; it is the reference for
\(\Delta E\). (b)~A single flipped spin, \(n=1\), \(B=4\). (c)~A
\(3\times3\) seed, \(n=9\), \(B=12\). (d)~The same seed with a protrusion,
\(n=10\), \(B=14\). The energy relative to the all-up state,
\(\Delta E=2JB-2|h|n\) from Eq.~\eqref{eq:ising-2d-DeltaE}, is given below
each panel.}
\label{fig:ising-seeds}
\end{figure}

Fig.~\ref{fig:ising-seeds} shows what happens on the way to the down
phase. Panel~(a) shows the all-up state, which is the reference with
\(\Delta E=0\). Panel~(b) shows a single flipped spin in the up background:
it has \(n=1\) and four broken bonds, \(B=4\), so \(\Delta E=8J-2|h|\),
which is positive for \(|h|<4J\). Panel~(c) shows a \(3\times3\) seed with
\(n=9\) and \(B=12\), so \(\Delta E=24J-18|h|\), and panel~(d) the same
seed with one extra spin on its edge, with \(n=10\) and \(B=14\), so
\(\Delta E=28J-20|h|\). For a weak field \(\Delta E\) increases from (a) to
(b) to (c) to (d), so each step along this sequence has \(\delta E>0\).
Although the all-down state is lower, the all-up state is therefore stuck:
every small step towards the down phase is uphill, so none of them happens
spontaneously.

The seeds in Fig.~\ref{fig:ising-seeds} are all reached by uphill steps,
but they are small. Panels~(b) and (c) are the cases \(\ell=1\) and
\(\ell=3\) of a compact \(\ell\times\ell\) square seed, for which
\(n=\ell^2\) and \(B=4\ell\), so that Eq.~\eqref{eq:ising-2d-DeltaE} gives
\begin{equation}
    \Delta E(\ell)=8J\ell-2|h|\ell^2 .
    \label{eq:ising-2d-square}
\end{equation}
The bond cost grows with the perimeter, linearly in \(\ell\), while the
field gain grows with the area, as \(\ell^2\), so for large enough seeds
the gain must win. For weak fields, \(|h|\ll J\), the seeds that matter
are many lattice spacings across, and \(\ell\) may be treated as
continuous. The slope
\begin{equation}
    \frac{d\Delta E}{d\ell}=8J-4|h|\ell
    \label{eq:ising-2d-slope}
\end{equation}
is positive for small \(\ell\) and negative for large \(\ell\), and
vanishes at
\begin{equation}
    \ell^*=\frac{2J}{|h|},
    \qquad
    \Delta E^*\equiv\Delta E(\ell^*)=\frac{8J^2}{|h|},
    \label{eq:ising-2d-critical}
\end{equation}
where \(\Delta E(\ell)\) is largest. On the lattice itself, enlarging an
\(\ell\times\ell\) seed to \((\ell+1)\times(\ell+1)\) is a step with
\begin{equation}
    \delta E=\Delta E(\ell+1)-\Delta E(\ell)=8J-2|h|(2\ell+1),
    \label{eq:ising-2d-step}
\end{equation}
which changes sign at \(\ell=\ell^*-\tfrac12\). Discreteness therefore
shifts the critical size by less than one lattice spacing, which is
negligible when \(\ell^*\gg1\). Growing a seed smaller than \(\ell^*\)
has \(\delta E>0\), so such seeds tend to shrink back; growing a seed
larger than \(\ell^*\) has \(\delta E<0\), so their growth is spontaneous.
Note that for \(\ell^*<\ell<2\ell^*\) the seed still has
\(\Delta E(\ell)>0\), yet each further step has \(\delta E<0\): what
matters for growth is the slope of \(\Delta E(\ell)\), not its sign.

The up phase is therefore stuck only against small fluctuations. To decay,
it must form a seed of size \(\ell^*\), which requires climbing by uphill
steps to the energy \(\Delta E^*\) above the all-up state. This is the
situation of the Kramers problem: small seeds form and dissolve on a time
much shorter than the time needed to reach \(\ell^*\), so between such
rare events the up phase samples its nearby configurations with Boltzmann
weights relative to the all-up state. The probability of finding a
critical seed at a given place is therefore of order
\begin{equation}
    P^*\sim e^{-\Delta E^*/\kB T}=e^{-8J^2/(|h|\kB T)} .
    \label{eq:ising-2d-probability}
\end{equation}
This probability is exponentially small, because \(\Delta E^*\) is large
compared with \(\kB T\) for weak fields; but once a seed exceeds
\(\ell^*\), its growth lowers the energy, and it grows until the down
phase fills the system.

Equation~\eqref{eq:ising-2d-probability} is the Arrhenius factor of
Eq.~\eqref{eq:kramers-rate}, with \(\Delta E^*\) in place of \(\Delta U\):
the all-up state plays the role of the metastable minimum, and the critical
seed that of the barrier top. The lattice does not, however, supply an
equation of motion for the seed size to which the Kramers formula could be
applied: \(\ell\) is discrete, and near \(\ell=0\) its landscape is not
smooth. We return to this point below.

The field quench therefore leaves the up phase metastable: it is stable
against small fluctuations, since every small seed is reached by uphill
steps, yet it is not the state of lowest energy. It decays only when a
rare thermal fluctuation carries a seed over the barrier \(\Delta E^*\) to
the critical size \(\ell^*\), after which the seed grows spontaneously.
This fluctuation-driven formation of a critical seed is nucleation. The
lattice has thus exposed all its ingredients: a metastable phase; a seed of
the stable phase whose boundary raises the energy and whose bulk lowers it;
a critical size, set by the balance of the two, beyond which the seed
grows; and a barrier that makes reaching that size rare.

The counting above assigns each seed its energy alone, which is exact only
as \(T\to0\). At finite temperature many different configurations have
nearly the same energy: a seed of given size can take many shapes, and its
boundary can be straight or kinked. If \(\Omega\) such configurations share
an energy \(E\), their combined Boltzmann weight is
\(\Omega\,e^{-E/\kB T}=e^{-(E-TS)/\kB T}\), with entropy
\(S=\kB\ln\Omega\), so what controls the probabilities is the free energy
\(E-TS\) rather than the energy. The boundary fluctuates, and the entropy
of its kinks lowers its cost per unit length from \(2J\) to the
interfacial tension \(\sigma(T)\); thermal flips inside both phases reduce
the magnetisation per site from 1 to \(m_{\mathrm{s}}(T)\), so that the
gain per unit area becomes \(2|h|m_{\mathrm{s}}(T)\). The critical size and
the barrier, now a free-energy barrier, are then set by \(\sigma\) and
\(m_{\mathrm{s}}\) rather than by \(J\) alone, but the competition between
boundary and bulk is unchanged. As \(T\to0\), \(\sigma\to2J\) and
\(m_{\mathrm{s}}\to1\), and the lattice results of
Eqs.~\eqref{eq:ising-2d-square} and \eqref{eq:ising-2d-critical} are
recovered.

The lattice argument identifies the barrier but stops short of a rate. It
is tied to one model at low temperature, and its seed size has no equation
of motion of the kind the escape theory requires. Both limitations are
removed by coarse-graining. Nothing in the competition between boundary
and bulk is specific to spins: whenever a phase is carried into a region
where another phase is preferred, the same balance controls how the new
phase forms, and the field-quenched magnet is the counterpart of a
supersaturated vapour or a supercooled liquid. We therefore describe the
seed as a compact droplet of the stable phase, characterised by a single
continuous coordinate, its radius \(R\); for the Ising magnet, \(\sigma\)
is the interfacial tension between the up and down phases and the driving
force is \(\delta g=2|h|m_{\mathrm{s}}\). The droplet description does
three things the lattice could not: it applies to any system, it uses free
energies valid at finite temperature, and, as we show, its radius obeys an
overdamped Langevin equation, so that the escape theory applies in full,
prefactor included.

Consider a system held in a metastable phase~1, such as a supersaturated
vapour, in which a second phase~2, such as the liquid, is thermodynamically
preferred. In a fluid such a state is prepared by a quench in temperature
or pressure, which plays the role of the field reversal in the magnet:
cooling a liquid below its melting point, or supersaturating a vapour by
rapid adiabatic expansion, as in a cloud chamber. Let \(g_1\) and \(g_2\) be the Gibbs free energies per unit
volume of the two phases at the same temperature and pressure. The Gibbs
free energy is the appropriate one because nucleation normally occurs at
fixed temperature and pressure, and the two phases must be compared under
the same external conditions; per molecule, \(g\) is the chemical
potential, so \(g_1>g_2\) means that molecules lower their chemical
potential by passing from phase~1 to phase~2. For the Ising magnet the
field \(h\) is held fixed in the same way, and the relevant free energy is
the magnetic analogue of the Gibbs free energy; the pair
\((-h,M=\sum_i s_i)\) plays the role of
\(({\rm pressure},\,{\rm volume})\). Phase~2 is preferred when
\begin{equation}
    \delta g \equiv g_1 - g_2 > 0 ,
    \label{eq:driving-force}
\end{equation}
the free energy gained per unit volume of phase~1 converted into phase~2.

The new phase cannot appear everywhere at once; it must first form as a
small droplet, and the reversible work \(\Delta G(R)\) to create one of
radius \(R\) plays the role of the potential. Two effects compete, exactly
as for the lattice seed: the interior of the droplet is in phase~2, which
lowers the free energy by \(\delta g\) per unit volume, while its surface
costs the interfacial tension \(\sigma\) per unit area, because particles
there have fewer like neighbours. Treating the interface as sharp and the
interior as bulk (the capillarity approximation),
\begin{equation}
    \Delta G(R)
    =
    4\pi R^2\sigma
    -
    \frac{4\pi}{3}R^3\,\delta g .
    \label{eq:nucleation-free-energy}
\end{equation}
In two dimensions, as for the Ising magnet, \(4\pi R^2\) and
\(\tfrac43\pi R^3\) become \(2\pi R\) and \(\pi R^2\). The surface term
dominates at small \(R\) and the volume term at large \(R\), so small
droplets tend to evaporate and large ones to grow.

The crossover, \(\Delta G'(R^*)=0\), gives the critical radius and the
barrier,
\begin{equation}
    R^* = \frac{2\sigma}{\delta g},
    \qquad
    \Delta G^* \equiv \Delta G(R^*) = \frac{16\pi\sigma^3}{3(\delta g)^2}.
    \label{eq:nucleation-barrier}
\end{equation}
At \(R^*\) the Laplace pressure \(2\sigma/R^*\) of the droplet balances the
bulk driving force \(\delta g\), which has the same dimensions, so droplets
smaller than \(R^*\) shrink on average and larger ones grow. The barrier
diverges as \(\delta g\to0\), the continuum counterpart of
\(\Delta E^*\propto1/|h|\) in Eq.~\eqref{eq:ising-2d-critical}. The
curvatures
\begin{equation}
    \Delta G''(0) = 8\pi\sigma,
    \qquad
    \Delta G''(R^*) = -8\pi\sigma
    \label{eq:curvatures}
\end{equation}
identify the parent phase \(R=0\) as the minimum and the critical nucleus
\(R^*\) as the barrier top. In the full space of droplet shapes, \(R^*\) is
a saddle, since shape fluctuations of the critical droplet raise its free
energy.

Equation~\eqref{eq:nucleation-barrier} shows that the barrier exists only
because of the cost of the interface: it vanishes as \(\sigma\to0\),
since \(\Delta G^*\propto\sigma^3\). A foreign surface, such as a dust grain, an insoluble particle or the
container wall, that the new phase wets (i.e., one it prefers to be in
contact with over the parent phase) lowers the barrier by replacing part
of that interface. Nucleation on such surfaces, called \emph{heterogeneous
nucleation}, is therefore faster by many orders of magnitude, 
which is why ordinary water freezes only a few degrees below
\(0\,^{\circ}\)C, whereas pure water in small droplets can be cooled to
about \(-38\,^{\circ}\)C.

A droplet changes size through many elementary events at its surface:
single particles attaching or detaching in a fluid, or boundary spins
flipping in a magnet. Each event is fast and changes the size by one unit,
whereas an appreciable change of \(R\) requires very many of them. The
radius is therefore a slow coordinate driven by many fast, weakly
correlated kicks from the remaining degrees of freedom, which act as a
thermal bath. It also has no inertia: the momentum of each surface event is
shared with the surroundings within a few molecular collision times, long
before the radius changes appreciably, so \(\dot R\) has no persistence.
Collective inertia, as for vapour bubbles in a liquid, would alter only the
prefactor, not the Arrhenius exponent. The radius therefore obeys
Eq.~\eqref{eq:overdamped-langevin} with \(X\to R\), \(U\to\Delta G\),
\(\gamma\to\gamma_R\) and \(\Theta\to T\),
\begin{equation}
    \gamma_R\,dR
    =
    -\Delta G'(R)\,dt
    +
    \sqrt{2\gamma_R\kB T}\,dW .
    \label{eq:radius-langevin}
\end{equation}
The friction \(\gamma_R\) measures how strongly the radius resists change
and is set by the attachment kinetics, which thermodynamics does not fix.
For simplicity we take \(\gamma_R\) to be independent of \(R\), so that
the noise in Eq.~\eqref{eq:radius-langevin} is additive and the mapping onto
Eq.~\eqref{eq:overdamped-langevin} is exact; this also avoids having to
choose a stochastic convention, which a size-dependent \(\gamma_R\) would
require. The noise amplitude is fixed by the friction through the
fluctuation--dissipation relation, because the same surface events that
resist changes of the radius also make it fluctuate; as a result, droplet
sizes are distributed in the steady state according to the Boltzmann
weight \(e^{-\beta\Delta G(R)}\).

The landscape has the geometry assumed in the escape problem: the parent
phase \(R=0\) is the metastable minimum, the critical nucleus \(R^*\) is
the barrier top, and \(\Delta U\to\Delta G^*\). Reaching sizes beyond
\(R^*\) is escape, since such droplets grow on average. A droplet that has
just reached \(R^*\) is equally likely to grow or to shrink, so the
relevant rate is that of crossing the barrier, Eq.~\eqref{eq:kramers-rate},
rather than that of first reaching \(R^*\). Nucleation is therefore the
escape problem already solved, with one difference: the parent phase sits
at the reflecting boundary \(R=0\) rather than at an interior minimum, so
only half of the Gaussian around it contributes. For constant
\(\gamma_R\), the Laplace argument with the curvatures of
Eq.~\eqref{eq:curvatures} then gives, for \(\beta\Delta G^*\gg1\), the
escape rate of a single droplet,
\begin{eqnarray}
    k_{\mathrm{nuc}}
    &\simeq&
    \frac{\sqrt{\Delta G''(0)\,|\Delta G''(R^*)|}}{\pi\gamma_R}\,
    e^{-\Delta G^*/\kB T}\nonumber\\
    &=&
    \frac{8\sigma}{\gamma_R}\,e^{-\Delta G^*/\kB T},
    \label{eq:nucleation-arrhenius}
\end{eqnarray}
and the nucleation rate per unit volume \(I\propto e^{-\Delta G^*/\kB T}\).

The robust content of Eq.~\eqref{eq:nucleation-arrhenius} is the exponent,
which is fixed by thermodynamics alone through
Eq.~\eqref{eq:nucleation-barrier}. The prefactor
\(\sqrt{\Delta G''(0)\,|\Delta G''(R^*)|}/(\pi\gamma_R)\), by contrast, is
not universal. It depends on the friction near the critical nucleus, on
the curvatures of \(\Delta G\), and on how the parent phase is represented
near \(R=0\), where a description in terms of a continuous radius is least
reliable. The Arrhenius picture also requires \(\beta\Delta G^*\gg1\); at
strong driving the barrier falls to the order of \(\kB T\) and the escape
description breaks down.

Before concluding, we note a subtlety that arises once the friction is
allowed to depend on the droplet size: the choice of stochastic
convention. The rate of Eq.~\eqref{eq:nucleation-arrhenius} was obtained
for a friction \(\gamma_R\) independent of the droplet size. This is
rarely true: a droplet changes size through events at its surface, so its
friction depends on \(R\). The noise in Eq.~\eqref{eq:radius-langevin} is
then multiplicative, and the equation is ambiguous until a stochastic
convention is specified. The ambiguity is not a technicality, because each
convention predicts a different stationary distribution of droplet sizes.

To see this, write Eq.~\eqref{eq:radius-langevin} with a size-dependent
friction as
\begin{equation}
    dR=a\,dt+b\,dW_c ,
    \label{eq:sde-c}
\end{equation}
with drift \(a=-\mathcal{M}\Delta G'\), noise amplitude
\(b=\sqrt{2\kB T\,\mathcal{M}}\) and mobility \(\mathcal{M}=1/\gamma_R\).
The subscript on \(dW_c\) specifies where in each time step the amplitude
multiplying the noise increment is evaluated: \(b\,dW_c\) stands for
\(b\bigl(R(t+c\,dt)\bigr)\,dW(t)\), with \(0\le c\le1\). Thus \(c=0\) is
It\^o, \(c=\tfrac12\) is Stratonovich and \(c=1\) is anti-It\^o
(H\"anggi--Klimontovich). Evaluating \(b\) later in the step correlates it
with the noise and adds the drift \(c\,bb'\), so the equivalent It\^o
equation has drift \(a+c\,bb'\), and the probability density \(P(R,t)\)
obeys
\begin{equation}
    \partial_t P
    =
    -\partial_R\Bigl[\bigl(a+c\,bb'\bigr)P
    -\tfrac12\,\partial_R\bigl(b^2P\bigr)\Bigr].
    \label{eq:fp-c}
\end{equation}
Denote by \(P_{\mathrm{s}}(R)\) the stationary distribution. With
reflecting walls its flux vanishes, so
\(\bigl(a+c\,bb'\bigr)P_{\mathrm{s}}
=\tfrac12\,\partial_R\bigl(b^2P_{\mathrm{s}}\bigr)\). Using
\(bb'=\tfrac12(b^2)'=\kB T\,\mathcal{M}'\) and
\(\tfrac12(b^2P_{\mathrm{s}})'
=\kB T\bigl(\mathcal{M}'P_{\mathrm{s}}+\mathcal{M}P_{\mathrm{s}}'\bigr)\),
this becomes
\begin{equation}
    \frac{P_{\mathrm{s}}'}{P_{\mathrm{s}}}
    =
    -\beta\Delta G'-(1-c)\,\frac{\mathcal{M}'}{\mathcal{M}} .
\end{equation}
Integrating, and using \(\mathcal{M}=1/\gamma_R\),
\begin{equation}
    P_{\mathrm{s}}(R)
    \propto
    \gamma_R(R)^{\,1-c}\,e^{-\beta\Delta G(R)} .
    \label{eq:stationary-c}
\end{equation}
Each convention therefore predicts a different stationary distribution of
droplet sizes:
\begin{itemize}
    \item It\^o (\(c=0\)):
          \(P_{\mathrm{s}}\propto\gamma_R(R)\,e^{-\beta\Delta G(R)}\);
    \item Stratonovich (\(c=\tfrac12\)):
          \(P_{\mathrm{s}}\propto\sqrt{\gamma_R(R)}\,e^{-\beta\Delta G(R)}\);
    \item anti-It\^o (\(c=1\)):
          \(P_{\mathrm{s}}\propto e^{-\beta\Delta G(R)}\).
\end{itemize}
Only the last is the Boltzmann weight that thermodynamics demands; in the
other two, the kinetics would change how often each size occurs, which is
impossible in equilibrium. The convention is in fact fixed by the
elementary events introduced above: a droplet grows by attachment, the
gain of one molecule or the flip of one boundary spin, and shrinks by
detachment, the reverse event. In equilibrium every attachment between
two neighbouring sizes is balanced by the reverse detachment (detailed
balance), which fixes the ratio of the two rates through the free-energy
difference between the sizes. For droplets of many particles the size
changes in steps small compared with its range, so the discrete jumps can
be replaced by a continuous radius; in this limit the detailed-balance
condition singles out the anti-It\^o convention.

To see that the anti-It\^o convention arises naturally, one derives the
Fokker--Planck equation for the radius from the microscopic master
equation for attachment and detachment, the \emph{Becker--D\"oring equation}. Its
continuum limit, the \emph{Zeldovich--Frenkel equation}, has the form
\(\partial_tP=\partial_R\bigl[\mathcal{M}\bigl(\Delta G'P+\kB T\,\partial_RP\bigr)\bigr]\),
in which the diffusion coefficient \(\kB T\,\mathcal{M}\) stands outside
the inner derivative, multiplying \(\partial_RP\) rather than being
differentiated together with \(P\). This placement identifies the
convention: it is precisely the anti-It\^o form of the Fokker--Planck
equation, Eq.~\eqref{eq:FP-antiIto}.

With it, \(P_{\mathrm{s}}\) remains \(e^{-\beta\Delta G(R)}\) even when the
friction depends on size, so the barrier, and hence the Arrhenius exponent
\(\Delta G^*\), are unchanged. The friction affects only the prefactor: in
the mean escape time it multiplies \(e^{\beta\Delta G}\), which is sharply
peaked at the barrier top, so only its value there matters, and
Eq.~\eqref{eq:nucleation-arrhenius} holds with \(\gamma_R\) replaced by
\(\gamma_R(R^*)\).

The Arrhenius law itself would survive even a wrong choice of
convention. Eq.~\eqref{eq:stationary-c} can be written as a Boltzmann
weight in an effective free energy,
\begin{equation}
    P_{\mathrm{s}}\propto e^{-\beta\Delta G_{\mathrm{eff}}},
    \,\,
    \Delta G_{\mathrm{eff}}(R)=\Delta G(R)-(1-c)\,\kB T\ln\gamma_R(R),
\end{equation}
so a convention other than anti-It\^o adds to the landscape a term equal to
\(\kB T\) times the logarithm of the friction. A physical friction varies
only as a power of \(R\), for instance \(\gamma_R\propto R^2\) for
interface-limited growth, so this term changes by just a few \(\kB T\)
across the barrier, negligible against \(\Delta G^*\gg\kB T\). The barrier
height, and with it the exponent \(\Delta G^*/\kB T\), is therefore the same
in every convention; the rate keeps its Arrhenius form
\(k_{\mathrm{nuc}}\propto e^{-\Delta G^*/\kB T}\) whichever reading is
used. What changes is the prefactor, which acquires a factor
\(\bigl[\gamma_R(R^*)/\gamma_R(R\to0)\bigr]^{-(1-c)}\), a power of the ratio
of the frictions at the barrier and in the parent phase. The convention is
thus invisible in the exponent, which thermodynamics fixes, but not in the
prefactor, which the kinetics fixes; only the anti-It\^o reading, selected
by detailed balance, gives the prefactor correctly.

The two questions posed at the outset can now be answered. A metastable
phase survives because every small seed of the new phase is uphill in free
energy: its boundary costs more than its bulk gains. It decays only when a
rare thermal fluctuation assembles a seed of critical size, and the rate
of such fluctuations is the Arrhenius factor
\(e^{-\Delta G^*/\kB T}\), with a barrier
\(\Delta G^*\propto\sigma^3/(\delta g)^2\) that diverges as the driving
force vanishes. This is why weakly supercooled water, or a magnet in a
weakly reversed field, can stay in the wrong phase almost indefinitely,
and why nucleation, once the driving force is large enough to bring
\(\Delta G^*\) down to tens of \(\kB T\), sets in abruptly. As for the
second question, the rate separates cleanly into two parts. Its exponent
is fixed by thermodynamics alone, through \(\sigma\) and \(\delta g\), and
is insensitive to the kinetics, even to the stochastic convention used for
a size-dependent friction. Its prefactor is set by the kinetics: by the
friction at the critical nucleus, by the curvatures of the landscape, and,
when the friction depends on size, by the anti-It\^o convention that
detailed balance selects.
\subsection{Kinetic Ising Models}

The previous section followed one way in which a new phase appears.
After a field quench the up phase is metastable: every small step
towards the down phase is uphill, and the system decays only when a rare
thermal fluctuation assembles a seed of critical size \(\ell^*\),
Eq.~\eqref{eq:ising-2d-critical}. This is not the only possibility, and
the field is not the only thing that can be quenched. In a
\emph{temperature quench} at \(h=0\), a magnet prepared in its
disordered state above \(T_c\) is cooled suddenly below it. The
disordered state is then \emph{unstable}, with downhill steps available
everywhere, and ordering begins at once without any seed; this is
\emph{spinodal instability}. Which route follows depends on whether
the state left behind is metastable or unstable, not on what was
quenched. Either way, the system soon breaks up into \emph{domains},
regions in which the spins point predominantly up or predominantly down,
separated by \emph{domain walls}. The domains then grow and merge, the
typical domain size increasing with time, a process called
\emph{coarsening}. Throughout, the evolution is carried from its very first step
by the same microscopic moves, individual spins flipping or neighbouring
spins exchanging, one event at a time: the fluctuations that assemble a
critical seed, the growth of an unstable fluctuation, and the coarsening
that follows are all sequences of such moves.

The Ising Hamiltonian, Eq.~\eqref{eq:ising}, does not describe these
moves. It assigns an energy to every configuration and, through the
Boltzmann weight, fixes all equilibrium properties, but it generates no
motion: the spins are classical and discrete, with no conjugate momentum
and no equation of motion. For nucleation this was enough, because the
energies of the seeds determined the barrier and the dynamics entered
only through an effective equation for the droplet radius. To follow the system in time in general, we need an explicit rule for
how the spins change. Such a rule is necessarily an effective
description: the spins are kept, while the environment and the fast
microscopic, ultimately quantum, processes by which a spin actually
reverses are replaced by random transitions occurring at given rates.
The description is thus coarse in time rather than in space: it
resolves every spin, but not the details of each flip.

The dynamics come from the surroundings. Each spin exchanges energy
with a heat bath, such as the lattice vibrations, and these exchanges
are fast and effectively random on the time scale of the spins. The
natural description is therefore stochastic: a configuration \(C\)
changes to \(C'\) at a rate \(R(C'|C)\), and the probability
\(P(C,t)\) evolves by a master equation, Eq.~\eqref{eq:master} below.
The moves triggered by the bath are independent random events. In a
short interval \(dt\) a given move occurs with probability proportional
to \(dt\), so two independent moves occur together only with
probability of order \(dt^2\), and the configuration changes by one
elementary move at a time.

This is the \emph{kinetic Ising model}: the Ising Hamiltonian together
with a stochastic, typically Markovian, spin dynamics described by a
master equation, whose rates, for a system in contact with a heat bath,
are chosen to obey detailed balance with respect to the Boltzmann
distribution. It is the simplest model of the dynamics with the correct
physics, and it does not depend on the process being studied: the same
rules describe nucleation, spinodal instability, the coarsening that
follows them, and equilibrium fluctuations.

What counts as an elementary move depends on the order parameter. In a
magnet the total magnetisation \(M=\sum_i s_i\) need not be conserved: a
spin can reverse by exchanging energy with the lattice, so the
elementary move is a single flip, \(s_i\to-s_i\). This is
\emph{Glauber dynamics}---the dynamics implicit in the field quench. In a
binary alloy or a lattice gas, the two species correspond to
\(s_i=\pm1\) through the occupation \(n_i=(1+s_i)/2\), and the number of
atoms of each kind is fixed. An atom can move only by swapping with a
neighbouring atom of the other kind, so the elementary move is an
exchange of two neighbouring antiparallel spins, which conserves \(M\)
locally: magnetisation can be transported but never created or
destroyed. This is \emph{Kawasaki dynamics}. The exchange is a single
elementary event involving a pair of sites, not two independent flips
that happen to coincide, which would occur with probability of order
\(dt^2\) and would not conserve \(M\).

The rates cannot be chosen freely. The system must eventually reach
equilibrium at its temperature \(T\), so the Boltzmann distribution
\(P_{\mathrm{eq}}\propto e^{-H/\kB T}\) must be a stationary solution of
the master equation. The simplest way to guarantee this is detailed
balance: in equilibrium every move is balanced by its reverse, so the
ratio of the forward and backward rates of each move is
\(e^{-\Delta E/\kB T}\), where \(\Delta E\) is the energy change of the
move. This fixes the ratio but not the rates themselves, and the
remaining freedom is real: \emph{Glauber's choice}, the \emph{Metropolis rule} and
others all satisfy it.

Restricting the dynamics to one elementary move at a time loses none of
the physics that matters. Single flips connect every configuration to
every other, and neighbouring exchanges connect every configuration to
every other with the same magnetisation, so with detailed balance the
system always reaches the Boltzmann distribution, over all
configurations under Glauber dynamics and over those with the given
\(M\) under Kawasaki dynamics. Collective changes, such as the growth of
a droplet, the motion of a domain wall or the transport of
magnetisation, are sequences of these local moves. Large-scale
behaviour depends only on the energies and on what is conserved, so
moves of several spins at once would change the time scale but not the
physics; and one move at a time is exactly what Monte Carlo simulations
implement.

We now construct the standard transition rates of the kinetic Ising
model.
We begin with single spin flips. At equilibrium the probability of a configuration $C$ is
\begin{equation}
  P_{\rm eq}(C) = \frac{e^{-\beta H(C)}}{Z},\qquad \beta=\frac{1}{\kB T},
\end{equation}
where $C$ denotes a particular configuration. For $N$ Ising spins there are
$2^{N}$ possible configurations. As the system relaxes it must approach
$P_{\rm eq}(C)$. The evolution of $P(C,t)$ is governed by the master equation
\begin{equation}
  \frac{dP(C)}{dt} = \sum_{C'} R(C|C')\,P(C') - \sum_{C'} R(C'|C)\,P(C),
  \label{eq:master}
\end{equation}
where $R(C|C')$ is the (yet to be determined) transition rate from $C'$ to
$C$.
Physically, the memory lives in the
heat bath, and the Markov property holds because the bath relaxes much
faster than the spins: by the time a spin is ready to flip again, the
bath has forgotten the previous flip. The dynamics is thus a
continuous-time Markov jump process on the $2^N$ configurations.

For single spin flips, let $C_{s_i}$ denote a configuration with spin
$s_i\in\{+1,-1\}$ at site $i$, and $C_{-s_i}$ the same configuration with
that spin reversed, e.g.,
\begin{equation*}
  C_{s_3}:\ \fivechain{1}{1}{1}{1}{1}\,,\qquad
  C_{-s_3}:\ \fivechain{1}{1}{-1}{1}{1}\,,
\end{equation*}
where the boxed spin (site $i=3$) is the one in focus, i.e., the spin being flipped.
Detailed balance then reads
\begin{equation}
  R(C_{s_i}|C_{-s_i})\,P_{\rm eq}(C_{-s_i})
  = R(C_{-s_i}|C_{s_i})\,P_{\rm eq}(C_{s_i}).
  \label{eq:db}
\end{equation}

To evaluate this condition we need the energy. Write the Ising
Hamiltonian as
\begin{equation}
  H = -\sum_{\langle lk\rangle} J_{lk}\, s_l s_k,
\end{equation}
where the sum runs over the pairs of interacting sites $l$ and $k$, each
pair counted once. This form is deliberately general: the sites may sit
on any lattice or graph, in any dimension, and the couplings $J_{lk}$
may be restricted to nearest neighbours or extend further, be uniform or
vary from bond to bond. The only structure used below is that $H$ is a
sum of pair interactions, so that a given spin $s_i$ enters only through
its couplings $J_{ik}$ to the sites $k$ it interacts with, i.e.,
through the \emph{local field}
\begin{equation}
  h_i=\sum_k J_{ik}\,s_k
\end{equation}
exerted on it by those spins, with $J_{ik}=0$ for sites not coupled to
$i$ and $J_{ii}=0$. An external field $h_{\rm ext}$ would add a
term $-h_{\rm ext}\sum_l s_l$ to $H$; the couplings are unchanged, and
the local field on every spin becomes $h_i+h_{\rm ext}$. We take
$h_{\rm ext}=0$.

Denote by $w_i \equiv R(C_{-s_i}|C_{s_i})$ the rate of the flip
$s_i\to -s_i$, and by $w_{-i}$ the rate of the reverse flip. From
Eq.~\eqref{eq:db},
\begin{equation}
  \frac{w_{-i}}{w_i}
  = \frac{P_{\rm eq}(C_{s_i})}{P_{\rm eq}(C_{-s_i})}
  = \frac{\exp\!\big[\beta\sum_{\langle lk\rangle} J_{lk}\, s_l s_k\big]_{s_i}}
         {\exp\!\big[\beta\sum_{\langle lk\rangle} J_{lk}\, s_l s_k\big]_{-s_i}} ,
\end{equation}
where the subscripts $s_i$ and $-s_i$ indicate that the sum is evaluated
in configuration $C_{s_i}$ and $C_{-s_i}$ respectively. Separating the
terms that involve site $i$,
\begin{equation}
  \sum_{\langle lk\rangle} J_{lk}\, s_l s_k
  = \sum_{\substack{\langle lk\rangle\\ l,k\neq i}} J_{lk}\, s_l s_k
  + s_i h_i ,
\end{equation}
the part not involving $s_i$ cancels between numerator and denominator,
leaving
\begin{equation}
  \frac{w_{-i}}{w_i}
  = \frac{e^{+\beta s_i h_i}}{e^{-\beta s_i h_i}} .
  \label{eq:ratio_exp}
\end{equation}

This ratio can be turned into an explicit rate with a simple identity:
For $\varepsilon\in\{+1,-1\}$,
\begin{equation}
  e^{a\varepsilon} = \cosh a + \varepsilon \sinh a
  = \cosh a\,\big[1+\varepsilon\tanh a\big].
\end{equation}
Applying it with $\varepsilon = s_i$ and $a=\beta h_i$, the $\cosh$
factors cancel in Eq.~\eqref{eq:ratio_exp}:
\begin{equation}
  \frac{w_{-i}}{w_i}
  = \frac{1 + s_i \tanh(\beta h_i)}{1 - s_i \tanh(\beta h_i)} .
\end{equation}
This fixes the form of $w_i$ up to an overall constant, which sets the
time scale:
\begin{equation}
  w_i \propto 1 - s_i \tanh(\beta h_i).
\end{equation}
The rate can be written in terms of the energy change of the flip. Since
reversing $s_i$ changes the energy by $\Delta E=2s_ih_i$, and $s_i=\pm1$,
we have $s_i\tanh(\beta h_i)=\tanh(\beta s_ih_i)=\tanh(\beta\Delta E/2)$, so that
\begin{equation}
  w_i \propto 1-\tanh\!\Big(\frac{\beta\Delta E}{2}\Big).
\end{equation}
The constant is fixed by the energy-conserving flips. When $\Delta E=0$
neither orientation of the spin is favoured, so after an update it
should be equally likely to point up or down; this requires the rate
$\tfrac12$, and hence
\begin{equation}
  w_i = \frac{1}{2}\big[1 - s_i \tanh(\beta h_i)\big]
      = \frac{1}{1+e^{\beta\Delta E}} .
  \label{eq:glauber_general}
\end{equation}
The second form has a simple meaning. Hold all other spins fixed and
consider spin $i$ alone in the field $h_i$ of its neighbours. It has
two possible states, its present one, with energy $E_0$, and the
flipped one, with energy $E_0+\Delta E$. If the spin were in equilibrium
with the heat bath in this fixed field, the Boltzmann weights of the two
states would give the probability of finding it flipped as
\begin{equation}
  \frac{e^{-\beta(E_0+\Delta E)}}{e^{-\beta E_0}+e^{-\beta(E_0+\Delta E)}}
  = \frac{e^{-\beta\Delta E}}{1+e^{-\beta\Delta E}}
  = \frac{1}{1+e^{\beta\Delta E}} ,
\end{equation}
where $E_0$ cancels, so that only the energy difference matters. This is
exactly $w_i$. Glauber dynamics is therefore a \emph{heat-bath} rule: at
each update the spin forgets its previous value and is drawn afresh from
its own equilibrium distribution in the current field of its neighbours.

The same normalisation gives the expected behaviour at high
temperature. As $\beta\to0$, $w_i\to\tfrac12$ for every spin, whatever
its neighbours: at infinite temperature the coupling is irrelevant, every
flip is effectively energy-conserving, and an update turns each spin
over with probability one half, independently of its surroundings and
of its past.

The general rate [Eq.~\eqref{eq:glauber_general}] simplifies considerably in one dimension. Consider the
\emph{Ising chain}: spins $s_i$ on the sites of a line, each coupled
only to its two neighbours $i\pm1$ with the same coupling $J$, so that
$H=-J\sum_i s_is_{i+1}$, with periodic boundary conditions so that every
site is equivalent. The local field at site $i$ is then $h_i=J(s_{i-1}+s_{i+1})$,
so the rate becomes
\begin{equation}
  w_i = \frac{1}{2}\left[1 - s_i \tanh\!\left(2\beta J\,
        \frac{s_{i-1}+s_{i+1}}{2}\right)\right].
\end{equation}
Since $\tanh(\varepsilon x)=\varepsilon\tanh x$ for
$\varepsilon\in\{0,-1,+1\}$, and $(s_{i-1}+s_{i+1})/2$ takes exactly these
values, we obtain
\begin{equation}
  w_i = \frac{1}{2}\left[1 - \frac{\gamma}{2}\, s_i\,
  (s_{i-1}+s_{i+1})\right],\quad \gamma = \tanh(2\beta J).
  \label{eq:glauber1d}
\end{equation}

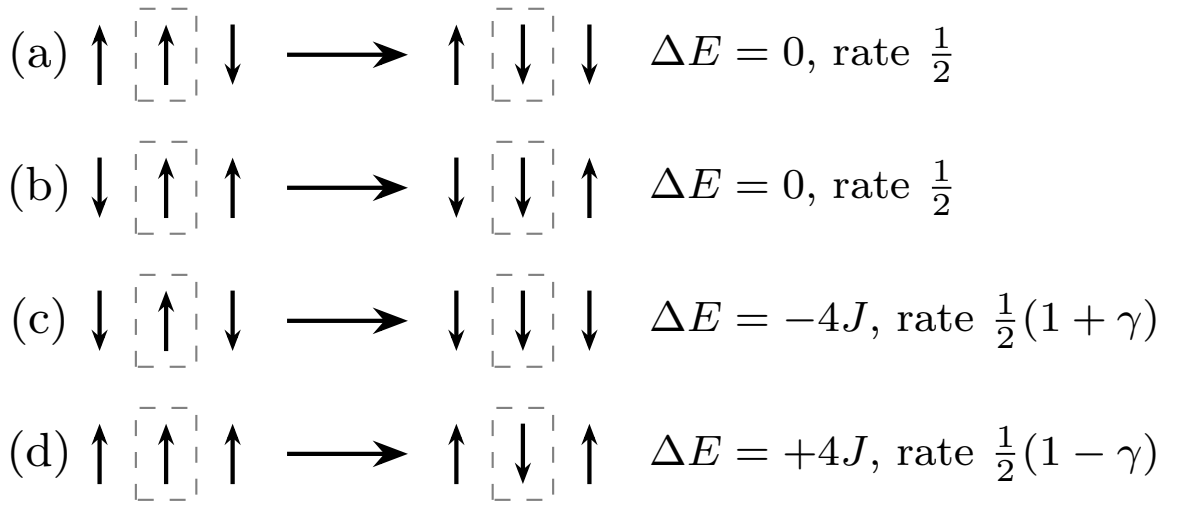
\begin{figure}[t]
\centering
\begin{tikzpicture}[scale=0.8]
  \gchain{0}{0}{1}{1}{-1}
  \draw[-{Stealth},thick] (1.55,0) -- (2.55,0);
  \gchain{2.95}{0}{1}{-1}{-1}
  \node[font=\footnotesize,anchor=west] at (4.4,0)
    {$\Delta E=0$, rate $\tfrac12$};
  \node at (-0.5,0) {(a)};
  \gchain{0}{-1.1}{-1}{1}{1}
  \draw[-{Stealth},thick] (1.55,-1.1) -- (2.55,-1.1);
  \gchain{2.95}{-1.1}{-1}{-1}{1}
  \node[font=\footnotesize,anchor=west] at (4.4,-1.1)
    {$\Delta E=0$, rate $\tfrac12$};
  \node at (-0.5,-1.1) {(b)};
  \gchain{0}{-2.2}{-1}{1}{-1}
  \draw[-{Stealth},thick] (1.55,-2.2) -- (2.55,-2.2);
  \gchain{2.95}{-2.2}{-1}{-1}{-1}
  \node[font=\footnotesize,anchor=west] at (4.4,-2.2)
    {$\Delta E=-4J$, rate $\tfrac12(1+\gamma)$};
  \node at (-0.5,-2.2) {(c)};
  \gchain{0}{-3.3}{1}{1}{1}
  \draw[-{Stealth},thick] (1.55,-3.3) -- (2.55,-3.3);
  \gchain{2.95}{-3.3}{1}{-1}{1}
  \node[font=\footnotesize,anchor=west] at (4.4,-3.3)
    {$\Delta E=+4J$, rate $\tfrac12(1-\gamma)$};
  \node at (-0.5,-3.3) {(d)};
\end{tikzpicture}
\caption{\small Glauber flip of the boxed spin in the one-dimensional chain,
with the rate of Eq.~\eqref{eq:glauber1d}.
(a), (b) The spin sits at a domain wall: the flip moves the wall by one
site and the energy is unchanged.
(c) Both neighbours are opposite: two domain walls annihilate and the
energy decreases.
(d) Both neighbours are aligned: two domain walls are created and the
energy increases.}
\label{fig:glauber}
\end{figure}

The rate takes only three values, according to the neighbours of the
flipping spin (Fig.~\ref{fig:glauber}). In the chain a domain wall is
simply a bond between two antiparallel neighbours. If both neighbours are
aligned with it, $s_i(s_{i-1}+s_{i+1})=+2$: the flip creates two domain
walls,
raises the energy by $4J$, and occurs at rate $\tfrac12(1-\gamma)$. If
the spin sits at a domain wall, with one neighbour aligned and one not,
$s_i(s_{i-1}+s_{i+1})=0$: the flip moves the wall by one site, leaves the
energy unchanged, and occurs at rate $\tfrac12$. If both neighbours are
opposite to it, $s_i(s_{i-1}+s_{i+1})=-2$: the flip removes two domain
walls, lowers the energy by $4J$, and occurs at rate
$\tfrac12(1+\gamma)$. Creating and removing a pair of domain walls are reverse moves, and the
ratio of their rates, $(1+\gamma)/(1-\gamma)=e^{4\beta J}$, is exactly
what detailed balance requires.

The rate of Eq.~\eqref{eq:glauber1d} has the following symmetries:
\begin{itemize}
  \item \emph{locality}: it depends only on the spin and its two
        neighbours;
  \item \emph{translation invariance}: it has the same form at every
        site;
  \item \emph{left--right symmetry}: it is unchanged under
        $s_{i-1}\leftrightarrow s_{i+1}$;
  \item \emph{up--down symmetry}: it is unchanged when every spin is
        reversed, $s_j\to-s_j$.
\end{itemize}
Glauber's rate is not the only one compatible with detailed balance.
The most general rate with the same symmetries also allows a term
$s_{i+1}s_{i-1}$ and an overall factor $\alpha$. Adding these to
Eq.~\eqref{eq:glauber1d},
\begin{equation}
  w_i = \frac{\alpha}{2}\Bigl[1 - \frac{\kappa}{2}\, s_i(s_{i+1}+s_{i-1})
        + \delta\, s_{i+1}s_{i-1}\Bigr],
  \label{eq:general_rate}
\end{equation}
with $-1\le\delta\le 1$, $\alpha>0$, and $\kappa$ to be determined
by detailed balance.
[The bounds on $\delta$ follow from $w_i\ge0$. When the spin sits at a
domain wall, $s_{i-1}s_{i+1}=-1$ and the $\kappa$ term is absent, so
$w_i=\frac{\alpha}{2}(1-\delta)$, which requires $\delta\le1$. When its
neighbours are aligned, $s_{i-1}s_{i+1}=+1$, and with $\kappa$ fixed by
detailed balance below, the rate becomes
$\frac{\alpha}{2}(1+\delta)(1\mp\gamma)$, which requires $\delta\ge-1$
since $0\le\gamma<1$.]
The flip-reversed rate is
\begin{equation}
  w_{-i} = \frac{\alpha}{2}\Bigl[1 + \frac{\kappa}{2}\, s_i(s_{i+1}+s_{i-1})
        + \delta\, s_{i+1}s_{i-1}\Bigr],
\end{equation}
so the ratio of rates is 
\begin{equation}
  \frac{w_{-i}}{w_i}
  = \frac{1+\delta s_{i+1}s_{i-1} + \frac{\kappa}{2}s_i(s_{i+1}+s_{i-1})}
         {1+\delta s_{i+1}s_{i-1} - \frac{\kappa}{2}s_i(s_{i+1}+s_{i-1})} .
  \label{eq:ratio_delta}
\end{equation}

For the chain, flipping $s_i$ changes the energy by
\begin{equation}
  \Delta E = H(C_{-s_i}) - H(C_{s_i}) = 2J s_i(s_{i+1}+s_{i-1}),
\end{equation}
and detailed balance [Eq.~\eqref{eq:ratio_exp}] requires:
\begin{align}
  \frac{w_{-i}}{w_i} = \frac{P_{\rm eq}(C_{s_i})}{P_{\rm eq}(C_{-s_i})}
  = e^{+\beta\Delta E}= \exp\!\bigl[2\beta J s_i(s_{i+1}+s_{i-1})\bigr].
  \label{eq:db_1d}
\end{align}
If the neighbours are anti-aligned, $s_{i+1}=-s_{i-1}$, both sides equal
unity and no constraint arises. The non-trivial constraint comes from
aligned neighbours, $s_{i+1}=s_{i-1}=\sigma\in\{+1,-1\}$, for which
$s_{i+1}s_{i-1}=1$ and $s_i(s_{i+1}+s_{i-1}) = 2\sigma s_i$.
Substituting into Eq.~\eqref{eq:ratio_delta},
\begin{equation}
  \frac{w_{-i}}{w_i} = \frac{(1+\delta)+\kappa\,\sigma s_i}
                            {(1+\delta)-\kappa\,\sigma s_i},
\end{equation}
which, by detailed balance [Eq.~\eqref{eq:db_1d}], must equal
$e^{2\beta J s_i(s_{i+1}+s_{i-1})}=e^{4\beta J\sigma s_i}$. For $\sigma s_i=+1$ this gives
\begin{align}
&  \frac{1+\delta+\kappa}{1+\delta-\kappa} = e^{4\beta J} ,\nonumber\\
\implies& \kappa = (1+\delta)\tanh(2\beta J)=(1+\delta)\gamma .
  \label{eq:gamma_delta}
\end{align}
The case $\sigma s_i=-1$ gives the same condition. For $\delta=0$ (and $\alpha=1$)
this reduces to Glauber's result $\kappa=\gamma=\tanh(2\beta J)$,
Eq.~\eqref{eq:glauber1d}.

Detailed balance therefore leaves a one-parameter family of rates, of
which Glauber's choice ($\delta=0$, $\alpha=1$) is only one member.
Another familiar member is the \emph{Metropolis rule}, the standard
Monte~Carlo prescription: a proposed flip that changes the energy by
$\Delta E$ is accepted with probability
\begin{equation}
  w(\Delta E)=\min\!\bigl(1,\,e^{-\beta\Delta E}\bigr),
  \label{eq:metropolis}
\end{equation}
so that neutral and downhill flips, for which
$e^{-\beta\Delta E}\ge1$, are always accepted once proposed, while
uphill flips are accepted with the Boltzmann factor $e^{-\beta\Delta E}$. The ratio of forward to backward
rates is $e^{-\beta\Delta E}$, so detailed balance is satisfied, but the
individual rates differ from Glauber's.

To locate this rule within the general family, we list the three possible
environments of a flipped spin in the chain and their rates from
Eq.~\eqref{eq:general_rate}:

\begin{itemize}
  \item {One adjacent domain wall} ($s_{i-1}s_{i+1}=-1$):
        flipping $s_i$ shifts the wall to the other bond,
        $\Delta E=0$ (neutral),
        $w_i=\frac{\alpha}{2}(1-\delta)$;
  \item {Two adjacent domain walls}
        ($s_{i-1}=s_{i+1}=-s_i$):
        flipping $s_i$ annihilates both walls,
        $\Delta E=-4J$ (downhill),
        $w_i=\frac{\alpha}{2}(1+\delta+\kappa)$;
  \item {No adjacent domain walls}
        ($s_{i-1}=s_{i+1}=+s_i$):
        flipping $s_i$ creates two walls,
        $\Delta E=+4J$ (uphill),
        $w_i=\frac{\alpha}{2}(1+\delta-\kappa)$.
\end{itemize}

Metropolis assigns probability~$1$ to both neutral and downhill moves,
so the first two cases must carry the same rate.  Equating them gives
\begin{equation}
  \frac{\alpha}{2}(1-\delta) = \frac{\alpha}{2}(1+\delta+\kappa)
  \quad\Longrightarrow\quad
  \kappa=-2\delta .
  \label{eq:metropolis_equal}
\end{equation}
The overall factor $\alpha$ is then fixed by requiring these rates to
equal~$1$:
\begin{equation}
  \frac{\alpha}{2}(1-\delta)=1 \quad\Longrightarrow\quad
  \alpha = \frac{2}{1-\delta}.
\end{equation}
Inserting $\kappa=-2\delta$ into the detailed-balance condition
$\kappa=(1+\delta)\gamma$ [Eq.~\eqref{eq:gamma_delta}] yields
$-2\delta=(1+\delta)\gamma$, hence
\begin{equation}
  \delta=-\frac{\gamma}{2+\gamma},\qquad
  \kappa=\frac{2\gamma}{2+\gamma},\qquad
  \alpha=\frac{2+\gamma}{1+\gamma}.
\end{equation}
Since $0\le\gamma<1$, $\delta$ lies between $-\frac13$ and $0$, well
within the allowed range $[-1,1]$.  Substituting these values, the three
rates become
\[
  w_i^{\text{neutral}} = w_i^{\text{downhill}} = 1,
  \qquad
  w_i^{\text{uphill}} = \frac{1-\gamma}{1+\gamma}=e^{-4\beta J},
\]
recovering the standard Metropolis prescription directly.  Expressed in
explicit form, the Metropolis rate for the chain is therefore
\begin{equation}
  w_i^{\rm M}=\frac{2+\gamma-\gamma\,s_i(s_{i-1}+s_{i+1})
               -\gamma\,s_{i-1}s_{i+1}}{2(1+\gamma)}.
  \label{eq:metropolis1d}
\end{equation}

Glauber and Metropolis dynamics are thus two points on the same line,
$\delta=0$ and $\delta=-\gamma/(2+\gamma)$, differing only in $\delta$ and
in the unit of time (set by $\alpha$). They share the same equilibrium state and the same
large-scale behaviour, but they are used for different purposes.
Metropolis is the usual choice in Monte Carlo simulations: it is the
simplest rule to implement, it accepts every move that does not raise the
energy, and it relaxes up to twice as fast as Glauber's rule, which
rejects half of the neutral moves. Glauber's rate is the usual choice in
analytic work, because it is a smooth function of the energy change and,
in one dimension, linear in the neighbouring spins---which, as we shall
see, is what makes the chain exactly solvable. It is also the natural
choice when the dynamics is meant to model a heat bath, since it resets
each spin according to its local equilibrium.

In the limit $\beta\to0$, we have $\gamma=\tanh(2\beta J)\to0$ and hence
$\kappa\to0$, so a member of the family with a fixed $\delta$ reduces to
$w_i\to\frac{\alpha}{2}\bigl(1+\delta\,s_{i-1}s_{i+1}\bigr)$. At
infinite temperature the energy no longer couples the spins, so the
rate should be the same, $\alpha/2$, whatever the neighbours. Only
$\delta=0$ gives this. For any fixed $\delta\neq0$ the rate is raised or
lowered by $\delta\,s_{i-1}s_{i+1}$ according to whether the two
neighbours are parallel or antiparallel, so the dynamics ``remembers''
an interaction that no longer has any energetic meaning. The Metropolis
rate avoids this, because its $\delta=-\gamma/(2+\gamma)$ itself
vanishes as $\beta\to0$, but it has a different defect: its rate tends
to $1$ for every spin, so every proposed flip is accepted. In a
simulation that updates the spins in a fixed order, each spin is then
simply reversed at every sweep, and its successive values are perfectly
anticorrelated rather than independent, as they should be at infinite
temperature. Glauber's rate, which tends to $\tfrac12$, is the only one
of these whose infinite-temperature limit is a fair coin for each spin.

\begin{figure}[h]
\centering
\begin{tikzpicture}[scale=0.85]
  \kchain{0}{0}{1}{1}{-1}{1}
  \draw[-{Stealth},thick] (2.1,0) -- (3.1,0);
  \kchain{3.5}{0}{1}{-1}{1}{1}
  \node[font=\footnotesize,anchor=west] at (5.5,0) {$\Delta E=0$};
  \node at (-0.5,0) {(a)};
  \kchain{0}{-1.1}{-1}{1}{-1}{-1}
  \draw[-{Stealth},thick] (2.1,-1.1) -- (3.1,-1.1);
  \kchain{3.5}{-1.1}{-1}{-1}{1}{-1}
  \node[font=\footnotesize,anchor=west] at (5.5,-1.1) {$\Delta E=0$};
  \node at (-0.5,-1.1) {(b)};
  \kchain{0}{-2.2}{-1}{1}{-1}{1}
  \draw[-{Stealth},thick] (2.1,-2.2) -- (3.1,-2.2);
  \kchain{3.5}{-2.2}{-1}{-1}{1}{1}
  \node[font=\footnotesize,anchor=west] at (5.5,-2.2) {$\Delta E=-4J$};
  \node at (-0.5,-2.2) {(c)};
  \kchain{0}{-3.3}{1}{1}{-1}{-1}
  \draw[-{Stealth},thick] (2.1,-3.3) -- (3.1,-3.3);
  \kchain{3.5}{-3.3}{1}{-1}{1}{-1}
  \node[font=\footnotesize,anchor=west] at (5.5,-3.3) {$\Delta E=+4J$};
  \node at (-0.5,-3.3) {(d)};
\end{tikzpicture}
\caption{\small Kawasaki exchange of the boxed antiparallel pair.
(a), (b) The outer spins are equal: the number of domain walls is unchanged
and the energy is conserved.
(c) Three domain walls become one: the energy decreases.
(d) One domain wall becomes three: the energy increases.}
\label{fig:kawasaki}
\end{figure}
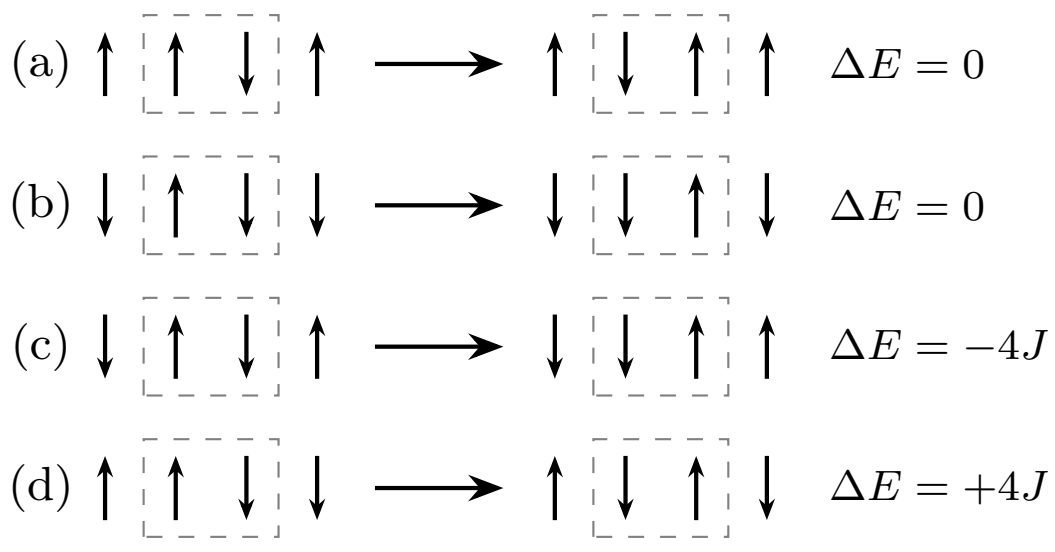

We turn now very briefly to conserved dynamics, again for the Ising
chain. In Kawasaki dynamics two neighbouring spins
change \emph{simultaneously} by exchanging their values---$
  \uparrow\downarrow \;\longrightarrow\; \downarrow\uparrow$---so that the magnetisation remains the same. Only an antiparallel pair can
change; exchanging a parallel pair does nothing. Exchanging the pair $(s_i,s_{i+1})$ affects only the
three bonds $(i-1,i)$, $(i,i+1)$ and $(i+1,i+2)$, and the outcome depends on
the outer spins $s_{i-1}$ and $s_{i+2}$, as shown in
Fig.~\ref{fig:kawasaki}.

In the chain, each of the three affected bonds has energy $-J$ if its
spins are aligned and $+J$ if it is a domain wall. Let
$w_{3\to1}$ denote the rate of an exchange that reduces the number of domain
walls on these bonds from three to one, as in Fig.~\ref{fig:kawasaki}(c),
and let $w_{1\to3}$ denote the rate of the reverse exchange,
Fig.~\ref{fig:kawasaki}(d). An exchange that leaves the energy
unchanged, as in (a) and (b), must by detailed balance occur at the same
rate as its reverse. Going from three domain walls to one lowers the energy from $3J$
to $-J$, so detailed balance requires $w_{3\to1}/w_{1\to3}=e^{4\beta J}$.
This constraint only fixes the ratio; the specific choice
$w_{3\to1}=\tfrac12(1+\tanh2\beta J)$ and
$w_{1\to3}=\tfrac12(1-\tanh2\beta J)$ corresponds to the heat-bath
(Glauber) prescription, where the rates are proportional to the
Boltzmann weights of the final states.  Any other pair satisfying the
same ratio (e.g., a Metropolis-like $w_{3\to1}=1$, $w_{1\to3}=e^{-4\beta J}$)
would be equally valid under detailed balance.

All three cases are captured by
the rate of exchanging the pair $(i,i+1)$,
\begin{multline}
  w_i(s_i,s_{i+1}) = \frac{1}{2}\left[1-\frac{\gamma}{2}
     \big(s_{i-1}s_i + s_{i+1}s_{i+2}\big)\right]\\
  \times \frac{1}{2}\big(1-s_i s_{i+1}\big),
  \qquad \gamma=\tanh 2\beta J .
  \label{eq:kawasaki}
\end{multline}
The second factor, put in by hand, equals $1$ for an antiparallel pair
and vanishes when the two spins are the same, so that only antiparallel
pairs are exchanged. In the first factor,
$s_{i-1}s_i+s_{i+1}s_{i+2}$ equals $+2$, $0$ and $-2$ for cases (d),
(a,b) and (c) of Fig.~\ref{fig:kawasaki}, giving the rates
$\tfrac12(1-\gamma)$, $\tfrac12$ and $\tfrac12(1+\gamma)$ respectively.
Eq.~\eqref{eq:kawasaki} is thus the rate of the kinetic Ising chain
with Kawasaki dynamics. With the single-flip and exchange rates in hand,
one may now ask how they make the magnetisation evolve.
\subsection{Glauber Dynamics}

The kinetic Ising model with single-spin-flip Glauber rates, often
called the \emph{Glauber--Ising model}, is the simplest stochastic
dynamics of a magnet that relaxes to the Boltzmann distribution. Glauber
introduced it and solved it exactly for the chain; since then it has
served as the reference model of non-conserved dynamics, against which
Monte Carlo simulations, mean-field theories and renormalisation-group
treatments of critical dynamics (model~A in the classification of
Hohenberg and Halperin) are compared. In this section we ask what the
Glauber rate implies for the magnetisation after a quench. The answer
depends on where the final temperature lies relative to the
\emph{critical temperature} $T_c$, below which the equilibrium state is
ordered, with a non-zero spontaneous magnetisation at zero field, and
above which it is disordered.

We follow the system after a \emph{temperature quench}: it is prepared
in equilibrium at a temperature \(T_i\), and at \(t=0\) the temperature
is changed suddenly to \(T_f\). The rates change at once but the
configuration does not, and the system then relaxes towards equilibrium
at \(T_f\), one elementary move at a time. We treat three cases of
increasing realism and decreasing tractability: the \emph{Curie--Weiss}
model, with \emph{all-to-all} coupling, where mean-field theory is exact
and the three regimes of a quench, above, at and below $T_c$, can be
followed in full; the Ising chain, for which $T_c=0$, which can be
solved exactly and shows critical slowing down without any
approximation; and local mean-field theory, which extends the
mean-field step to short-range lattices, and whose limitations we
expose by comparing it with the chain.

All three start from the same exact equation. The average of spin $i$
is $\langle s_i\rangle = \sum_C s_i\, P(C)$. In a time step $\Delta t$,
\begin{equation}
  s_i(t+\Delta t) =
  \begin{cases}
    -s_i(t) & \text{with probability } w_i\,\Delta t,\\
    \phantom{-}s_i(t) & \text{with probability } 1-w_i\,\Delta t ,
  \end{cases}
\end{equation}
so the spin changes by $-2s_i$ at rate $w_i$, and
$\langle\Delta s_i\rangle = -2\langle s_i w_i\rangle\,\Delta t$. Dividing
by $\Delta t$,
\begin{equation}
  \frac{d\langle s_i\rangle}{dt} = -2\,\langle s_i\, w_i\rangle ,
  \label{eq:dsdt}
\end{equation}
which holds for any single-flip rate on any lattice. It does not by itself
close, because $w_i$ depends on the neighbours of $i$. This reflects a
general point: the dynamics is Markovian for the full configuration $C$,
Eq.~\eqref{eq:master}, but not, in general, for reduced quantities such
as the magnetisation of a single spin, whose evolution depends on spins
that are not being followed. A closed equation for $\langle s_i\rangle$
needs either a property of the model that reduces
$\langle s_iw_i\rangle$ to single-spin averages, or an approximation
that does so.

We begin with the Curie--Weiss model. The general Hamiltonian can be
written in terms of the local field $h_i$,
\begin{equation}
  H = -\sum_{i<k} J_{ik}\, s_i s_k = -\frac{1}{2}\sum_i h_i s_i,
  \quad h_i = \sum_k J_{ik}\, s_k ,
\end{equation}
where the factor $\tfrac12$ corrects for double counting. In the
Curie--Weiss model, the coupling is all-to-all: every spin
interacts with every other spin with the same strength, however far
apart they are, so there is no lattice and no notion of distance. Since
there are $N(N-1)/2\simeq N^2/2$ pairs, the coupling of each pair must be
scaled as $J/N$ to keep the energy proportional to $N$:
\begin{equation}
  H = -\frac{J}{N}\sum_{i<k} s_i s_k ,
\end{equation}
and we use units with $J=1$. Each spin then feels a total coupling of
order one, shared equally among all the others, and the local field is
the same for every spin, $h_i=\frac1N\sum_k s_k$, the instantaneous
magnetisation of the whole system.

For the Curie--Weiss model the Glauber rate,
Eq.~\eqref{eq:glauber_general}, is
\begin{equation}
  w_i = \frac{1}{2}\left[1 - s_i \tanh\!\left(\beta\hat m\right)\right],
  \qquad \hat m \equiv \frac{1}{N}\sum_k s_k .
\end{equation}
Because every spin interacts with every other, the field on each spin is
the same quantity, the instantaneous magnetisation $\hat m$; this is a
property of the model, not an approximation. The \emph{mean-field
approximation} is the further step of replacing this fluctuating
variable by its ensemble average,
\begin{equation}
  \hat m \;\to\; m \equiv \langle \hat m\rangle = \frac{1}{N}\sum_k \langle s_k\rangle ,
\end{equation}
i.e.,  $\langle s_i\tanh(\beta\hat m)\rangle \approx \langle s_i\rangle\tanh(\beta m)$,
which neglects fluctuations of $\hat m$. For the Curie--Weiss model this is
exact as $N\to\infty$: $\hat m$ is a sum over all $N$ spins, so its
fluctuations about $m$ vanish as $1/\sqrt N$ by the law of large numbers
(away from the critical point). Using $s_i^2=1$, Eq.~\eqref{eq:dsdt}
becomes
\begin{equation}
  \frac{d\langle s_i\rangle}{dt} = -\big[\langle s_i\rangle - \tanh(\beta m)\big],
\end{equation}
and averaging over $i$,
\begin{equation}
  \frac{dm}{dt} = -m + \tanh(\beta m),
  \label{eq:mf}
\end{equation}
which is therefore exact for the Curie--Weiss model as $N\to\infty$.
The price is that the model has no notion of distance: it describes how
order grows in the system as a whole, but not domains. On short-range lattices the field on a spin, $h_i=\sum_k J_{ik}s_k$, is
a sum over its few neighbours only, its fluctuations do not vanish, and
replacing it by its average $\sum_k J_{ik}\langle s_k\rangle$ is a
genuine approximation, \emph{local mean-field theory}, to which we
return after the exact solution of the chain.

\begin{figure}[h]
\centering
\begin{tikzpicture}[xscale=1.9,yscale=1.9]
  \draw[->,gray] (-1.55,0) -- (1.65,0) node[right,black] {$m$};
  \draw[->,gray] (0,-1.15) -- (0,1.3) node[above,black] {$y$};
  \draw[very thick,gray] (-1.15,-1.15) -- (1.15,1.15)
    node[above left,font=\footnotesize,black] {$y=m$};
  \draw[thick,densely dotted,domain=-1.5:1.5,samples=80]
    plot (\x,{tanh(0.5*\x)});
  \node[font=\footnotesize,anchor=west] at (1.5,0.55)
    {$\beta<\beta_c$};
  \draw[thick,dashed,domain=-1.5:1.5,samples=80]
    plot (\x,{tanh(\x)});
  \node[font=\footnotesize,anchor=west] at (1.5,0.88)
    {$\beta=\beta_c=1$};
  \draw[thick,domain=-1.5:1.5,samples=80]
    plot (\x,{tanh(2*\x)});
  \node[font=\footnotesize,anchor=west] at (1.5,1.06)
    {$\beta>\beta_c$};
  \fill (0,0) circle (0.035);
  \node[font=\footnotesize,anchor=south east] at (-0.03,0.02) {$m^*=0$};
  \fill (0.9575,0.9575) circle (0.035);
  \fill (-0.9575,-0.9575) circle (0.035);
  \draw[dotted,gray] (0.9575,0.9575) -- (0.9575,0);
  \draw[dotted,gray] (-0.9575,-0.9575) -- (-0.9575,0);
  \node[font=\footnotesize,below] at (0.9575,0) {$m^*(\beta)$};
  \node[font=\footnotesize,below] at (-0.9575,0) {$-m^*(\beta)$};
\end{tikzpicture}
\caption{\small Fixed points of Eq.~\eqref{eq:mf}, which satisfy
$m=\tanh(\beta m)$, are the crossings of the straight line
$y=m$ (thick grey) with the curve $y=\tanh(\beta m)$, whose slope at the
origin is $\beta$; the curves are drawn for $\beta=\tfrac12$, $1$ and
$2$. For $\beta<\beta_c$ (dotted) the only crossing is $m^*=0$. At
$\beta=\beta_c=1$ (dashed) the curve is tangent to the line at the
origin. For $\beta>\beta_c$ (solid) $m^*=0$ becomes unstable and two
stable ordered states $\pm m^*(\beta)$ appear.}
\label{fig:selfconsistency}
\end{figure}
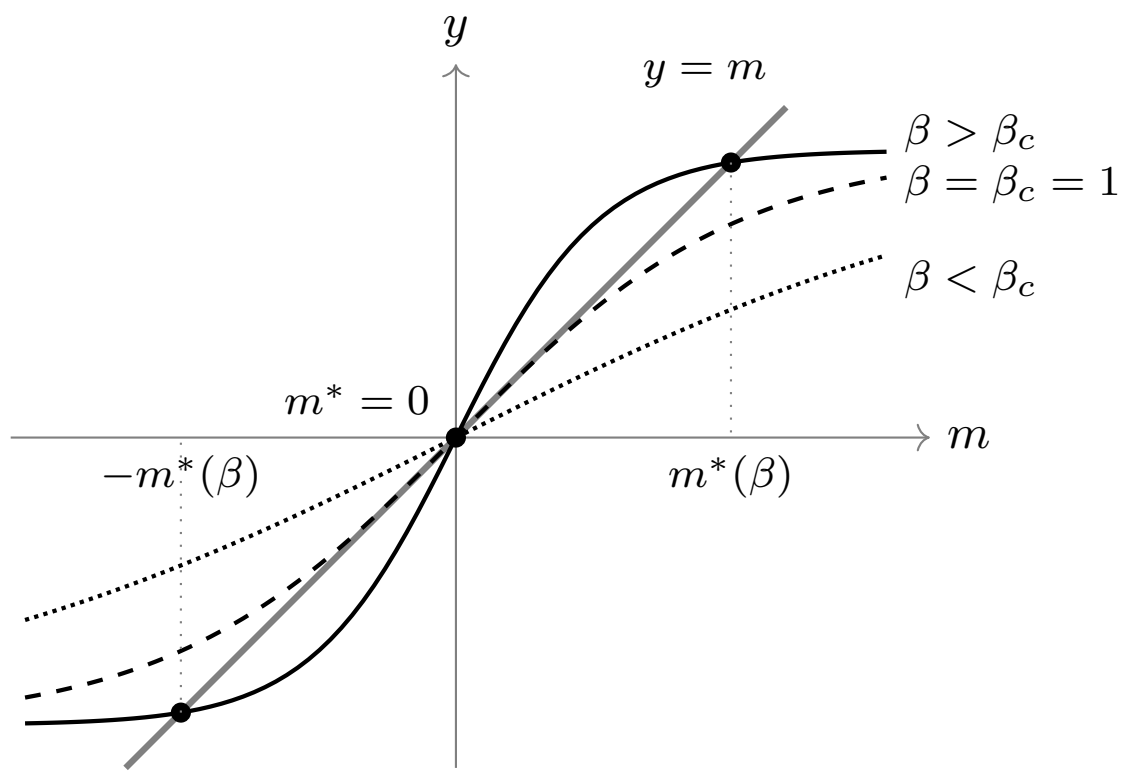

The fixed points of Eq.~\eqref{eq:mf} determine how the magnetisation
relaxes after a quench (Fig.~\ref{fig:selfconsistency}).

The fixed points satisfy $m^*=\tanh(\beta m^*)$. For small $m$,
$\tanh(\beta m)\simeq\beta m$, so the \emph{disordered state} $m^*=0$ is stable
for $\beta<1$ and unstable for $\beta>1$. The mean-field critical point
is therefore $\beta_c=1$, where the thermal energy $1/\beta$ just equals
the coupling $J=1$: at higher temperatures, thermal agitation outweighs
the coupling and the system stays disordered, at lower ones the coupling
wins and it orders.
Expanding $\tanh(\beta m) = \beta m - \tfrac{1}{3}(\beta m)^3 + \mathcal{O}(m^5)$,
\begin{equation}
  \frac{dm}{dt} \equiv f(m)
  = -(1-\beta)\,m - \frac{1}{3}(\beta m)^3 + \mathcal{O}(m^5),
  \label{eq:fm}
\end{equation}
so that close to $\beta=1$ the \emph{ordered states} are $\pm m^*(\beta)$
with $m^*(\beta)\simeq[3(\beta-1)/\beta^3]^{1/2}$. The expansion requires
$m$ to be small. This holds at late times after a quench to
$T_f\ge T_c$, once $m$ has decayed, but on the ordered side only close
to $T_c$: the ordered states, where the linear and cubic terms balance,
are small only when $\beta-1$ is. Three regimes follow, according to the
final temperature $T_f$ of the quench: the \emph{supercritical}
($T_f>T_c$), the \emph{subcritical} ($T_f<T_c$) and the \emph{critical}
($T_f=T_c$).

The two sides of the transition differ qualitatively, not just
quantitatively. In the supercritical regime, $T_f>T_c$ ($\beta<1$), only $m^*=0$ exists, and
it is stable. Linearizing Eq.~\eqref{eq:fm},
\begin{equation}
  m(t) \sim e^{-(1-\beta)t},
\end{equation}
an exponential decay with relaxation time $\tau=(1-\beta)^{-1}$. The
quench is undone: whatever the initial magnetisation, including its
sign, it decays and is forgotten, and the system returns to disorder.

In the subcritical regime, $T_f<T_c$ ($\beta>1$), the quench is instead
completed. The state $m^*=0$ has become unstable, and the system leaves
it in two stages. First a small initial magnetisation is amplified,
$m\sim m_0e^{(\beta-1)t}$, the mean-field form of a \emph{spinodal
instability}, with no barrier to cross and no seed to form. Then $m$
saturates at one of the ordered states $\pm m^*$, whose stability
follows from
\begin{align}
  f'(m^*) &= (\beta-1) - \beta^3 m^{*2} \nonumber\\
  &= (\beta-1) - 3(\beta-1) = -2(\beta-1) < 0,\label{eq:fprime}
\end{align}
so that both are approached exponentially, $|m-m^*|\sim e^{-2(\beta-1)t}$.
Which of the two is reached is decided by the sign of $m_0$: the initial
bias is not forgotten but selects the ordered state, and the up--down
symmetry is broken. A system started exactly at $m_0=0$ remains there as
$N\to\infty$, balanced on the unstable fixed point; at finite $N$ a
fluctuation tips it one way or the other.

What the two regimes share is a relaxation time that diverges as
$T_f\to T_c$, as $(1-\beta)^{-1}$ from above and $[2(\beta-1)]^{-1}$ from
below, so that at $T_c$ itself the exponential decay must give way to
something slower. In this critical case, $T_f=T_c$ ($\beta=1$), the
linear term vanishes and $dm/dt\simeq -\tfrac13 m^3$. Hence
\begin{equation}
  m(t) = \left[m_0^{-2} + \tfrac{2}{3}\, t\right]^{-1/2}
  \simeq \left(\tfrac{2}{3}\, t\right)^{-1/2},
\end{equation}
for $t \gg 3/(2 m_0^2)$, independent of $m_0$. The decay is a power law
rather than an exponential: this is \emph{critical slowing down}.

The subcritical results above rely on the expansion, Eq.~\eqref{eq:fm},
and so hold only near $T_c$. For arbitrary $T_f<T_c$ no expansion is
needed. Linearizing Eq.~\eqref{eq:mf} about $m^*=\tanh(\beta m^*)$, with
$m=m^*+\delta$ and $\mathrm{sech}^2(\beta m^*)=1-m^{*2}$, gives
$\dot\delta=-\lambda\delta$ with
\begin{equation}
  \lambda = 1-\beta\big(1-m^{*2}\big).
  \label{eq:lambda_sub}
\end{equation}
Close to $T_c$, $m^{*2}\simeq3(\beta-1)$ is small, and
$\lambda\simeq1-\beta+3(\beta-1)=2(\beta-1)$, which vanishes at $T_c$
and agrees with Eq.~\eqref{eq:fprime}. At low temperature $\beta$ is
large, so $\beta m^*\gg1$ and $m^*=\tanh(\beta m^*)$ is exponentially
close to $1$, with $1-m^{*2}\simeq4e^{-2\beta}$; the term
$\beta(1-m^{*2})$ then vanishes and $\lambda\to1$.

At $T_f=0$ itself, $\tanh(\beta m)\to{\rm sgn}(m)$, where
${\rm sgn}(m)=+1$ for $m>0$, $-1$ for $m<0$, and $0$ for $m=0$, and
Eq.~\eqref{eq:mf} becomes $\dot m=-m+{\rm sgn}(m)$. This can be solved
along the whole trajectory, not only near the fixed points:
\begin{equation}
  m(t) = {\rm sgn}(m_0)\big[\,1-(1-|m_0|)\,e^{-t}\big],
\end{equation}
so $m$ approaches $\pm1$ at rate $\lambda=1$, the $T_f\to0$ limit of
Eq.~\eqref{eq:lambda_sub}. The initial amplification stage has
disappeared: its rate $\beta-1$ diverges as $T_f\to0$, so any $m_0\neq0$,
however small, is pushed at once onto the approach to $\pm1$; while
$m_0=0$ stays put, since ${\rm sgn}(0)=0$.

The subcritical regime is thus, in a precise sense, a quench to zero
temperature: for every $T_f<T_c$ the flow has the same structure, an
unstable fixed point at $m=0$ and two stable ones at $\pm m^*$
approached exponentially. With no thermal noise in
Eq.~\eqref{eq:mf}, temperature only lowers $m^*$ and $\lambda$ from
$1$. The equivalence fails near $T_c$, where $\lambda\to0$. This is the mean-field form of a general principle of phase-ordering
kinetics: below $T_c$ the late-time dynamics is governed by its
zero-temperature limit, temperature merely renormalising quantities such
as $m^*$, so that a quench to $T_f=0$ is representative of any quench
into the ordered phase away from $T_c$.

The infinite one-dimensional chain, by contrast, can be treated
exactly, without any mean-field step. For the chain we insert the Glauber rate,
Eq.~\eqref{eq:glauber1d},
\begin{equation}
  w_i = \frac{1}{2}\left[1-\gamma\, s_i\,\frac{s_{i-1}+s_{i+1}}{2}\right],
  \qquad \gamma=\tanh(2\beta J),
\end{equation}
into Eq.~\eqref{eq:dsdt}:
\begin{align}
  \frac{d\langle s_i\rangle}{dt}
  &= -\left\langle s_i\left[1-\gamma\, s_i\,
     \frac{s_{i-1}+s_{i+1}}{2}\right]\right\rangle \nonumber\\
  &= -\langle s_i\rangle
     + \frac{\gamma}{2}\big[\langle s_{i-1}\rangle + \langle s_{i+1}\rangle\big].
  \label{eq:chain}
\end{align}
Unlike the mean-field case this equation is linear and closed, so it can
be solved exactly. No spatial average and no approximation is involved.
The equation closes because the one-dimensional rate is linear in the
neighbouring spins, so $\langle s_iw_i\rangle$ involves only
single-spin averages. For the initial condition
$\langle s_i(0)\rangle = \delta_{i,0}$ the solution is
\begin{equation}
  \langle s_i(t)\rangle = I_i(\gamma t)\, e^{-t},
\end{equation}
where $I_i$ is the \emph{modified Bessel function of the first kind}.
To see this, write $\langle s_i\rangle=e^{-t}u_i$, which removes the
$-\langle s_i\rangle$ term and leaves
$\dot u_i=\tfrac{\gamma}{2}(u_{i-1}+u_{i+1})$ with $u_i(0)=\delta_{i,0}$.
The function $u_i=I_i(\gamma t)$ satisfies both, since
$dI_n(u)/du=\tfrac12[I_{n-1}(u)+I_{n+1}(u)]$ and $I_n(0)=\delta_{n,0}$.
Using $I_i(u)\simeq e^{u}/\sqrt{2\pi u}$ for large $u$,
\begin{equation}
  \langle s_i(t)\rangle \sim
  \begin{cases}
    (2\pi\gamma t)^{-1/2}\, e^{-(1-\gamma)t}, & T\neq 0\ (\gamma<1),\\[2pt]
    (2\pi t)^{-1/2} \propto t^{-1/2}, & T=0\ (\gamma=1).
  \end{cases}
\end{equation}
The decay factor defines the relaxation time $\tau=1/(1-\gamma)$. As
$T\to0$, $\gamma\to1$ and $\tau\simeq\tfrac12e^{4\beta J}$ diverges: this
is critical slowing down at the critical point $T_c=0$ of the chain, and
at $T=0$ itself the decay becomes the power law $t^{-1/2}$.

In the language of the Curie--Weiss model, every quench of the chain to
$T>0$ is supercritical: since $T_c=0$, there is no ordered phase, and
hence neither a spinodal instability nor nucleation. A uniform initial
magnetisation $m_0$ stays uniform, and setting $\langle s_i\rangle=m(t)$
for all $i$ in Eq.~\eqref{eq:chain} gives $\dot m=-(1-\gamma)m$, so that
it simply decays, $m(t)=m_0e^{-(1-\gamma)t}$, exactly and at the same
rate $1-\gamma$. The only critical quench is to $T=0$ itself, and it
differs from the mean-field one. At the mean-field critical point, the
magnetisation decays, if only as a power law---$m\sim t^{-1/2}$; at $T=T_c=0$,
in the chain it does not decay at all, since with $\gamma=1$ the total
$\sum_i\langle s_i\rangle$ is conserved and a uniform magnetisation stays
at $m_0$. The $t^{-1/2}$ found above refers to something else: the
magnetisation at a single site when an initially localised magnetisation
spreads out diffusively, its total remaining fixed. Physically, a chain
quenched to $T=0$ breaks into domains whose walls random-walk and
annihilate on contact, so that the domains \emph{coarsen} without bound
while the average magnetisation keeps its initial value, zero for a
disordered start.

A second view of the same behaviour comes from the continuum limit.
With $m_i\equiv\langle s_i\rangle$, Eq.~\eqref{eq:chain} can be
rearranged as
\begin{align}
  \dot m_i &= -m_i + \frac{\gamma}{2}\big[m_{i-1}+m_{i+1}\big] \nonumber\\
  &= -(1-\gamma)\,m_i
     + \frac{\gamma}{2}\big[m_{i-1}-2m_i+m_{i+1}\big].
\end{align}
The bracket is the discrete second derivative. In the continuum limit
(unit lattice spacing) the local magnetisation $m_i(t)$ becomes a smooth
field $m(x,t)$, to be distinguished from the spatially averaged $m(t)$ of
the mean-field theory, and
\begin{equation}
  \frac{\partial m(x,t)}{\partial t}
  = -(1-\gamma)\,m(x,t) + \frac{\gamma}{2}\,\frac{\partial^2 m(x,t)}{\partial x^2},
\end{equation}
a diffusion equation with diffusion coefficient $\gamma/2$ and a decay
term. Writing $m=e^{-(1-\gamma)t}u$ removes the decay term and leaves the
ordinary diffusion equation for $u$, so for a localised initial
magnetisation, $m(x,0)=\delta(x)$, the solution is
\begin{equation}
  m(x,t)=\frac{e^{-(1-\gamma)t}}{\sqrt{2\pi\gamma t}}
  \exp\left(-\frac{x^2}{2\gamma t}\right),
  \label{eq:chain_continuum_sol}
\end{equation}
and for a general initial profile $m(x,t)$ is the convolution of
$m(x,0)$ with this kernel. At $x=0$, it reproduces the long-time form of
the exact lattice solution above. Eq.~\eqref{eq:chain_continuum_sol}
displays the two effects of Glauber dynamics on the chain separately.
The Gaussian factor spreads the magnetisation over a width
$\sqrt{\gamma t}$, which is the random walk of domain walls, and the
exponential factor makes it decay uniformly at the rate $1-\gamma$, which
is the thermal creation of domain walls that destroys order. Critical
slowing down at $T_c=0$ is thus simply the disappearance of this decay
term as $T\to0$, which leaves only the diffusive spreading.

The chain is solvable because its rate is linear in the neighbouring
spins. On a lattice in $d\geq2$ this is no longer so, and an
approximation is unavoidable. To see what it involves we need to be
precise about what is being averaged. Two different averages appear in
these notes. The \emph{ensemble average} $\langle\cdot\rangle$ is taken
over the probability $P(C,t)$ of configurations and keeps the site
index: $m_i=\langle s_i\rangle$ is the mean of spin $i$ over many
realisations of the dynamics. Every equation of motion we have written,
starting from Eq.~\eqref{eq:dsdt}, is for an ensemble average. The
\emph{spatial average} $\frac1N\sum_k$ is taken over all sites and
gives a single number for the whole system. The mean-field
approximation is a separate step: it replaces the average of a
nonlinear function of the spins by the same function of their
averages, $\langle\tanh(\beta h_i)\rangle\to\tanh(\beta\langle
h_i\rangle)$, and so neglects correlations between spins.

The treatments differ in how they combine these. The The \emph{global} mean-field
theory takes both the ensemble and the spatial average, and is exact
for all-to-all coupling. The chain takes only the ensemble average and
needs no approximation. On a lattice in $d\geq2$ the rate contains the
$\tanh$ of a sum of several spins, so the equation for
$\langle s_i\rangle$ does not close. The \emph{local} mean-field theory
applies the mean-field step but keeps the site index, replacing each
neighbouring spin by its own ensemble average $m_k=\langle s_k\rangle$
rather than by the system average. It neglects correlations between
neighbours but retains the spatial variation of the magnetisation.
With the general rate, Eq.~\eqref{eq:glauber_general},
Eq.~\eqref{eq:dsdt} becomes
\begin{equation}
  \frac{dm_i}{dt}
  = -m_i + \tanh\!\Big(\beta\sum_k J_{ik}\,m_k\Big).
  \label{eq:glauber_lmf}
\end{equation}
For a uniform coupling $J$ between nearest neighbours, the sum is $J$
times the sum of $m_k$ over the neighbours of $i$.

In one dimension Eq.~\eqref{eq:glauber_lmf} can be compared directly
with the exact Eq.~\eqref{eq:chain}, which shows concretely what the
mean-field step neglects. Both start from Eq.~\eqref{eq:dsdt} with the
Glauber rate and differ only in how the average of the $\tanh$ is
taken. In the chain the field is $J(s_{i-1}+s_{i+1})$, and for spins
$\pm1$ the identity
\begin{equation}
  \tanh\!\big(\beta J(s_{i-1}+s_{i+1})\big)
  = \frac{\gamma}{2}\,(s_{i-1}+s_{i+1})
\end{equation}
holds configuration by configuration. It is the origin of the linear
rate, Eq.~\eqref{eq:glauber1d}, and its average gives
Eq.~\eqref{eq:chain} exactly. Mean field instead linearises the
$\tanh$ only for small $m$, and the difference is one of coupling:
\begin{align}
  \frac{dm_i}{dt} &= -m_i+\frac{\gamma}{2}\,(m_{i-1}+m_{i+1}),
  \nonumber\\
  \frac{dm_i}{dt} &\simeq -m_i+\beta J\,(m_{i-1}+m_{i+1}),
\end{align}
exact and mean-field respectively.
Since $\tfrac12\tanh(2x)<x$ for every $x>0$, local mean-field theory
overestimates the coupling between neighbours. The consequence is
visible in the uniform mode. In the exact equation it relaxes at rate
$1-\gamma$, positive at every $T>0$, so the disordered state is always
stable, $T_c=0$, and the rate vanishes only at $T=0$, the critical
slowing down found above. In the mean-field equation it relaxes at
rate $1-2\beta J$, which becomes negative below $\kB T=2J$, so local
mean-field theory predicts a spurious transition at $\kB T_c=zJ$, with
$z=2$ the number of neighbours in the chain. The overestimate is the
effect of the neglected correlations: neighbouring spins are
correlated, their fluctuations partly cancel inside the $\tanh$, and
the exact average is weaker than the $\tanh$ of the averages. The two
equations also differ at $T=0$. The exact one becomes the diffusive
equation $dm_i/dt=-m_i+\tfrac12(m_{i-1}+m_{i+1})$ found above, whereas
the mean-field one becomes $dm_i/dt=-m_i+{\rm sgn}(m_{i-1}+m_{i+1})$,
which is nonlinear and drives every site to $\pm1$. In $d\geq2$ no
exact counterpart of Eq.~\eqref{eq:chain} exists, and
Eq.~\eqref{eq:glauber_lmf} is the approximation one is forced to use.

We conclude this section by noting how special the exact solution of
the chain is. Even among single-flip rates it singles out Glauber's
choice. With the Metropolis rate, Eq.~\eqref{eq:metropolis1d}, the
equation for $\langle s_i\rangle$ acquires a term
$\delta\,\langle s_{i-1}s_is_{i+1}\rangle$, which couples the
magnetisation to a three-spin correlation, and it no longer closes; only
for $\delta=0$ does this term vanish. Unlike the Glauber chain, Kawasaki
dynamics admits no exact calculation, even in one dimension. The
exchange rate depends on the spins surrounding both members of the
pair, so the equation for $\langle s_i\rangle$ involves averages of
products of three and more spins, the equations for those involve still
larger products, and at any finite temperature the hierarchy of moment
equations never closes. Mean-field theory is also less direct than for
Glauber dynamics. Its global form carries no information. The total
magnetisation is conserved by every exchange, so $dm/dt=0$ identically,
whereas everything that happens under Kawasaki dynamics is a
redistribution of magnetisation in space, which a description in terms
of $m$ alone cannot see. A local form, which does follow this
redistribution, leads to considerably more involved equations.

\subsection{TASEP: Periodic Boundaries}

The exclusion process is a minimal lattice model of transport in which particles interact only through excluded volume, so that each site holds at most one particle. It was first proposed in 1968 (MacDonald, Gibbs and Pipkin) to model ribosomes translating mRNA along a one-dimensional chain, and was introduced independently in 1970 by Spitzer, who gave it its name, as a many-particle extension of the lattice random walk on general lattices. The name of the \emph{Totally ASymmetric Exclusion Process (TASEP)} spells out its rules (Fig.~\ref{fig:moves}): the hopping is \emph{asymmetric}, since particles prefer one direction over the other; \emph{totally} so, since they move only in that direction and never back; and it obeys \emph{exclusion}, since no two particles can occupy the same site.
\begin{center}
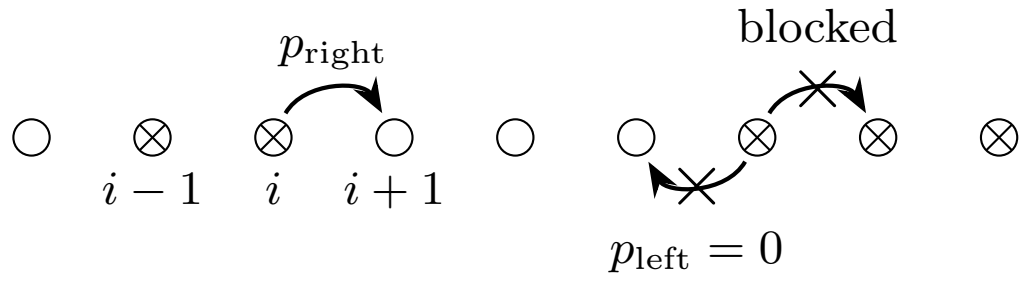
\begin{minipage}{\linewidth}\centering
\begin{tikzpicture}[scale=0.8]
  \foreach \x/\o in {0/0,1/1,2/1,3/0,4/0,5/0,6/1,7/1,8/1}{%
    \draw (\x,0) circle (0.15);
    \ifnum\o=1
      \draw (\x-0.106,-0.106) -- (\x+0.106,0.106);
      \draw (\x-0.106,0.106) -- (\x+0.106,-0.106);
    \fi}
  \node[below=3pt] at (1,0) {\small $i-1$};
  \node[below=3pt] at (2,0) {\small $i$};
  \node[below=3pt] at (3,0) {\small $i+1$};
  \draw[-{Stealth},thick] (2.1,0.2) to[bend left=60]
    node[above]{\small $p_{\text{right}}$} (2.9,0.2);
  \draw[-{Stealth},thick] (6.1,0.2) to[bend left=60]
    node[midway]{\Large $\times$}
    node[above=5pt]{\small blocked} (6.9,0.2);
  \draw[-{Stealth},thick] (5.9,-0.2) to[bend left=60]
    node[midway]{\Large $\times$}
    node[below=5pt]{\small $p_{\text{left}}=0$} (5.1,-0.2);
\end{tikzpicture}
\captionof{figure}{\small Allowed and forbidden moves in TASEP ($\bigcirc$ empty site, $\bigotimes$ occupied site). A particle at site $i$ hops to site $i+1$ with probability $p_{\text{right}}$ only if $i+1$ is empty; a hop onto an occupied site is blocked by exclusion, and hops to the left are forbidden ($p_{\text{left}}=0$).}
\label{fig:moves}
\end{minipage}\end{center}
TASEP can be defined on multi-lane systems, networks and higher-dimensional lattices, but its one-dimensional version is by far the most studied, since it is there that exact solutions are available. One-dimensional TASEP has become a cornerstone of nonequilibrium statistical mechanics: it sustains a steady current and breaks detailed balance, yet remains exactly solvable. Simple as it is, it shows boundary-driven phase transitions in one dimension, and it captures a wide range of transport phenomena, from molecular motors to road traffic.

To understand the rules and general vocabulary of the TASEP dynamics, consider a one-dimensional lattice of $L$ sites, each either empty or occupied by a single particle. A pair of adjacent sites $(i,i+1)$ is called a \emph{bond}. Two boundary conditions are commonly used. With \emph{periodic} boundary conditions, the sites are arranged on a ring, there are $K=L$ bonds, and the number of particles $N<L$ is conserved. With \emph{open} boundary conditions, particles enter at site $1$ with probability (or rate) $\alpha$ if it is empty, and leave from site $L$ with probability (or rate) $\beta$ if it is occupied; counting the entry bond $(0,1)$ and the exit bond $(L,L+1)$ along with the $L-1$ bulk bonds, there are $K=L+1$ bonds, and $N$ fluctuates. It is now important to specify the exact update mechanism. Two immediate mechanisms come to mind:
\begin{itemize}
\item \emph{Parallel updating at fixed discrete time steps:} at each time step $t\to t+\Delta t$, all bonds are updated simultaneously, using the configuration at time $t$. Every particle whose right neighbour is empty hops to the right with probability $p$, independently of the others; with open boundaries, entry and exit are attempted in the same step with probabilities $\alpha$ and $\beta$. A particle cannot move into a site vacated in the same step; it may do so only at the next step.
\item \emph{Random sequential updating at fixed discrete time steps:} at each time step $t\to t+\Delta t$, a single bond is chosen uniformly at random from the $K$ bonds. For a bulk bond $(i,i+1)$, if site $i$ is occupied and site $i+1$ is empty, the particle hops to the right; otherwise nothing happens. For the entry (exit) bond, a particle is inserted at site $1$ (removed from site $L$) with probability $\alpha$ ($\beta$), provided the site is empty (occupied). Choosing $\Delta t=1/K$ ensures that every bond is attempted once per unit time on average. $K$ consecutive attempts constitute one \emph{Monte Carlo sweep}, the natural unit of time. Since bonds are chosen at random, a given bond is not necessarily picked exactly once per sweep. The probability that it is picked $k$ times is
\begin{equation*}
  P(k)={}^{K}C_{k}\left(\frac1K\right)^k\left(1-\frac1K\right)^{K-k}
  \;\xrightarrow{K\to\infty}\;\frac{1^k e^{-1}}{k!},
\end{equation*}
so for large $K$ a bond is missed with probability $P(0)=e^{-1}\approx 0.37$ and picked more than once with probability $1-P(0)-P(1)=1-2e^{-1}\approx 0.26$.
\end{itemize}
Both schemes define discrete-time Markov chains: the configuration at $t+\Delta t$ depends only on that at $t$ and on freshly drawn random numbers, with no memory of earlier history. Henceforth, we shall consider only sequential updating.

Random sequential updating at fixed discrete time steps, with exponentially distributed rather than fixed time increments, becomes the \emph{continuous-time Poisson update}, which we explain as follows. Each bond $(i,i+1)$ carries an independent clock that fires at random times with rate $\lambda$ ($\lambda=1$ in TASEP). When the clock of bond $(i,i+1)$ fires, the particle at site $i$ hops to $i+1$ if site $i$ is occupied and site $i+1$ is empty; otherwise nothing happens. Equivalently, every particle whose right neighbour is empty hops at rate $\lambda$, while a blocked particle stays put.

The defining assumption is that in a small interval $dt$ a clock fires with probability $\lambda\,dt$, independently of how long ago it last fired. Let $P_k(t)$ be the probability of exactly $k$ firings up to time $t$. There are $k$ firings at time $t+dt$ either if there were $k$ at time $t$ and none in $dt$, or if there were $k-1$ and one in $dt$; two firings in $dt$ occur with probability $\mathcal{O}(dt^2)$ and drop out. Thus
\begin{equation}
  P_k(t+dt) = P_k(t)\,(1-\lambda\,dt) + P_{k-1}(t)\,\lambda\,dt,
\end{equation}
\begin{equation}
  \Rightarrow\; \frac{dP_k}{dt} = -\lambda P_k + \lambda P_{k-1},
  \qquad P_k(0) = \delta_{k0}.
\end{equation}
Substituting $P_k = Q_k\,e^{-\lambda t}$ removes the decay term, and integrating the resulting equation successively gives $Q_k=(\lambda t)^k/k!$. Therefore
\begin{equation}
  P_k(t) = \frac{(\lambda t)^k}{k!}\,e^{-\lambda t},
\end{equation}
the Poisson distribution with mean $\mathbb{E}[k]=\lambda t$, which gives the update its name.

In particular, starting just after a firing, the waiting time $\tau$ to the next firing exceeds $t$ exactly when no firing occurs in $[0,t]$, so $P(\tau>t)=P_0(t)=e^{-\lambda t}$. Hence $P(\tau\le t)=1-e^{-\lambda t}$, and the probability density is $ \frac{d}{d\tau}\left(1 - e^{-\lambda\tau}\right) = \lambda e^{-\lambda\tau}$: the exponential distribution, the unique continuous \emph{memoryless} distribution, i.e.,
\begin{equation*}
  P(\tau > t+s \mid \tau > s) = P(\tau > t),
\end{equation*}
so a bond that has already waited a time $s$ is no closer to firing than a fresh one. Memorylessness, a property of the waiting-time random variable, makes TASEP a Markov process, a property of the stochastic process as a whole: its future depends only on the current configuration, not on its history.

Considering independent clocks $j$ with rates $\lambda_j$ and waiting times $\tau_j$, the first firing occurs at $T=\min_j\tau_j$, and the probability that none has fired by time $t$ is
\begin{equation}
  P(T>t)=\prod_j P(\tau_j>t)=\prod_j e^{-\lambda_j t}=e^{-\Lambda t},
\end{equation}
with $\Lambda\equiv\sum_j\lambda_j$; so the first firing occurs at rate $\Lambda$, and it belongs to clock $j$ with probability $\lambda_j/\Lambda$. On a ring, the $L$ bond clocks all have rate $\lambda$, so $\Lambda=L\lambda$ and each firing is equally likely to belong to any bond: the Poisson update is equivalent to choosing a bond uniformly at random at each event, which connects it to the random sequential update introduced earlier. With open boundaries the rates differ ($\lambda$ in the bulk, $\alpha$ at the entry, $\beta$ at the exit), and the equivalence requires choosing each event with probability $\lambda_j/\Lambda$, where $\lambda_j\in\{\lambda,\alpha,\beta\}$ and $\Lambda=(L-1)\lambda+\alpha+\beta$. 

Since both select bonds with the same probabilities, the discrete-time and continuous-time random sequential updates produce the same sequence of moves and hence agree in their stationary and long-time behaviour; they differ only in how time is counted. One may therefore wonder why the seemingly more complicated Poisson update is needed at all. There are several reasons. First, many realistic systems, such as ribosomes, molecular motors and vehicles, hop asynchronously at random times with fixed rates, which forces exponentially distributed waiting times. Second, for infinite systems the discrete-time update is not even defined, since a bond cannot be picked uniformly at random from infinitely many. Third, on short time scales, of order $1/L$ or less, the deterministic time increments of the discrete-time update distort fluctuations and correlations, whereas the Poisson update remains exact at all time scales.

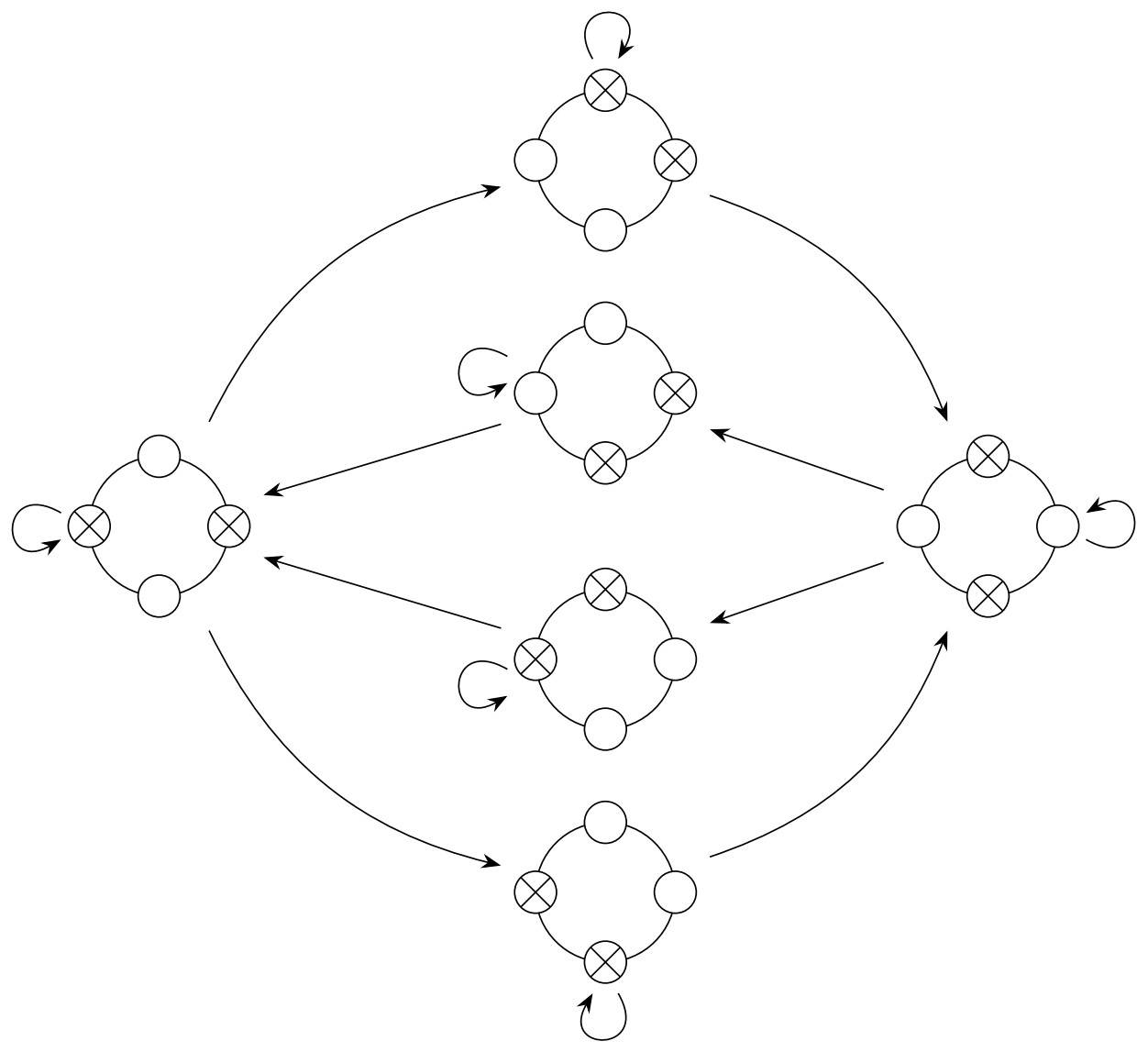
\begin{figure}[h]\centering
\scalebox{0.8}{%
\begin{tikzpicture}[x=1.15cm,y=0.6cm,>={Stealth[length=5pt]}]
  \node (B13) at (6.5,0)    {\ring1010};
  \node (B24) at (0,0)      {\ring0101};
  \node (A23) at (3.5,2.0)  {\ring0110};
  \node (A14) at (3.5,-2.0) {\ring1001};
  \node (A12) at (3.5,5.5)  {\ring1100};
  \node (A34) at (3.5,-5.5) {\ring0011};
  \draw[->] (B13) -- (A23);
  \draw[->] (B13) -- (A14);
  \draw[->] (A23) -- (B24);
  \draw[->] (A14) -- (B24);
  \draw[->] (B24) to[bend left=25] (A12);
  \draw[->] (A12) to[bend left=25] (B13);
  \draw[->] (B24) to[bend right=25] (A34);
  \draw[->] (A34) to[bend right=25] (B13);
  \begin{scope}[looseness=7]
    \draw[->] (A12.north) ++(-0.1,-0.05) to[out=120,in=60] ++(0.2,0);
    \draw[->] (A23.west) ++(0.05,0.55) to[out=150,in=210] ++(0,-0.4);
    \draw[->] (A14.west) ++(0.05,-0.15) to[out=150,in=210] ++(0,-0.4);
    \draw[->] (A34.south) ++(0.1,0.05) to[out=-60,in=-120] ++(-0.2,0);
    \draw[->] (B13.east) ++(-0.05,-0.2) to[out=-30,in=30] ++(0,0.4);
    \draw[->] (B24.west) ++(0.05,0.2) to[out=150,in=210] ++(0,-0.4);
  \end{scope}
\end{tikzpicture}}
\caption{\small TASEP transitions for $L=4$, $N=2$ under the discrete-time random sequential update. All ${}^{4}C_{2}=6$ configurations on a ring ($\bigotimes$: occupied site, $\bigcirc$: empty site; hopping clockwise). Every configuration has as many incoming as outgoing transitions; self loops denote the probability of remaining in the same configuration.}
\label{fig:ring}
\end{figure}

Next, in the rest of this section, let us confine ourselves to periodic boundary conditions. Consider an illustrative Markov chain for TASEP with $K=L=4$ and $N=2$ under the discrete-time random sequential update. The possible configurations (and the transitions from one to another) are shown in Fig.~\ref{fig:ring}.
We note a few important features of this chain: (i) it is irreducible, since every configuration can be reached from every other; (ii) it is finite and irreducible, hence all its states are positive recurrent and it has a unique stationary distribution; and (iii) it is aperiodic, since a self loop gives its state period 1, and period is a class property. (Without the self loops the chain would in fact be periodic with period 4.) Together, (i)--(iii) ensure that the stationary distribution is also the limiting distribution. In continuous time (Poisson update), the self loops play no role. A clock on a blocked bond still fires, but the configuration does not change; such a firing is not a jump of the Markov chain, and its rate would enter the master equation once as a gain and once as a loss and cancel. Staying in a configuration is then simply the waiting time until the next actual jump, which is exponentially distributed. Moreover, periodicity cannot arise in continuous time, since these waiting times can be arbitrarily short or long, so every reachable configuration has a positive probability at every time $t>0$. Irreducibility and finiteness alone therefore guarantee that the unique stationary distribution is also the limiting one.

Mathematically, the evolution of the probability $p(C,t)$ of finding the system in configuration $C$ at time $t$ is governed by the master equation
\begin{equation}
  \frac{{dp}(C,t)}{dt} = \sum_{C'\neq C} R(C\,|\,C')\,p(C',t) - \sum_{C'\neq C} R(C'\,|\,C)\,p(C,t), \label{eq:tpbcmc}
\end{equation}
where $R(C\,|\,C')$ denotes the transition rate from configuration $C'$ to configuration $C$. For TASEP on a ring, $R(C\,|\,C')=1$ if $C$ can be obtained from $C'$ by a single allowed hop, and $R(C\,|\,C')=0$ otherwise. In the steady state, $d{p}(C,t) /dt=0$ for every $C$, i.e., the total flow into each configuration balances the flow out of it; in steady state we denote the distribution as $p_s(C)$. We note that the rates in  Eq.~\eqref{eq:tpbcmc} are independent of time: the hopping rule is the same at all times and depends only on the configuration. This time-homogeneity, together with the Markov property, is what allows a time-independent steady state.

We guess that all configurations are equally probable, i.e.,
\begin{equation}
  p_s(C)=\frac{1}{{}^LC_{N}}.
\end{equation}
(The configuration $C$ should not be confused with the $C$ in the binomial coefficient ${}^{L}C_{N}$.)
The guess is correct if it satisfies the steady-state condition
\begin{equation}
  \sum_{C'\neq C} R(C\,|\,C')\,p_s(C') = \sum_{C'\neq C} R(C'\,|\,C)\,p_s(C)
  \label{eq:R=R}
\end{equation}
for every configuration $C$. Since $p_s$ is constant, it cancels from both sides, and since each $R$ is either $0$ or $1$, we have to show that the number of configurations leading to $C$ in a single hop equals the number of configurations reachable from $C$ in a single hop.

To count these, note that the particles in any configuration $C$ group into \emph{blocks} of consecutively occupied sites, separated by empty sites. A segment of such a configuration is shown below:
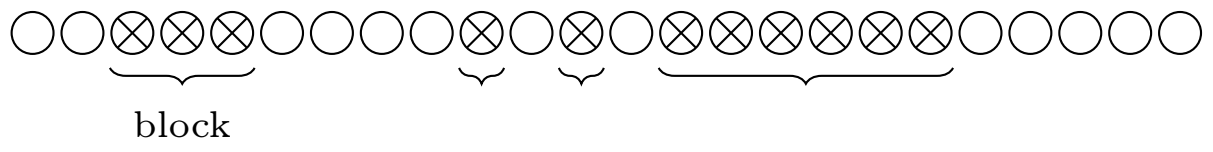
\begin{center}
\begin{tikzpicture}[x=0.33cm,y=0.33cm]
  \foreach \o [count=\x] in {0,0,1,1,1,0,0,0,0,1,0,1,0,1,1,1,1,1,1,0,0,0,0,0}{%
    \draw (\x,0) circle (0.42);
    \ifnum\o=1
      \draw (\x-0.3,-0.3) -- (\x+0.3,0.3);
      \draw (\x-0.3,0.3) -- (\x+0.3,-0.3);
    \fi}
  \foreach \a/\b in {3/5,10/10,12/12,14/19}{%
    \draw[decorate,decoration={brace,mirror,amplitude=3pt}]
      (\a-0.45,-0.7) -- (\b+0.45,-0.7);}
  \node[below] at (4,-1.2) {\scriptsize block};
\end{tikzpicture}
\end{center}
Inside a block no particle can move, since the site to its right is occupied. A configuration $C'$ leading to $C$ must therefore coincide with $C$ except that the left-most particle of one block is displaced one site to the left, into the empty site preceding that block; hence there are exactly as many such $C'$ as there are blocks. Similarly, a configuration $C'$ reachable from $C$ coincides with $C$ except that the right-most particle of one block is displaced one site to the right, into the empty site following it; hence there are again exactly as many such $C'$ as there are blocks. On a ring every block has exactly one left end and one right end, so the two counts coincide, and the global balance condition~\eqref{eq:R=R} holds. Since the chain is irreducible, $p_s$ is the unique stationary distribution.

Detailed balance, $R(C\,|\,C')\,p_s(C')=R(C'\,|\,C)\,p_s(C)$, is nevertheless violated: particles never hop backwards, so whenever $R(C\,|\,C')=1$ we have $R(C'\,|\,C)=0$, and detailed balance would require $p_s(C')=0$, which is impossible. The steady state therefore carries a nonzero probability current and is not an equilibrium distribution; TASEP always gives rise to a nonequilibrium steady state (NESS).

Now let us find how the particle distribution changes under the dynamics. The relevant variable is the \emph{occupation number} $n_i$, which equals $1$ if site $i$ is occupied and $0$ if it is empty. In the ensemble picture, consider $M$ copies of the ring, all with the same $L$ and $N$, and let $m_i(t)$ be the number of copies in which site $i$ is occupied at time $t$. The probability that site $i$ is occupied is then
\begin{equation}
  \rho_i(t)=\lim_{M\to\infty}\frac{m_i(t)}{M},
\end{equation}
and the ensemble average of the occupation number is
\begin{equation}
  \langle n_i(t)\rangle = 1\times \rho_i(t) + 0\times\left[1-\rho_i(t)\right] = \rho_i(t).
\end{equation}
Since a site holds at most one particle, $\rho_i(t)$ is also the mean number of particles at site $i$, i.e., the \emph{local density}. In terms of the configuration probabilities, $\langle n_i(t)\rangle=\sum_C n_i(C)\,p(C,t)$, where $n_i(C)$ is the occupation of site $i$ in configuration $C$; that is, $\rho_i(t)$ is the marginal of $p(C,t)$ at site $i$. Averaging the local density over the ring gives the \emph{global density}, which is fixed by particle conservation at all times:
\begin{equation}
  \frac1L\sum_{i=1}^{L}\rho_i(t)=\frac{1}{L}\avg{\sum_{i=1}^{L}n_i}=\frac{N}{L}\equiv\rho.
\end{equation}
In the steady state, $p(C,t)\to p_s(C)$, and we denote the corresponding steady-state local density by $\rho_{i,s}\equiv\lim_{t\to\infty}\rho_i(t)$, and the steady-state average by $\langle n_i\rangle_s\equiv \rho_{i,s}$. Then
\begin{equation}
  \rho_{i,s}=\sum_C n_i(C)\,p_s(C)=\frac{{}^{L-1}C_{N-1}}{{}^{L}C_{N}}=\frac{N}{L}=\rho,
\end{equation}
since the sum counts the configurations with site $i$ occupied, i.e., with the remaining $N-1$ particles distributed over the other $L-1$ sites. In the steady state, the local density thus equals the global density at every site. This uniformity is a manifestation of the translation invariance of the ring, together with the uniqueness of the steady state; away from the steady state, the local densities may differ from site to site, but their average always remains $\rho$.

Let us now find the temporal evolution of the occupation number. In the random sequential update, a single bond is chosen in each time step $\Delta t$, so the occupation $n_i$ of site $i$ can change only if one of the two bonds containing site $i$ is chosen. Consider first the bond $(i,i+1)$:
\begin{center}
\small
\renewcommand{\arraystretch}{1.4}\begin{tabular}{c|c|c|c|c}
time & \multicolumn{4}{c}{\shortstack{Bond [adjacent pair $(i,i+1)$]\\ configuration update}} \\
\hline
$t$          & $\overset{i}{\bigcirc}\;\overset{i+1}{\bigcirc}$ & $\bigcirc\;\bigotimes$ & $\bigotimes\;\bigcirc$ & $\bigotimes\;\bigotimes$ \\
\hline
$t+\Delta t$ & $\bigcirc\;\bigcirc$ & $\bigcirc\;\bigotimes$ & $\bigcirc\;\bigotimes$ & $\bigotimes\;\bigotimes$ \\
\end{tabular}
\end{center}
Out of the four possible bond configurations, only $\bigotimes\bigcirc$ changes: the particle at site $i$ hops to site $i+1$. With $n=0$ for a vacant site ($\bigcirc$) and $n=1$ for an occupied site ($\bigotimes$), the update of site $i$ can be written compactly as
\begin{equation}
  n_i(t+\Delta t) = n_i(t)\,n_{i+1}(t),
\end{equation}
i.e., site $i$ remains occupied only if it was occupied and its right neighbour was blocking the hop.

Similarly, for the bond $(i-1,i)$:
\begin{center}
\small
\renewcommand{\arraystretch}{1.4}\begin{tabular}{c|c|c|c|c}
time & \multicolumn{4}{c}{\shortstack{Bond [adjacent pair $(i-1,i)$]\\ configuration update}} \\
\hline
$t$          & $\overset{i-1}{\bigcirc}\;\overset{i}{\bigcirc}$ & $\bigcirc\;\bigotimes$ & $\bigotimes\;\bigcirc$ & $\bigotimes\;\bigotimes$ \\
\hline
$t+\Delta t$ & $\bigcirc\;\bigcirc$ & $\bigcirc\;\bigotimes$ & $\bigcirc\;\bigotimes$ & $\bigotimes\;\bigotimes$ \\
\end{tabular}
\end{center}
Again only $\bigotimes\bigcirc$ changes, now with the particle at site $i-1$ hopping into site $i$. Site $i$ is occupied after the update if it was already occupied, or if it was empty and site $i-1$ was occupied:
\begin{equation}
  n_i(t+\Delta t) = n_i(t) + n_{i-1}(t)\,\big[1-n_i(t)\big].
\end{equation}
If neither bond is chosen, 
\begin{equation}
  n_i(t+\Delta t) = n_i(t).
\end{equation}
Since each of the $L$ bonds is chosen with probability $1/L=\Delta t$, the three cases occur with probabilities
\begin{equation*}
  \underbrace{\Delta t,}_{\text{bond }(i,i+1)}\quad
  \underbrace{\Delta t,}_{\text{bond }(i-1,i)}\quad
  \underbrace{1-2\Delta t.}_{\text{neither}}
\end{equation*}
These probabilities are exact for the fixed-step update, since only one bond is chosen per step. In the Poisson (continuous-time) update they hold only up to corrections of order $\Delta t^2$, arising from the possibility of two firings within the same interval $\Delta t$. Weighting the three outcomes by their probabilities and averaging over the ensemble,
\begin{align}
  \avg{n_i(t+\Delta t)} ={}& \Delta t\,\avg{n_i n_{i+1}}
  + \Delta t\,\avg{n_i + n_{i-1}(1-n_i)} \nonumber\\
  &+ (1-2\Delta t)\,\avg{n_i},\label{eq:nitasepm}
\end{align}
where all occupation numbers on the right are evaluated at time $t$. Rearranging, in the limit $\Delta t\to0$,
\begin{equation}
   \frac{d\avg{n_i(t)}}{dt}=\frac{d\rho_i(t)}{dt} = J_{i-1,i}-J_{i,i+1},
\end{equation}
where $J_{i-1,i}\equiv \avg{n_{i-1}(1-n_i)}$ and $J_{i,i+1}\equiv\avg{n_i(1-n_{i+1})}$ are the particle currents between $i-1$ and $i$, and between $i$ and $i+1$, respectively.

In steady state, of course, $J_{i-1,i}=J_{i,i+1}=J$ (say), and explicit calculation of $\avg{n_{i-1}(1-n_i)}$ yields:
\begin{equation}
  J=\avg{n_{i-1}(1-n_i)}_s=\sum_C n_{i-1}(C)\,\big[1-n_i(C)\big]\,p_s(C).
\end{equation}
The product $n_{i-1}(1-n_i)$ equals $1$ only if site $i-1$ is occupied and site $i$ is empty, and $0$ otherwise. The sum therefore counts the configurations in which site $i-1$ holds a particle and site $i$ is empty, with the remaining $N-1$ particles distributed over the other $L-2$ sites:
\begin{equation}
  J=J_{i-1,i}=\frac{{}^{L-2}C_{N-1}}{{}^{L}C_{N}}=\frac{N(L-N)}{L(L-1)}.
\end{equation}
The same result follows for $J_{i,i+1}$ from $\avg{n_i(1-n_{i+1})}_s$, as it should. In terms of the density $\rho=N/L$,
\begin{equation}
  J=\frac{N\left(1-\frac{N}{L}\right)L}{L(L-1)}=\rho(1-\rho)\,\frac{L}{L-1}\;\xrightarrow{L\to\infty}\;\rho(1-\rho).
\end{equation}
Neighbouring sites are weakly anticorrelated,
\begin{equation}
  \avg{n_{i-1}n_i}_s-\avg{n_{i-1}}_s\avg{n_i}_s=-\frac{\rho(1-\rho)}{L-1},
\end{equation}
since fixing one particle leaves slightly fewer for the remaining sites. This correlation vanishes as $L\to\infty$, which is why the \emph{mean-field} result, $J=\rho(1-\rho)$, becomes exact in this limit. Recall that in the mean-field approximation, each particle is assumed to feel the same average background field created by all the other particles, independently of where they actually are. Correlations between sites are thereby neglected, i.e., $\avg{n_{i-1}n_i}_s=\avg{n_{i-1}}_s\avg{n_i}_s$.

In the large-$L$ limit, the current as a function of the density is $J=\rho(1-\rho)$. It vanishes for an empty ring ($\rho=0$, no particles to carry it) and for a full ring ($\rho=1$, no room to hop), and it is maximal where
\begin{equation}
  \frac{dJ}{d\rho} = 1-2\rho = 0 \;\Rightarrow\; \rho = \frac12 \;\Rightarrow\;
   J=J_{\rm max}=\frac14\,.
\end{equation}
Physically, $J=\rho(1-\rho)$ is the product of the supply of particles and the supply of empty sites to hop into: adding particles increases the number of carriers but reduces the room to move, and the two effects balance at $\rho=\tfrac12$.
 
One may ask whether the ring exhibits a phase transition. A nonequilibrium steady state has no Boltzmann distribution and hence no \emph{free energy} in the equilibrium sense. Its defining macroscopic property is instead the stationary current $J$, and in driven lattice gases $J$ is commonly used as the analogue of the free energy: phase transitions show up as non-analyticities of $J$, just as they do in the free energy at equilibrium. A phase transition further requires a \emph{control parameter}, which is tuned through a critical value, and an \emph{order parameter}, which responds to it. On the ring, the only parameters are the hopping rate $\lambda$, which merely sets the time scale [$J=\lambda\rho(1-\rho)$], and the density $\rho=N/L$. The density is thus the control parameter, and with this identification the ring shows no phase transition, for two immediate reasons. First, $J=\rho(1-\rho)$ is an analytic function of $\rho$ on the whole interval $0\le\rho\le1$: it has no kink and no discontinuity in any derivative, and the maximum at $\rho=\tfrac12$ is a smooth extremum, not a critical point. Second, there is no candidate for an order parameter: $\rho$ is fixed by particle conservation, and the unique steady state is uniform and translation invariant at every density, so different regions of the ring can never settle at different densities, and no symmetry is spontaneously broken. Phase transitions in TASEP require breaking the translation invariance, for instance by open boundaries, where the entry and exit rates $\alpha$ and $\beta$ serve as control parameters and the bulk density, no longer fixed, becomes the order parameter.
\subsection{TASEP: Open Boundaries}

We now turn our attention exclusively to open boundary conditions: particles are injected at site $1$ at rate $\alpha$ if it is empty, hop through the bulk at rate $\lambda=1$ subject to exclusion, and are removed from site $L$ at rate $\beta$ if it is occupied (Fig.~\ref{fig:open}).
\begin{center}
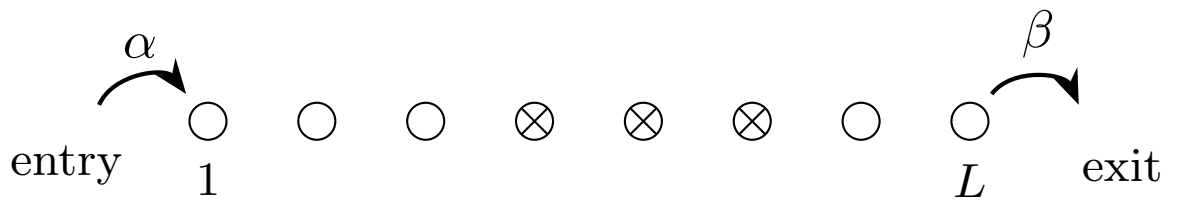
\begin{minipage}{\linewidth}\centering
\begin{tikzpicture}[scale=0.72]
  \foreach \x/\o in {1/0,2/0,3/0,4/1,5/1,6/1,7/0,8/0}{%
    \draw (\x,0) circle (0.17);
    \ifnum\o=1
      \draw (\x-0.12,-0.12) -- (\x+0.12,0.12);
      \draw (\x-0.12,0.12) -- (\x+0.12,-0.12);
    \fi}
  \node[below] at (1,-0.22) {\small $1$};
  \node[below] at (8,-0.22) {\small $L$};
  \draw[-{Stealth},thick] (0,0.15) to[bend left=60] node[above]{$\alpha$} (0.8,0.25);
  \draw[-{Stealth},thick] (8.2,0.25) to[bend left=60] node[above]{$\beta$} (9,0.15);
  \node at (-0.3,-0.4) {\small entry};
  \node at (9.4,-0.4) {\small exit};
\end{tikzpicture}
\captionof{figure}{\small TASEP with open boundaries ($\bigotimes$ occupied site, $\bigcirc$ empty site): particles enter at site $1$ at rate $\alpha$, hop to the right at rate $1$, and leave from site $L$ at rate $\beta$.}
\label{fig:open}
\end{minipage}\end{center}
The number of particles is no longer conserved, and the boundaries break the translation invariance of the ring: the first and last sites are each coupled on one side to a reservoir instead of a neighbouring site, so not all sites are equivalent. On the ring, translation invariance together with the uniqueness of the steady state forced every site to have the same density, and the block-counting argument showed that all configurations are equally probable. Both simplifications are now lost, as we show next.

As on the ring, the open chain is a finite Markov chain, but its state space now consists of all $2^L$ configurations: entries and exits connect configurations with different particle numbers, so the chain cannot be restricted to a fixed $N$. For $\alpha,\beta>0$ the chain is irreducible: from any configuration the system can be emptied through the exit, and any configuration can then be built up through the entry. Being finite and irreducible, it is positive recurrent with a unique stationary distribution. It is also aperiodic: In the discrete-time random sequential update, each step picks one bond at random; if the chosen bond cannot fire, the configuration stays unchanged, so every configuration with a blocked bond has a self-loop, i.e., a nonzero probability of returning to itself in a single step. Every configuration has such a bond: the entry needs site $1$ to be empty, whereas a hop on the bond $(1,2)$ needs it to be occupied, so at least one of the two is always blocked, implying existence of a self-loop. In continuous time, periodicity cannot arise anyway. The stationary distribution is therefore also the limiting one.

\begin{figure}[h!]\centering
\begin{tikzpicture}[>={Stealth[length=5pt]},x=1cm,y=1cm]
  \newcommand{\site}[2]{%
    \draw (#1,0) circle (0.15);
    \ifnum#2=1
      \draw (#1-0.105,-0.105) -- (#1+0.105,0.105);
      \draw (#1-0.105,0.105) -- (#1+0.105,-0.105);
    \fi}
  \newcommand{\conf}[3]{\tikz[baseline=-0.5ex]{\site{0}{#1}\site{0.4}{#2}\site{0.8}{#3}}}
  \node (c000) at (2.6,0)   {\conf000};
  \node (c100) at (0,1.7)   {\conf100};
  \node (c010) at (2.6,1.7) {\conf010};
  \node (c001) at (5.2,1.7) {\conf001};
  \node (c110) at (0,3.4)   {\conf110};
  \node (c101) at (2.6,3.4) {\conf101};
  \node (c011) at (5.2,3.4) {\conf011};
  \node (c111) at (2.6,5.1) {\conf111};
  \foreach \y/\n in {0/0,1.7/1,3.4/2,5.1/3}{\node[anchor=east] at (-0.9,\y) {\small $N=\n$};}
  \draw[->] (c100) -- node[above]{\small $1$} (c010);
  \draw[->] (c010) -- node[above]{\small $1$} (c001);
  \draw[->] (c110) -- node[above]{\small $1$} (c101);
  \draw[->] (c101) -- node[above]{\small $1$} (c011);
  \draw[->] (c000) -- node[pos=0.35,left]{\small $\alpha$} (c100);
  \draw[->] (c010) -- node[pos=0.2,left]{\small $\alpha$} (c110);
  \draw[->] (c001) -- node[pos=0.2,left]{\small $\alpha$} (c101);
  \draw[->] (c011) -- node[pos=0.35,right]{\small $\alpha$} (c111);
  \draw[->] (c001) -- node[pos=0.35,right]{\small $\beta$} (c000);
  \draw[->] (c101) -- node[pos=0.2,right]{\small $\beta$} (c100);
  \draw[->] (c011) -- node[pos=0.2,right]{\small $\beta$} (c010);
  \draw[->] (c111) -- node[pos=0.35,left]{\small $\beta$} (c110);
\end{tikzpicture}
\caption{\small Continuous-time Markov chain of TASEP with open boundaries for $L=3$: all $2^3=8$ configurations ($\bigotimes$ occupied site, $\bigcirc$ empty site), arranged in rows of fixed particle number $N$. Horizontal arrows are bulk hops at rate $1$, which conserve $N$; entries (rate $\alpha$, possible when site $1$ is empty) raise $N$ by one, and exits (rate $\beta$, possible when site $3$ is occupied) lower it by one. In the corresponding discrete-time chain, each configuration would also carry a self loop.}
\label{fig:openchain}
\end{figure}
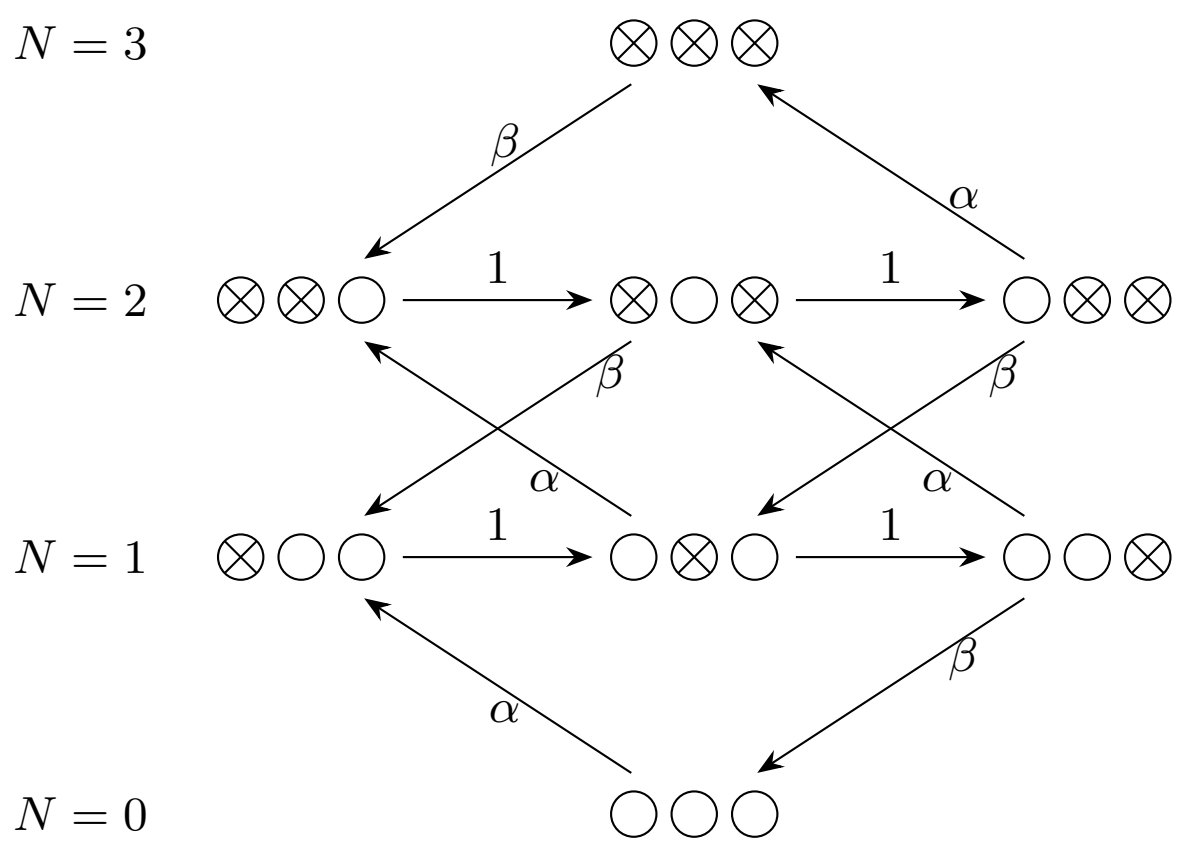

This stationary distribution, however, is no longer uniform, because the block-counting argument fails for blocks touching the boundaries. To see this explicitly, let's take $L=3$ and $\alpha=\beta=1$, whose full Markov chain is shown in Fig.~\ref{fig:openchain}, and label configurations by their occupation numbers $(n_1n_2n_3)$. 

Since every allowed move then has unit rate, the master equation~\eqref{eq:tpbcmc} with $\dot{p}=0$ reads, for each configuration $C$,
\begin{equation}
  \sum_{C'\to C}p_s(C') = (\text{number of moves out of } C)\times p_s(C),
\end{equation}
where the sum runs over the configurations $C'$ leading into $C$: the outflow carries the common factor $p_s(C)$, since all moves out of $C$ start from $C$, whereas each inflow carries the probability of its own starting configuration. For example, for $C=100$, the only move out is a hop on the bond $(1,2)$, so the right-hand side is $1\times p_s(100)$, while the configurations leading into $100$ are $000$ (by an entry) and $101$ (by an exit), so the left-hand side is $p_s(000)+p_s(101)$; hence $p_s(100)=p_s(000)+p_s(101)$. Solving the eight such equations gives the stationary weights
\begin{align}
  &p_s(000):p_s(001):p_s(010):p_s(011)\nonumber\\
  &:p_s(100):p_s(101):p_s(110):p_s(111)\nonumber\\
  &\qquad=1:1:2:1:3:2:3:1,
\end{align}
normalized by their sum, $14$, which are not uniform even though all rates are equal; in general, the stationary distribution is not even a product of single-site distributions. Moreover, since particles never hop backwards, detailed balance is violated as on the ring, and the steady state is a nonequilibrium steady state rather than an equilibrium distribution.

With translation invariance broken, the density $\rho_i=\avg{n_i}$ may vary from site to site. The current, however, does not vary: in the steady state particles cannot accumulate anywhere, so every bond, including the entry and the exit, carries the same stationary current $J$. In a long system, the boundaries affect the density mainly near the ends, and far from them the system looks locally homogeneous, with a well-defined bulk density $\rho_{\rm b}$. As we shall see, the boundaries nevertheless control which bulk density is selected. Throughout, we therefore work in the large-$L$ limit, which is essential: only there is the bulk well separated from the boundaries, and only there can $J$ and $\rho_{\rm b}$ develop non-analyticities---as functions of $\alpha$ and $\beta$---that signal phase transitions. For any finite $L$, the stationary probabilities of the finite, irreducible Markov chain are ratios of polynomials in the rates (\emph{Markov chain tree theorem}), with a denominator that never vanishes for $\alpha,\beta>0$; the current and the density profile are, therefore, analytic functions of $\alpha$ and $\beta$, and every transition is smoothed out.

To find the density profile, we write down its evolution equations. In the bulk ($2\le i\le L-1$), site $i$ is affected only by the bonds $(i-1,i)$ and $(i,i+1)$, exactly as on the ring, so
\begin{equation}
  \frac{d\rho_i}{dt}=\avg{n_{i-1}(1-n_i)}-\avg{n_i(1-n_{i+1})}.\label{eq:tase1}
\end{equation}
At the boundaries, the entry and exit bonds replace the missing neighbours. A particle enters site $1$ at rate $\alpha$ provided the site is empty, and leaves site $L$ at rate $\beta$ provided it is occupied:
\begin{align}
  \frac{d\rho_1}{dt}&=\alpha\avg{1-n_1}-\avg{n_1(1-n_2)},\label{eq:tase2}\\
  \frac{d\rho_L}{dt}&=\avg{n_{L-1}(1-n_L)}-\beta\avg{n_L}.\label{eq:tase3}
\end{align}
This set of equations [Eqs.~\eqref{eq:tase1}--\eqref{eq:tase3}] is exact but not closed, since they involve the two-point correlations $\avg{n_in_{i+1}}$, whose equations in turn involve three-point correlations, and so on. The open chain can in fact be solved exactly, by the technically involved \emph{matrix product method}. We here follow a simpler route. On the ring, the exact steady state showed that correlations between neighbouring sites vanish as $L\to\infty$; here we {assume} that they can be neglected, i.e., we make the mean-field approximation $\avg{n_in_{i+1}}\approx \rho_i\rho_{i+1}$, which gives
\begin{equation}
  \frac{d\rho_i}{dt}=\rho_{i-1}(1-\rho_i)-\rho_i(1-\rho_{i+1}).
\end{equation}
As we shall see, this simple closure turns out to be remarkably successful.

In the steady state, the time derivatives vanish. From here on, $\rho_i$ denotes the steady-state density---$\rho_{i,s}$ in the notation used for the ring---and we drop the subscript $s$ for brevity. The equal currents on all bonds then read, in mean field,
\begin{equation}
  J=\alpha(1-\rho_1)=\rho_i(1-\rho_{i+1})=\beta \rho_L,
\end{equation}
for $i=1,\dots,L-1$. The boundary terms take the same form as the bulk ones if we introduce two fictitious sites,
\begin{equation}
  \rho_0\equiv\alpha,\qquad \rho_{L+1}\equiv1-\beta,
\end{equation}
since then $\alpha(1-\rho_1)=\rho_0(1-\rho_1)$ and $\beta \rho_L=\rho_L(1-\rho_{L+1})$. For $\alpha,\beta\le1$ these have a direct interpretation: the entry acts as a reservoir of density $\alpha$ to the left of site $1$, and the exit as a reservoir of density $1-\beta$ to the right of site $L$, both coupled to the system by bonds of unit rate. In general, however, $\alpha$ and $\beta$ are rates, not probabilities, and may exceed $1$, so that $\rho_0$ and $\rho_{L+1}$ are merely effective boundary values rather than densities; only the densities $\rho_1,\dots,\rho_L$ of the real sites must lie in $[0,1]$. The steady-state condition $J=\rho_i(1-\rho_{i+1})$ for $i=0,\dots,L$ then becomes
\begin{equation}
  \rho_{i+1}=f(\rho_i)\equiv1-\frac{J}{\rho_i},
  \qquad i=0,\dots,L.
  \label{eq:map}
\end{equation}
This is a discrete dynamical system in which the site index plays the role of time: starting from $\rho_0=\alpha$, the map generates the density profile, and $J$ must be chosen such that after $L+1$ steps it arrives at $\rho_{L+1}=1-\beta$. To understand its solutions, we first recall some basic facts about one-dimensional maps.

Consider a map $x_{n+1}=g(x_n)$, where $x_n$ takes values in a phase space $\mathcal{P}\subseteq\mathbb{R}$ and $g:\mathcal{P}\to\mathcal{P}$, and where the discrete time $n$ runs over the non-negative integers, $n=0,1,2,\dots$, starting from an initial value $x_0$. The sequence $x_0,x_1,x_2,\dots$ generated in this way is called the \emph{trajectory} of $x_0$. A point $x^*$ is a \emph{fixed point} if $x^*=g(x^*)$: a trajectory that reaches it stays there. To decide whether nearby trajectories approach it, write $x_n=x^*+\delta x_n$ and expand:
\begin{align}
  x^*+\delta x_{n+1}&=g(x^*+\delta x_n)\nonumber\\
  &=g(x^*)+g'(x^*)\,\delta x_n+\mathcal{O}(\delta x_n^2).
\end{align}
Since $g(x^*)=x^*$, to linear order
\begin{equation}
  \delta x_{n+1}=g'(x^*)\,\delta x_n
  \;\Rightarrow\;
  \delta x_n=\big[g'(x^*)\big]^n\,\delta x_0.
\end{equation}
Hence the fixed point is \emph{stable} if $|g'(x^*)|<1$, since deviations shrink geometrically and nearby trajectories are attracted to $x^*$, and \emph{unstable} if $|g'(x^*)|>1$, since deviations grow and nearby trajectories are repelled from it. A trajectory approaching a stable fixed point thus consists of a \emph{transient}, during which it is still away from $x^*$, followed by its long-time behaviour, in which it remains arbitrarily close to $x^*$. The case $|g'(x^*)|=1$ is \emph{marginal}: the linear analysis is inconclusive, and the approach, if any, is slower than geometric.

Returning to the density map~\eqref{eq:map}, the iteration index is now the site index, running over the finite range $0,1,\dots,L+1$. The densities $\rho_1,\dots,\rho_L$ of the real sites must lie in $[0,1]$, whereas the boundary values $\rho_0=\alpha$ and $\rho_{L+1}=1-\beta$ need not. Note also that $f$ does not map all of $[0,1]$ into itself, since $f(\rho)<0$ for $\rho<J$; physically admissible profiles are therefore restricted to the part of $[0,1]$ that the iteration does not leave, as we shall see below.

A fixed point $\rho^*=1-J/\rho^*$ satisfies
\begin{equation}
  {\rho^*}^2-\rho^*+J=0
  \;\Rightarrow\;\rho^*=
  \rho_\pm=\frac12\pm\sqrt{\frac14-J}.
\end{equation}
Real fixed points exist only for $J\le\frac14$, which is the same maximal current $J_{\max}=\frac14$ as on the ring. The two fixed points are related by \emph{particle--hole symmetry}. Whenever a particle hops from site $i$ to $i+1$, a hole (empty site) hops from $i+1$ to $i$; the holes therefore perform a TASEP of their own, moving to the left at the same rate, with density $1-\rho_i$ at site $i$. Exchanging particles and holes, $n_i\to1-n_i$, while reversing the direction of the lattice, $i\to L+1-i$, thus maps TASEP onto itself, with the roles of $\alpha$ and $\beta$ interchanged: holes enter at the right end at rate $\beta$ and leave at the left end at rate $\alpha$. In particular, the fixed-point condition $J=\rho^*(1-\rho^*)$ is unchanged under $\rho^*\to1-\rho^*$, so the fixed points of the map come in pairs $\rho_\pm$ with
\begin{equation}
  \rho_++\rho_-=1,\qquad \rho_+\rho_-=J,
\end{equation}
and a uniform profile at either fixed point carries the current $J=\rho_\pm(1-\rho_\pm)$, as on a ring of the same density. Their stability follows from $f'(\rho)=J/\rho^2$ with $J=\rho_+\rho_-$:
\begin{equation}
  f'(\rho_+)=\frac{\rho_-}{\rho_+}<1,\qquad
  f'(\rho_-)=\frac{\rho_+}{\rho_-}>1,
\end{equation}
for $J<\frac14$. Hence $\rho_+$ is stable and $\rho_-$ is unstable:
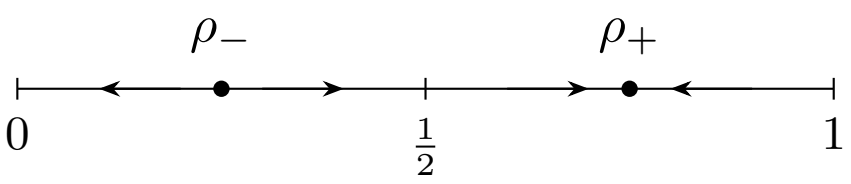
\begin{center}\begin{minipage}{\linewidth}\centering
\begin{tikzpicture}[scale=0.9,>={Stealth[length=4pt]}]
  \draw (0,0) -- (6,0);
  \foreach \x/\l in {0/0,3/\tfrac12,6/1}{%
    \draw (\x,0.08) -- (\x,-0.08) node[below]{\small $\l$};}
  \fill (1.5,0) circle (0.06) node[above=3pt]{$\rho_-$};
  \fill (4.5,0) circle (0.06) node[above=3pt]{$\rho_+$};
  \draw[->] (1.2,0) -- (0.6,0); \draw[->] (1.8,0) -- (2.4,0);
  \draw[->] (3.6,0) -- (4.2,0); \draw[->] (5.4,0) -- (4.8,0);
\end{tikzpicture}
\end{minipage}\end{center}
Under iteration from left to right, any starting density $\rho_0>\rho_-$ is therefore attracted to $\rho_+$, while a profile can stay near $\rho_-$ only if it starts exactly there.

Beyond the stability of its fixed points, the map~\eqref{eq:map} can in fact be solved exactly. Writing $J=\rho_+\rho_-$ and subtracting $\rho_-$ and $\rho_+$, respectively, from both sides, we arrive at
\begin{align}
  \rho_{i+1}-\rho_-&=1-\frac{\rho_+\rho_-}{\rho_i}-\rho_-,\\
  \rho_{i+1}-\rho_+&=1-\frac{\rho_+\rho_-}{\rho_i}-\rho_+.
\end{align}
Multiplying numerator and denominator of their ratio by $\rho_i$, and using $1-\rho_-=\rho_+$ and $1-\rho_+=\rho_-$,
\begin{eqnarray}
  \frac{\rho_{i+1}-\rho_-}{\rho_{i+1}-\rho_+}
  =\frac{\rho_i(1-\rho_-)-\rho_+\rho_-}{\rho_i(1-\rho_+)-\rho_+\rho_-}
  =\frac{\rho_+(\rho_i-\rho_-)}{\rho_-(\rho_i-\rho_+)}.\qquad
\end{eqnarray}
In terms of $x_i\equiv(\rho_i-\rho_-)/(\rho_i-\rho_+)$, this is the linear map $x_{i+1}=(\rho_+/\rho_-)\,x_i$, whose solution is $x_i=(\rho_+/\rho_-)^i x_0$, with $x_0$ fixed by the initial value $\rho_0=\alpha$. Hence
\begin{equation}
  {\frac{\rho_i-\rho_-}{\rho_i-\rho_+}=\left(\frac{\rho_+}{\rho_-}\right)^{i}\frac{\rho_0-\rho_-}{\rho_0-\rho_+}}
  \label{eq:profile}
\end{equation}
Since $\rho_+/\rho_->1$, $|x_i|$ grows as $e^{i/\xi}$ with
\begin{equation}
  \xi=\frac{1}{\ln(\rho_+/\rho_-)},
\end{equation}
so a profile starting from any $\rho_0>\rho_-$ approaches $\rho_+$ exponentially, over a characteristic number of sites $\xi$. (A profile starting below $\rho_-$ instead decreases and soon leaves the physical interval $[0,1]$; although Eq.~\eqref{eq:profile} formally describes it, it does not correspond to an admissible density profile.) The length $\xi$, measured in lattice spacings, sets the width of the \emph{boundary layers}: the regions near the entry or the exit where the density crosses over from the value imposed by the reservoir to its bulk value $\rho_{\rm b}$. Beyond a distance of a few $\xi$ from the boundary, the density equals $\rho_{\rm b}$ up to exponentially small corrections. As $J\to\frac14$, $\rho_+\to \rho_-$ and $\xi$ diverges, so the boundary layers spread over the whole system.

With the profile in hand, we can now determine the steady state by matching it to both boundary conditions, $\rho_0=\alpha$ and $\rho_{L+1}=1-\beta$, for large $L$. Three phases emerge.

\emph{High-density phase:} If $\alpha>\rho_-$, the profile starting from $\rho_0=\alpha$ is attracted to the stable fixed point $\rho_+$. In the language of the map, the boundary layer near the entry is the transient of the trajectory, lasting a few $\xi$ iterations, while the bulk corresponds to its long-time behaviour: for $L\gg\xi$, the trajectory spends almost all of its $L+1$ steps at $\rho_+$, which is therefore the bulk density. Since the trajectory cannot leave the vicinity of a stable fixed point again, it must still be at $\rho_+$, up to exponentially small corrections, at the right end, so that $\rho_{L+1}=1-\beta=\rho_+$. Hence $\beta=1-\rho_+=\rho_-$, and
\begin{equation}
  \rho_{\rm b}=1-\beta,\qquad J=\beta(1-\beta).
\end{equation}
Since $\rho_-=\frac12-\sqrt{\frac14-J}<\frac12$ for $J<\frac14$, this phase requires $\beta<\frac12$; and since the trajectory must start above $\rho_-=\beta$ to approach $\rho_+$, it further requires $\alpha>\beta$. The bulk density is thus set by the exit, which acts as a bottleneck, while the entry affects the profile only within the boundary layer.

\emph{Low-density phase:} If instead $\alpha\le\rho_-$, the profile cannot approach $\rho_+$; since a profile starting below $\rho_-$ leaves $[0,1]$, the only admissible possibility is that it stays at the unstable fixed point $\rho_-$, which requires it to start there, $\rho_0=\alpha=\rho_-$, up to an exponentially small deviation. In the language of the map, the trajectory must linger near the unstable fixed point for almost all of its $L+1$ steps, and yet leave it at the end to reach $\rho_{L+1}=1-\beta$; a trajectory starting exactly at $\rho_-$ would never leave. 

To see how precisely this must hold, write $\rho_i=\rho_-+\delta_i$. As long as $\rho_i$ stays close to $\rho_-$, Eq.~\eqref{eq:profile} gives
\begin{equation*}
  \delta_i\simeq\left(\frac{\rho_+}{\rho_-}\right)^{i}\delta_0=e^{i/\xi}\,\delta_0 ,
\end{equation*}
so a small initial deviation $\delta_0=\alpha-\rho_-$ is amplified by a factor $e^{i/\xi}$ after $i$ sites. At the exit, however, the profile must reach $\rho_{L+1}=1-\beta$, which in general differs from $\rho_-$ by an amount of order one; hence $e^{L/\xi}\delta_0\sim1$, i.e.,
\begin{equation*}
  \delta_0=\alpha-\rho_-\sim e^{-L/\xi}.
\end{equation*}
A larger $\delta_0$ would make the profile leave $\rho_-$ well before the exit, and a smaller one would leave it unable to reach $1-\beta$; with this choice, the deviation becomes appreciable only during the last few $\xi$ sites, which form the boundary layer at the exit. The current $J$ adjusts accordingly, and as $L\to\infty$ one recovers $\alpha=\rho_-$ exactly. 

The bulk density and current are therefore
\begin{equation}
  \rho_{\rm b}=\alpha,\qquad J=\alpha(1-\alpha).
\end{equation}
Since $\alpha=\rho_-<\frac12$, this phase requires $\alpha<\frac12$. Moreover, the other fixed point is $\rho_+=1-\rho_-=1-\alpha$. On leaving $\rho_-$, the trajectory either increases towards $\rho_+$, which it never exceeds, or decreases; in either case it can reach the required end value $\rho_{L+1}=1-\beta$ only if $1-\beta<\rho_+=1-\alpha$, i.e., $\beta>\alpha$. For $\beta<\alpha$ the exit value lies above $\rho_+$, no such trajectory exists, and the system is instead in the high-density phase. The bulk density is thus set by the entry, which now acts as the bottleneck, while the exit affects the profile only within its boundary layer. The same result follows from particle--hole symmetry: The holes perform a TASEP with entry rate $\beta$ and exit rate $\alpha$; in their high-density phase---which is low-density phase of the particles---their bulk density is $1-\alpha$, so the particle density is $\rho_{\rm b}=\alpha$. 

\emph{Maximal-current phase:} If $\alpha\ge\frac12$ and $\beta\ge\frac12$, neither boundary can impose its density: for $J<\frac14$ the only solutions are the high- and low-density ones, which require $\beta=\rho_-$ and $\alpha=\rho_-$, respectively, and both are impossible since $\rho_-<\frac12$. Nor can the current exceed $\frac14$ for large $L$. For $J>\frac14$, the map has no fixed points; moreover, since $\rho^2-\rho+J=(\rho-\frac12)^2+(J-\frac14)\ge J-\frac14$, we have for $0<\rho\le1$:
\begin{equation}
  \rho_{i+1}=f(\rho_i)=\rho_i-\frac{\rho_i^2-\rho_i+J}{\rho_i}\le\rho_i-\left(J-\tfrac14\right),
\end{equation}
so the density would decrease by at least $J-\frac14$ at every step. Starting from $\rho_1\le1$, after $k$ steps $\rho_{1+k}\le1-k(J-\frac14)$, which becomes negative once $k>1/(J-\frac14)$; at most about $1/(J-\frac14)$ consecutive sites can thus have densities in $[0,1]$. Keeping all $L$ densities in $[0,1]$, therefore, requires $L\lesssim1/(J-\frac14)$, i.e., $J\le\frac14+\mathcal{O}(1/L)$. In the large-$L$ limit, the current therefore takes its maximal value $J=\frac14$, at which ${\rho^*}^2-\rho^*+J=(\rho^*-\frac12)^2$ has a double root; i.e., the two fixed points merge,
\begin{equation}
\rho_+=\rho_-=\tfrac12=\rho_{\rm b}\qquad (\textrm{at}~J=J_{\rm max}=\tfrac14),
\end{equation}
and the merged fixed point is marginal, $f'(\tfrac12)=J/(\tfrac12)^2=1$. Writing $\rho_i=\frac12+\delta_i$, the map $\rho_{i+1}=1-1/(4\rho_i)$ gives exactly $1/\delta_{i+1}=1/\delta_i+2$, so that $\delta_i=\delta_0/(1+2i\delta_0)\simeq1/(2i)$: the bulk density is approached algebraically rather than exponentially, consistent with the divergence of $\xi$.

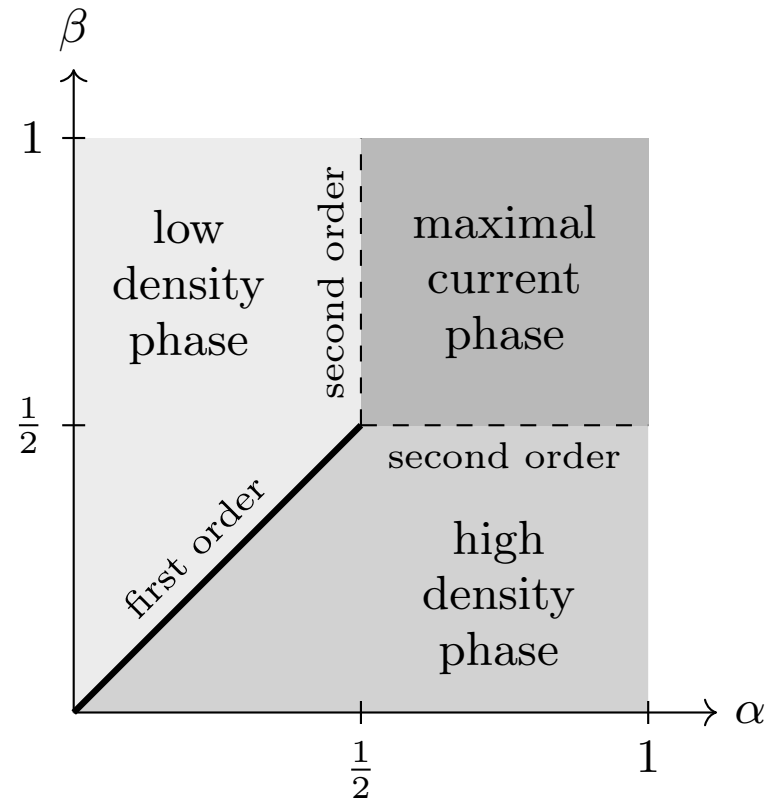
\begin{figure}[h!]
{\centering
\begin{tikzpicture}[scale=3.8]
  \fill[gray!15] (0,0) -- (0,1) -- (0.5,1) -- (0.5,0.5) -- cycle;   
  \fill[gray!35] (0,0) -- (0.5,0.5) -- (1,0.5) -- (1,0) -- cycle;   
  \fill[gray!55] (0.5,0.5) -- (0.5,1) -- (1,1) -- (1,0.5) -- cycle; 
  \draw[->] (0,0) -- (1.12,0) node[right]{$\alpha$};
  \draw[->] (0,0) -- (0,1.12) node[above]{$\beta$};
  \foreach \t in {0.5,1}{%
    \draw (\t,0.02) -- (\t,-0.02);
    \draw (0.02,\t) -- (-0.02,\t);}
  \draw[very thick] (0,0) -- (0.5,0.5)
    node[midway,sloped,above,font=\scriptsize] {first order};
  \draw[dashed] (0.5,0.5) -- (0.5,1)
    node[midway,sloped,above,font=\scriptsize] {second order};
  \draw[dashed] (0.5,0.5) -- (1,0.5)
    node[midway,below,font=\scriptsize] {second order};
  \node[align=center,font=\small] at (0.2,0.74) {low\\density\\phase};
  \node[align=center,font=\small] at (0.74,0.2) {high\\density\\phase};
  \node[align=center,font=\small] at (0.75,0.75) {maximal\\current\\phase};
  \node[below] at (0.5,-0.02) {\small $\tfrac12$};
  \node[left] at (-0.02,0.5) {\small $\tfrac12$};
  \node[below] at (1,-0.02) {\small $1$};
  \node[left] at (-0.02,1) {\small $1$};
\end{tikzpicture}\par}
\caption{\small Phase diagram of the TASEP with open boundaries. The solid line $\alpha=\beta<\tfrac12$ is a line of
first-order transitions; the dashed lines are lines of second-order
transitions.}
\label{fig:phase}
\end{figure}

The resulting mean-field phase diagram is shown in Fig.~\ref{fig:phase}. Unlike the ring, the open system thus shows genuine phase transitions, with $\alpha$ and $\beta$ as control parameters, the bulk density $\rho_{\rm b}$ playing the role of an order parameter, and the current $J$ that of a free energy.

To see this, fix $\beta<\frac12$ and let $\alpha$ increase through $\beta$, i.e., cross the line from the low-density side ($\alpha<\beta$) to the high-density side ($\alpha>\beta$). On the low-density side, $J=\alpha(1-\alpha)$, which tends to $\beta(1-\beta)$ as $\alpha\to\beta^-$; on the high-density side, $J=\beta(1-\beta)$, independently of $\alpha$. Hence $J$ is continuous across the line. Its slope, however, drops abruptly: $\partial J/\partial\alpha=1-2\alpha\to1-2\beta>0$ as $\alpha\to\beta^-$, whereas $\partial J/\partial\alpha=0$ for $\alpha>\beta$, so that $J$ has a kink. At the same time, the order parameter jumps upwards, from $\rho_{\rm b}=\alpha\to\beta$ as $\alpha\to\beta^-$ to $\rho_{\rm b}=1-\beta$ for $\alpha>\beta$. The slope of $J$ thus decreases, and the bulk density increases, by the same amount $1-2\beta$; both jumps vanish as $\beta\to\frac12$, where the line of first-order transitions ends. This is a \emph{first-order} transition.

On the coexistence line itself, $\rho_0=\alpha=\rho_-$ and $\rho_{L+1}=1-\beta=\rho_+$ are both fixed points: the trajectory lingers near $\rho_-$, crosses over to $\rho_+$ within a few $\xi$ sites, and remains there, so that a low-density and a high-density region are separated by a \emph{domain wall}, i.e., a narrow region, a few $\xi$ sites wide, across which the density jumps from $\rho_-$ to $\rho_+$. It is also called a \emph{shock}, since on large scales it is a sharp density front. In mean field, the position of the wall is undetermined.

Similarly, fixing $\beta>\frac12$ and letting $\alpha$ increase through $\frac12$, i.e., cross the dashed line from the low-density side ($\alpha<\frac12$) to the maximal-current side ($\alpha>\frac12$). On the low-density side, $J=\alpha(1-\alpha)$, which tends to $\frac14$ as $\alpha\to\frac12^-$; in the maximal-current phase, $J=\frac14$, independently of $\alpha$. Hence $J$ is continuous across the line. So is its slope: $\partial J/\partial\alpha=1-2\alpha\to0$ as $\alpha\to\frac12^-$, matching $\partial J/\partial\alpha=0$ for $\alpha>\frac12$, so that $J$ has no kink. Only its curvature changes abruptly, rising from $\partial^2J/\partial\alpha^2=-2$ to $0$. Correspondingly, the order parameter varies continuously, from $\rho_{\rm b}=\alpha\to\frac12$ as $\alpha\to\frac12^-$ to $\rho_{\rm b}=\frac12$ for $\alpha>\frac12$, with only its slope dropping from $\partial\rho_{\rm b}/\partial\alpha=1$ to $0$. This is a \emph{second-order} transition into the maximal-current phase. The same holds at $\beta=\frac12$ for $\alpha>\frac12$, with the roles of $\alpha$ and $\beta$ interchanged, the high-density side replacing the low-density one.

On the dashed lines themselves, both adjacent phases give $J=\frac14$ and $\rho_{\rm b}=\frac12$, and the lines belong to the maximal-current phase. Approaching them, the transition announces itself through a diverging length scale: on the low-density side $\rho_\pm\to\frac12$, and the width of the boundary layer,
\begin{equation}
  \xi=\frac{1}{\ln[(1-\alpha)/\alpha]}\simeq\frac{1}{4(\frac12-\alpha)},
\end{equation}
diverges as $\alpha\to\frac12$, until the exponential approach to the bulk density becomes algebraic. A diverging length is the hallmark of a continuous transition; the same holds on the high-density side, with $\beta$ in place of $\alpha$. The three phases meet at the point $(\frac12,\frac12)$, where the first-order line ends.

Finally, we return to the validity of the mean-field approximation. Unlike on the ring, where it became exact as $L\to\infty$, for open boundaries it remains an approximation; yet the exact (matrix product) solution confirms its phase diagram, together with the bulk densities and currents of all three phases. Mean field fails only in finer details, such as the exponent of the algebraic approach to the bulk density in the maximal-current phase, and the wandering of the domain wall on the coexistence line.

To summarize, TASEP illustrates in its simplest form how phases and phase transitions arise in nonequilibrium steady states. Even on the ring the steady state carries a current and violates detailed balance, but its uniform steady state leaves no room for distinct phases, and no transition occurs. With open boundaries, the entry and exit rates determine the state of the entire bulk, giving low-density, high-density and maximal-current phases separated by first- and second-order transitions. These \emph{nonequilibrium phase transitions}, known more specifically as \emph{boundary-induced phase transitions}, differ from their equilibrium counterparts in three respects: they occur in one dimension with short-range interactions, which equilibrium forbids, and are made possible by the drive acting together with the boundaries; the boundaries, whose influence the current carries through the whole system, control the bulk phase, whereas in equilibrium they cannot do so in the thermodynamic limit; and, lacking a free energy, the transitions are characterized by the non-analyticities of the current, with the bulk density as order parameter.

\begin{center}
	$\symking\symqueen\symrook\symknight\symbishop\sympawn\epsdice{1}\,\epsdice{2}\,\epsdice{3}\,\epsdice{4}\,\epsdice{5}\,\epsdice{6}\,\spadesuit\heartsuit\clubsuit\diamondsuit$
\end{center}
\newpage
\clearpage
\onecolumngrid

\acknowledgments

\emph{The author is indebted to Saikat Ghosh for the initial motivation and enthusiasm that sparked and shaped these lecture notes. The author is also thankful to all those IIT Kanpur students who audited or credited the one-semester course---partially or fully---upon which these notes are based. The author regrets not maintaining a comprehensive record of all references consulted during preparation. Consequently, the selected references listed below represent only a minimal set of authoritative introductory materials on the subject, to whose authors and publishers the author is greatly indebted. During the preparation of these notes, various generative AI tools (including Anthropic Claude, DeepSeek, Google Gemini, OpenAI ChatGPT, and other large language models) were consulted to assist with language refinement, stylistic editing, creating figures, LaTeX formatting, proofreading, improving readability, and verifying and streamlining steps of mathematical calculations. All mathematical content, derivations, proofs, conceptual frameworks, and pedagogical aspects were envisaged by the author, who retains full responsibility for any errors or omissions in the content.}

\begin{center}
	$\symking\symqueen\symrook\symknight\symbishop\sympawn\epsdice{1}\,\epsdice{2}\,\epsdice{3}\,\epsdice{4}\,\epsdice{5}\,\epsdice{6}\,\spadesuit\heartsuit\clubsuit\diamondsuit$
\end{center}

\renewcommand{\bibsection}{\section*{Selected References}}

\begin{center}
	$\symking\symqueen\symrook\symknight\symbishop\sympawn\epsdice{1}\,\epsdice{2}\,\epsdice{3}\,\epsdice{4}\,\epsdice{5}\,\epsdice{6}\,\spadesuit\heartsuit\clubsuit\diamondsuit$
\end{center}
\clearpage
\twocolumngrid
\end{document}